\documentclass[letterpaper,11pt]{article}
\usepackage[english]{babel}
\usepackage{mathtools}
\input{headers}

\title{On Weighted Graph Sparsification by Linear Sketching\footnote{This work was supported in part by NSF awards CCF-1763514, CCF-1934876, and CCF-2008304.}}

\newcommand*\samethanks[1][\value{footnote}]{\footnotemark[#1]}
\author{
  Yu Chen\thanks{University of Pennsylvania.
  Email: {{\small{\texttt{\{chenyu2,sanjeev,huanli\}@cis.upenn.edu}}}}}
  \and Sanjeev Khanna\samethanks\and Huan Li\samethanks}
\date{}

\begin{document}

\maketitle

\thispagestyle{empty}

\begin{abstract}
    A seminal work of~[Ahn-Guha-McGregor, PODS'12] showed that one can compute a cut sparsifier
  of an {\em unweighted} undirected graph by taking a near-linear number of linear measurements on the graph.
  Subsequent works also studied computing other graph sparsifiers using linear sketching,
  and obtained near-linear upper bounds for spectral sparsifiers~[Kapralov-Lee-Musco-Musco-Sidford, FOCS'14]
  and first non-trivial upper bounds for spanners~[Filtser-Kapralov-Nouri, SODA'21]. %
  All these linear sketching algorithms, however, only work on {\em unweighted} graphs,
  and are extended to weighted graphs by weight grouping,
  a non-linear operation not implementable in, for instance, general turnstile streams.

  In this paper, we initiate the study of {\em weighted} graph sparsification by linear sketching
  by investigating a natural class of linear sketches
  that we call {\em incidence sketches},
  in which each measurement is a linear combination of the weights of edges incident on a {\em single} vertex.
  This class captures {\em all} aforementioned linear sketches for unweighted sparsification.
  It also covers linear sketches implementable in the {\em simultaneous communication model},
  where edges are distributed across $n$ machines.
  Our results are:

  \begin{enumerate}
    \item \textbf{Weighted cut sparsification:} We give an algorithm that computes a
      $(1 + \eps)$-cut sparsifier using
      $\Otil(n \eps^{-3})$\footnote{$\Otil$ hides $\polylog(n,\eps^{-1},\frac{\wmax}{\wmin})$ factors,
      where $\wmax$ and $\wmin$ are the largest and smallest edge weights.}
      linear measurements,
      which is nearly optimal.
      This also implies a turnstile streaming algorithm with $\Otil(n \eps^{-3})$ space.
      Our algorithm is achieved by building a so-called ``weighted edge sampler'' for each vertex.
    \item \textbf{Weighted spectral sparsification:} We give an algorithm that computes
      a $(1 + \eps)$-spectral sparsifier using
      $\Otil(n^{6/5} \eps^{-4})$ linear measurements.
      This also implies a turnstile streaming algorithm with $\Otil(n^{6/5} \eps^{-4})$ space.
      Key to our algorithm is a novel analysis of how the effective resistances
      change under {\em vertex sampling}.
      Complementing our algorithm,
      we then prove a superlinear lower bound of $\Omega(n^{21/20-o(1)})$ measurements
      for computing some $O(1)$-spectral sparsifier using incidence sketches.
    \item \textbf{Weighted spanner computation:}
      We first show that
      any $o(n^2)$ linear measurements can only recover a spanner of stretch that in general depends
      {\em linearly} on $\frac{\wmax}{\wmin}$. %
      We thus focus on %
      graphs with $\frac{\wmax}{\wmin}=O(1)$ and study the stretch's dependence on $n$.
      On such graphs,
      the algorithm in~[Filtser-Kapralov-Nouri, SODA'21] can
      obtain a spanner of stretch
      $\Otil(n^{\frac{2}{3}(1-\alpha)})$ using
      $\Otil(n^{1+\alpha})$ measurements for any $\alpha\in[0,1]$.
      We prove that, for incidence sketches, this tradeoff is {\em optimal up to an
      $n^{o(1)}$ factor for all $\alpha < 1/10$}.
  \end{enumerate}

  We prove both our lower bounds by
  analyzing the ``effective resistances'' in certain {\em matrix-weighted} graphs,
  where we develop a number of new tools for reasoning about such graphs --
  most notably
  (i) a matrix-weighted analog of the widely used {\em expander decomposition} of ordinary graphs, and
  (ii) a proof that a random {\em vertex-induced subgraph} of a matrix-weighted expander is also an expander.
  We believe these tools are of independent interest.

\end{abstract}

\clearpage

\thispagestyle{empty}

\tableofcontents

\thispagestyle{empty}

\newpage

\pagenumbering{arabic}

\section{Introduction}\label{sec:intro}

Graph sparsification is a process that reduces the number of edges in a dense graph significantly
while preserving certain useful properties. Besides being an interesting problem in its own right,
graph sparsification has also been used as a fundamental building block in many modern graph algorithms
such as 
maximum flow and minimum cut algorithms~\cite{BenczurK15,Sherman13,KelnerLOS14,Peng16},
solvers for graph structured linear systems~\cite{ST14,CohenKMPPRX14,JambulapatiS21},
and graph clustering~\cite{ChenSWZ16}.

In addition to designing fast algorithms in the classic computational model,
a rich body of work has also studied graph sparsification by {\em linear sketching}.
In this setting, we can only access the input graph by taking {\em linear measurements},
each of which returns a linear combination of the edge weights,
and the goal is then to compute a sparsifier of the input graph using as few measurements as possible.
To state the previously known results in this setting,
let us first recall the definitions of three extensively studied graph sparsifiers
that we will study in this work.

\begin{definition}[Cut sparsifiers]
  Given a weighted graph $G = (V,E,w)$ and a parameter $\eps \in (0,1)$, another weighted graph $H = (V,F,w')$ with $F \subseteq E$ is called
  a {\em $(1 + \eps)$-cut sparsifier} of $G$
  if for every cut $(S,V-S)$, its weight $w_G(S,V-S)$ in $G$ and its weight $w_H(S,V-S)$ in $H$ satisfy
  that $(1 - \eps) w_G(S,V-S) \leq w_H(S,V-S) \leq (1 + \eps) w_G(S,V-S)$.
\end{definition}

\begin{definition}[Spectral sparsifiers]
  Given a weighted graph $G = (V,E,w)$ with Laplacian matrix $L_G$ and a parameter $\eps \in (0,1)$,
  another weighted graph $H = (V,F,w')$ with Laplacian matrix $L_H$ and $F \subseteq E$ is called
  a {\em $(1 + \eps)$-spectral} sparsifier of $G$ if
  for every vector $x\in\mathbb{R}^{n}$ we have
  $(1 - \eps) x^T L_G x \leq x^T L_H x \leq (1 + \eps) x^T L_G x$.
\end{definition}

\begin{definition}[Spanners]
  Given a weighted graph $G = (V,E,w)$ and a parameter $ t \ge 1$
  (called the {\em stretch}), 
  another weighted graph $H = (V,F,w')$ with $F \subseteq E$ is called
  a {\em $t$-spanner} of $G$ if for every vertex pair $u,v$, the shortest path length $d_G(u,v)$\footnote{Note that
  in a weighted graph, the length of a path is defined as the total edge weight along the path.}
  between $u,v$ in $G$ and the shortest path length $d_H(u,v)$ in $H$ satisfy
  $d_G(u,v)\leq d_H(u,v) \leq t\cdot d_G(u,v)$.
\end{definition}

A seminal work by Ahn, Guha, and McGregor~\cite{AhnGM12} showed
that one can compute a $(1 + \eps)$-cut sparsifier of an unweighted graph using
$\Otil(n\eps^{-2})$ linear measurements,
which is nearly optimal.
Then subsequent works by Kapralov, Lee, Musco, Musco, and Sidford~\cite{KapralovLMMS14}
and Kapralov, Mousavifar, Musco, Musco, Nouri, Sidford, and Tardos~\cite{KapralovMMMNST20}
showed that one can also compute a $(1 + \eps)$-spectral sparsifier of an unweighted graph
using $\Otil(n\eps^{-2})$ linear measurements.
Finally, a recent work by Filtser, Kapralov, and Nouri~\cite{FiltserKN21} showed that
one can compute an $\Otil(n^{\frac{2}{3}})$-spanner of an unweighted graph using $\Otil(n)$ linear measurements.
\cite{FiltserKN21} also showed that one can compute an $\Otil(n^{\frac{2}{3}(1-\alpha)})$-spanner
of an unweighted graph using $\Otil(n^{1+\alpha})$ linear measurements for any $\alpha\in [0,1]$,
and conjectured that this tradeoff might be close to optimal.

In all of these works~\cite{AhnGM12,KapralovLMMS14,KapralovMMMNST20,FiltserKN21},
the authors also showed that their linear sketching algorithms can also be applied
to computing graph sparsifiers of weighted graphs in {\em dynamic streams}, where the input graph
is given by a stream of {\em insertions and deletions of weighted edges}, and the goal is to compute
a sparsifier of the input graph using a small amount of space. In all these works, this is achieved
by grouping the edge weights geometrically, and then applying the
linear sketching algorithm for {\em unweighted graphs} %
to the subgraph induced by edges in each weight group. Note that this approach crucially requires that
if an edge $e$ is inserted with weight $w_e$, then any subsequent deletion of the edge $e$
again reveals its weight and it must be {\em identical} to $w_e$.
This is crucial to ensuring that each edge $e$ is inserted into or deleted from the same geometric group.

However, the operation of grouping edges by weight is not linear,
and as a result, the above approach for extending unweighted sketches to weighted ones
is not implementable in the more general {\em turnstile streams},
where the graph is given as a stream of {\em arbitrary edge weight updates}.
Surprisingly, little seems to be known about graph sparsification in this setting.

\subsection{Our results}

In this paper, we initiate the study of {\em weighted} graph sparsification by linear sketching.
To state our results, we need to introduce some notation first.
Let $w\in\mathbb{R}_{\geq 0}^{\binom{n}{2}}$ denote the weights of the edges
of the input graph, where $w_e = 0$ means that there is no edge in the edge slot $e$.
A {\em linear sketch of $N = N(n)$ measurements} consists of a (random) {\em sketching matrix}
$\Phi \in \mathbb{R}^{N\times \binom{n}{2}}$, and a (randomized) {\em recovery algorithm} $\Acal$
that takes as input $\Phi w\in\mathbb{R}^{N}$ and outputs a sparsifier of the graph with edge weights $w$.
Note that, by definition, the linear sketch is {\em non-adaptive}.

We focus on a natural class of linear sketches that we call {\em incidence sketches},
in which each row of the sketching matrix $\Phi$ is supported\footnote{Recall that
the support of a vector is the set of indices at which it is non-zero.}
on edges incident on a {\em single} vertex
(which could be different for different rows).
This class captures all linear sketches
that are implementable in a
distributed computing setting,
where the edges are stored across $n$ machines such that machine $i$ has all edges incident on the $i^{\mathrm{th}}$ vertex
(a.k.a.{\em simultaneous communication model}).
Moreover, it also covers {\em all} aforementioned linear sketches used in previous works
for unweighted cut sparsification~\cite{AhnGM12},
spectral sparsification~\cite{KapralovLMMS17,KapralovMMMNST20}, and spanner computation~\cite{FiltserKN21}.

We now present our results for computing these three kinds of sparsifiers in weighted graphs.
When describing our results, we use $\wmax$ and $\wmin$ to denote the largest
and the smallest non-zero edge weights, respectively, and always assume $\wmax \geq 1 \geq \wmin$.
We also write $\Otil(\cdot)$ to hide $\polylog(n, \eps^{-1}, \frac{\wmax}{\wmin})$ factors.

\paragraph{Weighted cut sparsification.}{
  We design an incidence sketch with a near-linear number of measurements
  for computing a $(1 + \eps)$-cut sparsifier of a weighted graph.
  
  \begin{theorem}[Algorithm for weighted cut sparsification]
    \label{thm:cutalg}
    For any $\eps\in(0,1)$,
    there exists an incidence sketch
    with random sketching matrix $\Phi_1\in\mathbb{R}^{N_1\times \binom{n}{2}}$ satisfying
    $N_1\leq \Otil(n \eps^{-3})$ and
    a recovery algorithm $\Acal_1$, such that for any $w\in\mathbb{R}_{\geq 0}^{\binom{n}{2}}$,
    $\Acal_1(\Phi_1 w)$ returns, with probability $1 - \frac{1}{\poly(n)}$,
    a $(1+\eps)$-cut sparsifier of the graph with edge weights $w$.
  \end{theorem}
  
  Thus, we achieve a similar performance as the linear sketch for unweighted graphs given in~\cite{AhnGM12},
  which uses $O(n \eps^{-2} \poly (\log n))$ measurements.
  It is well known that even to detect the connectivity of a graph, $\Omega(n)$ linear measurements
  are needed. Therefore, our upper bound in Theorem~\ref{thm:cutalg} is nearly optimal
  in $n$. %

  Similar to~\cite{Indyk06,AhnGM12,KapralovLMMS14,KapralovMMMNST20,FiltserKN21},
  by using Nisan's well known pseudorandom number generator~\cite{Nisan92},
  we can turn our linear sketching algorithm to a low space streaming algorithm.
  \begin{corollary}[of Theorem~\ref{thm:cutalg}]
    There is a single pass turnstile streaming algorithm with $\Otil(n \eps^{-3})$ space
    that, at any given point of the stream, recovers a $(1+\eps)$-cut sparsifier of the current graph
    with high probability.
  \end{corollary}
  Note that in the turnstile model, the stream consists of arbitrary edge weight updates.
}

\paragraph{Weighted spectral sparsification.}{
  We design an incidence sketch with about $n^{6/5}$ measurements for computing
  a $(1 + \eps)$-spectral sparsifier of a weighted graph.

  \begin{theorem}[Algorithm for weighted spectral sparsification]
    \label{thm:spectralalg}
    For any $\eps\in(0,1)$,
    there exists an incidence sketch
    with random sketching matrix $\Phi_2\in\mathbb{R}^{N_2\times \binom{n}{2}}$ satisfying
    $N_2\leq \Otil(n^{6/5} \eps^{-4})$ and
    a recovery algorithm $\Acal_2$, such that for any $w\in\mathbb{R}_{\geq 0}^{\binom{n}{2}}$,
    $\Acal_2(\Phi_2 w)$ returns, with probability $1 - \frac{1}{\poly(n)}$,
    a $(1+\eps)$-spectral sparsifier of the graph with edge weights $w$.
  \end{theorem}
  
  Similar to the cut sparsification case, we have the following corollary:
  \begin{corollary}[of Theorem~\ref{thm:spectralalg}]
    There is a single pass turnstile streaming algorithm with $\Otil(n^{6/5} \eps^{-4})$ space
    that, at any given point of the stream, recovers a $(1+\eps)$-spectral sparsifier of the current graph
    with high probability.
  \end{corollary}
  We complement this result by showing that a superlinear number of measurements are indeed necessary
  for any incidence sketch to recover some $O(1)$-spectral sparsifier.
  
  \begin{restatable}[Lower bound for weighted spectral sparsification]{theorem}{spectrallb}
    \label{thm:spectrallb}
    There exist constants $\eps,\delta \in (0,1)$
    such that
    any incidence sketch of $N$ measurements that computes a $(1+\eps)$-spectral sparsifier
    with probability $\geq 1 - \delta$ on any $w$
    must satisfy $N \geq n^{21/20-o(1)}$.
  \end{restatable}
  Note that this is in sharp contrast to the unweighted case, where a near-linear number of incidence sketch measurements
  are sufficient for computing an $O(1)$-spectral sparsifier~\cite{KapralovLMMS17,KapralovMMMNST20}.
  Theorem~\ref{thm:spectrallb} also draws a distinction between
  spectral sparsification and cut sparsification, %
  as for the latter a near-linear number of measurements are enough even in the weighted case
  (by Theorem~\ref{thm:cutalg}).
}

\paragraph{Weighted spanner computation.}{
  We first show that any $o(n^2)$ linear measurements can only recover
  a spanner of stretch that in general depends {\em linearly} on $\frac{\wmax}{\wmin}$.
  This differs fundamentally from the case of cut or spectral sparsification,
  where we can recover an $O(1)$-sparsifier, whose error is completely independent of $\frac{\wmax}{\wmin}$,
  using a sublinear-in-$n^2$ number of measurements.
  Specifically, we prove the following proposition, whose proof appears in Appendix~\ref{sec:appendspanner}.

  \begin{proposition}\label{prop:spanner}
    Any linear sketch (not necessarily an incidence sketch) of $N$ measurements that
    computes an $o(\frac{\wmax}{\wmin})$-spanner with probability $\geq .9$ on any $w$
    must satisfy $N \geq \Omega(n^2)$.
  \end{proposition}
  This proposition is a consequence of that edge weights are {\em proportional} to the edge lengths.
  More specifically,
  consider a complete graph where we weight a uniformly random edge by $\wmin$
  and all other $\binom{n}{2} - 1$ edges by $\wmax$.
  Then, while we can ignore the $\wmin$-weight edge in an $O(1)$-cut or spectral sparsifier,
  we have to include it in any $o(\frac{\wmax}{\wmin})$-spanner,
  since otherwise the shortest path length between its two endpoints would have been blown up by
  at least a factor of $\frac{2\wmax}{\wmin}$.
  Now note that, as the weight of this edge is smaller than all other edges, in order to find it
  we have to essentially recover all entries of the edge weight
  vector $w$, which inevitably requires
  $\Omega(n^2)$ linear measurements\footnote{In our actual proof of the proposition,
  we have to add some random Gaussian noise to each edge's weight for the lower bound to carry out.}.

  In light of the above proposition, we turn our focus to graphs with $\frac{\wmax}{\wmin} = O(1)$,
  and study the stretch's optimal dependence on $n$.
  On such graphs, the approach in~\cite{FiltserKN21} is able to obtain
  the following tradeoff between the stretch of the spanner and the number of linear measurements needed.
  \begin{theorem}[Algorithm for weighted spanner computation~\cite{FiltserKN21}]
    \label{thm:spectralspanner}
    For any constant $\alpha\in[0,1]$,
    there exists an incidence sketch
    with random sketching matrix $\Phi_3\in\mathbb{R}^{N_3\times \binom{n}{2}}$ satisfying
    $N_3\leq \Otil(n^{1+\alpha})$ and
    a recovery algorithm $\Acal_3$, such that for any
    $w\in\mathbb{R}_{\geq 0}^{\binom{n}{2}}$ with
    $\frac{\wmax}{\wmin} \leq O(1)$,
    $\Acal_3(\Phi_3 w)$ returns, with probability $1 - \frac{1}{\poly(n)}$,
    an $\Otil(n^{\frac{2}{3}(1 - \alpha)})$-spanner of the graph with edge weights $w$.
  \end{theorem}
  
  \cite{FiltserKN21} also conjectured that on unweighted graphs, to obtain a spanner of stretch
  $O(n^{\frac{2}{3}-\eps})$ for any constant $\eps > 0$, a superlinear number of measurements are needed
  for any linear sketch (in other words, the tradeoff is optimal at $\alpha = 0$).
  We make progress on this question by showing that this is indeed true
  for a natural class of linear sketches (i.e. incidence sketches)
  on ``almost'' unweighted graphs (i.e. those with $\frac{\wmax}{\wmin} = O(1)$).
  In fact, we show that in such a setting, the tradeoff obtained in the above theorem is
  {\em optimal for all} $\alpha < 1/10$.
  
  \begin{restatable}[Lower bound for weighted spanner computation]{theorem}{spannerlb}
    \label{thm:spannerlb}
    For any constant $\alpha \in (0,1/10)$,
    there exist constants $C\geq 1, \delta \in (0,1)$
    such that
    any incidence sketch of $N$ measurements that computes
    an $o(n^{\frac{2}{3}(1 - \alpha)})$-spanner with probability $\geq 1 - \delta$
    on any $w$ with $\frac{\wmax}{\wmin} \leq C$ must satisfy $N\geq n^{1+\alpha-o(1)}$.
  \end{restatable}

}

\subsection{Roadmap}

The rest of this paper is structured as follows.

We start by giving an overview of the techniques used in proving our main results. 
Specifically, Section~\ref{sec:ovcutalg} gives an overview
of our algorithms for weighted cut and spectral sparsification.
Then in Section~\ref{sec:ovspectrallb}, we give an overview
of our lower bound for weighted spectral sparsification.
Note that we do {\em not} give a separate overview
of our lower bound for weighted spanner computation,
because the ideas are similar
to the ones described in Section~\ref{sec:ovspectrallb}. %

Section~\ref{sec:preli} contains some
preliminaries that we will rely on throughout. Then Sections~\ref{sec:algocut} and~\ref{sec:algospectral} present
in detail our algorithms and their analysis for
weighted cut and spectral sparsification, respectively.

The remaining Sections~\ref{sec:hdg}-\ref{sec:spanner} are
devoted to proving our lower bounds for weighted spectral
sparsification and spanner computation,
both of which rely on a number of new tools
for analyzing certain {\em matrix-weighted} graphs.
Section~\ref{sec:hdg} first defines such matrix-weighted graphs and sets up some notation.
Then Sections~\ref{sec:fewsmall},\ref{sec:hargd},\ref{sec:hared},\ref{sec:hdevs} present the new tools we develop
for analyzing them.
Finally, Sections~\ref{sec:lbspectral},\ref{sec:spanner}
prove our lower bounds for weighted spectral sparsification
and spanner computation, respectively.

\section{Overview of weighted cut and spectral sparsification algorithms}\label{sec:ovcutalg}

\subsection{Overview of the algorithm for weighted cut sparsification}\label{sec:cutoverview}

\paragraph{Recap of the unweighted cut sparsification algorithm in~\cite{AhnGM12}.}{
  At a high-level, the approach taken is to reduce cut sparsification to (repeatedly) recovering a spanning forest
  of a subgraph of the input graph, obtained by sampling edges
  {\em uniformly} at some rate $p\in(0,1)$ known beforehand. This task is then further reduced to the task of sampling an edge connecting $S$ to $\bar{S}$ for an arbitrary subset $S$ of vertices, as this can be used to create a spanning forest by growing connected components.  

  Now to implement this latter task, in the sketching phase, we apply an $\ell_0$-sampler sketch to the incidence vector
  of each vertex $u$ (i.e. each column of the edge-vertex incidence matrix) in the {\em sub-sampled graph}.
  Then in the recovery phase, in order to recover an edge going out of a vertex set $S$,
  we add up the sketches of the vertices inside $S$. By linearity, this summed sketch is taken
  over the sum of the incidence vectors of vertices inside $S$, and the latter contains
  exactly the edges going out of $S$, since the edges inside cancel out.
  As a result, we can recover an edge going out of $S$, and create a spanning forest of the sub-sampled graph.
   Note that this approach crucially utilizes the fact that the edges are sampled uniformly at a rate
  that is known {\em beforehand}.
  This means that we can sample all $\binom{n}{2}$ edge slots beforehand, and apply the linear sketch only
  to the sampled edge slots.
}

\paragraph{Our approach for weighted cut sparsification.}{
We also reduce the task to recovering a spanning forest in a sub-sampled graph.
However, the latter graph is now obtained by sampling edges {\em non-uniformly}.
Specifically, we need to recover a spanning forest in a subgraph obtained by
sampling each edge $e$ with probability $\min\setof{w_e p,1}$ for some
parameter $p\in(0,1)$ that is known beforehand.
Therefore, in order to apply the idea as in the unweighted case, we will now need to design
a variant of $\ell_0$-sampler that, given a vector $x\in\mathbb{R}_{\geq 0}^{N}$
and a parameter $p\in(0,1)$, recovers a nonzero entry of $x$ after each entry $i\in[N]$
is sampled with probability $\min\setof{x_i p, 1}$. %
We call such a sampler ``weighted edge sampler''.

Note that the edge weights are {\em not} known to us beforehand, so we cannot
sample the edge slots with our desired probabilities as in the unweighted graph case.
We instead build such a weighted edge sampler using
a {\em rejection sampling} process, in which we sample edges uniformly at $\Otil(1)$ geometric rates,
but use {\em $\ell_1$-samplers}
to try recovering edges at each rate, and only output a recovered edge $e$ if the sampling rate $\approx w_e p$.
We then show that with high probability, we can efficiently find a desired edge.

Roughly,
our analysis involves proving that
there exists a geometric rate $q$ such that,
after uniformly sampling edges at rate $q$,
the total weight of edges $e$ satisfying $w_e p\approx q$ accounts for a large portion
of that of all sampled edges. As a result, by using a few independent $\ell_1$-samplers,
we can find one such edge with high probability.

}

\subsection{Overview of the algorithm for weighted spectral sparsification}\label{sec:spectraloverview}

\newcommand\sq{\mathrm{sq}}

As in previous linear sketches for unweighted graphs~\cite{KapralovLMMS14,KapralovMMMNST20},
the key task is to recover edges with $\Omegatil(1)$ effective resistances (or in weighted case,
$\Omegatil(1)$-leverage scores), which we refer to as {\em heavy edges}.
The high-level idea used in previous works is to (i) compute, for each vertex pair $s,t$,
a set of vertex potentials $x_{s,t}\in\mathbb{R}^{n}$ induced by an electrical flow from $s$ to $t$,
and then (ii) apply an $\ell_2$-heavy hitter to $B_G x_{s,t} \in\mathbb{R}^{\binom{n}{2}}$ to try
recovering the edge $(s,t)$, where $B_G$ is the edge-vertex incidence matrix of $G$
(see Section~\ref{sec:gmat}).
They achieve (i) by simulating an iterative refinement process in~\cite{LiMP13}.
To achieve (ii),
they make a key observation that
\begin{align*}
  \norm{ B_G x_{s,t} }_2^2 = x_{s,t}^T B_G^T B_G x_{s,t} = x_{s,t}^T L_G x_{s,t}
\end{align*}
is the energy of $x_{s,t}$,
and the entry of $B_G x_{s,t}$ indexed by edge $(s,t)$
is $(B_G x_{s,t})_{(s,t)} = b_{s,t}^T x = x_s - x_t$.
Therefore by the energy minimization characterization of effective resistances
(Fact~\ref{fact:energymin}), 
whenever the effective resistance between $s,t$ is $b_{s,t}^T L_G^{\dag} b_{s,t} \geq \Omegatil(1)$,
we have
$$ (B_G x_{s,t})_{(s,t)}^2 \geq \Omegatil(1) \norm{ B_G x_{s,t} }_2^2, $$
and hence the entry $(s,t)$ is an $\ell_2$-heavy hitter.

However, when the graph is weighted,
we are only allowed to access the graph through linear measurements on its weight vector $w_G$.
As a result, we can only apply $\ell_2$-heavy hitters
to $W_G B_G x_{s,t}$, whose squared $\ell_2$-norm is $x_{s,t}^T B_G^T W_G^2 B_G x_{s,t}$.
Now notice that $B_G^T W_G^2 B_G$ is the Laplacian matrix of a ``squared'' graph (call it $G^{\sq}$),
which has the same edges as $G$, but whose edges are weighted by $w_e^2$ as opposed to $w_e$.
Therefore, we will be recovering
edges that are heavy in $G^{\sq}$ instead of in $G$ if we apply the same approach as in previous works.
Unfortunately, a heavy edge in $G$ is not in general heavy in $G^{\sq}$, since the energy on the edges
with very large weights will blow up when we square the edge weights
(i.e. $w_e^2(x_u - x_v)^2 \gg w_e(x_u - x_v)^2$),
and hence make the total energy
grow unboundedly.

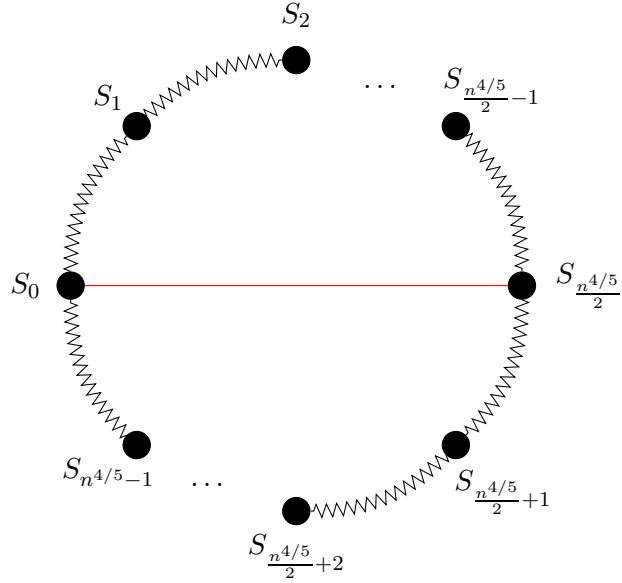
\begin{figure}[ht!]
  \centering
  \begin{tikzpicture}
    [scale = 0.5]%
    \node (1)[style={circle, fill=black, draw=black,}] at  (1.757, 1.757) {};
    \node (3)[style={circle, fill=black, draw=black,}] at  (1.757, 10.243){};
    \node (5)[style={circle, fill=black, draw=black,}] at  (10.243, 10.243){};
    \node (7)[style={circle, fill=black, draw=black,}] at  (10.243, 1.757){};
    \node (2)[style={circle, fill=black, draw=black,}] at  (0, 6){};
    \node (4)[style={circle, fill=black, draw=black,}] at (6, 12){};
    \node (6)[style={circle, fill=black, draw=black,}] at (12, 6){};
    \node (8)[style={circle, fill=black, draw=black,}] at (6, 0){};
    \node (9) at (1,1) {$S_{n^{4/5}-1}$};
    \node (10) at (-1.2,6) {$S_0$};
    \node (11) at (1,11) {$S_1$};
    \node (12) at (6,13.2) {$S_2$};
    \node (13) at (11.2,11.2) {$S_{\frac{n^{4/5}}{2}-1}$};
    \node (14) at (13.8,6) {$S_{\frac{n^{4/5}}{2}}$};
    \node (15) at (11.5,0.5) {$S_{\frac{n^{4/5}}{2}+1}$};
    \node (16) at (6,-1.2) {$S_{\frac{n^{4/5}}{2}+2}$};
    \node (17) at (8.3,11.3) {$\dots$};
    \node (18) at (3.7,0.7) {$\dots$};

    \draw[-,decoration={zigzag,segment length=4}] (1) [out=135,in=270] edge[decorate] (2);
    \draw[-,decoration={zigzag,segment length=4}] (2) [out=90,in=225] edge[decorate] (3);
    \draw[-,decoration={zigzag,segment length=4}] (3) [out=45,in=180] edge[decorate] (4);
    \draw[-,decoration={zigzag,segment length=4}] (5) [out=315,in=90] edge[decorate] (6);
    \draw[-,decoration={zigzag,segment length=4}] (6) [out=270,in=45] edge[decorate] (7);
    \draw[-,decoration={zigzag,segment length=4}] (7) [out=225,in=0] edge[decorate] (8);
    \draw[-,draw=red] (2) [out=0,in=180] to (6);
  \end{tikzpicture}
  \caption{A block cycle graph on $n$ vertices.
    Each $S_i$ represents a block of $n^{1/5}$ vertices connected by a clique
    and each zigzag represents the edges of a complete bipartite graph between adjacent blocks.
    The red edge represents the crossing edge.
  All edges along the cycle have weights $n^{2/5}$, and the crossing edge has weight $1$.}
  \label{fig:block}
\end{figure}

To see an intuitive example, suppose $G$ is a ``block cycle graph'' on $n$ vertices whose edges are
generated as follows (see also Figure~\ref{fig:block}):
\begin{enumerate}
  \item Partition the vertices into $n^{4/5}$ blocks $S_0,\ldots,S_{n^{4/5}-1}$,
    each with $n^{1/5}$ vertices.
  \item For each $0\leq i< n^{4/5}$, add on $S_i$
    a complete graph of $n^{1/5}$ vertices with edge weights $n^{2/5}$, i.e. $n^{2/5} K_{n^{1/5}}$.
  \item For each $0\leq i< n^{4/5}$, add on $(S_i,S_{i+1})$
    a complete bipartite graph of $2 n^{1/5}$ vertices with edge weights $n^{2/5}$
    and bipartition $(S_i,S_{i+1})$\footnote{Note that we consider $i+1$ as $0$ when $i=n^{4/5}-1$.},
    i.e. $n^{2/5} K_{n^{1/5},n^{1/5}}$.
  \item Finally, add a ``crossing edge'' $e^*$ of weight $1$ between a randomly chosen vertex pair $s,t$.
\end{enumerate}
We note that, in this construction, the crossing edge $e^*$ spans $\Omega(n^{4/5})$ consecutive blocks,
Note that, typically, the crossing edge $e^*$ spans $\Omega(n^{4/5})$ consecutive blocks,
and therefore has effective resistance (and also leverage score) $\Omega(1)$.

\begin{proposition}\label{prop:largeer}
  If $s\in S_i, t\in S_j$ such that $\min\setof{|i-j|,n^{4/5}-|i-j|} \geq \Omega(n^{4/5})$,
  then the effective resistance of $e^*$ satisfies $r_{e^*} \geq \Omega(1)$.
\end{proposition}

\begin{proof}%
  Let $s,t$ be the endpoints of the crossing edge $e^*$.
  By the energy minimization characterization of effective resistances
  (Fact~\ref{fact:energymin}), it suffices to show that there is a set of vertex potentials
  whose normalized energy with respect to $s,t$ is $O(1)$. %
  Specifically, consider the set of potentials $x\in\mathbb{R}^n$ such that
  $x_u = \frac{i}{n^{4/5}}$ for all $u\in S_i$.
  Then we have $x_s - x_t = \Theta(1)$,
  and the total energy is $\sum_{e=(u,v)} w_e(x_u - x_v)^2 = n^{6/5}\cdot n^{2/5} (\Theta(n^{-4/5}))^2 + \Theta(1)
  = \Theta(1)$, as desired.
\end{proof}

However, in the squared graph $G^{\sq}$, all edge weights along the cycle
are blown up by a factor of $n^{2/5}$, and thus $e^*$ only has
leverage score $O(n^{-2/5})$ in $G^{\sq}$. To recover in a vector $x$ every entry with $\ell_2$-contribution
$\geq n^{-2/5} \norm{x}_2^2$, one will need an $\Omega(n^{-2/5})$ factor
blowup in the number of linear measurements, resulting in a total of $n^{7/5}$ measurements needed to
recover $e^*$.

We can in fact improve the number of linear measurements needed for recovering $e^*$
to $\Otil(n^{6/5})$ using the {\em vertex sampling} trick, an idea first used in~\cite{FiltserKN21} for
sketching spanners.
Namely, consider sampling a vertex set $C\subset V$ by including each vertex with probability $n^{-1/5}/100$,
and looking at the vertex-induced subgraph $G^{\sq}[C]$.
Then one can show  that, conditioned on $e^*\in G^{\sq}[C]$,
with constant probability, the two endpoints of $e^*$ will be disconnected in $G^{\sq}[C]\setminus e^*$.
As a result, the leverage score of $e^*$ becomes $1$ in $G^{\sq}[C]$,
and we can recover $e^*$ by recovering heavy edges in $G^{\sq}[C]$, which, as will show,
can be done using $\Otil(|C|) \approx \Otil(n^{4/5})$ measurements.
Since $e^*\in G[C]$ with probability $\approx \frac{1}{n^{2/5}}$,
repeating this sampling process independently
for $\Otil(n^{2/5})$ times allows us to recover $e^*$ in at least one vertex induced subgraph.
This results in a linear sketch of $\Otil(n^{6/5})$ measurements.

What if we slightly increase each block's size to $n^{1/5+\delta}$
and decrease the edge weights along the cycle to $n^{2/5-3\delta}$?
While one can still verify that the crossing edge $e^*$
has leverage score $\Omega(1)$, applying the same vertex sampling process as above
will not disconnect the endpoints of $e^*$ with $\Omega(1)$ probability.
However, one can alternatively show that,
with constant probability,
the number of edges along the cycle reduces by a factor of $n^{2/5}$.
Since now the energy of each edge only blows up by a factor smaller than $n^{2/5}$ in $G^{\sq}$,
this will also make the leverage score of $e^*$ become $\Omega(1)$ in $G^{\sq}[C]$,
and thus we can apply the same linear sketch of $\Otil(n^{6/5})$ measurements.

The above warm-up seems to suggest that the sampling rate of $\approx n^{-1/5}$ is a sweet spot
for recovering heavy edges in any graphs with the block cycle structure.
Indeed, we prove a key vertex sampling lemma showing that in {\em any} weighted graph $G$,
a heavy edge $e$ in $G$ is also likely heavy in a vertex-induced subgraph of $G^{\sq}$ obtained by sampling vertices
at rate $\approx n^{-1/5}$.
This is proved by carefully analyzing the structures of the edges of different weights after vertex sampling,
and then explicitly constructing a set of vertex potentials with small total energy in the induced subgraph.
Finally, by integrating this lemma into an iterative refinement process in~\cite{LiMP13}
(as the authors did in~\cite{KapralovLMMS14,KapralovMMMNST20})
and a spectral sparsification algorithm in~\cite{Koutis14},
we are able to recover a spectral sparsifier of $G$ using $\Otil(n^{6/5})$
linear measurements.
We note that the latter step of using heavy edge recovery to build a spectral sparsifier
is also more involved than in the unweighted case and requires a few extra techniques;
we refer the reader to the overview at the beginning of Section~\ref{sec:algmainspectral}
and the discussion therein for more details.

We note that this method of recovering heavy edges by vertex sampling is inspired by the one
used in~\cite{KapralovLMMS14} for spanners.
However, for spectral sparsification,
the correctness of such a method follows from fairly different reasoning,
and the proof is arguably more involved.

\section{Overview of lower bound for weighted spectral sparsification}\label{sec:ovspectrallb}

In this section we give an overview of our lower bound for weighted spectral sparsification (Theorem~\ref{thm:spectrallb}).
We prove our lower bound on a family of hard instances
that turn out to have the exact same structure as the one in Figure~\ref{fig:block}, which
we used to illustrate the difficulty of recovering spectral sparsifiers in weighted graphs.
Specifically, our hard instances are %
weighted ``block cycle graphs'' plus an extra {\em crossing edge}
that is included with probability $1/2$.
In a block cycle graph, the vertices are partitioned into blocks that are arranged in a cyclic manner.
Each block is a complete graph, and the vertices of adjacent blocks are connected by a complete bipartite graph.
Here we draw all edge
weights from Gaussian distributions and permute the vertices uniformly at random.
On such graphs, for a suitable choice of edge weights,
computing a spectral sparsifier essentially boils down to detecting
the presence/absence
of the crossing edge.
We then show that for the latter task,
the success probability of any incidence sketch can be bounded by the
``effective resistance'' of the crossing edge
in a {\em matrix-weighted} graph, where the matrix weights are in turn determined
by the sketching matrix.

We note that this family of hard instances has a similar structure
to the ones conjectured in~\cite{FiltserKN21}.
However, instead of using Bernoulli distributions on the edges as suggested in~\cite{FiltserKN21},
we use Gaussian distributions, which makes it easier to build the connection to effective resistances.

In order to show that the effective resistance is small for any incidence sketch with a limited number of measurements,
we develop a number of new tools for analyzing such matrix-weighted graphs. Most importantly,
we present (i) a matrix-weighted analog of the widely used expander decomposition of ordinary
graphs~\cite{GoldreichR99,KannanVV04,SpielmanT04}, and (ii) a proof that a random vertex-induced subgraph
of a matrix-weighted expander is also an expander with high probability.
We highly recommend reading Section~\ref{sec:onerow} to get intuition on why these two techniques
are useful,
where we use the ordinary graph version of (i) and (ii) to prove a lower bound for a simple class of sketches.

The rest of this section is structured as follows.
In Section~\ref{sec:harddist} we describe the distribution from which we generate our hard instances.
In Section~\ref{sec:tver} we explain how we bound the success probability of an incidence sketch
by the effective resistance in a matrix-weighted graph.
In Section~\ref{sec:onerow} we prove, as a warm-up, a lower bound for a simple class of sketches,
where we only need to analyze the effective resistances in {\em ordinary graphs}.
Finally in Section~\ref{sec:multirow} we outline our proof for arbitrary incidence sketches,
which requires analyzing matrix-weighted graphs.
We note that some proofs in this section are deferred to Appendix~\ref{sec:apovlb}.

\subsection{The hard distribution}~\label{sec:harddist}

We first state how we generate the input weighted graph $G = (V,E,w)$.
Let $n$ be the number of vertices
and define $s \defeq n^{1/5}$ and
$\ell \defeq n^{4/5}$.
We choose a random permutation $\pi:1..n\to 1..n$
and construct a block cycle graph as follows.
The $i$-th block (where $0\leq i < \ell$) consists
of vertices $\pi(s i + 1),\ldots,\pi(s i + s)$.
For simplicity we denote
the $a$-th vertex in the $i$-th block
(i.e. $\pi(s i + a)$) as $u_{i,a}$.
The block index $i$ will always be modulo $\ell$ implicitly.
We then add a complete graph to each block, and a complete bipartite graph
between each pair of adjacent blocks.
Namely, for each $0\leq i < \ell$, we add a graph $G_{i}$ with edges connecting
$u_{i,a},u_{i,b}$ for all $a < b \in \setof{1,\ldots,s}$,
and add another bipartite graph $G_{i,i+1}$ with edges connecting
$u_{i,a},u_{i+1,b}$ for all $a,b \in \setof{1,\ldots,s}$.
Finally, with probability $1/2$,
we add an edge between
vertices $\pi(1)$ and $\pi(n/2+1)$ (assume $n$ is even).
We refer to this edge as the {\em crossing edge} with respect to $\pi$
and any other edge in $G$ as a {\em non-crossing edge} with respect to $\pi$.
We will omit ``with respect to $\pi$'' when the underlying permutation $\pi$ is clear.

We next describe how the edge weights are determined.
The weights of all non-crossing edges
are drawn independently from $\Ncal(8n^{2/5}, n^{4/5} \log^{-1} n)$
(the Gaussian distribution with mean $8n^{2/5}$ and
variance $n^{4/5} \log^{-1} n$). The weight of the crossing edge
is drawn from the standard Gaussian $\Ncal(0,1)$.
If the crossing edge has negative weight, we say the input is {\em invalid}, and
accept any sketch as a valid sketch.
Our goal will be to detect the presence/absence of the crossing edge with high probability.

In the following, we will call the conditional distribution on the presence of the crossing edge
the {\em \sfy distribution}, and call the conditional distribution on the absence of the crossing edge
the {\em \sfn distribution}.
We then show that with high probability, the effective resistance of the crossing edge is large,
and therefore any linear sketch for computing spectral sparsifiers must distinguish between
the two distributions with good probability.

\begin{proposition}\label{prop:erbig}
  With probability at least $1 - 1/n$,
  all non-crossing edges have weights in the range $[4 n^{2/5}, 12 n^{2/5}]$,
  and as a result the effective resistance
  between vertices $\pi(1)$ and $\pi(n/2+1)$ is at least $1/48$
  in a \sfn instance.
\end{proposition}

\begin{proposition}\label{prop:0.6}
  Any linear sketch that can compute a $1.0001$-spectral sparsifier with probability $0.9$
  can distinguish between the \sfy and \sfn distributions with probability $0.6$.
\end{proposition}

The first proposition follows from an application of the Chernoff bound (Theorem~\ref{thm:GaussianChernoff}).
The proof of the second proposition is deferred to Appendix~\ref{sec:apovlb}.

In the following, we will assume, for ease of our analysis, that the sketch will be given the permutation
$\pi$ {\em after} computing the linear sketch.
That is, the recovery algorithm $\Acal$ takes as input both $\Phi w$ and $\pi$.
We will show that even with this extra piece of information,
any incidence sketch with $n^{21/20-\eps}$ measurements for constant $\eps>0$ cannot distinguish between the \sfy
and \sfn distributions with high probability.

\subsection{A bound on the success probability via effective resistance}\label{sec:tver}

We first show that for our lower bound instance, any incidence sketch can be reduced to a more restricted class of linear sketches
by only increasing the number of measurements by an $O(\log n)$ factor.
Specifically,
let us fix an arbitrary orientation of the edges,
and consider sketches taken over the weighted signed edge-vertex incidence matrix $B^{w}\in\mathbb{R}^{\binom{n}{2}\times n}$,
where the latter is given by
\begin{align*}
  B_{eu}^w =
  \begin{cases}
    w_e & \text{$e\in E$ and $u$ is $e$'s head} \\
    -w_e & \text{$e\in E$ and $u$ is $e$'s tail} \\
    0 & \text{otherwise.} \\
  \end{cases}
\end{align*}
That is, the algorithm must choose a (random) sketching matrix $\Phi\in \mathbb{R}^{k\times \binom{n}{2}}$ with
the $e^{\mathrm{th}}$ column $\phi_e\in\mathbb{R}^{k}$ corresponding to the edge slot $e$.
The sketch obtained is then $\Phi B^w \in \mathbb{R}^{k\times n}$.
Notice that the total number of measurements in $\Phi B^w$ is $k n$,
as each vertex applies the sketching matrix $\Phi$ to its incident edges.
Let us call this class of sketch {\em signed sketches}.
By Yao's minimax principle~\cite{Yao77}, to prove a lower bound for distinguishing the \sfy and \sfn distributions,
it suffices to focus on deterministic sketches.
The proof of the proposition below appears in Appendix~\ref{sec:apovlb}.

\begin{proposition}[Reduction to signed sketches]
  \label{prop:B}
  Consider any incidence sketch of
  $N$ measurements
  with a deterministic sketching matrix $\Phi\in\mathbb{R}^{N\times \binom{n}{2}}$ and
  a recovery algorithm $\Acal$ that, given $\Phi w$ and $\pi$,
  distinguishes between the \sfy and \sfn distributions 
  with probability at least $0.6$.
  Then there exists a signed sketch with a sketching matrix $\Phi'\in\mathbb{R}^{k\times \binom{n}{2}}$,
  where $k = O(1)\cdot \max\{1, \frac{N\log n}{n}\}$,
  and a recovery algorithm $\Acal'$ that, given $\Phi' B^w$ and $\pi$,
  distinguishes between the \sfy and \sfn distributions with probability at least $0.55$.
\end{proposition}

Let us now fix a sketching matrix $\Phi\in\mathbb{R}^{k\times \binom{n}{2}}$
and aim to obtain an upper bound on the success probability of any signed sketch using $\Phi$.
For notational convenience, let us write $(\Phi B^w)_{\mathrm{yes}}$ to denote $\Phi B^w$ conditioned on
the presence of the crossing edge %
and $(\Phi B^w)_{\mathrm{no}}$ to denote $\Phi B^w$ conditioned on
the absence of the crossing edge. %
We will also write $(\Phi B^w)_{\pi,\mathrm{yes}}$ or $(\Phi B^w)_{\pi,\mathrm{no}}$
to denote an extra conditioning on the permutation being $\pi$ in addition
to the presence/absence of the crossing edge.
Then to bound the success probability of any signed sketch using $\Phi$ ,
it suffices to show that the total variation distance (TV-distance) between
$(\Phi B^w)_{\mathrm{yes}}$ and $(\Phi B^w)_{\mathrm{no}}$ is small.

To state our upper bound on the TV-distance, %
we need to first introduce some notation.
For an edge $(u,v)$, define $b_{uv}\in\mathbb{R}^{nk}$ by writing it as a block vector (with block size $k$)
as follows:
\begin{align}\label{eq:defbb}
  b_{uv} = 
  \begin{blockarray}{c@{}c@{}cl}
    \begin{block}{(c@{}c@{}c)l}
      & 0 \\
      & \vdots \\
      & \phi_{uv} & & \text{$u^{\mathrm{th}}$ block} \\
      & 0 \\
      & \vdots \\
      & -\phi_{uv} & & \text{$v^{\mathrm{th}}$ block} \\
      & 0 \\
      & \vdots \\
    \end{block}
  \end{blockarray}\in \mathbb{R}^{nk},
\end{align}
where $\phi_{uv}\in\mathbb{R}^{k}$ is the column of $\Phi$ corresponding to the edge slot $(u,v)$.
For a permutation $\pi$,
we then define $L_{\pi} = \sum_{\text{non-crossing $(u,v)$}} n^{4/5}\log^{-1} n\,b_{uv} b_{uv}^T$.
The following proposition is essentially a consequence of Theorem~\ref{thm:tvd}~\cite{DevroyeMR18},
which bounds the TV-distance between multivariate Gaussians with the same mean.
We give its proof in Appendix~\ref{sec:apovlb}.
Note that we use $\dag$ to denote taking the Moore-Penrose pseudoinverse of a matrix.

\begin{proposition}\label{prop:tver}
  For any permutation $\pi$ such that
  $b_{\pi(1)\pi(n/2+1)}$ is in the range\footnote{Recall that the range
  of a symmetric matrix is the linear span of its columns.}
  of $L_{\pi}$,
  \begin{align*}
    d_{\tv}((\Phi B^w)_{\pi,\mathrm{yes}},(\Phi B^w)_{\pi,\mathrm{no}}) \leq O(1)\cdot
    \min\setof{1, b_{\pi(1)\pi(n/2+1)} L_{\pi}^{\dag} b_{\pi(1)\pi(n/2+1)}}.
  \end{align*}
\end{proposition}

Our plan is then to show that $b_{\pi(1)\pi(n/2+1)} L_{\pi}^{\dag} b_{\pi(1)\pi(n/2+1)}$
is small on average for every choice of a signed sketch with $k = n^{1/20-\eps}$
for constant $\eps > 0$.

Note that if $k=1$ and each $\phi_{uv}\in \setof{0,1}$,
then $L_{\pi}$ is exactly the Laplacian matrix of the graph
(call it $\Hcal_{\pi}$) that is formed by the non-crossing edges
$(u,v)$ such that $\phi_{uv} = 1$,
where
each edge is weighted by $n^{4/5} \log^{-1} n$.
Therefore, $b_{\pi(1)\pi(n/2+1)} L_{\pi}^{\dag} b_{\pi(1)\pi(n/2+1)}$ is the effective resistance
between $\pi(1)$ and $\pi(n/2+1)$ in $\Hcal_{\pi}$, if $\phi_{\pi(1)\pi(n/2+1)} \neq 0$
(otherwise $b_{\pi(1)\pi(n/2+1)}$ is the zero vector).

In fact,
to get a quick intuition as to why we should expect the effective resistance
between $\pi(1)$ and $\pi(n/2+1)$
to be small,
let us assume $\phi_{uv} = 1$ for all edge slots $(u,v)$.
Then, for any permutation $\pi$,
$\Hcal_{\pi}$ is the graph formed by {\em all} non-crossing edges,
each weighted by $n^{4/5} \log^{-1} n$.
Note that these weights are about $n^{2/5}$ times larger than
the weights $\Theta(n^{2/5})$ in the actual input graph (Proposition~\ref{prop:erbig}).
As a result, the effective resistance between $\pi(1)$ and $\pi(n/2+1)$ is about $n^{2/5}$ times smaller
than the effective resistance between them in the input graph (the former roughly equals $n^{-2/5}$).

When $k > 1$, we can view $L$ as the Laplacian of a {\em matrix-weighted} graph (again, call it $\Hcal_{\pi}$)
formed by the non-crossing edges,
where each edge $(u,v)$ has a $k\times k$ matrix weight $n^{4/5}\log^{-1} n\, \phi_{uv} \phi_{uv}^T$.
Now $b_{\pi(1)\pi(n/2+1)} L_{\pi}^{\dag} b_{\pi(1)\pi(n/2+1)}$ can be seen as
the (generalized) effective resistance
between $\pi(1)$ and $\pi(n/2+1)$ in $\Hcal_{\pi}$.

\subsection{Warm-up: one-row signed sketches have small TV-distance}\label{sec:onerow}

As a warm-up, we show that for any signed sketch,
in the case that $k=1$ and the sketching matrix $\Phi$ has $0/1$ entries,
we have, for any constant $\eps > 0$,
\begin{align}\label{eq:dtvonerow}
  \expec{\pi}{d_{\tv}\kh{ (\Phi B^w)_{\pi,\yes}, (\Phi B^w)_{\pi,\no} }} \leq \frac{1}{n^{1/5 - O(\eps)}}.
\end{align}
By Proposition~\ref{prop:tver}, we know that
$d_{\tv}\kh{ (\Phi B^w)_{\pi,\yes}, (\Phi B^w)_{\pi,\no} }$
can be bounded by the effective resistance
between $\pi(1),\pi(n/2+1)$ in $\Hcal_{\pi}$
if $\phi_{\pi(1)\pi(n/2+1)} = 1$, and is zero otherwise.
Here $\Hcal_{\pi}$ is formed by the non-crossing edges whose $\phi_{uv} = 1$,
where each edge $(u,v)$ has scalar weight $n^{4/5}\log^{-1} n$.
We can focus on the $\Phi$'s whose number of nonzero entries is at least $n^{9/5 + \eps}$,
since otherwise
\begin{align*}
  \prob{\pi}{\phi_{\pi(1)\pi(n/2+1)} = 1} = \frac{\nnz(\Phi)}{\binom{n}{2}} \leq
  \frac{1}{n^{1/5-O(\eps)}},
\end{align*}
and we would already have our desired result~(\ref{eq:dtvonerow}).

Our proof of~(\ref{eq:dtvonerow}) will rely on
decomposing $\Hcal_{\pi}$ into expanders with large minimum degree.
Since $\Hcal_{\pi}$'s edges all have the same weight $n^{4/5}\log^{-1} n$,
it is more convenient to work with the {\em unweighted version of $\Hcal_{\pi}$},
which we denote by $H_{\pi}$.
We now briefly review the definition of unweighted expanders,
as well as state a known expander decomposition lemma that we will utilize.

\begin{definition}[Expander]
  An unweighted graph $H = (V,E)$ is a $\zeta$-expander for some $\zeta \in [0,1]$
  if its conductance is at least $\zeta$, namely,
  for every nonempty $S\subset V$, we have
  \begin{align*}
    |E(S,V-S)| \geq \zeta \cdot \min\setof{ \vol(S), \vol(V-S) },
  \end{align*}
  where $\vol(S)$ is the total degree of vertices in $S$.
\end{definition}

Note that in the lemma below, we slightly abuse the notion of ``regular graphs''.
Specifically, we will say a graph is regular if
its minimum vertex degree $d_{\min}$ is not much smaller than the average degree $d$.

\begin{restatable}[Almost regular expander decomposition, see e.g.~\cite{KapralovKTY21}]{lemma}{lemared}
  \label{lem:ared}
  Given an unweighted graph $H = (V,E)$ with average degree $d\geq 16$, there exists
  a subgraph $I = (U,F)$ where $U\subseteq V$ and $F\subseteq E$
  such that $I$ is a $\frac{1}{16\log n}$-expander with
  minimum degree $d_{\min} \geq \frac{d}{16}$.
\end{restatable}

We will also need the following lemma, which shows that
a random vertex-induced subgraph of an expander with large minimum degree is almost
certainly an expander.
We give the proof of this lemma in Section~\ref{sec:expwarmup}.
To the best of our knowledge, even this result was not known before.

\begin{restatable}[Expanders are preserved under vertex sampling]{lemma}{lemedvs}
  \label{lem:edvs}
  There exists a $\theta = \theta(n) = n^{o(1)}$ with the following property.
  Consider an unweighted $\frac{1}{16\log n}$-expander $H=(V,E)$ %
  with minimum degree $d_{\min} \geq 4\cdot 10^6 \cdot \theta(n)$.
  For any
  $s \geq \frac{4\cdot 10^6}{d_{\min}}\cdot \theta(n)\cdot n$,
  let $C\subseteq V$ be a uniformly random vertex subset of size $s$.
  Then with probability at least $1 - 1/n^7$,
  the vertex-induced subgraph $H[C]$ is a $\frac{1}{n^{o(1)}}$-expander
  with minimum degree at least $\frac{s}{2n}\cdot d_{\min}$.
\end{restatable}

\begin{proof}[Proof of~(\ref{eq:dtvonerow}) using Lemmas~\ref{lem:ared},\ref{lem:edvs}]
  As argued above we can assume w.l.o.g. that $\nnz(\Phi) \geq n^{9/5 + \eps}$.
  We want to obtain, for each edge slot $e$ satisfying $\phi_e = 1$,
  conditioned on $e$ being the crossing edge w.r.t. $\pi$,
  an upper bound (call it $u_e$) on the typical effective resistance
  between the endpoints of $e$ in the graph $\Hcal_{\pi}$.
  In other words, conditioned on $e$ being the crossing edge,
  $u_e$ should be an upper bound on the effective resistance between the endpoints of $e$
  in $\Hcal_{\pi}$ with high probability over $\pi$.
  Then the total variation distance between $(\Phi B^w)_{\yes}$ and $(\Phi B^w)_{\no}$
  can be bounded by
  \begin{align}\label{eq:dtvue}
      \expec{\pi}{d_{\tv}\kh{ (\Phi B^w)_{\pi,\yes}, (\Phi B^w)_{\pi,\no} } } \leq
      O(1)\cdot \frac{1}{\binom{n}{2}} \sum_{e: \phi_e = 1} u_e.
  \end{align}

  To obtain the $u_e$'s, let us define the unweighted graph
  $H_{\phi} = (V,E_{\phi})$
  where $E_{\phi}$ contains {\em all} edges $e$ whose $\phi_e = 1$
  (including the ones not present in the input graph, i.e. $|E_{\phi}| = \nnz(\Phi))$.
  Now consider the following process, %
  where we repeatedly delete an expander subgraph from $H_{\phi}$ and obtain
  $u_e$'s for the edges in the expander.
  \begin{enumerate}
    \item While $|E_{\phi}| \geq 10^9 n^{9/5+\eps}$:
      \begin{enumerate}
        \item Find a subgraph $I = (U,F)$ of $H_{\phi} = (V,E_{\phi})$ that is a
          $\frac{1}{16\log n}$-expander with minimum degree
          $d_{\min} \geq \frac{|E_{\phi}|}{8n}$
          (existence is guaranteed by Lemma~\ref{lem:ared}).
        \item For each edge $f\in F$,
          let $u_f\gets \kh{\frac{10^9 n^{9/5+\eps}}{|E_{\phi}|}}^2$.
          \label{step:2b}
        \item Delete the edges in $F$ from $H_{\phi}$ by letting $E_{\phi} \gets E_{\phi} \setminus F$.
      \end{enumerate}
    \item Let $u_f \gets 1$ for all $f$ in the remaining $E_{\phi}$.
  \end{enumerate}
  To show that $u_e$'s are valid upper bounds, let us consider a fixed iteration of the while loop.
  For $i=0,\ldots,n^{4/5}-1$, let $U_i$ denote the vertices in $I$ that are
  in the $i^{\mathrm{th}}$ block of the input block cycle graph:
  \begin{align*}
    U_i \defeq U\intersect \setof{\pi(n^{1/5} i + 1), \ldots, \pi(n^{1/5} i +  n^{1/5})}.
  \end{align*}
  Then by Chernoff bounds, with probability at least $1 - 1/n^5$ over the random choice of $\pi$,
  we have $|U_i| \geq \frac{|U|}{2 n^{4/5}} \geq \frac{4\cdot 10^6}{d_{\min}}\cdot |U|^{1+\eps}$.
  Then by invoking Lemma~\ref{lem:edvs}, %
  with probability at least $1 - 1/n^4$ over $\pi$,
  all vertex-induced subgraphs $I[U_i\union U_{i+1}]$ are
  $\frac{1}{n^{o(1)}}$-expanders with minimum degree at least $\frac{|E_{\phi}|}{16 n^{9/5}}$.
  Using this fact, we obtain the following claim, whose proof appears in Appendix~\ref{sec:apovlb}.
  \begin{claim}\label{claim:ue}
    For each edge $f\in F$,
    conditioned on $f$ being the crossing edge,
    with probability at least $1-1/n^2$ over $\pi$,
    the effective resistance between the endpoints of $f$ in $\Hcal_{\pi}$
    is at most $u_f$.
  \end{claim}
  Now let us divide the above process for obtaining $u_e$'s into $O(\log n)$ phases,
  where in phase $i \in \setof{1,\ldots,O(\log n)}$, we have $|E_{\phi}| \in (\binom{n}{2} / 2^i, \binom{n}{2}/2^{i-1}]$.
  Then we have
  \begin{align*}
    \sum_{e: \phi_e = 1} u_e =
    & \sum_{e: \phi_e = 1} \kh{\frac{10^9 n^{9/5+\eps}}{|E_{\phi}|}}^2
    \leq
    n^{O(\eps)}\cdot
    \sum_{i=1}^{O(\log n)} \frac{\binom{n}{2}}{2^i} \cdot \kh{\frac{2^i}{n^{1/5}}}^2 \\
    \leq & n^{8/5 + O(\eps)} \sum_{1}^{O(\log n)} 2^i \leq
    n^{9/5 + O(\eps)}
  \end{align*}
  where in the first line we have used $n^{O(\eps)}$ to hide the constant factors,
  and the last inequality holds since in the last phase we have $2^i\leq n^{1/5}$.
  Plugging this into~(\ref{eq:dtvue}) finishes the proof.
\end{proof}

\subsection{The general case: proof of Theorem~\ref{thm:spectrallb}}\label{sec:multirow}

Note that even though for $k = 1$, the TV-distance is $\Otil(n^{-1/5})$, this does not imply
that $k$ must be large for the TV-distance to become $\Omega(1)$.

By Proposition~\ref{prop:B},
in order to prove Theorem~\ref{thm:spectrallb},
it suffices to prove the following:

\begin{restatable}{theorem}{tvdsmall}
  \label{thm:tvdsmall}
  For any fixed sketching matrix
  $\Phi\in\mathbb{R}^{k\times \binom{n}{2}}$ where
  $k \leq n^{1/20 - \eps}$ for some constant $\eps > 0$,
  we have
  \begin{align*}
    \expec{\pi}{d_{\tv}\kh{ (\Phi B^w)_{\pi,\yes}, (\Phi B^w)_{\pi,\no} } } \leq o(1).
  \end{align*}
\end{restatable}

By Proposition~\ref{prop:tver}, our goal is to bound the ``effective resistance''
$b_{\pi(1)\pi(n/2+1)}L_{\pi}^{\dag} b_{\pi(1)\pi(n/2+1)}$
between
vertices $\pi(1),\pi(n/2+1)$
in the matrix-weighted graph $\Hcal_{\pi}$ consisting of the non-crossing edges,
where edge $(u,v)$
has matrix weight $n^{4/5} \log^{-1} n\, \phi_{uv}\phi_{uv}^T\in\mathbb{R}^{k\times k}$.
We will do so by (significantly) generalizing our previous approach based on expander decomposition
for ordinary graphs in Section~\ref{sec:onerow}.
Our approach for the $k=1$ case essentially consists of two steps:
(i) decomposing the graph $H_{\phi}$ into large expander subgraphs and
(ii) proving that a random vertex induced subgraph of an expander is still an expander.

First note that there does not appear to be a combinatorial analog of conductance in matrix-weighted graphs,
which suggests that we should define expanders in an algebraic way.
Let us first recall the algebraic characterization of expanders for ordinary, unweighted graphs.
The definition is based on eigenvalues of the {\em normalized Laplacian} of the graph,
which is given by $N = D^{-1/2} L D^{-1/2}$,
where $D$ is a diagonal matrix with $D_{uu}$ equal to the degree $d_u$ of $u$.
\begin{definition}[Algebraic definition of ordinary, unweighted expanders]
  \label{def:uwexp}
  An unweighted graph $H$ is a $\zeta$-expander for some $\zeta\in[0,1]$ if
  the smallest nonzero eigenvalue of its normalized Laplacian matrix $N$ is at least $\zeta$.
\end{definition}
By Cheeger's inequality~\cite{AlonM85},
for $\zeta \geq \Omegatil(1)$,
this definition translates to that
the graph $H$ is a union of vertex-disjoint combinatorial expanders, each with conductance $\Omegatil(1)$.
To come up with an analogous definition for matrix-weighted graphs,
let us first define their associated matrices formally.

\paragraph{Matrices associated with matrix-weighted graphs.}{
We consider a $k\times k$ matrix-weighted graph $H = (V,E)$ with $|V| = n$.
For each edge $(u,v)\in E$,
there is a vector $\phi_{uv}\in\mathbb{R}^{k}$, indicating that
$(u,v)$ is weighted by the $k\times k$ rank-$1$ matrix $\phi_{uv} \phi_{uv}^T$.

\begin{definition}[Degree matrices]
  For a vertex $u$, its generalized degree is given by
  \begin{align*}
    D_u = \sum_{u\sim v} \phi_{uv} \phi_{uv}^T\in\mathbb{R}^{k\times k}.
  \end{align*}
  We then define the $nk\times nk$ degree matrix $D$ as a block diagonal matrix
  (with block size $k\times k$),
  with the $u^{\mathrm{th}}$ block on the diagonal being $D_{uu} = D_u\in\mathbb{R}^{k\times k}$.
\end{definition}

\begin{definition}[Laplacian matrices]
  The Laplacian matrix is given by $L = \sum_{(u,v)\in E} b_{uv} b_{uv}^T$,
  where $b_{uv}$'s are defined in~(\ref{eq:defbb}).
\end{definition}

We will call $b_{uv}$ the {\em incidence vector} of edge $(u,v)$.
Note that the Laplacian matrix here differs from the connection Laplacian matrix~\cite{KyngLPSS16} which is also defined
to be a block matrix. In particular, the Laplacian matrix of a matrix-weighted graph is not necessarily block diagonally dominant
(bDD) (Definition 1.1 of~\cite{KyngLPSS16}).

\begin{definition}[Normalized Laplacian matrices]
  The normalized Laplacian matrix is given by $N\defeq D^{\dag/2} L D^{\dag/2}$.
  Equivalently, we have $N = \sum_{(u,v)\in E} D^{\dag/2} b_{uv} b_{uv}^T D^{\dag/2}$.
\end{definition}

We will call $D^{\dag/2} b_{uv}$ the {\em normalized incidence vector} of edge $(u,v)$.

The following proposition says that similar to scalar-weighted graphs,
the eigenvalues of the normalized Laplacian of a matrix-weighted graph are also between $[0,2]$.
The proof this proposition appears in Appendix~\ref{sec:aphdg} (in ``Proof of Proposition~\ref{prop:eigvalofN}'').

\begin{proposition}
  The eigenvalues of $N$ are between $[0,2]$.
\end{proposition}

}

Now, a first attempt might be to define matrix-weighted expanders to also be graphs
whose normalized Laplacians' nonzero eigenvalues are large,
and then try to decompose any matrix-weighted graph into large
expander subgraphs. However, we show that the latter goal may not be achievable in general, by presenting
in Appendix~\ref{sec:aphd}
a hard instance, for which any large subgraph has a small nonzero eigenvalue.

\paragraph{Our approach.}{
  In light of the hard instance, we loosen the requirement of being an expander
  by allowing small eigenvalues, but requiring instead that
  each edge, compared to the average, does not have too large ``contribution'' to the small eigenvectors.
  Formally, we want that every edge's normalized incidence vector has small
  (weighted) projection onto the bottom eigenspace.
  We will also need an analog of ``almost regularity'', which for ordinary, unweighted graphs says that
  the minimum degree is large.
  We give the formal definition of an almost regular matrix-weighted expander below.
  \begin{definition}[Almost regular matrix-weighted expanders]
    \label{def:hde1}
    For a $k\times k$ matrix-weighted graph $H$,
    let $\lambda_1\leq \ldots\leq\lambda_{nk}$ be the eigenvalues of its normalized Laplacian $N$, and
    let $f_1,\ldots,f_{nk} \in \mathbb{R}^{nk}$ be a set of corresponding orthonormal eigenvectors.
    We say $H$ is a $(\gamma,\zeta,\psi)$-almost regular expander if
    \begin{enumerate}
      \item ($\gamma$-almost regularity) For every vertex $u$ and every incident edge $(u,v)\in E$, we have
        \begin{align}\label{eq:arty}
          \phi_{uv}^T D_u^{\dag} \phi_{uv} \leq \frac{\gamma \cdot k}{n}.
        \end{align}
      \item ($(\zeta,\psi)$-expander) For every edge $(u,v)\in E$ we have
        \begin{align}\label{eq:expdef}
          \kh{D^{\dag/2} b_{uv} }^T
          \kh{ \sum_{i: \lambda_i\in (0,\zeta]} \frac{1}{\lambda_i} f_i f_i^T } D^{\dag/2} b_{uv} \leq
          \frac{\psi\cdot k^2}{n^2}.
        \end{align}
    \end{enumerate}
  \end{definition}
  The LHS of~(\ref{eq:arty}) is the so-called {\em leverage score} of $\phi_{uv}$ w.r.t. $D_u$
  (Definition~\ref{def:deflvg}).
  It is known that the sum of leverage scores equals the rank of the matrix:
  \begin{proposition}
    For any fixed vertex $u$,
    $\sum_{(u,v)\in E} \phi_{uv}^T D_{u}^{\dag} \phi_{uv} = \rank{D_u}$.
  \end{proposition}
  Since $D_u$ is a $k\times k$ matrix, we have $\rank{D_u}\leq k$.
  Therefore, in the case that $u$ has $\Omega(n)$ incident edges,~(\ref{eq:arty})
  is essentially saying that no incident edge's leverage score exceeds the average by too much.
  
  To get intuition for condition~(\ref{eq:expdef}),
  we need the following two results.
  The first theorem is proved in Section~\ref{sec:fewsmall}.
  The second proposition is proved in Appendix~\ref{sec:apovlb}.
  \begin{theorem}\label{thm:fewsmallintro}
    Let $H$ be a $k\times $k-matrix weighted graph that is $\gamma$-almost regular
    (in the sense of~(\ref{eq:arty})).
    Then for any $\zeta \in (0,1)$,
    the number of eigenvalues of its normalized Laplacian that are between $(0,\zeta]$ 
    is at most $\frac{\gamma \cdot k^2}{(1 - \zeta)^2}$.
  \end{theorem}
  \begin{proposition}\label{prop:ellavg}
    Let $\ell$ be the number of $\lambda_i$'s
    that are between $(0,\zeta]$. Then
    \begin{align}
      \sum_{(u,v)\in E}
      \kh{D^{\dag/2} b_{uv} }^T
      \kh{ \sum_{i: \lambda_i\in (0,\zeta]} \frac{1}{\lambda_i} f_i f_i^T } D^{\dag/2} b_{uv}
      = \ell.
    \end{align}
  \end{proposition}
  Therefore,
  in the case that $|E| = \Omega(n^2)$,~(\ref{eq:expdef}) is essentially saying that
  the LHS for every edge $(u,v)$ does not exceed the average by too much.

  We then show that every dense enough matrix-weighted graph can indeed be made into an expander
  by downscaling a small number of edges.
  To this end, let us define, for a {\em scaling} $s: E\to[0,1]$,
  the rescaled graph $H^s$, which is obtained from $H$
  by rescaling each edge $(u,v)$'s weight to $s_{uv}^2 \phi_{uv} \phi_{uv}^T$.
  The proof of the following theorem appears in Sections~\ref{sec:hargd} and~\ref{sec:hared}.
  \begin{theorem}\label{thm:edintro}
    There is an algorithm that,
    given any $k\times k$ matrix-weighted graph $H = (V,E)$ with $|E| \geq \Omega(n^2)$,
    outputs a scaling $s: E \to [0,1]$ such that
    \begin{enumerate}
      \item The rescaled graph $H^s$ is a $(\gamma,\zeta,\psi)$-almost regular expander
        for
        \begin{align*}
          \gamma = 8 \log n,\ \zeta = \frac{1}{\log n},\ \psi = 16 k^2 \log^3 n.
        \end{align*}
      \item The number of edges $(u,v)\in E$ with $s_{uv} < 1$ is $o(n^2)$.
    \end{enumerate}
  \end{theorem}
  We next show that almost-regular expanders are preserved under vertex sampling.
  However, we will now use a different notion of ``preservation''.
  To state our specific result, let us define some additional notations.
  For a vertex subset $C\subseteq V$, we write $L_{G[C]}$ to denote the Laplacian
  of the vertex-induced subgraph $G[C]$.
  We also let $D_{CC}$ be the submatrix of $D$ (the degree matrix of the original graph $H$)
  with rows and columns restricted to vertices in $C$,
  and let $(f_i)_{C}$ denote, for an eigenvector $f_i$, the vector $f_i$ with indices restricted to $C$.
  We then have the following theorem, whose proof appears in Section~\ref{sec:mtexpsamp}.
  \begin{theorem}\label{thm:expsampintro}
    There exists a $\theta = \theta(n) \leq n^{o(1)}$ with the following property.
    Let $H = (V,E)$ be a $k\times k$ matrix-weighted,
    $(\gamma,\zeta,\psi)$-almost regular expander where $\zeta\leq 1/\log n$.
    For an
    $s \geq 2\cdot 10^6 \gamma \psi \zeta^{-1} k^2 \theta(n)$,
    let $C\subseteq V$ be a uniformly random vertex subset of size $s$.
    Then with probability at least $1 - 1/n^5$, we have
    that
    \begin{enumerate}
      \item The null space of $D_{CC}^{\dag/2} L_{G[C]} D_{CC}^{\dag/2}$ is exactly
        the linear span of $\setof{(f_i)_C : \lambda_i = 0}$.
      \item For all vectors $x\in\mathbb{R}^{|C|k}$ such that
        $x^T (f_i)_C = 0,\forall i: \lambda_i = 0$,
        \begin{align}\label{eq:preservance}
          x^T \kh{ D_{CC}^{\dag/2} L_{G[C]} D_{CC}^{\dag/2} }^{\dag} x \leq
          n^{o(1)}\cdot
          x^T \kh{ \frac{n^2}{s^2} \sum_{i: \lambda_i\in(0,\zeta]} \frac{1}{\lambda_i} (f_i)_C (f_i)_C^T +
          \frac{n}{s}\cdot \frac{1}{\zeta} I } x.
        \end{align}
    \end{enumerate}
  \end{theorem}
  We argue that~(\ref{eq:preservance}) is roughly saying that
  the pseudoinverse of the subgraph $G[C]$ can be bounded by the pseudoinverse of the original
  graph that is (i) restricted to indices in $C$ and (ii) rescaled in a certain way.
  For technical reasons, on the LHS of~(\ref{eq:preservance}) we
  normalize the Laplacian of the vertex-induced subgraph using the degree matrix
  of the {\em original graph $H$}.
  As for the RHS,
  we can see it as a rescaled version of the pseudoinverse of $N$ restricted to $C$,
  by noting that
  \begin{align*}
    \kh{ N^{\dag} }_{CC} = \sum_{\lambda_i > 0} \frac{1}{\lambda_i} (f_i)_C (f_i)_C^T.
  \end{align*}
  Thus, on the RHS of~(\ref{eq:preservance}) we blow up the small eigenvalues
  quadratically in $1/(\text{sampling\ rate})$,
  but blow up the large eigenvalues linearly in $1/(\text{sampling\ rate})$.

  With these tools, we are finally able to prove Theorem~\ref{thm:spectrallb}.
  We present the proof in Section~\ref{sec:lbspectral}.
}

\subsubsection{Techniques for proving Theorems~\ref{thm:fewsmallintro},~\ref{thm:edintro},~\ref{thm:expsampintro}}

We now explain, at a very high level,
the techniques that we use to prove these three key theorems,
as well as their connections to previous works.
More details can be found in the subsequent sections.

\paragraph{Proof of Theorem~\ref{thm:fewsmallintro}.}{
  We consider the ``spectral embedding'' %
  induced by the bottom eigenvectors of the normalized Laplacian,
  which maps each vertex to a rectangular matrix.
  Such a spectral embedding may be seen as the matrix-weighted counterpart of the ones for scalar-weighted graphs,
  which map each vertex to a {\em vector}.
  The latter embeddings were previously used to prove higher-order Cheeger inequalities~\cite{LouisRTV12,LeeGT14}.
  We show that, for matrix-weighted graphs that are almost regular (in the sense of~(\ref{eq:arty})),
  the spectral embedding has vertex-wise bounded spectral norm (Lemma~\ref{lem:fsu2}),
  and as a result the number of bottom eigenvectors must be small (Theorem~\ref{thm:fewsmall}).
}

\paragraph{Proof of Theorem~\ref{thm:edintro}.}{
  Our proof consists of two steps:
  (i) decomposing the graph into an almost regular graph (Theorem~\ref{thm:hargd}), and
  (ii) decomposing the graph into an almost regular expander (Theorem~\ref{thm:hared}).
  In achieving (ii), we actually invoke (i) repeatedly to maintain the almost regularity of the graph.

  As noted above, the almost regularity condition~(\ref{eq:arty}) is essentially saying that
  no incident edge has leverage score too large comparing to the average.
  A similar task to (i) has in fact been investigated by a previous work~\cite{CohenLMMPS15}, 
  where the authors showed that given a set of vectors, one can,
  by downscaling a small number of them,
  make every vector have small leverage score comparing to the average.
  This result is achieved by an algorithm that iteratively downscales vectors
  with large leverage scores, while analyzing how each vector's leverage score changes in the process.
  While it is possible to directly invoke the result from~\cite{CohenLMMPS15} to get
  a large almost regular graph, its guarantee does not suffice for our purpose of smoothly incorporating (i) into (ii).
  In particular, since we will repeatedly invoke (i) in (ii), we need,
  in addition to that the number of rescaled edges is small,
  an extra bound on the number of completely deleted edges (i.e. those rescaled to $0$) that
  is proportional to the rank change of the degree matrix $D$.
  As a result, we design a more involved algorithm for obtaining the scaling
  as well as carry out a more careful analysis of the algorithm.

  Achieving (ii) turns out to be much more challenging.
  Although the LHS of~(\ref{eq:expdef}) may be seen as a leverage score,
  there is the intrinsic difficulty that whenever the edge weights change,
  so do the eigenvalues and eigenvectors of the normalized Laplacian,
  as well as the degree matrix itself (hence also the normalized incidence vectors).
  Thus it is not clear how the LHS of~(\ref{eq:expdef}) will change. %
  As a result, when trying to obtain a desired scaling,
  we have to use some global measure of progress. 
  This is in contrast to (i), where we can track the leverage score change of each edge {\em locally}.
  We resolve this issue
  by considering, as a potential function, the determinant of the normalized Laplacian restricted to the bottom eigenspace.
  In other words, our potential function is the product of the eigenvalues of $N$ that are between
  $(0,\zeta]$\footnote{Due to technical reasons, the actual potential function slightly differs from the one stated here.}.
  We show that, by a delicate global analysis of such a potential function, we are able to
  make the graph an expander by only downscaling a small number of edges.
}

\paragraph{Proof of Theorem~\ref{thm:expsampintro}.}{
  Our proof is motivated by the {\em approximate Gaussian elimination} of
  the Laplacian matrices of scalar-weighted graphs, which was previously used as an algorithmic tool for solving
  graph structured linear systems~\cite{KyngLPSS16,KyngS16}
  and building data structures for dynamically maintaining effective resistances~\cite{DurfeeKPRS17,LiZ18,LiPYZ20}.
  Our approach also relies 
  on analyzing matrix-valued martingales which played a key role in~\cite{KyngS16}.
  which have played key roles in constructing vertex/subspace sparsifiers~\cite{KyngS16,LiS18,ForsterGLPSY21}.

  Let us first briefly review the Gaussian elimination of the Laplacian matrix of a scalar-weighted graph.
  Roughly speaking, by eliminating the row and column of $L$ corresponding to a vertex $u$,
  we can obtain another Laplacian matrix $L'$ supported on $V\setminus\setof{u}$ whose pseudoinverse {\em equals}
  the pseudoinverse of the original $L$ restricted to $V\setminus\setof{u}$
  (i.e. $(L')^{\dag} = (L^{\dag})_{V\setminus\setof{u}V\setminus\setof{u}}$).
  Given a vertex subset $C\subseteq V$,
  one can also eliminate the vertices outside of $C$ one by one
  and get a Laplacian matrix $L''$ supported
  on $C$ with the same property that $(L'')^{\dag} = (L^{\dag})_{CC}$.
  The matrix $L''$ is referred to as the {\em Schur complement} of $L$ onto $C$.
  However, the graphs associated with $L'$ and $L''$ could be dense,
  which are inefficient for algorithm design.
  Therefore~\cite{KyngS16} showed that one can perform an {\em approximate} Gaussian elimination
  by, upon each elimination, implicitly sub-sampling the edges in $L'$.
  They then showed that we eventually get a good approximation to $L''$
  by analyzing a matrix-valued martingale induced by this process.

  We now explain how to apply this idea to prove Theorem~\ref{thm:expsampintro}.
  Since we are considering an induced subgraph $G[C]$ where $C$ is a uniformly random
  subset of size $s$, we can also view the process for choosing $C$
  as deleting a sequence of $n-s$ vertices from $V$ uniformly at random.
  Our goal is to compare the pseudoinverse of $G[C]$ with that of the original graph,
  therefore it suffices to compare it with the Schur complement of $L$ onto $C$.
  We will in fact do such a comparison upon the elimination of every vertex.
  That is, if we let $C_i$ be the set of remaining vertices at the $i^{th}$ step,
  then we want to compare the Laplacian of $G[C_i]$ with the Schur complement of $L$ onto $C_i$.
  At a high level,
  we do so by setting up a matrix-valued martingale,
  and show that it has good concentration
  when $G$ is a matrix-weighted expander (in the sense of Definition~\ref{def:hde1}).

}

\section{Preliminaries}\label{sec:preli}

\subsection{Matrices}

\begin{definition}[Pseudoinverse]
  Let $A$ be an $n\times n$ symmetric matrix.
  Let $\lambda_1,\lambda_2,\ldots,\lambda_n$
  be the eigenvalues of $A$ and let $v_1,v_2,\ldots,v_n$
  be orthonormal (i.e. unit and orthogonal)
  eigenvectors of $A$,
  then by the spectral theorem
  $A = \sum_{i=1}^{n} \lambda_i v_i v_i^T$.
  The pseudoinverse of $A$ is then defined as
  \begin{align}
    A^{\dag} \defeq \sum_{\lambda_i\neq 0}
    \frac{1}{\lambda_i} v_i v_i^{T}.
  \end{align}
\end{definition}

\begin{definition}[Matrix partial ordering]
  \label{def:matrixorder}
  For two $n\times n$ symmetric matrices $A,B$,
  we write $A\pleq B$ if for any vector $x\in\mathbb{R}^{n}$ we have
  $x^{T} A x \leq x^T B x$
  (i.e. $B - A$ is positive semi-definite).
\end{definition}

\begin{fact}[Properties of the matrix partial ordering]\label{fact:inverseordering}
  \quad
  \begin{enumerate}
    \item If $A \pleq B$ and $C \pleq D$, then $A + C \pleq B + D$.
    \item If $A \pleq B$ where $A,B\in\mathbb{R}^{n\times n}$,
      then for any $W$ with $n$ rows, $W^T A W \pleq W^T \BB W$.
    \item For two positive semidefinite matrices $A,B$
      that have the same null space,
      $A\pleq B$ implies that $B^{\dag} \pleq A^{\dag}$.
  \end{enumerate}
\end{fact}

\begin{theorem}[Matrix chernoff bound~\cite{Tropp12}]\label{thm:matrixchernoff}
  Let $X_1,\ldots,X_m\in\mathbb{R}^{n\times n}$ be independent random positive
  semidefinite matrices such that $\lambda_{\mathrm{max}}(X_i)\leq R, \forall i$
  and $\ex{\sum_{i=1}^{m} X_i} = I_{n\times n}$ where $I_{n\times n}$ is the
  $n\times n$ identity matrix. Then with probability at least
  $1 - 2n \exp\setof{-\frac{\eps^2}{3R}}$
  \begin{align}
    (1 - \eps) I \pleq \sum_{i=1}^{m} X_i \pleq (1 + \eps) I.
  \end{align}
\end{theorem}

\begin{definition}[Leverage scores]
  \label{def:deflvg}
  Let $a_1,\ldots,a_m \in \mathbb{R}^{n}$ and
  $A = \sum_{i=1}^{m} a_i a_i^T \in \mathbb{R}^{n\times n}$.
  The leverage score of $a_i$ w.r.t. $A$ is defined to be
  $\tau_i(A) = a_i^T A^{\dag} a_i$.
\end{definition}

\begin{fact}\label{thm:dual}
  Let $a_1,\ldots,a_m \in \mathbb{R}^{n}$ and
  $A = \sum_{i=1}^{m} a_i a_i^T \in \mathbb{R}^{n\times n}$.
  Also let $B = (a_1 ,\ldots , a_m )\in\mathbb{R}^{n\times m}$,
  so we have $A = B B^T$.
  Let $b\in\mathbb{R}^{n}$
  be a vector in the span of $a_1,\ldots,a_m$.
  Then we have
  \begin{align*}
    \min_{Bx = b} \norm{x}_2^2 = b^{T} (B B^T)^{\dag} b.
  \end{align*}
\end{fact}

\subsection{Multivariate Gaussian distributions}

\begin{definition}\label{def:multivargaussian}
  Let $\mu\in \mathbb{R}^{d}$ be a vector
  and $\Sigma\in\mathbb{R}^{d\times d}$ be a matrix.
  We say a random vector $x\in\mathbb{R}^d$ follows a multivariate Gaussian distribution
  with mean $\mu$ and covariance matrix $\Sigma$, denoted by $\mathcal{N}(\mu,\Sigma)$, if
  \begin{enumerate}
    \item Each $x_i$ distributes as $\Ncal(\mu_i,\Sigma_{ii})$,
      a univariate Gaussian with mean $\mu_i$ and variance $\Sigma_{ii}$. \label{cond:1}
    \item $\ex{(x_i - \mu_i)(x_j - \mu_j)} = \Sigma_{ij}$ for all pairs $i,j$.
      Or equivalently,
      \begin{align}
        \ex{(x - \mu) (x - \mu)^{T}} = \Sigma.
      \end{align}
      \label{cond:2}
  \end{enumerate}
\end{definition}

\begin{fact}
  The covariance matrix $\Sigma$ is symmetric and positive semi-definite.
\end{fact}

\begin{fact}\label{thm:sumgaussian}
  Let $x_1,x_2 \in \mathbb{R}^d$ be independent random vectors
  such that $x_1\sim \Ncal(\mu_1,\Sigma_1)$ and $x_2\sim \Ncal(\mu_2,\Sigma_2)$.
  Then $x_1 + x_2 \sim \Ncal(\mu_1 + \mu_2, \Sigma_1 + \Sigma_2)$.
\end{fact}

\begin{theorem}[$\ell_1$-distance between multivariate Gaussians with the same mean~\cite{DevroyeMR18}]
  \label{thm:tvd}
  Let $\mu\in\mathbb{R}^{d}$ and let
  $\Sigma_1,\Sigma_2\in\mathbb{R}^{d\times d}$ be positive semidefinite
  such that
  \begin{enumerate}
    \item $\Sigma_1$ and $\Sigma_2$ have the same null space.
    \item $\Sigma_2 = \Sigma_1 + w w^T$ for some $w\in\mathbb{R}^{d}$.
  \end{enumerate}
  Then we have
  \begin{align*}
    d_{TV} \kh{ \Ncal(\mu,\Sigma_1), \Ncal(\mu,\Sigma_2) }
    = \Theta\kh{ \min\setof{1, w^T \Sigma_1^{\dag} w} }. %
  \end{align*}
\end{theorem}

\begin{theorem}[Chernoff bound for univariate Gaussians, Theorem 9.3 of~\cite{MitzenmacherU2017}]
  \label{thm:GaussianChernoff}
  Let $X$ be a univariate Gaussian with mean $\mu$ and variance $\sigma^2 > 0$:
  $X\sim \Ncal(\mu,\sigma^2)$. Then for any $a > 0$
  \begin{align}
    \pr{ \sizeof{ X - \mu } \geq a \sigma } \leq 2 e^{-a^2/2}.
  \end{align}
\end{theorem}

\subsection{$\ell_2$-heavy hitter, $\ell_p$-sampler and $\ell_2$ estimation}

\begin{proposition}[$\ell_2$-heavy hitters~\cite{KapralovLMMS14}]\label{prop:heavyhitter}
  For any $\eta > 0$, there is a decoding algorithm $D$ and a distribution on
  matrices $A \in \mathbb{R}^{O(\eta^{-2} \polylog(N))\times N}$ such that,
  for any $x\in \mathbb{R}^{N}$, given $A x$, the algorithm $D$ returns a vector $\xtil$
  such that $\xtil$ has $O(\eta^{-2}\polylog(N))$ non-zeros and satisfies
  $$\norm{x - \xtil}_{\infty} \leq \eta \norm{x}_2$$ with high probability over the choice of $A$.
  The sketch $A x$ can be maintained and decoded in
  $O(\eta^{-2} \polylog(N))$ space.
\end{proposition}

Given a vector $x$ of size $N$ and a number $\delta > 0$, an $\ell_p$ sampler is an algorithm that output an index $i$ with probability 
$$
p_i \in (1 \pm N^{-c}) \frac{\sizeof{x_i}^p}{\norm{x}^p_p} \pm O(N^{-c})
$$
where $c$ is arbitrary constant. The algorithm may also output $\textsf{Fail}$ with probability at most $\delta$. 

For any $0 \le p \le 2$, there is a polylog size linear sketch for $\ell_p$-sampler.

\begin{proposition}[\cite{JowhariST11,CormodeF14}] \label{l0-sampler}  
    For any constant $c$ and $0<\delta<1$, there is a linear sketch for $\ell_0$-sampling with measurement size $O(\log^2 n \log 1/\delta)$.
\end{proposition}

\begin{proposition}[\cite{JayaramW21}] \label{l1-sampler}  
    For any $0 < p < 2$, any $\eps,\delta_1,\delta_2>0$ and any constant $c>0$, there is a linear sketch with measurement size $O(\log^2 N (\log \log N)^2 \log (1/\delta_1) + \eps^{-p} \log N \log^2(1/\delta_2)\log (1/\delta_1))$ such that given an $N$-dimensional vector $x$, with probability $(1-\delta_1)$, it can recover an index $i$, such that the probablity of outputting $i$ is 
$$
    p_i \in (1 \pm N^{-c}) \frac{\sizeof{x_i}^p}{\norm{x}^p_p} \pm O(N^{-c})
$$
    Moreover, if the sketch does output an index $i$, then it also recovers a value $x'_i$ such that $\sizeof{x_i}\le x'_i \le (1 + \eps)\sizeof{x_i}$, with probability $1-\delta_2$.
\end{proposition}

\begin{proposition}[\cite{KapralovLMMS14}] \label{l2-heavy-hitter}  
    For any $\eps>0$ and any constant $c>0$, there is a linear sketch of size $O(\eps^{-2} \polylog(N))$ such that given an $N$-dimensional vecotr $x$, with probability $N^{-c}$, we can recover a vector $x'$ such that $\norm{x-x'}_{\infty} \le \eps \norm{x}_2$.
\end{proposition}

\begin{proposition}[\cite{johnson1984extensions}] \label{l2-norm}
    For any $0<\delta<1$ and any constant $c>0$, there is a linear sketch of size $O(\delta^{-2} \log N)$ such that given an $N$-dimensional vector $x$, with probability $N^{-c}$, we can recover a vector $x'$ such that $\norm{x'}_2 \in (1 \pm \delta) \norm{x}_2$.
\end{proposition}

\subsection{Edge strengths and cut sparsifiers}
Given a graph $G$, a {\em $k$-strongly connected component} is a maximal vertex induced subgraph whose minimum cut size is at least $k$. Thus, all $k$-strongly connected components form a partition of the entire vertex set.
The following fact gives an equivalent way of obtaining the $k$-strongly connected components.
\begin{fact}\label{fact:removecut}
  Given any graph and integer $k$,
  consider a process
  where we iteratively remove (the edges across) an arbitrary cut of size strictly smaller than $k$,
  until there is no such cut left.
  Then the connected components in the resulting
  graph are the $k$-strongly connected components of the original graph.
\end{fact}

    The {\em strength} of an edge $e$ in the graph, denoted $k_e$, is the maximum value of $k$ such that a $k$-strong component of $G$ contains both endpoints of $e$. The weighted sum of the inverse of the strength of every edge in a graph is at most $n-1$.

\begin{claim} [\cite{BenczurK15}] \label{clm:sum_n-1} For any weighted graph $G=(V,F,w)$ on $n$ vertices, $\sum_{f\in F}\frac{w_f}{k_f}\leq n-1$.
\end{claim}

Given a graph $G$, an $(1 \pm \eps)$-cut sparsifier $G'$ is a subgraph of $G$ such that any cut in $G$ is preserved in $G'$ to within a factor of $(1 \pm \eps)$. The seminal result~\cite{BenczurK15} shows that, for any graph $G$, if we construct a graph $G'$ as follows: we include each edge $e$ in $G'$ with probability $p_e \ge \Omega(\frac{w_e \log n}{\eps^2 k_e})$, and give weight $\frac{w_e}{p_e}$ if it gets chosen, then $G'$ is a $(1 \pm \eps)$ cut sparsifier of $G$ with high probability.

\subsection{Graph matrices, leverage scores, and spectral sparsifiers}\label{sec:gmat}
Fix an arbitrary orientation of all possible $\binom{n}{2}$ edge slots of the graph.
Let $B \in \mathbb{R}^{\binom{n}{2} \times n}$ be the {\em edge-vertex incidence matrix} of an undirected, unweighted complete graph over $n$ vertices. That is, for every edge $e=(u,v)$ oriented from $u\to v$,
there is a row $b_e\in\mathbb{R}^{n}$ in $B$ corresponding to $e$ such that the column $u$ has value $1$,
the column $v$ has value $-1$, and all other columns have value $0$. 
We also write $b_e$ as $b_{u,v}$.
For a graph $G$,
we write $B_G\in\mathbb{R}^{\binom{n}{2}\times n}$ to denote the matrix
obtained from $B$ by zeroing out rows corresponding to absent edges in $G$.

Given any weighted graph $G$, let $W_G \in \mathbb{R}^{\binom{n}{2} \times \binom{n}{2}}$ be the diagonal matrix whose diagonal entries are the weights of the edges corresponding to them, i.e. $(W_{G})_{ee} = w_e$. If an edge $e$ is not present in the graph, then
$(W_G)_{ee} = 0$. The {\em Laplacian matrix} of $G$ is given by $L_G = B_G^{\top} W_G B_G = (W_G^{1/2} B_G)^{\top} (W_G^{1/2} B_G)$.
Notice that for unweighted graphs, we have $L_G = B_G^T B_G$.

For any edge $e$, %
its {\em leverage score} $\tau_e$ is given by $\tau_e = (\sqrt{w_e} b_e)^{\top} L^{\dagger}_G (\sqrt{w_e} b_e) = w_e b_e^{\top} L^{\dagger}_G b_e$, where $L^{\dagger}_G$ is the Moore-Penrose pseudoinverse of the Laplacian matrix $L_G$.
\begin{fact}
  The sum of the leverage scores of all edges $\sum_e \tau_e = \mathrm{rank}(W_G^{1/2} B_G) \le n-1$. 
\end{fact}
In general, for {\em any} matrix $C\in\mathbb{R}^{m\times n}$ whose $i^{\mathrm{th}}$ row is denoted by
$c_i\in\mathbb{R}^{n}$,
we define the leverage score of the $i^{th}$ row by $\tau_i = c_i^T (C^T C)^{\dag} c_i$,
and we once again have $\sum_{i=1}^{m} \tau_i = \rank{C}$.

If we view the graph $G$ as an electrical network where each edge $e$ has resistance $1/w_e$,
then the {\em effective resistance} between the two vertices $s,t$ is given by $r_{s,t} = b_{s,t}^{\top} L^{\dagger}_G b_{s,t}$.
For an edge $e = (s,t)$, we also define its effective resistance as $r_e = r_{s,t}$.
Thus, the leverage score $\tau_e = w_e r_e$.

If we inject $f$ units of electrical flow into a vertex $s$ and extract $1$ unit from a vertex $t$,
then $f L^{\dagger} b_{s,t} \in \mathbb{R}^{n}$ is referred to as {\em the set of vertex potentials induced by the electrical flow}.
The relation between vertex potentials of the electrical flow is characterized by Ohm's Law.

\begin{fact}[Ohm's Law]
  Let $x = f L^{\dag} b_{s,t}\in\mathbb{R}^{n}$ be the set of vertex potentials when we send
  $f$ units of electrical flow
  from $s$ to $t$. Then
  we have $x_s - x_t = f (b_{s,t} L^{\dag} b_{s,t})$.
  Moreover, for any edge $e = (u,v)$ with $x_u \geq x_v$,
  the flow on this edge is in the direction of $u\to v$ and has amount exactly
  $w_e (x_u - x_v)$.
\end{fact}

It is also known that the vertex potentials induced by an electrical flow minimizes the total energy.
Specifically, for an arbitrary set of vertex potentials $x\in \mathbb{R}^{n}$,
we define its {\em energy} to be $x^T L_G x = \sum_{e = (u,v)\in G} w_e (x_u - x_v)^2$,
and define its {\em normalized energy with respect to vertices $s,t$} to be $\frac{x^T L_G x}{(x^T b_{s,t})^2}$
(i.e. the energy divided by $(x_s - x_t)^2$).
Then we have:
\begin{fact}\label{fact:energymin}
  The vertex potentials induced by an electrical flow from $s$ to $t$ minimizes
  the normalized energy with respect to $s,t$.
  That is, for any $f > 0$, we have
  $$ f L^{\dag} b_{s,t} \in \argmin_{x\in\mathbb{R}^{n}} \frac{x^T L_G x}{(x^T b_{s,t})^2}, $$
  and thus, by plugging in $x = L^{\dag} b_{s,t}$,
  the smallest normalized energy w.r.t. $s,t$ is
  \begin{align*}
    \min_{x\in\mathbb{R}^{n}} \frac{x^T L_G x}{(x^T b_{s,t})^2} = 
    \frac{ b_{s,t} L_G^{\dag} L_G L_G^{\dag} b_{s,t}}{(b_{s,t}^T L_G^{\dag} b_{s,t})^2} = 
    \frac{b_{s,t} L_G^{\dag} b_{s,t}}{(b_{s,t}^T L_G^{\dag} b_{s,t})^2}  =
    \frac{1}{{b_{s,t}^T L_G^{\dag} b_{s,t}}},
  \end{align*}
  exactly $1$ over the effective resistance between $s,t$.
\end{fact}

Given a graph $G$ with Laplacian matrix $L_G$, a $(1 \pm \eps)$-spectral sparsifier $G'$ is a graph with Laplacian matrix $L_{G'}$ such that for any $x \in \mathbb{R}^{n}$, $x^{\top} L_G' x \in (1 \pm \eps) x^{\top} L_G x$. In other words, $\frac{1}{1+\eps} L_G \preceq L_{G'} \preceq (1+\eps) L_G$ (recall Definition~\ref{def:matrixorder}). %
We also use $L_G \approx_{1+\eps} L_{G'}$ to denote the same relation between $L_G$ and $L_{G'}$.
For two scalars $a,b\geq 0$, we also write $a\approx_{1+\eps} b$ to denote $\frac{a}{1+\eps} \leq b \leq (1 + \eps) a$.

If we sample each edge of a graph with probability proportional to its leverage score or larger,
and reweight it accordingly,
then with high probability we get a spectral sparsifier.
In fact, this sampling process gives a good spectral approximation for any $C\in\mathbb{R}^{m\times n}$

\begin{theorem} [\cite{SpielmanS11,Tropp12}] \label{clm:lev-sample}
  Let $\eps > 0$.
  Given a matrix $C \in \mathbb{R}^{m \times n}$, let $p_1,\ldots,p_m\in\mathbb{R}$ be such that
  $1 \geq p_i \geq \min\setof{1, 100 \tau_i \eps^{-2}\log n}$ for all $i \in [m]$.
  Let $\tilde{W}\in\mathbb{R}^{m\times m}$ be a diagonal matrix such that
  $\tilde{W}_{ii} = 1/p_i$ with probability $p_i$ and $\tilde{W}_{ii} = 0$ otherwise.
  Then with high probability,
  $$
  C^T \tilde{W} C \approx_{\eps} C^T C.
  $$
\end{theorem}

Recall that when $C = W_G^{1/2} B_G$, we have $L_G = C^T C$.
Thus if we sample each edge $e$ with probability $p_e = \min\setof{1, 100 \tau_e \eps^{-2} \log n}$
and reweight it to $w_e / p_e$ if sampled, we get, by the claim above
and that $\sum_{e} \tau_e \leq n - 1$, a $(1+\eps)$-spectral sparsifier of $O(n\eps^{-2} \log n)$ edges.
We could also do an oversampling, where we sample each edge $e$ with some probability
$p_e \geq \min\setof{1, 100 \tau_e \eps^{-2} \log n}$ and also reweight it to $w_e/p_e$ if sampled,
and then we get a $(1 + \eps)$-spectral sparsifier of $O(\sum_{e} p_e)$ edges.

\section{A linear sketching algorithm for weighted cut sparsification}\label{sec:algocut}

In this section, we present a linear sketch
with $\tilde{O}(n \eps^{-3})$ measurements that computes a $(1+\eps)$-cut sparsifier of a weighted graph.
In particular, our linear sketch will be an incidence sketch.
Our algorithm is obtained by generalizing the cut sparsifier algorithm of~\cite{RubinsteinSW18},
which is an $O(\log n)$ round cut query algorithm that works only for unweighted graphs.
We first describe the weighted generalization of the algorithm of~\cite{RubinsteinSW18}
in a model-oblivious manner,
and then show how to implement it by linear sketching.

\subsection{A model oblivious algorithm for weighted cut sparsification}

\begin{figure}
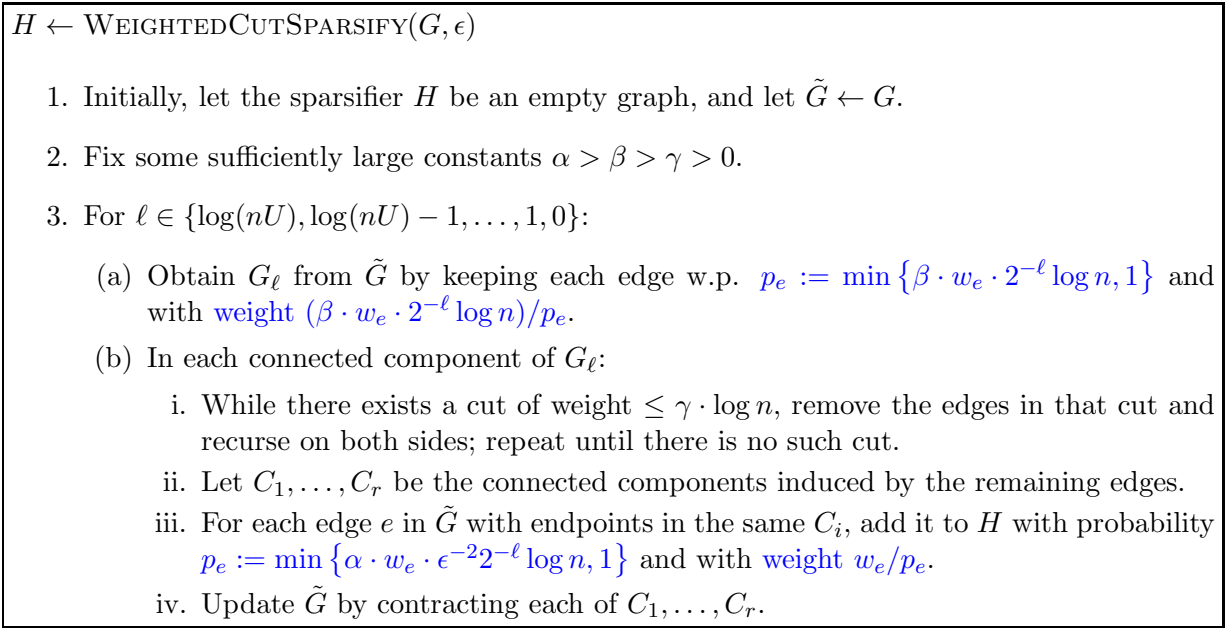

\begin{algbox}
    $H\gets \WGM(G, \eps)$\\

    \begin{enumerate}
      \item Initially, let the sparsifier $H$ be an empty graph, and let $\Gtil \gets G$.
      \item Fix some sufficiently large constants $\alpha > \beta > \gamma > 0$. \label{line:abc}
        \item For $\ell \in\setof {\log (nU), \log (nU) - 1, \ldots, 1,0}$:
            \begin{enumerate}
                \item Obtain $G_{\ell}$ from $\Gtil$ by keeping each edge w.p.
                {\color{blue} $p_e := \min\setof{\beta \cdot w_e\cdot 2^{-\ell} \log n,1}$} and with
                {\color{blue} weight $(\beta \cdot w_e\cdot 2^{-\ell} \log n)/p_e$}. \label{line:sample3}
                \item In each connected component of $G_{\ell}$:
                    \begin{enumerate}
                        \item While there exists a cut of weight $\leq \gamma \cdot \log n$,
                            remove the edges in that cut and recurse on both sides; repeat until there is no such cut.
                        \item Let $C_1,\ldots, C_r$ be the connected components induced by the remaining edges.
                        \item For each edge $e$ in $\Gtil$ with endpoints in the same $C_i$,
                            add it to $H$ with probability
                            {\color{blue} $p_e:= \min\setof{\alpha \cdot w_e\cdot \eps^{-2} 2^{-\ell} \log n,1}$}
                            and with {\color{blue} weight $w_e/p_e$}. \label{line:sample4}
                          \item Update $\Gtil$ by contracting each of $C_1,\ldots,C_r$. \label{line:contraction2}
                    \end{enumerate}
            \end{enumerate}
    \end{enumerate}
\end{algbox}
\caption{Model oblivious algorithm for weighted cut sparsification.}
\label{fig:wcutalg}
\end{figure}

We present a model-oblivious algorithm $\WGM$ for weighted cut sparsification in Figure~\ref{fig:wcutalg},
which generalizes the idea presented in~\cite{RubinsteinSW18} for unweighted graphs.
Here, we assume the edge weights of the input graph are between $[1,U]$ for some $U\geq 1$ that is known
to us.
We characterize the performance of the algorithm in the lemma below.

\begin{lemma}\label{prop:cutsparsifier}
  Fix $\alpha = 800,\beta = 400,\gamma = 100$ at Line~\ref{line:abc} of the algorithm $\WGM$.
  Let $\eps$ be an arbitrary number in $(0,1)$.
  Then,
  the algorithm $\WGM$ outputs, with high probability, a $(1+\eps)$-cut sparsifier of the input graph $G$
  with $O(n \eps^{-2} \log n)$ edges.
\end{lemma}

To prove the proposition, we will need the following
lemma.

\begin{lemma}\label{lem:strengthl}
  In each iteration $\ell$, with high probability:
    \begin{enumerate}
        \item The edges in $\Gtil$ that survive the contractions have strength at most $2^{\ell-1}$. \label{item:strengthupw}
        \item The edges in $\Gtil$ that are within the components $C_1,\ldots,C_r$ have strength
          $[2^{\ell-4},2^{\ell}]$. \label{item:strengthlbw}
    \end{enumerate}
\end{lemma}
\begin{proof}
  We show that at the end of each iteration $\ell$,
  the edges within $C_1,\ldots,C_r$ have strength $\geq 2^{\ell-4}$,
  and the edges between them have strength at most $2^{\ell-1}$.
  Then by noting that all edges in the graph have strength at most $n U$,
  the lemma follows by an induction on $\ell$.

  Consider iteration $\ell$ of the for loop.
  Let $V_1,\ldots,V_{t}$ be the partition of $\Gtil$ into
  maximal $2^{\ell-4}$-strongly connected components.
  Then by Fact~\ref{fact:removecut} there exists a way to arrive at these components by starting from the entire graph $\Gtil$
  and iteratively removing a cut with size $< 2^{\ell-4}$.
  By applying a Chernoff bound and a union bound over the sequence of (at most $n-1$) cuts
  that we remove in this process,
  we have that after sampling, each of these cuts has size at most $50 \log n$.
  As a result, the partition $C_1,\ldots,C_r$ of $G_{\ell}$ into maximal $100 \log n$-strongly connected components is a refinement
  of $V_1,\ldots,V_{t}$.
  This implies that all edges within $C_1,\ldots,C_r$ have strength $\geq 2^{\ell-4}$.

  Now consider the partition $V_1,\ldots,V_{s}$ of $\Gtil$ into maximal $2^{\ell-1}$-strongly connected components.
  After sampling at Line~\ref{line:sample3}, these components still have min cut $\geq 100 \log n$ with high probability.
  Thus, the partition $C_1,\ldots,C_r$ of $G_{\ell}$ into maximal $100\log n$-strongly connected components
  is a coarsening of $V_1,\ldots,V_{s}$. This implies that all edges going across different $C_i$'s
  have strength at most $2^{\ell-1}$. This finishes the proof of the lemma.
\end{proof}

\begin{proof}[Proof of Lemma~\ref{prop:cutsparsifier}]
  By Lemma~\ref{lem:strengthl}, whenever we sample an edge $e$ of the graph, we sample it
  with probability $p_e \geq \min\setof{1, 50 w_e k_e^{-1} \eps^{-2} \log n }$ and re-weight it to $w_e / p$.
  Moreover, in the last iteration $\ell = 0$, all edges are within the components $C_1,\ldots,C_r$
  (since each edge's strength is at least $1$). Therefore ultimately all edges get sampled in our algorithm.
  As a result, we get with high probability a $(1+\eps)$-cut sparsifier. %
  On the other hand, the probability $p_e$ with which we sample $e$ also satisfies
  $p_e \leq 800 w_e k_e^{-1} \eps^{-2} \log n$, and thus we get a sparsifier
  with $O(n\eps^{-2}\log n)$ edges.
\end{proof}

\subsection{Implementation by linear sketching}

Now we show how to implement the algorithm by linear sketching. The implementation is motivated by the techniques first used in~\cite{AhnGM12a}. Note that it suffices to implement the two edge sampling processes at Lines~\ref{line:sample3},~\ref{line:sample4} and the contraction operations at Line~\ref{line:contraction2}.

We note that the implementation of both sampling processes can be seen as the following task:
we first independently generate a uniformly random real number $R_e \in [0,1)$ for each edge slot $e$,
and then recover all edges satisfying $R_e < w_e p$ for some given $p$ (which in iteration $\ell$ equals $\beta 2^{-\ell} \log n$
for the first process and $\alpha \eps^{-2} 2^{-\ell} \log n$ for the second process).
Here we can generate the $R_e$'s offline, but have to recover the sampled edges using linear sketching.
We achieve the latter by repeatedly finding a spanning forest formed by the sampled edges.
We will show that the sampled edges can be found by performing $\Otil(1)$ iterations of spanning forest recovery.

We shall first show how to recover a spanning forest formed by the sampled edges via linear sketching.
To this end, we need a linear sketching subroutine that we call {\em weighted edge sampler},
with the following guarantee.

\begin{lemma} \label{clm:wei_l0}
  Let $R_1,\ldots,R_N$ be $N$ numbers independently and uniformly at random
  generated from $[0,1)$, let $c > 0$ be an arbitrary constant,
  and let $p \in (0,1)$ be a parameter.
  There exists a linear sketch of $\polylog(N,1/p)$ measurements with the following guarantee.
  For any vector $w\in\mathbb{R}^{N}$, %
  if there exists an entry $e$ such that $R_e \le w_e p$ and $w_e > N^{-c}$,
  then with high probability,
  the sketch recovers an index $e'$ such that $R_{e'} \leq (1 + \eps) w_{e'} p$
  along with a $(1+\eps)$-approximate estimate of $w_{e'}$.
\end{lemma}

To find a spanning forest of the sampled edges (those with $R_e < w_e p$),
we apply the weighted edge sampler sketch to the incidence vectors of the vertices in $G$. Specifically, we fix an arbitrary orientation of each of the $\binom{n}{2}$ potential edges.  Then for a vertex $u\in G$, we consider its incidence vector $b_u \in \mathbb{R}^{\binom{n}{2}}$ given by \begin{align}
  (b_u)_e=
    \begin{cases}
        w_e & \text{$e\in G$, $u$ is $e$'s head} \\
        -w_e & \text{$e\in G$, $v$ is $e$'s tail} \\
        0 & \text{$e\notin G$ or $e$ does not touch $u$.}
    \end{cases}
\end{align}

Let $t = \polylog(n)$. For any $i \in [t]$,
let $A_i$ be an independently generated weighted edge sampler sketching matrix.
For each $i\in [t]$ and vertex $u\in G$, we compute the sketch $A_i b_u$.
Thus we make $\Otil(n)$ measurements in total. 
Now using the sketches we have taken,
we recover a spanning forest of sampled edges via a $\Otil(1)$-round process as follows.

In the first round, for each vertex $u$, we find an arbitrary outgoing edge using the weighted edge sampler sketch $A_1 b_u$.  We then find all connected components induced by these edges, and add up the sketches $A_2 b_u$ of vertices within the same component.  Note that the edges within the same component cancel out in the summation, so the resulting sketches are in fact taken over the outgoing edges of each component. As a result, in the next round we are able to find an outgoing edge of each component.
We then proceed similarly in the $i^{\mathrm{th}}$ round using sketches $A_{i} b_u$'s.
Since in each round, the number of components is at least reduced by a factor of $2$, we can find a spanning forest in $O(\log n)$ rounds of this process.

In order to iteratively find $\Otil(1)$ edge-disjoint spanning forests,
each time we find one,
we ``delete'' the found edges from the other linear sketches, and restart the $O(\log n)$-round process above.
Note that however, since we do not have the exact weights of the edges (Lemma~\ref{clm:wei_l0} only gives approximation of them),
we do not delete the found edges completely, bur rather decrease each of their weights by an $\Omega(1)$ factor.

Finally, to implement the contraction operations at Line~\ref{line:contraction2}, we once again add the sketches of the vertices within each contracted component, just as we did in finding spanning forests. Then starting from the next iteration, the sketches work for the contracted graph $\Gtil$.

We conclude this subsection by proving that the sampled edges can be recovered
by $\Otil(1)$ edge-disjoint spanning forests.

\begin{lemma}
  The edges in $G_{\ell}$ can be found by $O(\log^2 n)$ edge-disjoint spanning forests.
\end{lemma}
\begin{proof}
  We first show that the edges in $G_{\ell}$ all have low strength.
  \begin{claim}
    Every edge in the graph $G_{\ell}$ has strength $O(\log n)$.
  \end{claim}
  \begin{proof}
    By Lemma~\ref{lem:strengthl}, we know that at the beginning of iteration $\ell$,
    all edges in $\Gtil$ have strength at most $2^{\ell}$.
    This means that there is a way of removing all edges in the graph
    by iteratively removing a cut of size $\leq 2^{\ell}$.
    Then after sampling at Line~\ref{line:sample3}, these cuts all have size $O(\log n)$.
    Thus it follows that all edges in $G_{\ell}$ have strength $O(\log n)$.
  \end{proof}
  Let $C > 0$ be a constant such that all edges in $G_{\ell}$ have strength $\leq C \log n$.
  Then this claim implies that $G_{\ell}$ is {\em uniformly sparse}, in the sense that
  for any vertex induced subgraph $G[S]$ where $S\subseteq V$,
  the number of edges in $G[S]$ is $(|S| - 1) C \log n$.
  Indeed, each edge in $G[S]$ has strength only smaller than in $G$,
  and thus all edges in $G[S]$ can be removed by iteratively (for at most $|S|-1$ times) removing a cut
  of size $C \log n$.
  This in particular means that at any point,
  a spanning forest contains an $1 / (C \log n)$ fraction of the total remaining edges.
  Thus $O(\log^2 n)$ edge-disjoint spanning forests recover all edges in $G_{\ell}$.
\end{proof}

\subsection{Proof of Lemma~\ref{clm:wei_l0}}

Roughly, we will simulate the non-uniform sampling process, where we want to sample
each $e$ with probability $w_e p$, by sampling
the elements uniformly, but at different geometric rates.
We will then essentially implement a {\em rejection sampling} process.
Specifically, when subsampling all elements at some uniform rate $q$, we use $\ell_1$-samplers to recover
a few elements that are sampled. We will then check if any one of the sampled elements $e$
satisfies $w_e p \approx q$. If so, we will output this element; otherwise, we go the next sampling
rate and repeat this step. We will show that with high probability, we will successfully recover
a desired element.

\let\oldell\ell
\renewcommand*{\ell}{\gamma}

\begin{proof}[Proof of Lemma~\ref{clm:wei_l0}]
  We will analyze the linear sketch given in Figure~\ref{fig:wl0}.
  The basic idea of the algorithm is as follows: for each $0 \le j \le \ell$, we maintain $\polylog(N,\frac{1}{p})$ number of $\ell_1$-sketches given by Proposition~\ref{l1-sampler} with failure probability $\delta_1,\delta_2 = N^{-100}$ that work for $w^j$ where $w^j$ is the vector generated by sampling each entry of $w$ with probability $2^{-j}$ (i.e. $q$ in the overview above). For each sampled element $e$, we check if $w_e p$ is indeed larger than $2^{-j}$, if so, $e$ indeed gets sampled and we can output $e$, otherwise $e$ might not get sampled, and we discard $e$. We prove that whenever there are elements that get sampled, we will find one of them with high probability.

\begin{figure}
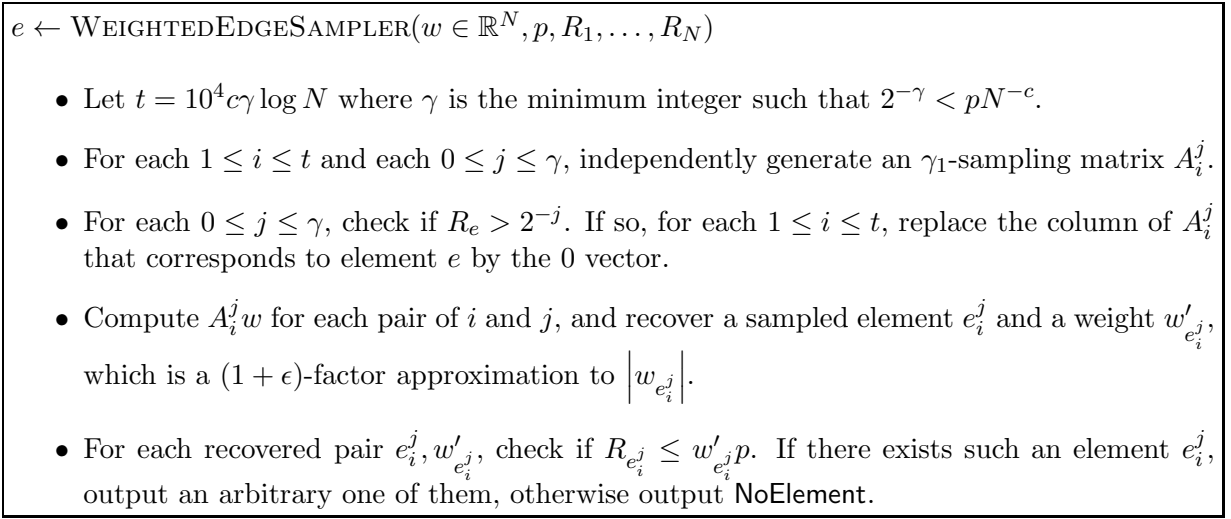

\begin{algbox}
  $e\gets \WL(w\in \mathbb{R}^{N},p, R_1,\ldots,R_N)$\\

    \begin{itemize}
        \item Let $t = 10^4 c \ell \log N$
          where $\ell$ is the minimum integer such that $2^{-\ell}<p N^{-c}$.
        \item For each $1 \le i \le t$ and each $0 \le j \le \ell$, independently generate an $\ell_1$-sampling
          matrix $A_i^j$.
        \item For each $0 \le j \le \ell$, check if $R_e > 2^{-j}$. If so, for each $1 \le i \le t$, replace the column of $A_i^j$ that corresponds to element $e$ by the $0$ vector.
        \item Compute $A_i^j w$ for each pair of $i$ and $j$, and recover a sampled element $e_i^j$ and a weight $w'_{e_i^j}$, which is a $(1+\eps)$-factor approximation to $\sizeof{w_{e_i^j}}$.
        \item For each recovered pair $e_i^j,w'_{e_i^j}$, check if $R_{e_i^j} \le w'_{e_i^j} p $. If there exists such an element $e_i^j$, output an arbitrary one of them, otherwise output $\mathsf{No Element}$.
    \end{itemize}
\end{algbox}
\caption{Weighted edge sampler.}
\label{fig:wl0}
\end{figure}

    The analysis is conditioned on the event that none of the $\ell_1$-samplers fails, which is a high probability event. Since the algorithm only uses $\ell_1$ samplers, we can without loss of generality assume that each element has positive weight. Note that the algorithm uses $t \ell$ number of $\ell_1$ sampling matrices in total, so the total number of measurements of this sketch is $\polylog(N,U,\frac{1}{p})$. Moreover, by Proposition~\ref{l1-sampler}, if the algorithm outputs an element $e$, we have $R_e \le w'_e \le (1+\eps) w_e p$. %
    it is sufficient to prove that if there exists $e$ such that $R_e \le w_e p$, with high probability, we will not output $\textsf{No Element}$.

For any $1 \le k \le \ell$, let $S_k$ be the set of elements $e$ such that $2^{-k} \le w_e p < 2^{-k+1}$, and let $S_0$ be the set of elements $e$ such that $w_e p \ge 1$. For any $0 \le j \le \ell$, let $S_k^j$ be the set of elements $e \in S_k$ such that $R_e \le 2^{-j}$. For any $j$ and $k$, let $n_k = \sizeof{S_k}$, $n_k^j=\sizeof{S_k^j}$ and $W_k^j$ be the total weight of elements in $S_k^j$. The following claim follows from Chernoff bound.

\begin{claim} \label{clm:sam_con}
    With high probability, for any $0 \le j,k \le \ell$, if $n_k > 1000 \cdot 2^j \log N$, then $\sizeof{2^j n_k^j - n_k} < n_k/2$, otherwise $n_k^j < 1500 \log N$.
\end{claim}

\begin{proof}
    For each element $e \in n_k$, the probability that $e \in n_k^j = 2^{-j}$, so the size of $n_k^j$ is the sum of $\sizeof{n_k}$ independent random $0/1$ variables each with expectation $2^{-j}$. The expected size of $n_k^j$ is $2^{-j} n_k$. If $\sizeof{n_k} > 1000 \cdot 2^j \log N$, by Chernoff bound, the probability that $\sizeof{2^j n_k^j - n_k} > n_k/2$ is at most $< 2 e^{\frac{n_k}{2^{j+4}}} < N^{-50}$.

    If $\sizeof{n_k} \le 1000 \cdot 2^j \log N$, the probability that $\sizeof{n_k^j} \ge 1500 \log N$ is at most the probability of the case when $n_k = 1000 \cdot 2^j \log N$. So the probability is less than $N^{-50}$.
\end{proof}

Let $k^*$ be the index that maximizes $\frac{n_{k^*}}{2^{k^*}}$. If $n_{k^*} > 1000 \cdot 2^{k^*} \log N$, by Claim~\ref{clm:sam_con}, $n_{k^*}^{k^*}> \frac{n_{k^*}}{2^{k^*+1}}$. Since each element in $S_{k^*}$ has weight at least $2^{-k^*}/p$, $W_{k^*}^{k^*} > \frac{n_{k^*}}{2^{2k^*+1}p}$. On the other hand, for any $k > k^*$, we have $n_k^{k^*} < \max \{ 1500 \log N, \frac{3n_k}{2^{k^*+1}}\}$ by Claim~\ref{clm:sam_con}. Since each element in $S_k$ has weight at most $2^{-k+1}/p$, we have $W_k^{k^*} < \max\{ \frac{3000 \log N}{2^k p} , \frac{3n_k}{2^{k+k^*+1}}\}$. As $n_{k^*} > 1000 \cdot 2^{k^*} \log N$, $\frac{3000 \log N}{2^k p} < \frac{3 n_{k^*}}{2^{k+k^*} p} < 3 W_{k^*}^{k^*}$. Also, by definition of $k^*$, $\frac{n_{k^*}}{2^{k^*}} \ge \frac{n_k}{2^k}$, which means $\frac{3n_k}{2^{k+k^*+1}} \le \frac{3n_{k^*}}{2^{2k^*+1}} < 3 W_{k^*}^{k^*}$. Thus, we have $W_k^{k^*} < 3W_{k^*}^{k^*}$. Note that for any $1 \le i \le t$, $e_i^{k^*}$ is obtained by an $\ell_1$ sampler from $\cup_{k=0}^{\ell} S_k^{k^*}$, so with probability at least $\frac{1}{3\ell + 3}$, $e_i^{k^*}$ is an element in $S_k$ such that $k \le k^*$. In this case, $R_e \le 2^{-k^*} \le w_e p$, and so the algorithm will output an element that gets sampled. Since $t=10^4 \ell \log N$, there exists such an $i$ with high probability.

If $n_{k^*} \le 1000 \cdot 2^{k^*} \log N$, then for any $k$, $\frac{n_k}{2^k} \le 1000 \log N$. Let $e$ be a maximum weight element such that $R_e < w_e p$. Suppose $e \in S_k$, then we have $R_e < w_e p < 2^{-k+1}$. Let $k'$ be the largest index such that $e \in S_k^{k'}$, we have $k' \ge k-1$. For any $e'$ such that $e' \in S_{k''}$ with $k''< k$, by definition of $e$ and $k$, $R_{e'} > w_{e'} p \ge 2 ^ {-k''} \ge 2^{-k+1} \ge 2^{k'}$, which means $e' \notin S_{k''}^{k'}$. So $S_{k''}^k = \emptyset$ for any $k''<k$. On the other hand, for any $k'' \ge k$, $n_{k''}^{k'} < \max\{ 1500 \log N , \frac{3n_{k''}}{2^{k'}+1}\}$ by Claim~\ref{clm:sam_con}. Since any element in $S_{k''}$ has weight at most $2^{-k''+1}/p$, $W_{k''}^{k'} < \{ \frac{3000 \log N}{2^{k''} p} , \frac{3n_{k''}}{2^{k''+k'+1}}\}$. Since $w_e \ge \frac{1}{2^k p}$ and $k'' \ge k$, $\frac{3000 \log N}{2^{k''} p} \le 3000 \log N w_e$. Moreover, since $\frac{n_{k''}}{2^{k''}} \le 1000 \log N$, $\frac{3n_{k''}}{2^{k''+k'+1}} \le \frac{3000 \log N}{2^{k'+1} p} \le 3000 \log N w_e$. So $W_{k''}^{k'} < 3000 \log N w_e$ for any $k'' \ge k$, which means for any $1 \le i \le t$, $e_i^{k'} = e$ with probability at least $\frac{1}{3000 (\ell+1) \log N}$. Since $t = 10^4 \ell \log N$, with high probability, there exists one $i$ such that $e_i^{k'} = e$ and the algorithm will output an element that gets sampled.

So in both cases, the algorithm succeeds with high probability.
\end{proof}

\let\ell\oldell

\section{A linear sketching algorithm for weighted spectral sparsification}\label{sec:algospectral}

\let\oldomega\omega
\renewcommand*{\omega}{\eta}

\newcommand{\supi}[1]{(#1)}

In this section, we present a linear sketch with $\tilde{O}(n^{6/5}\eps^{-4})$ measurements
that computes a $(1+\eps)$-spectral sparsifier of a weighted graph.
Our linear sketch will be an incidence sketch.

This section is structured as follows.
First in Section~\ref{sec:vertexsample} we prove a key vertex sampling lemma,
which says that a heavy edge in $G$ is likely heavy in a random vertex-induced
subgraph of $G^{\sq}$.
Then in light of this lemma,
in Section~\ref{sec:heavyedge}
we present our linear sketch for recovering heavy edges in $G$, where we also assume black-box
access to two other linear sketches for sparsifying and recovering heavy edges in $G^{\sq}$, respectively.
Next in Section~\ref{sec:algmainspectral}, we present our main linear sketching algorithm
for weighted spectral sparsification using the heavy edge recovery sketch in Section~\ref{sec:heavyedge}.
Finally in Section~\ref{sec:sparsifyG^2} we present the linear sketches
for sparsifying and recovering heavy edges in $G^{\sq}$
that we invoke in Section~\ref{sec:heavyedge}.

\subsection{A vertex sampling lemma}\label{sec:vertexsample}

In this subsection we prove a key vertex sampling lemma.
This lemma will enable us to recover heavy edges in $G$ by subsampling vertices at rate $\approx n^{-1/5}$
and then recovering edges in the vertex-induced subgraph of $G^{\sq}$.

\begin{lemma}[Vertex sampling lemma]\label{lem:vertex-sample}
  Let $\omega\in (0,1)$.
  Given a weighted graph $G$,
  let $C$ be a vertex set obtained by including each vertex in $G$ with probability $\frac{\omega}{100n^{1/5}}$
  independently.
  For any edge $e$ in $G$ with leverage score $w_e b_e^T L_G^{\dag} b_e \geq \omega$,
  conditioned on $e\in G[C]$,
  with probability at least $.1$,
  its leverage score in $G^{\sq}[C]$ satisfies
  $w_e^2 b_e^T L_{G^{\sq}[C]}^{\dag} b_e \geq 1/1000$.
\end{lemma}

Roughly, our proof of the lemma proceeds as follows: (i) group vertices according to their potentials
induced by an electrical flow between the endpoints of $e$ in $G$;
(ii) analyze the structure of the edges in the vertex-induced subgraph based on their weights
and the potential difference between their endpoints; (iii)
explicitly construct a set of vertex potentials in $G^{\sq}$ that certifies
the heaviness of the edge $e$.

\begin{proof}[Proof of Lemma~\ref{lem:vertex-sample}]
  Let $e=(s,t)$ and without loss of generality, assume $w_e=1$
  (since we could always scale all edge weights simultaneously without changing any leverage scores).
  Since the leverage score $\tau_e \geq \omega$,
  we also have that the effective resistance $b_e^{T} L_G^{\dagger} b_e \geq \omega$.
  We use the electrical network view of the graph $G$, and let $x = \frac{L_G^{\dag} b_e}{b_e^T L_G^{\dag} b_e} \in\mathbb{R}^{n}$
  be the set of vertex potentials induced by an 
  electrical flow from $s$ to $t$
  of $\frac{1}{b_e^{T} L_G^{\dagger} b_e}\leq \frac{1}{\omega}$ units.
  We also assume without loss of generality $x_t=0$ (since we could always shift all vertex potentials by $x_t$ otherwise).
  Since $b_e^{T} L_G^{\dagger} b_e$ is the effective resistance between $s$ and $t$, we have $x_s=1$ by Ohm's law.
  Moreover, by Fact~\ref{fact:energymin},
  the normalized energy of $x$ w.r.t. $s,t$ satisfies that
  \begin{align*}
    x^T L_G x  =
    \frac{1}{{b_e^T L_G^{\dag} b_e}} \leq 1/\omega.
  \end{align*}

  We now partition the vertices other than $s$ and $t$
  into $n^{4/5}$ groups based on their potentials.
  Specifically, the $i^{\mathrm{th}}$ group $S_i$ contains all vertices satisfying
  $(i-1)\cdot n^{-4/5}\leq x_u \leq i\cdot n^{-4/5}$,
  where we break ties arbitrarily.
  For an edge $f = (u,v)$, we say $f$ passes through $S_i$ if
  $x_u \leq (i-1)\cdot n^{-4/5} < i\cdot n^{-4/5} \leq x_v$ or
  $x_v \leq (i-1)\cdot n^{-4/5} < i\cdot n^{-4/5} \leq x_u$.
  \begin{claim}
    For any $i\in [n^{4/5}]$,
    the total weight of edges that pass through $S_i$ is at most
    $n^{4/5} / \omega$.
    \label{claim:totalw}
  \end{claim}
  \begin{proof}
    Consider an edge $f = (u,v)$ that passes through $S_i$,
    and assume without loss of generality $x_u \geq x_v$.
    By Ohm's law, the flow on edge $f$ is in the direction $u\to v$ and has amount
    $w_f (x_u - x_v) \geq w_f n^{-4/5}$. Since the total amount of flow across the cut
    \begin{align*}
      \kh{\setof{t}\union S_1\union\ldots \union S_{i},\ S_{i+1}\union\ldots \union S_{n^{4/5}}\union\setof{s}}
    \end{align*}
    is at most $1/\omega$, we have
    \begin{align*}
      \sum_{\text{$f$: $f$ passes through $S_i$}} w_f n^{-4/5} \leq 1/\omega,
    \end{align*}
    which means that the total weight of such edges is at most $n^{4/5}/\omega$.
  \end{proof}

  We now consider what happens when we look at a vertex-induced subgraph $G[C]$
  where $C$ is obtained by including each vertex (other than $s,t$)
  with probability $\frac{\omega}{100 n^{1/5}}$, and then also including $s,t$.
  We say an edge $f = (u,v)$ is {\em intermediate} if $\setof{u,v}\intersect\setof{s,t} = \emptyset$.
  We say a group $S_i$ is {\em good} if (i)
  none of the vertices in $S_i$ gets sampled in $C$, and (ii) all intermediate edges that pass through $S_i$ and have
  both endpoints in $C$ have weight at most $n^{2/5}$.
  We say $S_i$ is {\em bad} otherwise.
  \begin{claim}\label{claim:allgood}
    With probability at least $2/3$, the number of good $S_i$'s with $i\in(\frac{1}{4}n^{4/5},\frac{3}{4}n^{4/5}]$
    is at least $n^{4/5}/20$.
  \end{claim}
  \begin{proof}
    First, by Markov's inequality,
    at least $.8$ fraction of the $S_i$'s with $i\in(\frac{1}{4}n^{4/5},\frac{3}{4}n^{4/5}]$
    have size at most $10 n^{1/5}$.
    For any fixed $S_i$ with $|S_i|\leq 10n^{1/5}$,
    we have
    \begin{align*}
      \ex{\sizeof{S_i\intersect C}} \leq \frac{|S_i| {\omega}}{100 n^{1/5}} \leq \frac{{\omega}}{10}
      \leq \frac{1}{10}.
    \end{align*}
    Therefore, once again by Markov's inequality,
    the probability that (i) happens for any fixed $S_i$ with $|S_i|\leq 10n^{1/5}$ is at least $.9$.

    On the other hand, by Claim~\ref{claim:totalw},
    the total number of edges with weight $> n^{2/5}$ that pass through $S_i$
    is at most $n^{2/5}/\omega$.
    The probability of any intermediate edge belonging to $G[S]$ is $\kh{\frac{{\omega}}{100 n^{1/5}}}^2 =
    \frac{\omega^2}{10000 n^{2/5}}$.
    These combined give us that the expected number of intermediate edges with weight $> n^{2/5}$ that pass through $S_i$
    and have both endpoints in $C$ is at most $1/10000$.
    Now an application of Markov's inequality gives that
    (ii) happens for $S_i$ with probability at least $1 - 1/10000$.

    Therefore, by a union bound, each $S_i$ with $|S_i|\leq 10n^{1/5}$
    is good with probability at least $.89$.
    Thus the expected number of bad $S_i$'s with $|S_i|\leq 10 n^{1/5}$
    is at most $.11 n^{4/5}$.
    By Markov's inequality, the number of
    bad $S_i$'s with $|S_i|\leq 10 n^{1/5}$
    is at most $.33 n^{4/5}$ with probability $\geq 2/3$.
    Thus, with probability at least $2/3$,
    the number of good $S_i$'s with $i\in(\frac{1}{4}n^{4/5},\frac{3}{4}n^{4/5}]$
    is at least $.4 n^{4/5} - .33 n^{4/5} \geq .05 n^{4/5}$, as desired.
  \end{proof}

  We will now construct a set of vertex potentials (call it $y\in\mathbb{R}^{|C|}$) in $G[C]$ from $x$.
  We will show that the energy of the new set of potentials is small in $G^{\sq}[C]$ (i.e. even with edge weights squared),
  but the potential difference between $s,t$ is still large, which result in
  a small normalized energy w.r.t. $s,t$, and thus certify the ``heaviness'' of edge $(s,t)$ in $G^{\sq}[C]$.

  Specifically, we obtain $y$ by ``collapsing'' the vertex potentials within each bad $S_i$,
  so that the intermediate edges that do not pass through any good $S_i$'s will have both endpoints getting the same potential.
  To take care of the edges incident on $s$ or $t$ that span less than $1/4$ fraction
  of the groups, we will also
  collapse the vertex potentials in the range $[0,1/4]$ and $[3/4,1]$.
  Precisely, $y_u$ is given as follows for each $u\in C$:
  \begin{enumerate}
    \item If $x_u \geq 3/4$, then set $y_u \gets 1$.
    \item Otherwise, suppose $x_u\in S_i$ for some $i\leq \frac{3}{4} n^{4/5}$.
      Count the number of good $S_j$'s with $j\in (\max\setof{i,\frac{1}{4}n^{4/5}}, \frac{3}{4} n^{4/5}]$ and let
      $k$ denote that number. Then set $y_u \gets 1 - k n^{-4/5}$.
  \end{enumerate}
  \begin{claim}\label{claim:yx}
    $y$ satisfies the following properties:
    \begin{enumerate}
      \item For any $u,v$, $|y_u - y_v| \leq |x_u - x_v|$. \label{item:uv}
      \item With probability at least $2/3$,
        $y_s - y_t \geq 1/20$. \label{item:st}
      \item With probability at least $.8$,
        $y^T L_{G^{\sq}[C]} y \leq 2$. \label{item:yty}
    \end{enumerate}
  \end{claim}
  \begin{proof}
    Note that for any $i$, $C\intersect S_i \neq \emptyset$ implies that $S_i$ is bad.
    Therefore, by our construction, all vertices within the same bad $S_i$ will end up having
    the same potentials in $y$. As a result, for any $u,v$,
    $|y_u - y_v|$ equals $n^{-4/5}$ times the number of good groups $S_i$ between
    $x_u$ and $x_v$ such that $i\in(\frac{1}{4} n^{4/5}, \frac{3}{4} n^{4/5}]$, which implies~\ref{item:uv}.

    By Claim~\ref{claim:allgood}, with probability $2/3$, the number of good $S_i$'s
    with $i\in (\frac{1}{4} n^{4/5}, \frac{3}{4} n^{4/5}]$ is at least $n^{4/5}/20$.
    Thus we have~\ref{item:st}.

    We then prove~\ref{item:yty}.
    First note that for edges that do not pass through any good $S_i$, both their endpoints
    have the same potential in $y$. So the total energy contributed by these edges is zero.
    Now the remaining edges can be divided into two types: (A)
    edges that are incident on $s$ or $t$; (B) intermediate edges
    that pass through some good $S_i$.
    By the definition of good $S_i$'s, edges of type (B) have weight at most $n^{2/5}$ each.

    For edges of type (A), they are of the form $(s,u)$ or $(v,t)$. If $x_u \geq 3/4$ or
    $x_v \leq 1/4$, then once again both endpoints of the edge have the same potential in $y$,
    and the energy contribution from such edges is zero.
    We thus focus on the edges of type (A) such that $x_u < 3/4$ or $x_v > 1/4$,
    and refer to those edges as type (A').
    For any type (A') edge $f = (a,b)$,
    we have $|x_a - x_b|\geq 1/4$,
    and thus by Ohm's law the amount of flow on $f$ is at least $w_f/4$.
    Since the total amount of flow going out of $s$ or going into $t$
    is upper bounded by $1/\omega$,
    the total weight of type (A') edges is at most $8/\omega$.
    This means that the number of type (A') edges $f$ with $w_f > 1$ is at most $8/\omega$.
    Thus the expected number of such edges in $G[C]$ is at most
    $(8/\omega) \frac{\omega}{100 n^{1/5}} \leq \frac{1}{10}$.
    By Markov's inequality, none of such edges is in $G[C]$ with probability at least $.9$
    (call this event $\Ecal_1$).
    As for any type (A') edge $f = (a,b)$ with $w_f \leq 1$, we have
    $w_f^2 (y_a - y_b)^2 \leq w_f (x_a - x_b)^2$.
    This combined with the fact that each $f$ belongs to $G[C]$ with probability $\frac{\omega}{100 n^{1/5}}$,
    the expected contribution of type (A') edges with weight at most $1$ to $y^T L_{G^{\sq}[C]} y$ is at most
    $\frac{\omega}{100 n^{1/5}}\cdot x^T L_G x \leq \frac{1}{100 n^{1/5}} \leq \frac{1}{100}$.
    Thus this contribution does not exceed $1$ with probability $.99$ (call this event $\Ecal_2$).

    Finally we consider type (B) edges. Since their weights are at most $n^{2/5}$ each, squaring the edge weights
    blows up the energy on them by at most a factor of $n^{2/5}$.
    On the other hand, since these are intermediate edges,
    the probability that any such edge belongs to $G[C]$ is
    $\kh{ \frac{\omega}{100 n^{1/5}} }^2 = \frac{\omega^2}{10000n^{2/5}}$.
    Therefore, the expected total contribution of type (B) edges to $y^T L_{G^{\sq}[C]} y$ is at most
    $\frac{\omega^2}{10000 n^{2/5}}\cdot n^{2/5} x^T L_G x \leq \frac{\omega}{10000} \leq \frac{1}{10000}$.
    Thus this contribution does not exceed $1$ with probability $1 - 1/10000$
    (call this event $\Ecal_3$).

    By a union bound, $\Ecal_1,\Ecal_2,\Ecal_3$ simultaneously happen with probability at least $.8$,
    in which case we have $y^T L_{G^{\sq}[C]} y \leq 2$.
  \end{proof}

  By Claim~\ref{claim:yx} and a union bound over the events in the claim,
  we have with probability at least $.1$ that
  the normalized energy of $y$ with respect to $s,t$ in $G^{\sq}[C]$
  is $\frac{y^T L_{G^{\sq}[C]} y}{(y_s - y_t)^2} \leq 800$.
\end{proof}
\subsection{Recovery of heavy edges}\label{sec:heavyedge}

Armed with the vertex sampling lemma,
we are now ready to design a linear sketch to recover all heavy edges in $G$.
In doing so, we will also need to invoke two other linear sketches
for sparsifying and recovering heavy edges in $G^{\sq}$, respectively.
We summarize their performance in the two lemmas below,
and prove them later in Section~\ref{sec:sparsifyG^2}.
We note that the sketch designed in Lemma~\ref{lem:shv} is basically a direct application of $\ell_2$-heavy hitters.
The sketch designed in Lemma~\ref{lem:sparsifyG^2} is essentially a reduction from sparsification to heavy edge recovery,
which will be very similar to our main linear sketching algorithm for weighted spectral sparsification
in the next subsection (Section~\ref{sec:algmainspectral}).

\begin{restatable}{lemma}{sqsparsify}
  For any parameter $\eps_2 > 0$ and any integer $n$,
  there exists a linear sketch with sketching matrix
  $\SSQ\in \mathbb{R}^{n \eps_2^{-4} \polylog(n,\eps_2^{-1},\frac{\wmax}{\wmin})\times \binom{n}{2}}$
  and recovery algorithm $\SQR$
  such that,  given an input graph $G$ of $n$ vertices with weight vector $w_G$,
  $\SQR(\SSQ w_G)$ returns
  a $(1+\eps_2)$-spectral sparsifier of $G^{\sq}$ with high probability.
  \label{lem:sparsifyG^2}
\end{restatable}

\begin{restatable}{lemma}{lemshv}
  For any parameters $\omega_3,\eps_3\in(0,1)$ and any integer $n$,
  there exists a linear sketch with sketching matrix
  $\SSV\in\mathbb{R}^{n \omega_3^{-1} \eps_3^{-2} \polylog(n)\times \binom{n}{2}}$ and recovery algorithm
  $\SVR$ such that, for an input graph $G$ of $n$ vertices with weight vector $w_G$ and
  another graph $\Gtil$,
  $\SVR(\SSV w_G,\Gtil)$ recovers a set of edges $F$ in $G$ along with estimates of their weights $\wtil_f$'s
  such that with high probability
  \begin{enumerate}
    \item $F$ contains all edges $e$ satisfying
      \begin{align*}
        \frac{((w_G)_e b_e^T L_{\Gtil}^{\dag} b_e)^2}{ b_e^T L_{\Gtil}^{\dag} L_{G^{\sq}} L_{\Gtil}^{\dag} b_e }
        \geq \omega_3.
      \end{align*}
    \item All edges $f\in F$ satisfy $\frac{1}{1 + \eps_3} (w_G)_f \leq \wtil_f \leq (1 + \eps_3) (w_G)_f$.
  \end{enumerate}
  \label{lem:shv}
\end{restatable}
We remark that to understand the first guarantee of the above lemma, one should think of $\Gtil$
as a good spectral sparsifier of $G^{\sq}$, so that the numerator $((w_G)_e b_e^T L_{\Gtil}^{\dag} b_e)^2 \approx
((w_G)_e b_e^T L_{G^{\sq}}^{\dag} b_e)^2$
and the denominator
$b_e^T L_{\Gtil}^{\dag} L_{G^{\sq}} L_{\Gtil}^{\dag} b_e \approx
b_e^T L_{G^{\sq}}^{\dag} b_e$, and thus the LHS
$\approx (w_G)_e^2 b_e^T L_{G^{\sq}}^{\dag} b_e$, the leverage score of $e$ in $G^{\sq}$.
That is, this guarantee is essentially saying that all heavy edges in $G^{\sq}$ will be recovered.

We now describe the linear sketch for recovering heavy edges in $G$ in Figure~\ref{fig:hvs},
and characterize its performance in the lemma below.

\begin{figure}[!htbp]
\begin{algbox}
  $\HVS(G,\omega_1,\eps_1)$

  \quad

  \begin{enumerate}
    \item Let $t = \ceil{10000 \omega_1^{-2} n^{2/5}\log n}$ and
      let $V_1,\ldots,V_t$ be vertex subsets, each obtained by
      including each vertex in $G$ with probability $\frac{\omega_1}{100 n^{1/5}}$ independently.
      
      \hfill \textcolor{blue}{(Subsample the vertices sufficiently many times to cover every edge)}
    \item For each $i\in [t]$, let
      $\SSQ_i \in \mathbb{R}^{|V_i| \eps_2^{-4} \polylog(|V_i|,\eps_2^{-1}, \frac{\wmax}{\wmin})\times \binom{|V_i|}{2}}$
      be a sketching matrix with error $\eps_2 = .01$ (Lemma~\ref{lem:sparsifyG^2}),
      and let $\SSV_i\in\mathbb{R}^{|V_i|\omega_3^{-1}\eps_3^{-2}\polylog(|V_i|)\times \binom{|V_i|}{2}}$
      with $\omega_3 = 1/2000$ and $\eps_3 = \eps_1/100$ (Lemma~\ref{lem:shv}).
    \item Concatenate the following sketches as $\SHV w_G$:
      \begin{enumerate}
        \item $\SSQ_1 w_{G[V_1]}, \ldots, \SSQ_t w_{G[V_t]}$, where
          $w_{G[V_i]}\in\mathbb{R}^{\binom{|V_i|}{2}}$ is the weight vector of the vertex induced subgraph $G[V_i]$.

          \hfill \textcolor{blue}{(Create a sparsification sketch for each vertex-induced subgraph)}
        \item $\SSV_1 w_{G[V_1]}, \ldots, \SSV_t w_{G[V_t]}$.

      \hfill \textcolor{blue}{(Create a heavy edge sketch for each vertex-induced subgraph)}
      \end{enumerate}
  \end{enumerate}

  \quad

  $\HVR(\SHV w_G)$ %

  \begin{enumerate}
    \item For each $i\in[t]$:
      \begin{enumerate}
        \item Use Lemma~\ref{lem:sparsifyG^2} to
          recover from $\SSQ_i w_{G[V_i]}$ a sparsifier $\Gtil_i$ of $G^{\sq}[V_i]$.

          \hfill \textcolor{blue}{($\Gtil_i$ is a $1.01$-spectral sparsifier of $G^{\sq}[V_i]$)}
        \item Use the recovery algorithm from Lemma~\ref{lem:shv} to recover from
          $\SSV_i w_G[V_i]$ and $\Gtil_i$ a set $F$ of edges along with their estimated weights $\wtil_f$'s,
          and mark all edges in $F$ as heavy.
      \end{enumerate}

      \hfill \textcolor{blue}{(Feed the sparsifier $\Gtil_i$ to the recovery algorithm
      to get all heavy edges)}
    \item Return all edges marked heavy along with the estimates
      of their weights (if for some edge there are multiple estimates of its weight, pick
      an arbitrary one).
  \end{enumerate}
\end{algbox}
\caption{Linear sketch for recovering heavy edges in $G$.}
\label{fig:hvs}
\end{figure}

\begin{restatable}%
  {lemma}{heavyedge}
  For any parameters $\omega_1 \in (n^{-4/5}\log n,1)$, $\eps_1\in(0,1)$ and integer $n$,
  there exists a linear sketch with sketching matrix
  $\SHV\in\mathbb{R}^{n^{6/5}\omega_1^{-1} \eps_1^{-2}
    \polylog (n, \frac{\wmax}{\wmin})
  \times \binom{n}{2}}$
  and recovery algorithm $\HVR$
  such that,
  for an input graph $G$ of $n$ vertices with weight vector $w_G$,
  $\HVR(\SHV w_G)$ recovers a set $F$ of edges in $G$ along with estimates of their weights
  $\wtil_f$'s
  such that with high probability
  \begin{enumerate}
    \item All edges $e$ whose leverage score in $G$ satisfy $(w_G)_e b_e^T L_{G}^{\dag} b_e \geq \omega_1$
      belong to $F$. \label{item:g1}
    \item All edges $f\in F$ satisfy $\frac{1}{1 + \eps_1} (w_G)_f \leq \wtil_f \leq (1 + \eps_1) (w_G)_f$.
      \label{item:g2}
  \end{enumerate}
  \label{lem:heavyedge}
\end{restatable}

We now prove Lemma~\ref{lem:heavyedge} 
using Lemmas~\ref{lem:vertex-sample},\ref{lem:sparsifyG^2},\ref{lem:shv}.

\begin{proof}[Proof of Lemma~\ref{lem:heavyedge}
  using Lemmas~\ref{lem:vertex-sample},\ref{lem:sparsifyG^2},\ref{lem:shv}]
  \textbf{Number of linear measurements.}{
    We observe that with high probability, $|V_i|\leq 100 \omega_1 n^{4/5}$ for all $i\in[t]$, and
    thus each $\SSQ_i w_G[V_i] \in \mathbb{R}^{\omega_1 n^{4/5}\polylog(n,\frac{\wmax}{\wmin})}$
    and each $\SSV_i w_G[V_i] \in \mathbb{R}^{\omega_1 n^{4/5} \eps_1^{-2}\polylog(n)}$.
    Therefore, the total number of linear measurements is bounded by
    \begin{align*}
        t\cdot \omega_1 n^{4/5} \eps_1^{-2}\polylog(n,\frac{\wmax}{\wmin}) \leq
        n^{6/5} \omega_1^{-1} \eps_1^{-2} \polylog(n,\frac{\wmax}{\wmin}).
    \end{align*}
  }

  \textbf{Guarantee~\ref{item:g1}.}{
      Consider fixing any edge $e$ with $w_e b_e^T L_G^{\dag} b_e \geq \omega_1$.
      By Lemma~\ref{lem:vertex-sample},
      for each $i\in [t]$, with probability $\frac{\omega_1^2}{1000 n^{2/5}}$,
      we have $e\in G[V_i]$ and $w_e^2 b_e^T L_{G^{\sq}[V_i]}^{\dag} b_e\geq 1/1000$ (call this event $\Ecal_i$).
      Therefore at least one of $\Ecal_1,\ldots,\Ecal_t$ happens with probability
      $\geq 1 - (1 - \frac{\omega_1^2}{1000 n^{2/5}})^{t} \geq 1 - 1/n^{-10}$.
      Whenever an $\Ecal_i$ happens, using the fact that $\Gtil_i$ is a $1.01$-spectral sparsifier
      of $G^{\sq}[V_i]$ (by Lemma~\ref{lem:sparsifyG^2}),
      we have
      \begin{align*}
        (w_G)_e b_e^T L^{\dag}_{\Gtil_i} b_e \geq \frac{1}{1.01}
        (w_G)_e b_e^T L^{\dag}_{G^{\sq}[V_i]} b_e
      \end{align*}
      and
      \begin{align*}
        b_e^T L_{\Gtil_i}^{\dag} L_{G^{\sq}[V_i]} L^{\dag}_{\Gtil_i} b_e \leq 1.01
        b_e^T L_{\Gtil_i}^{\dag} L_{\Gtil_i} L^{\dag}_{\Gtil_i} b_e =
        b_e^T L^{\dag}_{\Gtil_i} b_e \leq 1.01^2
        b_e^T L^{\dag}_{G^{\sq}[V_i]} b_e,
      \end{align*}
      and thus
      \begin{align*}
        \frac{((w_G)_e b_e^T L^{\dag}_{\Gtil_i} b_e)^2}{b_e^T L_{\Gtil_i}^{\dag} L_{G^{\sq}[V_i]} L^{\dag}_{\Gtil_i} b_e}
        \geq
        1.01^{-3} (w_G)_e^2 b_e^T L_{G^{\sq}[V_i]}^{\dag} b_e \geq 1.01^{-3}/1000 \geq 1/2000.
      \end{align*}
      By Lemmas~\ref{lem:shv},
      with high probability, $e$ is among the recovered edges. Therefore by a union bound
      over all such edges $e$, we have the desired result.
  }

  \textbf{Guarantee~\ref{item:g2}.}{
      This follows directly from Lemma~\ref{lem:shv}.
  }
\end{proof}

\subsection{Main algorithm for weighted spectral sparsification}\label{sec:algmainspectral}

We now show how to use the heavy edge recovery sketch in the previous section to obtain a
spectral sparsifier of $G$.

We first briefly summarize the main ideas.
The first idea is to use the iterative refinement process in~\cite{LiMP13}
as in the previous works on unweighted graphs~\cite{KapralovLMMS14,KapralovMMMNST20}.
That is, we consider, for some large $\alpha = \poly(\wmax,n),\ t = \polylog(\eps^{-1},\frac{\wmax}{\wmin},n)$
and a constant $\beta\in(0,1)$,
the sequence of graphs
\begin{align*}
  G + \alpha K_n, G + \beta \alpha K_n, G + \beta^2 \alpha K_n, \ldots,
  G + \beta^t \alpha K_n,
\end{align*}
which have the properties that
\begin{enumerate}
  \item $\alpha K_n$ is an $O(1)$-spectral sparsifier of $G + \alpha K_n$.
  \item $G + \beta^k \alpha K_n$ is an $O(1)$-spectral sparsifier of $G + \beta^{k+1} \alpha K_n$ for all $k\geq 0$.
  \item $G + \beta^t \alpha K_n$ is a $(1 + \eps)$-spectral sparsifier of $G$.
\end{enumerate}
The idea is then to iteratively obtain a sparsifier of each of these graphs,
where we use the sparsifier of $G + \beta^k \alpha K_n$ to guide the sparsification
of $G + \beta^{k+1} \alpha K_n$ (in particular, we use the sparsifier of the former 
to estimate the effective resistances and leverage scores in the latter).
Thus it boils down to how to sparsify $G^{\supi{k+1}} := G + \beta^{k+1} \alpha K_n$ using heavy edge recovery.

\begin{remark}
  We remark that, when given access to a heavy edge recovery sketch,
  the sparsification of $G^{\supi{k+1}}$
  is relatively easy to achieve in the unweighted case, for the following reason.
  Consider, in an unweighted graph,
  an edge $e = (s,t)$ with effective resistance (thus also leverage score)
  $r_e$, and let $x_{s,t}\in\mathbb{R}^{n}$ be the set of vertex potentials induced by an electrical flow
  from $s$ to $t$ and assume w.l.o.g. $x_s - x_t = 1$. By Fact~\ref{fact:energymin},
  we have $x^T L_G x = 1/r_e$.
  Now notice that we can also assume $x_u \in [0,1]$ for all $u$,
  since letting $x_u\gets 1$ for all $x_u > 1$ and $x_v \gets 0$ for all $x_v < 0$
  can only decrease the total energy.

  This means that the energy $(x_s - x_t)^2 = 1$ on edge $e=(s,t)$ is the largest among all edges,
  and thus if we sample all edges uniformly at rate $\approx r_e$, the total energy will be
  $\Otil(1)$ with high probability by Chernoff bounds. This implies that in the latter subsampled graph,
  $e$ (if sampled) becomes heavy with high probability.
  Since $r_e$ is exactly (up to an $O(\log n)$ factor) the probability with which we want to sample $e$,
  we can apply the heavy edge sketch to subgraphs of $G$ obtained by sampling edges
  at geometrically decreasing rates, and then try to recover each edge $e$
  from the subgraph with sampling rate $\approx r_e$.

  However, for weighted graphs, the energy on some other edges of very large weights can be unboundedly big.
  Thus concentration bounds no longer give us high success probability of recovering an edge $e$
  when sampling all edges uniformly at rate $\approx \tau_e = w_e r_e$, even though
  the energy reduces significantly in expectation.
\end{remark}

To sparsify $G^{\supi{k+1}}$,
we will utilize the spectral sparsification framework in~\cite{Koutis14},
which is itself model oblivious.
The framework works as follows:
\begin{enumerate}
  \item Fix some constant $p\in (0,1)$.
  \item While the number of edges in the graph is $> n \eps^{-2} \polylog(n)$:
    \begin{enumerate}
      \item Find all edges whose leverage score $\geq \omega := \eps^2/\polylog(n)$,
        and call these edges $F$.
      \item Sample each edge not in $F$ with probability
        $p$, and multiply its weight by $1/p$ if sampled.
    \end{enumerate}
\end{enumerate}
Notice that since the leverage scores of all edges sum up to at most $n-1$,
the total number of edges in $F$ is at most $n\eps^{-2} \polylog(n)$.
Thus, in each while loop iteration, the number of edges decreases by a constant factor,
and as a result there can be at most $O(\log n)$ iterations.
Then using Theorem~\ref{clm:lev-sample}, we have that the final graph
is a $(1 + \eps)$-spectral sparsifier of $G^{\supi{k+1}}$.
Notice that here, the first step in the while loop is exactly the recovery of heavy edges.

We now describe the difficulty that arises in implementing the above process
using linear sketching {\em non-adaptively}, and our way around it.
First, let $E_0 \supseteq E_1\supseteq\ldots\supseteq E_{O(\log n)}$ be such that
$E_0 = \binom{V}{2}$ and $E_{i+1}$ is obtained by subsampling each edge slot in $E_i$
with probability $p$ (the constant fixed at the first step of the above process).
We apply the heavy edge recovery sketch to each $G^{\supi{k+1}}[E_i]$.
We then implement each iteration of the while loop in the above process.%

At first, we recover all heavy edges in $G^{\supi{k+1}}$ using the sketch of $G^{\supi{k+1}}[E_0] = G^{\supi{k+1}}$,
and call these edges $F_0$.
We would like to sample each edge in $G^{\supi{k+1}}[\binom{V}{2}\setminus F_0]$ with probability $p$,
and multiply its weight by $1/p$ if sampled.
Then in the next iteration, we would want to recover heavy edges in the latter subsampled graph.
That is, we would like to have a sketch
of the graph $(1/p) G^{\supi{k+1}}[E_1\setminus F_0] + G^{\supi{k+1}}[F_0]$. However, we only have a sketch
of $G^{\supi{k+1}}[E_1]$. By linearity, we can multiply it by $1/p$
and add to it $G^{\supi{k+1}}[F_0\setminus E_1]$ and get a sketch
of $(1/p)G^{\supi{k+1}}[E_1] + G^{\supi{k+1}}[F_0\setminus E_1]$.
Nonetheless, this sketch is still not taken on our desired graph
$(1/p) G^{\supi{k+1}}[E_1\setminus F_0] + G^{\supi{k+1}}[F_0]$,
since the weights of the edges in $E_1\intersect F_0$ in the former graph
are larger than in the latter by a factor of $1/p$. We say these edges are {\em overweighted} by $1/p$,
and call the former graph {\em overweighted} graph.

One might hope to further subtract from the sketch
$(1/p - 1)G^{\supi{k+1}}[E_1\intersect F_0]$ to bring down the weights of the overweighted edges by a factor of $(1/p)$.
However, notice that we do {\em not} have the exact weights of the edges in $F_0$
from our heavy edge recovery sketch.
Rather, we only have some estimates of their weights.
Moreover,
while in the second iteration we are only looking to subtract
edges that are overweighted by $1/p$,
in subsequent iterations, we might need to subtract edges that
are overweighted by $\poly(n)$,
which means that our weight estimates for such edges must have inverse polynomial accuracy for the subtraction to work.

Our way around this issue is to repeatedly {\em re-estimate} the weights of the overweighted edges.
Specifically, we show that the edges that are overweighted the most must be heavy in the overweighted graph.
Thus we can apply the heavy edge recovery sketch to the overweighted graph, get estimates
of the weights of these edges, and bring their weights down by a factor of $(1/p)$.
We then repeatedly apply this step $O(\log n)$ times (where we re-estimate the weights each time)
until there are no overweighted edges. Since we only bring down the edge weights by a constant factor each time
and always re-estimate the weights once changed,
we will never have too large an error.

We now present in Figure~\ref{fig:wss} our main algorithm for weighted spectral sparsification,
which invokes the heavy edge recovery sketch in Section~\ref{sec:heavyedge}.
Specifically, for each graph $G^{(k)}$ in the iterative refinement process,
we apply independent heavy edge recovery sketches to subgraphs of $G^{(k)}$
obtained by sampling edges at geometrically decreasing rates.
Then in the recovery step, we first simulate the iterative refinement process
using an outer for loop of $k$, and then implement the framework from~\cite{Koutis14}
in an inner for loop of $i$.
Inside each iteration of the inner for loop, we start with the sketch of an overweighted graph $Z$,
and then gradually bring down the weights of the overweighted edges by repeatedly recovering heavy
edges in the current (overweighted) graph and subtracting a constant fraction of their weights.
Finally when there are no overweighted edges left, we recover the heavy edges in the resulting graph
and then go to the next sampling rate.

\begin{figure}[!htbp]
  \vspace{-30pt}
\begin{algbox}
  $\WSS(G,\eps)$

  \quad

  \begin{enumerate}
    \item Let $t = \ceil{10000 \log (10^6 \eps^{-1} \frac{\wmax}{\wmin}\cdot n^{10})}$,
      and then for each $k = 0,1,2,\ldots,t$:
      \begin{enumerate}
        \item %
          Let $E_0^{\supi{k}}\supseteq E_1^{\supi{k}}\supseteq\ldots\supseteq E_t^{\supi{k}}$
          be edge subsets %
          where $E_0^{\supi{k}} = \binom{V}{2}$ and $E_i^{\supi{k}}$
          is obtained by sub-sampling each edge slot in $E_{i-1}^{\supi{k}}$
          with probability $(1+1/1000)^{-1}$.
        \item For each pair $0\leq i,j\leq t$, use Lemma~\ref{lem:heavyedge} to generate a sketching matrix
          $(\SHV)_{i,j}^{\supi{k}} \in \mathbb{R}^{O(n^{6/5}\omega_1^{-1} \eps_1^{-2}
          \polylog(n,\frac{\wmax}{\wmin}))\times \binom{n}{2}}$
          with $\omega_1 = \eps^2/(10^{12} t^2 \log n)$,
          $\eps_1 = \eps / (10^6 t)$. %
      \end{enumerate}
    \item For each $0\leq k,i,j \leq t$,
      let $G^{\supi{k}} \gets G + (1+1/10^4)^{-k} 10^6 \wmax n^5 K_n$,
      and compute the sketch
      $(\SHV)_{i,j}^{\supi{k}} w_{G^{\supi{k}}[E^{\supi{k}}_{i}]}$, where $w_{G^{\supi{k}}[E^{\supi{k}}_{i}]}\in\mathbb{R}^{\binom{n}{2}}$ is the weight vector
      of $G^{\supi{k}}[E^{\supi{k}}_i]$. %

      \hfill \textcolor{blue}{(Take sufficiently many independent heavy edge sketches on
      each subsampled graph)}
    \item Concatenate these sketches as $\SSS w_G$.
  \end{enumerate}

  \quad

  $H^{\supi{t}} = \WSR(\SSS w_G)$

  \begin{myEnumerate}
  \item Initially, let $H^{\supi{0}} \gets 10^6 \wmax n^5 K_n$.
    \hfill \textcolor{blue}{($H^{\supi{0}}$ is a $1.001$-spectral sparsifier of $G^{\supi{0}}$)}
    \item For $k = 1,2,\ldots,t$:
      \begin{myEnumerate}
      \item Let $H^{\supi{k}} \gets \emptyset$.
        \hfill \textcolor{blue}{($H^{\supi{k}}$ will be a $(1+\eps/1000)$-spectral sparsifier of $G^{\supi{k}}$)}
      \item Set $c_e\gets 0$ for all $e\in \binom{V}{2}$.
        \hfill \textcolor{blue}{($c_e$ will be s.t. $e$ is added to $H^{\supi{k}}$ when $i = c_e$)}
        \item For $i = 0,1,\ldots,t$:
          \begin{myEnumerate}
          \item Let $Z\gets (1 + 1/1000)^i G^{\supi{k}}[E_i^{\supi{k}}] + H^{\supi{k}}[\binom{V}{2}\setminus E^{\supi{k}}_i]$.

            \hfill \textcolor{blue}{($Z$ records the graph on which our linear sketches are currently taken)}
          \item Compute sketches $s_j := (\SHV)^{\supi{k}}_{i,j} w_Z, j\in[0,t]$,
            where $w_Z$ is $Z$'s weight vector.
          \item For each $f\in H^{\supi{k}}\intersect E_i^{\supi{k}}$, let
              $\delta_f \gets i - c_f$, and let $\delta_f \gets 0$ for all other edges.

              \hfill \textcolor{blue}{($(w_Z)_f$ needs to be brought down by a factor of $(1+1/1000)^{\delta_f}$)}
            \item Let $j\gets 0$. Then while $\exists f: \delta_f > 0$, do the following:
              \begin{itemize}
                \item Use Lemma~\ref{lem:heavyedge} to recover from $s_j$ a set $F$ of edges and then let $j\gets j+1$.
                \item For each $f\in F$ such that $\delta_f > 0$, let $\wtil_f$ be the estimate of its weight:
                  \begin{itemize}
                    \item $Z\gets Z - (1 - (1+1/1000)^{-1}) \wtil_f f$.
                    \item $s_{j'}\gets s_{j'} - (1 - (1+1/1000)^{-1}) (\SHV)^{\supi{k}}_{i,j'} (\wtil_f \chi_f)$
                      for all $j'\in [0,t]$.
                    \item $\delta_f \gets \delta_f - 1$.

                      \hfill \textcolor{blue}{(Bring down $(w_Z)_f$ by $(1+1/1000)$ and update all sketches accordingly)}
                  \end{itemize}
              \end{itemize}
            \item Use Lemma~\ref{lem:heavyedge} to recover from $s_j$ a set $F^*$ of edges.
            \item For each edge $f\in F^*$ with estimated weight $\wtil_f$ such that
              $\wtil_f b_f^T L_{H^{\supi{k-1}}}^{\dag} b_f \geq 8 \omega_1$
              and $f$ is not already in $H^{\supi{k}}$,
              add $f$ to $H^{\supi{k}}$ with weight $\wtil_f$,
              and let $c_f\gets i$.

              \hfill \textcolor{blue}{(Add recovered heavy edges to $H^{\supi{k}}$,
              then go to the next sampling rate)}
          \end{myEnumerate}
      \end{myEnumerate}
  \end{myEnumerate}
\end{algbox}
\caption{Linear sketch for weighted spectral sparsification.}
\label{fig:wss}
\end{figure}

The performance of our main linear sketching algorithm for weighted graph sparsification
is characterized in Theorem~\ref{thm:sparsifyG}.

\begin{restatable}{theorem}{sparsifyG}
  For any parameter $\eps > 0$ and any integer $n$,
  there exists a linear sketch with sketching matrix
  $\SSS\in \mathbb{R}^{n^{6/5} \eps^{-4} \polylog(n,\eps^{-1},\frac{\wmax}{\wmin})\times \binom{n}{2}}$
  and recovery algorithm $\WSR$
  such that,  given an input graph $G$ of $n$ vertices with weight vector $w_G$,
  $\WSR(\SSS w_G)$ returns
  a $(1+\eps)$-spectral sparsifier of $G$ with high probability.
  \label{thm:sparsifyG}
\end{restatable}

Before proving the theorem, we shall first give some useful intermediate lemmas.
The following proposition directly follows from the definition of spectral sparsifiers.

\begin{proposition}
  For any two graphs $G_1,G_2$ whose weight vectors satisfy
  that $(w_{G_1})_e \approx_{1+\eps} (w_{G_2})_e$
  for all $e\in \binom{V}{2}$,
  $G_1$ is a $(1 + \eps)$-spectral sparsifier of $G_2$.
\end{proposition}

This proposition then immediately implies the following two lemmas.

\begin{lemma}\label{label:basecaseG}
  $10^6 \wmax n^5 K_n$ is a $1.001$-spectral sparsifier of $G^{\supi{0}}$.
\end{lemma}

\begin{lemma}
  For all $k\geq 1$, $G^{\supi{k-1}}$ is a $1.001$-spectral sparsifier of $G^{\supi{k}}$.
\end{lemma}

\begin{lemma}
  $G^{\supi{t}}$ is a $(1 + \eps/2)$-spectral sparsifier of $G$.
\end{lemma}
\begin{proof}
  By definition
  \begin{align*}
    L_{G^{\supi{t}}} = & L_G + L_{(1+10^{-4})^{-t} 10^6 \wmax n^5 K_n} \\
    \pleq & L_G + .1 \eps {\wmin} n^{-5} L_{K_n},
  \end{align*}
  where the last line follows from our choice of $t$.
  Thus, the largest eigenvalue of the second term is bounded by $.1\eps \wmin n^{-4}$.
  By standard lower bounds on the second smallest eigenvalue, %
  the second smallest eigenvalue of $L_{G}$ is at least $\wmin / n^2$.
  Therefore we have
  \begin{align*}
    - .1\eps L_{G} \pleq L_{G^{\supi{t}}} - L_{G} \pleq .1 \eps L_{G},
  \end{align*}
  which implies that $G^{\supi{t}}$ is a $(1 + \eps/2)$-spectral sparsifier of $G$.
\end{proof}

Fix an iteration of the outer for loop of $k$.
Then for an iteration of the inner for loop of $i$,
let $H_i^{\supi{k}}$ be the $H^{\supi{k}}$ at the {\em beginning} of the iteration,
and let $F_i^{\supi{k}}$ be the edges in $H_i^{\supi{k}}$.
Define graph $$J_i^{\supi{k}} := (1 + 1/1000)^{{i}} G^{\supi{k}}[E_i^{\supi{k}} \setminus F_i^{\supi{k}}] +
\sum_{\ell=0}^{{i-1}} (1+1/1000)^{\ell} G^{\supi{k}}[F_{\ell+1}^{\supi{k}}\setminus F_{\ell}^{\supi{k}}].$$
Notice that $F^{\supi{k}}_0 = \emptyset$ and $J_0^{\supi{k}} = G^{\supi{k}}$.
Also by the way we are assigning values to $c_f$ in the algorithm,
we have, at the beginning of the for loop (of $i$) iteration, $f \in F^{\supi{k}}_{c_f + 1}\setminus F^{\supi{k}}_{c_f}$
for all $f\in F^{\supi{k}}_i$.
Thus we also have
\begin{align}\label{eq:Jdef2G}
  J_i^{\supi{k}} := (1 + 1/1000)^i G^{\supi{k}}[E_i^{\supi{k}} \setminus F_i^{\supi{k}}] +
  \sum_{f\in F_i^{\supi{k}}} (1 + 1/1000)^{c_f} (w_{G^{\supi{k}}})_f f.
\end{align}
\begin{lemma}\label{lem:inductionkG}
  Suppose $H^{\supi{k-1}}$ is a $1.001$-spectral sparsifier of $G^{\supi{k-1}}$.
  Then with high probability, for all $0\leq i < t$,
  \begin{enumerate}
    \item After the while loop inside the $i^{\mathrm{th}}$ iteration terminates,
      for all $f$,
      $$\frac{1}{1+\eps/10000} (w_{J_{i}^{\supi{k}}})_f \leq (w_{Z})_f \leq (1 + \eps/10000) (w_{J_{i}^{\supi{k}}})_f.$$
      \label{item:whileG}
    \item For all $f\in F^{\supi{k}}_{i+1}\setminus F^{\supi{k}}_i$, %
      $$\frac{1}{1+\eps/(10^6 t)} (w_{J_{i+1}^{\supi{k}}})_f \leq
      (w_{H^{\supi{k}}_{i+1}})_f
      \leq (1 + \eps / (10^6 t)) (w_{J_{i+1}^{\supi{k}}})_f.$$
      \label{item:wtili+1G}
    \item All edges in $F^{\supi{k}}_{i+1}$ have leverage scores in $J^{\supi{k}}_{i}$ at least $4\omega_1$.
      \label{item:o4G}
    \item $J^{\supi{k}}_{i+1}$ is a $(1 + \eps/(10^4 t))$-spectral sparsifier of $J^{\supi{k}}_i$.
      \label{item:jkG}
  \end{enumerate}
\end{lemma}
\begin{proof}
  We prove all statements of this lemma by induction on $i$.
  For $i = 0$, since $\delta_f = 0$ for all $f$,
  the while loop will not execute.
  Thus throughout this iteration we have $Z = J^{\supi{k}}_0 = G^{\supi{k}}$. %
  This immediately gives~\ref{item:whileG}.
  By Lemma~\ref{lem:heavyedge}, the $F^*$ we recover in this iteration
  contains all edges whose leverage score in $G^{\supi{k}}$ is at least $\omega_1$,
  and all edges in $F^*$ have weight estimates satisfying~\ref{item:wtili+1G}.
  Since $H^{\supi{k-1}}$ is a $1.001$-spectral sparsifier of $G^{\supi{k-1}}$,
  and $G^{\supi{k-1}}$ is in turn a $1.001$-spectral sparsifier of of $G^{\supi{k}}$,
  we have that $H^{\supi{k-1}}$ is a $1.003$-spectral sparsifier of $G^{\supi{k}}$.
  As a result, we know that,
  at the last step of the for loop iteration,
  all edges with leverage score at least $\geq 10\omega_1$ in $G^{\supi{k}}$ will be added to $H^{\supi{k}}$,
  and all edges added to $H^{\supi{k}}$ have leverage score at least $\geq 4\omega_1$ in $G^{\supi{k}}$,
  so we have~\ref{item:o4G}.
  This means that
  $J^{\supi{k}}_{1}$ is obtained by sampling a set of edges in $J^{\supi{k}}_0$ whose leverage scores in $J^{\supi{k}}_0$ are
  at most $10 \omega_1$ with probability $(1+1/1000)^{-1}$,
  and multiply their weights by $(1+1/1000)$
  if sampled. Using Theorem~\ref{clm:lev-sample}, we have~\ref{item:jkG}.

  We now do an inductive step. Suppose all four statements hold
  for iterations $0,1,\ldots,i-1$ where $1 < i < t$.
  We show that they also hold for iteration $i$.
  We first need to analyze the while loop inside iteration $i$.
  Let us number a while loop iteration by the value of $j$ at the {\em end} of the iteration.
  \begin{claim}\label{claim:whileapxG}
    At the end of while loop iteration $j$ where $j \leq t$, we have
    for all $f\in E_i^{\supi{k}}\intersect F_i^{\supi{k}}$
    $$\frac{1}{(1+2\eps_1)^j} \cdot (1+1/1000)^{\delta_f} (w_{J_{i}^{\supi{k}}})_f \leq (w_{Z})_f \leq
    (1 + 2\eps_1)^{j} (1+1/1000)^{\delta_f} (w_{J_{i}^{\supi{k}}})_f.$$
        \label{item:weightZG}
  \end{claim}
  \begin{proof}
    We prove this claim by an induction on $j$.
    First we show that the statement is true for $j=0$ at the beginning of while loop iteration $1$.
    Here all $f\in E_i^{\supi{k}}\intersect F_i^{\supi{k}}$ satisfy
    that $(w_Z)_f = (1+1/1000)^{i} (w_{G^{\supi{k}}})_f$.
    Since we set $\delta_f \gets i - c_f$ before the while loop,
    and by~(\ref{eq:Jdef2G}) $(w_{J^{\supi{k}}_i})_f = (1 + 1/1000)^{c_f} (w_{G^{\supi{k}}})_f$,
    we have
    $(w_Z)_f = (1 + 1/1000)^{\delta_f} (w_{J^{\supi{k}}_i})_f$, as desired.

    Now suppose the statement is true
    at the end of iteration $j-1$ where $1 < j \leq t$.
    We then show that the statement is also true at the end of iteration $j$.
    Let $Z_0$ be the $Z$ before our updates to $Z$ in iteration $j$ and let $Z_1$ be the $Z$ after our updates.
    By Lemma~\ref{lem:heavyedge}, all edges recovered $f\in F$ have their estimated edge weights
    $\wtil_f \in [\frac{1}{1+\eps_1} (w_{Z_0})_f, (1 + \eps_1) (w_{Z_0})_f]$.
    Therefore after our updates, we have for any $f\in F$ such that $\delta_f > 0$ that
    $(w_{Z_1})_f \in [\frac{1}{1+2\eps_1} (1+1/1000)^{-1} (w_{Z_0})_f, (1 + 2\eps_1) (1+1/1000)^{-1} (w_{Z_0})_f]$,
    and $(w_{Z_1})_f = (w_{Z_0})_f$ for other edges $f$.
    Since we let $\delta_f\gets \delta_f - 1$ for such edges, and do not change the $\delta_f$'s of other edges,
    we have our desired statement for $j$.
 \end{proof}
  \begin{claim}
    The while loop terminates after at most $t$ iterations.
  \end{claim}
  \begin{proof}
    It suffices to show that $\max_f {\delta_f}$ decreases by $1$ in each while loop iteration.
    Since $\delta_f \leq t$ for any $f$, this will imply that there can be at most $t$ iterations.
    Then it boils down to showing that for all $f^*$ with $\delta_{f^*} = \max_f \delta_f$,
    $f^*$ belongs to the recovered edge set $F$.
    Since $f^*\in F^{\supi{k}}_i$, by~\ref{item:o4G} of our induction hypothesis,
    the leverage score of $f^*$ in $J_{i-1}^{\supi{k}}$ is at least $4\omega_3$.
    Notice that by~\ref{item:jkG} of our induction hypothesis,
    $J^{\supi{k}}_{i-1}$ is a $(1+1/1000)$-spectral sparsifier of $G^{\supi{k}}$.
    Then using the fact that $H^{\supi{k-1}}$ is a $1.003$-spectral sparsifier of $G^{\supi{k}}$
    (which we proved at the beginning of the proof of this lemma), we have that
    $H^{\supi{k-1}}$ is a $1.005$-spectral sparsifier of $J_{i-1}^{\supi{k}}$.

    By Claim~\ref{claim:whileapxG}, we have at the beginning of each while loop that, for all $f$,
    \begin{align}\label{eq:wzf1G}
      (w_Z)_f \in &
      [\frac{1}{(1+2\eps_1)^t}(1+1/1000)^{\delta_f} (w_{J^{\supi{k}}_i})_f,(1 + 2\eps_1)^t (1 + 1/1000)^{\delta_f} (w_{J^{\supi{k}}_i})_f] \notag\\
      \subseteq &
      [\frac{1}{1.01}(1+1/1000)^{\delta_f} (w_{J^{\supi{k}}_i})_f,1.01 (1 + 1/1000)^{\delta_f} (w_{J^{\supi{k}}_i})_f].
    \end{align}
    Since $\delta_{f^*} \geq \delta_f$ for all $f$,
    the above implies
    \begin{align}
      L_{Z} \pleq {1.03} (1 + 1/1000)^{\delta_{f^*}} L_{J_{i-1}^{\supi{k}}}.
      \label{eq:ZHG}
    \end{align}
    By inverting both sides, we then get
    $$L_{Z}^{\dag} \pgeq 1.03^{-1} (1 + 1/1000)^{-\delta_{f^*}} L_{J_{i-1}^{\supi{k}}}^{\dag}.$$
    Combining this with~(\ref{eq:wzf1G}),
    the leverage score of $f^*$ in $Z$ satisfies
    \begin{align*}
      (w_Z)_{f^*} b_{f^*} L_{Z}^{\dag} b_{f^*} \geq
      \frac{1}{1.01\cdot 1.03} (w_{J_i^{\supi{k}}})_{f^*} b_{f^*}^T L_{J^{\supi{k}}_{i-1}}^{\dag} b_{f^*} \geq
      \frac{4\omega_3}{1.01\cdot 1.03} \geq \omega_3,
    \end{align*}
    as desired.
  \end{proof}

  By Claim~\ref{claim:whileapxG}, after the while loop terminates,
  we have that for all $f$,
  \begin{align*}
    (w_Z)_f \in & [\frac{1}{(1+2\eps_1)^t} (w_{J^{\supi{k}}_i})_f, (1 + 2\eps_1)^t (w_{J^{\supi{k}}_i})_f] \\
    \subseteq & [\frac{1}{(1+\eps/10000)} (w_{J^{\supi{k}}_i})_f, (1 + \eps/10000) (w_{J^{\supi{k}}_i})_f],
  \end{align*}
  and thus we have~\ref{item:whileG}.
  This also implies that $Z$ is a $(1 + \eps/10000)$-spectral sparsifier of $J^{\supi{k}}_i$,
  and as result, for each edge $f$, its leverage scores in $Z$ and $J^{\supi{k}}_i$ are within
  a $(1+\eps/10000)^2 < 1.01$ factor of each other.

  For all edges in $E^{\supi{k}}_i\setminus F^{\supi{k}}_i$, their weights in $Z$ equal exactly
  their weights in $J^{\supi{k}}_i$,
  therefore by Lemma~\ref{lem:heavyedge},
  all edges recovered in $F^*$ not in $F_i^{\supi{k}}$ have weight estimates satisfying~\ref{item:wtili+1G}.

  Notice that by~\ref{item:jkG} of our induction hypothesis,
  $J^{\supi{k}}_i$ is a $(1+1/1000)$-spectral sparsifier of $G^{\supi{k}}$.
  Then using the fact that $H^{\supi{k-1}}$ is a $1.003$-spectral sparsifier of $G^{\supi{k}}$
  (which we proved at the beginning of this proof),
  we have that $H^{\supi{k-1}}$ is a $1.01$-spectral sparsifier of $Z$.
  Thus, at the last step of the for iteration,
  all edges added to $H^{\supi{k}}$ have leverage score at least $\geq 5\omega_1$ in $Z$,
  and all edges with leverage score $\geq 9\omega_1$ in $Z$ will be added to $H^{\supi{k}}$.
  Thus we also know that all edges added to $H^{\supi{k}}$ have leverage score $\geq 4\omega_1$ in $J^{\supi{k}}_i$
  (which gives~\ref{item:o4G}),
  and all edges with leverage score $\geq 10\omega_1$ in $J^{\supi{k}}_i$ will be added to $H^{\supi{k}}$.

  The above reasoning also implies that
  $J^{\supi{k}}_{i+1}$ is obtained by sampling a set of edges in $J^{\supi{k}}_i$ whose leverage score
  is at most $10 \omega_1$ with probability $(1+1/1000)^{-1}$,
  and multiply their weights by $(1+1/1000)$
  if sampled. Using Theorem~\ref{clm:lev-sample}, we have~\ref{item:jkG}.
\end{proof}

\begin{proof}[Proof of Theorem~\ref{thm:sparsifyG}]
  \textbf{Number of linear measurements.}{
    Notice that each $(\SHV)_{i,j}^{\supi{k}} w_{G^{\supi{k}}[E^{\supi{k}}_{i}]}\in\mathbb{R}^{n^{6/5}
    \omega_1^{-1}\eps_1^{-2}\polylog(n,\frac{\wmax}{\wmin})}$,
    so the total number of linear measurements is bounded by
    \begin{align*}
      t^3 n^{6/5} \omega_1^{-1}\eps_1^{-2}\polylog(n,\frac{\wmax}{\wmin}) \leq
      n^{6/5} \eps^{-4} \polylog(n,\frac{\wmax}{\wmin},\eps^{-1}).
    \end{align*}
  }

  \textbf{Spectral sparsifier guarantee.}
  By Lemma~\ref{label:basecaseG},
  $H^{\supi{0}}$ is a $1.001$-spectral sparsifier of $G^{\supi{0}}$.
  We then show that whenever $H^{\supi{k-1}}$ is a $1.001$-spectral sparsifier of $G^{\supi{k-1}}$,
  $H^{\supi{k}}$ is a $(1 + \eps/1000)$-spectral sparsifier of $G^{\supi{k}}$ with high probability.
  Notice that inside each iteration of the outermost for loop of $k$,
  for $i = t$, we have that with high probability
  $E^{\supi{k}}_t = \emptyset$. This means that $J^{\supi{k}}_t$ consists of solely edges
  in $F^{\supi{k}}_t$. Thus by Lemma~\ref{lem:inductionkG}, $H^{\supi{k}}_t$ is a
  $(1+\eps/(10^6t))$-spectral sparsifier of $J^{\supi{k}}_t$.
  Also by Lemma~\ref{lem:inductionkG}, $J^{\supi{k}}_t$ is a $(1+\eps/(10^4 t))^t$-spectral sparsifier 
  of $G^{\supi{k}}$. These combined imply that
  $H^{\supi{k}}$ is a $(1 + \eps/1000)$-spectral sparsifier of $G^{\supi{k}}$.
  Now applying an induction on $k$,
  we have that $H^{\supi{t}}$ is a $(1 + \eps/1000)$-spectral sparsifier of $G^{\supi{t}}$.
  Since $G^{\supi{t}}$ is a $(1 + \eps/2)$-spectral sparsifier of $G$,
  $H^{\supi{t}}$ is a $(1 + \eps)$-spectral sparsifier of $G$, as desired.
\end{proof}

\subsection{Sparsification of \texorpdfstring{$G^{\sq}$}{}}\label{sec:sparsifyG^2}

\subsubsection{Sparsification of \texorpdfstring{$G^{\sq}$}{} by heavy edge recovery}\label{sec:similar}

We first give in Figure~\ref{fig:sqs} the linear sketch for sparsifying $G^{\sq}$ using the recovery of heavy edges
in Lemma~\ref{lem:shv}.
We will then prove Lemma~\ref{lem:shv} later in Section~\ref{sec:shv}.
The ideas for the former linear sketch are the same as the ones we used in Section~\ref{sec:algmainspectral},
since both are about how to sparsify a graph by repeatedly recovering heavy edges.

\begin{figure}[!htbp]
\begin{algbox}
  $\SQS(G,\eps_2)$ %

  \quad

  \begin{enumerate}
    \item Let $t = \ceil{10000 \log (10^6 \eps_2^{-1} \frac{\wmax}{\wmin}\cdot n^{10})}$,
      and then for each $k = 0,1,2,\ldots,t$:
      \begin{enumerate}
        \item %
          Let $E_0^{\supi{k}}\supseteq E_1^{\supi{k}}\supseteq\ldots\supseteq E_t^{\supi{k}}$ be subsets of edge slots
          where $E_0^{\supi{k}} = \binom{V}{2}$ and $E_i^{\supi{k}}$ is obtained by sub-sampling each edge slot in $E_{i-1}^{\supi{k}}$ with probability
          $(1+1/1000)^{-2}$.
        \item For each pair $0\leq i,j\leq t$,
          let $(\SSV)^{\supi{k}}_{i,j}\in \mathbb{R}^{n \omega_3^{-1}\eps_3^{-2}\polylog(n)\times \binom{n}{2}}$ be a sketching matrix
          in Lemma~\ref{lem:shv} with $\omega_3 = \eps_2^2/(10^{12} t^2\log n)$, $\eps_3 = \eps_2/(10^6t)$.
      \end{enumerate}
    \item For each $k,i,j \in [0,t]$,
      let $G^{\supi{k}} \gets G + {(1+10^{-4})^{-k} 10^6 \wmax n^5} K_n$
      and compute the sketch
      $(\SSV)_{i,j}^{\supi{k}} w_{G^{\supi{k}}[E^{\supi{k}}_{i}]}$
      where $w_{G^{\supi{k}}[E^{\supi{k}}_{i}]}\in\mathbb{R}^{\binom{n}{2}}$ is 
      weight vector of the subgraph $G^{\supi{k}}[E^{\supi{k}}_i]$. %
    \item Concatenate these sketches as $\SSQ w_G$.
  \end{enumerate}

  \quad

  $\SQR(\SSQ w_G)$

  \begin{myEnumerate}
  \item Initially, let $H^{\supi{0}} \gets 10^6 \wmax n^5 K_n$ (where $(H^{\supi{0}})^{\sq}$ is a $1.001$-spectral sparsifier of $(G^{\supi{0}})^{\sq}$).
    \item For $k = 1,2,\ldots,t$:
      \begin{myEnumerate}
      \item Let $H^{\supi{k}} \gets \emptyset$ (where $(H^{\supi{k}})^{\sq}$ will be a $(1+\eps_2/1000)$-spectral sparsifier of $(G^{\supi{k}})^{\sq}$).
        \item Set $c_e\gets 0$ for all $e\in \binom{V}{2}$
          (where $c_e$ will be such that $e$ is added to $H^{\supi{k}}$ when $i = c_e$).
        \item For $i = 0,1,\ldots,t$:
          \begin{myEnumerate}
          \item Let $Z\gets (1 + 1/1000)^{i} G^{\supi{k}}[E_i^{\supi{k}}] + H^{\supi{k}}[\binom{V}{2}\setminus E_i^{\supi{k}}]$
            ($Z$ is to record
            what graph our linear sketches are taken over).
          \item Compute sketches $s_j:=(\SSV)^{\supi{k}}_{i,j} w_Z, j\in[0,t]$, where $w_Z$ is the weight vector of $Z$.
          \item For each $f\in H^{\supi{k}}\intersect E_i^{\supi{k}}$, let
              $\delta_f \gets i - c_f$, and let $\delta_f \gets 0$ for all other edges.
            \item Let $j\gets 0$. Then while $\exists f: \delta_f > 0$, do the following:
              \begin{myEnumerate}
              \item Use Lemma~\ref{lem:shv} to recover from $s_j$ and $(H^{\supi{k-1}})^{\sq}$ an edge set $F$, and $j\gets j+1$.
                \item For each $f\in F$ such that $\delta_f > 0$, let $\wtil_f$ be the estimate of its weight:
                  \begin{myEnumerate}
                    \item $Z\gets Z - (1 - (1 + 1/1000)^{-1}) \wtil_f f$.
                    \item $s_{j'} \gets s_{j'} - (1 - (1 + 1/1000)^{-1})(\SSV)^{\supi{k}}_{i,j'} (\wtil_f \chi_f)$
                      for all $j\leq j' \leq t$
                    \item $\delta_f \gets \delta_f - 1$.
                  \end{myEnumerate}
              \end{myEnumerate}
            \item Use Lemma~\ref{lem:shv} to recover from $s_t$ and $(H^{\supi{k-1}})^{\sq}$ a set $F^*$ of edges.
            \item For each edge $f\in F^*$ with estimated weight $\wtil_f$ such that
              $\wtil_f^2 b_f^T L_{(H^{\supi{k-1}})^{\sq}}^{\dag} b_f \geq 8 \omega_3$
              and $f$ is not already in $H^{\supi{k}}$,
              add $f$ to $H^{\supi{k}}$ with weight $\wtil_f$,
              and let $c_f\gets i$.
          \end{myEnumerate}
      \end{myEnumerate}
    \item Return $(H^{\supi{t}})^{\sq}$.
  \end{myEnumerate}
\end{algbox}
\caption{Linear sketch for sparsifying $G^{\sq}$.}
\label{fig:sqs}
\end{figure}

The performance of the linear sketch is characterized by Lemma~\ref{lem:sparsifyG^2}.

\sqsparsify*

The proof of Lemma~\ref{lem:sparsifyG^2} will also be largely similar to that
of Theorem~\ref{thm:sparsifyG} in Section~\ref{sec:algmainspectral}, However, we still include the proof here for completeness.

As in Section~\ref{sec:algmainspectral}, we first prove some useful intermediate lemmas.

\begin{lemma}\label{label:basecase}
  $(10^6 \wmax n^5 K_n)^{\sq}$ is a $1.001$-spectral sparsifier of $(G^{\supi{0}})^{\sq}$.
\end{lemma}

\begin{proof}
  Let $\Pi\in\mathbb{R}^{n\times n}$ be the projection matrix on the $n-1$-dimensional subspace
  orthogonal to the all-one vector. Then $\Pi$ has $n-1$ eigenvalues of $1$ and one eigenvalue of $0$
  (with eigenvector being the all-one vector).
  It is known that $L_{K_n} = n \Pi$, so $L_{(10^6\wmax n^5 K_n)^{\sq}} = 10^{12} \wmax^2 n^{11} \Pi$,
  and has all $n-1$ nonzero eigenvalues equal to $10^{12} \wmax^2 n^{11}$.
  On the other hand, notice that by expanding
  \begin{align*}
    L_{(G^{\supi{0}})^{\sq}} = & L_{(G + 10^6 \wmax n^5 K_n)^{\sq}} \\ =
    & L_{G^{\sq}} + 2\cdot 10^6 \wmax n^5 L_G + 10^{12} \wmax^2 n^{10} L_{K_n}.
  \end{align*}
  By standard bounds on the largest eigenvalue of the Laplacian matrix of a weighted graph,
  the largest eigenvalue of an $n$-vertex graph with maximum weight $\wmax$
  is at most $n\wmax$.
  Therefore,
  the largest eigenvalue of $L_{(G^{\supi{0}})^{\sq} - (10^6 \wmax n^5 K_n)^{\sq}}$ is at most $3\cdot 10^6 \wmax^2 n^6$.
  This combined with $L_{(G^{\supi{0}})^{\sq}} \geq L_{(10^6 \wmax n^5 K_n)^{\sq}}$
  shows that
  \begin{align*}
    - \frac{1}{10^5 n^5} L_{(G^{\supi{0}})^{\sq}} \pleq 
    L_{(G^{\supi{0}})^{\sq}} - L_{(10^6 \wmax n^5 K_n)^{\sq}} \pleq
    \frac{1}{10^5 n^5} L_{(G^{\supi{0}})^{\sq}}.
  \end{align*}
  This implies that
  \begin{align*}
    \frac{1}{1.001} L_{(G^{\supi{0}})^{\sq}} \pleq
    L_{(10^6 \wmax n^5 K_n)^{\sq}} \pleq 1.001 L_{(G^{\supi{0}})^{\sq}}
  \end{align*}
  as desired.
\end{proof}

\begin{lemma}
  For all $k\geq 1$, $(G^{\supi{k-1}})^{\sq}$ is a $1.001$-spectral sparsifier of $(G^{\supi{k}})^{\sq}$.
\end{lemma}
\begin{proof}
  By definition
  \begin{align*}
    L_{(G^{\supi{k-1}})^{\sq}} = & L_{(G + (1+10^{-4})^{-(k-1)} 10^6 \wmax n^5 K_n)^{\sq}} \\
    = & L_{G^{\sq}} + 2(1 + 10^{-4})^{-(k-1)} 10^6 \wmax n^5 L_G + (1+10^{-4})^{-2(k-1)} 10^{12} \wmax^2 n^{10} L_{K_n}
  \end{align*}
  and
  \begin{align*}
    L_{(G^{\supi{k}})^{\sq}} = & L_{(G + (1+10^{-4})^{-k} 10^6 \wmax n^5 K_n)^{\sq}} \\
    = & L_{G^{\sq}} + 2(1 + 10^{-4})^{-k} 10^6 \wmax n^5 L_G + (1+10^{-4})^{-2k} 10^{12} \wmax^2 n^{10} L_{K_n}.
  \end{align*}
  The second terms of the above two expressions are $(1+10^{-4})$-sparsifiers of each other,
  and the third terms of the above two expressions are $(1+10^{-4})^2$-sparsifiers of each other.
  Therefore, $L_{(G^{\supi{k-1}})^{\sq}}$ is a $1.001$-spectral sparsifier of $L_{(G^{\supi{k}})^{\sq}}$.
\end{proof}

\begin{lemma}
  $(G^{\supi{t}})^{\sq}$ is a $(1 + \eps_2/2)$-spectral sparsifier of $G^{\sq}$.
\end{lemma}
\begin{proof}
  By definition
  \begin{align*}
    L_{(G^{\supi{t}})^{\sq}} = & L_{(G + (1+10^{-4})^{-t} 10^6 \wmax n^5 K_n)^{\sq}} \\
    = & L_{G^{\sq}} + 2(1 + 10^{-4})^{-t} 10^6 \wmax n^5 L_G + (1+10^{-4})^{-2t} 10^{12} \wmax^2 n^{10} L_{K_n} \\
    \pleq & L_{G^{\sq}} + .1 \eps_2 \frac{\wmin^2}{\wmax} n^{-5} L_G +
    .001 \eps_2^2 \frac{\wmin^4}{\wmax^2} n^{-10} L_{K_n},
  \end{align*}
  where the last line follows from our choice of $t$.
  Thus, the largest eigenvalue of the sum of the last two terms is bounded by $.2\eps_2 \wmin^2 n^{-4}$.
  By standard lower bounds on the second smallest eigenvalue, %
  the second smallest eigenvalue of $L_{G^{\sq}}$ is at least $\wmin^2 / n^2$.
  Therefore we have
  \begin{align*}
    - .2\eps_2 L_{G^{\sq}} \pleq L_{(G^{\supi{t}})^{\sq}} - L_{G^{\sq}} \pleq .2 \eps_2 L_{G^{\sq}},
  \end{align*}
  which implies that $(G^{\supi{t}})^{\sq}$ is a $(1 + \eps_2/2)$-spectral sparsifier of $G^{\sq}$.
\end{proof}

Fix an iteration of the outer for loop of $k$.
Then for an iteration of the inner for loop of $i$,
let $H_i^{\supi{k}}$ be the $H^{\supi{k}}$ at the {\em beginning} of the iteration,
and let $F_i^{\supi{k}}$ be the edges in $H_i^{\supi{k}}$.
Define graph $$J_i^{\supi{k}} := (1 + 1/1000)^i G^{\supi{k}}[E_i^{\supi{k}} \setminus F_i^{\supi{k}}] +
\sum_{\ell=0}^{i-1} (1+1/1000)^{\ell} G^{\supi{k}}[F_{\ell+1}^{\supi{k}}\setminus F_{\ell}^{\supi{k}}].$$
Notice that $F^{\supi{k}}_0 = \emptyset$ and $J_0^{\supi{k}} = G^{\supi{k}}$.
Also by the way we are assigning values to $c_f$ in the algorithm,
we have, at the beginning of the for loop (of $i$) iteration, $f \in F^{\supi{k}}_{c_f + 1}\setminus F^{\supi{k}}_{c_f}$
for all $f\in F^{\supi{k}}_i$.
Thus we also have
\begin{align}\label{eq:Jdef2}
  J_i^{\supi{k}} := (1 + 1/1000)^i G^{\supi{k}}[E_i^{\supi{k}} \setminus F_i^{\supi{k}}] +
  \sum_{f\in F_i^{\supi{k}}} (1 + 1/1000)^{c_f} (w_{G^{\supi{k}}})_f f.
\end{align}
\begin{lemma}\label{lem:inductionk}
  Suppose $(H^{\supi{k-1}})^{\sq}$ is a $1.001$-spectral sparsifier of $(G^{\supi{k-1}})^{\sq}$.
  Then with high probability, for all $0\leq i < t$,
  \begin{enumerate}
    \item After the while loop inside the $i^{\mathrm{th}}$ iteration terminates,
      for all $f$,
      $$\frac{1}{1+\eps_2/10000} (w_{J_{i}^{\supi{k}}})_f \leq (w_{Z})_f \leq (1 + \eps_2/10000) (w_{J_{i}^{\supi{k}}})_f.$$
      \label{item:while}
    \item For all $f\in F^{\supi{k}}_{i+1}\setminus F^{\supi{k}}_i$, %
      $$\frac{1}{1+\eps_2/(10^6 t)} (w_{J_{i+1}^{\supi{k}}})_f \leq
      (w_{H^{\supi{k}}_{i+1}})_f
      \leq (1 + \eps_2 / (10^6 t)) (w_{J_{i+1}^{\supi{k}}})_f.$$
      \label{item:wtili+1}
    \item All edges in $F^{\supi{k}}_{i+1}$ have leverage scores in $(J^{\supi{k}}_{i})^{\sq}$ at least $4\omega_3$.
      \label{item:o4}
    \item $(J^{\supi{k}}_{i+1})^{\sq}$ is a $(1 + \eps_2/(10^4 t))$-spectral sparsifier of $(J^{\supi{k}}_i)^{\sq}$.
      \label{item:jk}
  \end{enumerate}
\end{lemma}
\begin{proof}
  We prove all statements of this lemma by induction on $i$.
  For $i = 0$, since $\delta_f = 0$ for all $f$,
  the while loop will not execute.
  Thus throughout this iteration we have $Z = J^{\supi{k}}_0 = G^{\supi{k}}$. %
  This immediately gives~\ref{item:while}.
  Since $(H^{\supi{k-1}})^{\sq}$ is a $1.001$-spectral sparsifier of $(G^{\supi{k-1}})^{\sq}$,
  and $(G^{\supi{k-1}})^{\sq}$ is in turn a $1.001$-spectral sparsifier of of $(G^{\supi{k}})^{\sq}$,
  we have that $(H^{\supi{k-1}})^{\sq}$ is a $1.003$-spectral sparsifier of $(G^{\supi{k}})^{\sq}$.
  Therefore by letting $\Gtil = (H^{\supi{k-1}})^{\sq}$, we have
  \begin{align}\label{eq:bygtil}
    & \frac{((w_{G^{\supi{k}}})_e b_e^T L_{\Gtil}^{\dag} b_e)^2}{ b_e^T L_{\Gtil}^{\dag} L_{(G^{\supi{k}})^{\sq}} L_{\Gtil}^{\dag} b_e }
    \approx_{1.003^4}
    \frac{((w_{G^{\supi{k}}})_e b_e^T L_{(G^{\supi{k}})^{\sq}}^{\dag} b_e)^2}{ b_e^T L_{(G^{\supi{k}})^{\sq}}^{\dag} b_e }
    = (w_{G^{\supi{k}}})_e^2 b_e^T L_{(G^{\supi{k}})^{\sq}}^{\dag} b_e,
  \end{align}
  where in the first step we have used
  \begin{align*}
    (w_{G^{\supi{k}}})_e b_e^T L_{\Gtil}^{\dag} b_e \approx_{1.003} (w_{G^{\supi{k}}})_e b_e^T L_{(G^{\supi{k}})^{\sq}}^{\dag} b_e
  \end{align*}
  and
  \begin{align*}
    b_e^T L_{\Gtil}^{\dag} L_{(G^{\supi{k}})^{\sq}} L_{\Gtil}^{\dag} b_e \approx_{1.003}
    b_e^T L_{\Gtil}^{\dag} L_{\Gtil} L_{\Gtil}^{\dag} b_e = 
    b_e^T L_{\Gtil}^{\dag} b_e \approx_{1.003}
    b_e^T L_{(G^{\supi{k}})^{\sq}}^{\dag} b_e.
  \end{align*}
  Therefore by Lemma~\ref{lem:shv}, the $F^*$ we recover in this iteration
  contains all edges whose leverage score in $(G^{\supi{k}})^{\sq}$ is at least $1.1\omega_3$,
  and all edges in $F^*$ have weight estimates satisfying~\ref{item:wtili+1}.
  Also, at the final step of this for loop iteration,
  since $(H^{\supi{k-1}})^{\sq}$ is a $1.003$-spectral sparsifier of $(G^{\supi{k}})^{\sq}$,
  all edges with leverage score $\geq 10 \omega_3$ in $(G^{\supi{k}})^{\sq}$ will be added to $H^{\supi{k}}$,
  and all edges added to $H^{\supi{k}}$ have leverage score at least $\geq 4\omega_3$ in $(G^{\supi{k}})^{\sq}$,
  so we have~\ref{item:o4}.
  This means that
  $(J^{\supi{k}}_{1})^{\sq}$ is obtained by sampling a set of edges in $(J^{\supi{k}}_0)^{\sq}$ whose leverage scores in $(J^{\supi{k}}_0)^{\sq}$ are
  at most $10 \omega_3$ with probability $(1+1/1000)^{-2}$,
  and multiply their weights by $(1+1/1000)^2$
  if sampled. Using Theorem~\ref{clm:lev-sample}, we have~\ref{item:jk}.

  We now do an inductive step. Suppose all four statements hold
  for iterations $0,1,\ldots,i-1$ where $1 < i < t$.
  We show that they also hold for iteration $i$.
  We first need to analyze the while loop inside iteration $i$.
  Let us number a while loop iteration by the value of $j$ at the {\em end} of the iteration.
  \begin{claim}\label{claim:whileapx}
    At the end of while loop iteration $j$ where $j \leq t$, we have
    for all $f\in E_i^{\supi{k}}\intersect F_i^{\supi{k}}$
    $$\frac{1}{(1+2\eps_3)^j} \cdot (1+1/1000)^{\delta_f} (w_{J_{i}^{\supi{k}}})_f \leq (w_{Z})_f \leq
    (1 + 2\eps_3)^{j} (1+1/1000)^{\delta_f} (w_{J_{i}^{\supi{k}}})_f.$$
        \label{item:weightZ}
  \end{claim}
  \begin{proof}
    We prove this claim by an induction on $j$.
    First we show that the statement is true for $j=0$ at the beginning of while loop iteration $1$.
    Here all $f\in E_i^{\supi{k}}\intersect F_i^{\supi{k}}$ satisfy
    that $(w_Z)_f = (1+1/1000)^{i} (w_{G^{\supi{k}}})_f$.
    Since we set $\delta_f \gets i - c_f$ before the while loop,
    and by~(\ref{eq:Jdef2}) $(w_{J^{\supi{k}}_i})_f = (1 + 1/1000)^{c_f} (w_{G^{\supi{k}}})_f$,
    we have
    $(w_Z)_f = (1 + 1/1000)^{\delta_f} (w_{J^{\supi{k}}_i})_f$, as desired.

    Now suppose the statement is true
    at the end of iteration $j-1$ where $1 < j \leq t$.
    We then show that the statement is also true at the end of iteration $j$.
    Let $Z_0$ be the $Z$ before our updates to $Z$ in iteration $j$ and let $Z_1$ be the $Z$ after our updates.
    By Lemma~\ref{lem:shv}, all edges recovered $f\in F$ have their estimated edge weights
    $\wtil_f \in [\frac{1}{1+\eps_3} (w_{Z_0})_f, (1 + \eps_3) (w_{Z_0})_f]$.
    Therefore after our updates, we have for any $f\in F$ such that $\delta_f > 0$ that
    $(w_{Z_1})_f \in [\frac{1}{1+2\eps_3} (1+1/1000)^{-1} (w_{Z_0})_f, (1 + 2\eps_3) (1+1/1000)^{-1} (w_{Z_0})_f]$,
    and $(w_{Z_1})_f = (w_{Z_0})_f$ for other edges $f$.
    Since we let $\delta_f\gets \delta_f - 1$ for such edges (those with $\delta_f > 0$),
    and do not change the $\delta_f$'s of other edges,
    we have our desired statement for $j$.
 \end{proof}
  \begin{claim}
    The while loop terminates after at most $t$ iterations.
  \end{claim}
  \begin{proof}
    It suffices to show that $\max_f {\delta_f}$ decreases by $1$ in each while loop iteration.
    Since $\delta_f \leq t$ for any $f$, this will imply that there can be at most $t$ iterations.
    Then it boils down to showing that for all $f^*$ with $\delta_{f^*} = \max_f \delta_f$,
    $f^*$ belongs to the recovered edge set $F$.
    Since $f^*\in F^{\supi{k}}_i$, by~\ref{item:o4} of our induction hypothesis,
    the leverage score of $f^*$ in $(J_{i-1}^{\supi{k}})^{\sq}$ is at least $4\omega_3$.
    Notice that by~\ref{item:jk} of our induction hypothesis,
    $(J^{\supi{k}}_{i-1})^{\sq}$ is a $(1+1/1000)$-spectral sparsifier of $(G^{\supi{k}})^{\sq}$.
    Then using the fact that $(H^{\supi{k-1}})^{\sq}$ is a $1.003$-spectral sparsifier of $(G^{\supi{k}})^{\sq}$
    (which we proved at the beginning of the proof of this lemma), we have that
    $(H^{\supi{k-1}})^{\sq}$ is a $1.005$-spectral sparsifier of $(J_{i-1}^{\supi{k}})^{\sq}$.

    By Claim~\ref{claim:whileapx}, we have at the beginning of each while loop that, for all $f$,
    \begin{align}\label{eq:wzf1}
      (w_Z)_f \in &
      [\frac{1}{(1+2\eps_3)^t}(1+1/1000)^{\delta_f} (w_{J^{\supi{k}}_i})_f,(1 + 2\eps_3)^t (1 + 1/1000)^{\delta_f} (w_{J^{\supi{k}}_i})_f] \notag\\
      \subseteq &
      [\frac{1}{1.01}(1+1/1000)^{\delta_f} (w_{J^{\supi{k}}_i})_f,1.01 (1 + 1/1000)^{\delta_f} (w_{J^{\supi{k}}_i})_f].
    \end{align}
    Since $\delta_{f^*} \geq \delta_f$ for all $f$,
    the above implies
    \begin{align}
      L_{Z^{\sq}} \pleq {1.03} (1 + 1/1000)^{2 \delta_{f^*}} L_{(J_{i-1}^{\supi{k}})^{\sq}}
      \pleq {1.04} (1 + 1/1000)^{2 \delta_{f^*}} L_{(H^{\supi{k-1}})^{\sq}}, \label{eq:ZH}
    \end{align}
    where the second inequality follows from that $(H^{\supi{k-1}})^{\sq}$ is a
    $1.005$-spectral sparsifier of $(J^{\supi{k}}_{i-1})^{\sq}$.
    Let $\Gtil = (H^{\supi{k-1}})^{\sq}$.
    Now in order to show that $f^*$ will be recovered as an edge in $F^*$, by Lemma~\ref{lem:shv}, it suffices
    to show
    \begin{align}\label{eq:show}
      \frac{ ( (w_{Z})_{f^*} b_{f^*}^T L_{\Gtil}^{\dag} b_{f^*} )^2 }
      {b_{f^*}^T L_{\Gtil}^{\dag} L_{Z^{\sq}} L_{\Gtil}^{\dag} b_{f^*} } \geq \omega_3.
    \end{align}
    By~(\ref{eq:ZH}), the denominator satisfies
    \begin{align*}
      b_{f^*}^T L_{\Gtil}^{\dag} L_{Z^{\sq}} L_{\Gtil}^{\dag} b_{f^*}
      \leq {1.04} (1 + 1/1000)^{2 \delta_{f^*}} b_{f^*}^T L_{\Gtil}^{\dag} b_{f^*}.
    \end{align*}
    Therefore, the LHS of~(\ref{eq:show}) is at least
    $$(w_Z)_{f^*}^2(1 + 1/1000)^{-2 \delta_{f^*}} b_{f^*}^T L_{\Gtil}^{\dag} b_{f^*}\geq
    1.005^{-1}(w_Z)_{f^*}^2 (1 + 1/1000)^{-2 \delta_{f^*}} b_{f^*}^T L_{(J^{\supi{k}}_{i-1})^{\sq}}^{\dag} b_{f^*},
    $$
    where the inequality follows from that $(H^{\supi{k-1}})^{\sq}$ is a $1.005$-spectral sparsifier
    of $(J^{\supi{k}}_{i-1})^{\sq}$.
    Finally, using~(\ref{eq:wzf1}) and that the leverage score of $f^*$ is at least $2\omega_3$ in $(J^{\supi{k}}_{i-1})^{\sq}$,
    we have that the above is at least $\omega_3$, proving~(\ref{eq:show}).
  \end{proof}
  By Claim~\ref{claim:whileapx}, after the while loop terminates,
  we have that for all $f$,
  \begin{align*}
    (w_Z)_f \in & [\frac{1}{(1+2\eps_3)^t} (w_{J^{\supi{k}}_i})_f, (1 + 2\eps_3)^t (w_{J^{\supi{k}}_i})_f] \\
    \subseteq & [\frac{1}{(1+\eps_2/10000)} (w_{J^{\supi{k}}_i})_f, (1 + \eps_2/10000) (w_{J^{\supi{k}}_i})_f],
  \end{align*}
  and thus we have~\ref{item:while}.
  This also implies that $Z^{\sq}$ is a $(1 + \eps_2/10000)^2$-spectral sparsifier of $(J^{\supi{k}}_i)^{\sq}$,
  and as result, for each edge $f$, its leverage scores in $Z^{\sq}$ and $(J^{\supi{k}}_i)^{\sq}$ are within
  a $(1+\eps_2/10000)^3 < 1.01$ factor of each other.

  For all edges in $E^{\supi{k}}_i\setminus F^{\supi{k}}_i$, their weights in $Z$ equal exactly
  their weights in $J^{\supi{k}}_i$,
  therefore by Lemma~\ref{lem:shv},
  all edges recovered in $F^*$ not in $F_i^{\supi{k}}$ have weight estimates satisfying~\ref{item:wtili+1}.

  Notice that by~\ref{item:jk} of our induction hypothesis,
  $(J^{\supi{k}}_i)^{\sq}$ is a $(1+1/1000)$-spectral sparsifier of $(G^{\supi{k}})^{\sq}$.
  Then using the fact that $(H^{\supi{k-1}})^{\sq}$ is a $1.003$-spectral sparsifier of $(G^{\supi{k}})^{\sq}$
  (which we proved at the beginning of this proof),
  we have that $(H^{\supi{k-1}})^{\sq}$ is a $1.01$-spectral sparsifier of $Z^{\sq}$.
  By letting $\Gtil = (H^{\supi{k-1}})^{\sq}$
  we have, similar to~(\ref{eq:bygtil}),
  \begin{align}
    & \frac{((w_{Z})_e b_e^T L_{\Gtil}^{\dag} b_e)^2}{ b_e^T L_{\Gtil}^{\dag} L_{Z^{\sq}} L_{\Gtil}^{\dag} b_e }
    \approx_{1.01^4}
    \frac{((w_{Z})_e b_e^T L_{Z^{\sq}}^{\dag} b_e)^2}{ b_e^T L_{Z^{\sq}}^{\dag} b_e }
    = (w_{Z})_e^2 b_e^T L_{Z^{\sq}}^{\dag} b_e.
  \end{align}
  Therefore by Lemma~\ref{lem:shv}, the $F^*$ we recover in this iteration
  contains all edges whose leverage scores in $(Z)^{\sq}$ are at least $1.5\omega_3$.
  Also, at the last step of the for iteration, since $(H^{\supi{k-1}})^{\sq}$ is a $1.01$-spectral sparsifier of $Z^{\sq}$,
  all edges added to $H^{\supi{k}}$ have leverage scores at least $\geq 5\omega_3$ in $(Z)^{\sq}$,
  and all edges with leverage scores $\geq 9\omega_3$ in $(Z)^{\sq}$ will be added to $H^{\supi{k}}$.
  Thus we also know that all edges added to $H^{\supi{k}}$ have leverage scores at least $4\omega_3$ in $(J^{\supi{k}}_i)^{\sq}$
  (which gives~\ref{item:o4}),
  and all edges with leverage scores $\geq 10\omega_3$ in $(J^{\supi{k}}_i)^{\sq}$ will be added to $H^{\supi{k}}$,

  The above reasoning also implies that
  $(J^{\supi{k}}_{i+1})^{\sq}$ is obtained by sampling a set of edges in $(J^{\supi{k}}_i)^{\sq}$ whose leverage score
  is at most $10 \omega_3$ with probability $(1+1/1000)^{-2}$,
  and multiply their weights by $(1+1/1000)^2$
  if sampled. Using Theorem~\ref{clm:lev-sample}, we have~\ref{item:jk}.
\end{proof}
\begin{proof}[Proof of Lemma~\ref{lem:sparsifyG^2}]
  \textbf{Number of linear measurements.}{
    Notice that each $(\SSV)_{i,j}^{\supi{k}} w_{G^{\supi{k}}[E^{\supi{k}}_{i}]}\in\mathbb{R}^{n \omega_3^{-1}\eps_3^{-2}\polylog(n)}$,
    so the total number of linear measurements is bounded by
    \begin{align*}
      t^3 n \omega_3^{-1}\eps_3^{-2}\polylog(n) \leq n \eps_2^{-4} \polylog(n,\frac{\wmax}{\wmin},\eps_2^{-1}).
    \end{align*}
  }

  \textbf{Spectral sparsifier guarantee.}
  By Lemma~\ref{label:basecase},
  $(H^{\supi{0}})^{\sq}$ is a $1.001$-spectral sparsifier of $(G^{\supi{0}})^{\sq}$.
  We then show that whenever $(H^{\supi{k-1}})^{\sq}$ is a $1.001$-spectral sparsifier of $(G^{\supi{k-1}})^{\sq}$,
  $(H^{\supi{k}})^{\sq}$ is a $(1 + \eps_2/1000)$-spectral sparsifier of $(G^{\supi{k}})^{\sq}$ with high probability.
  Notice that inside each iteration of the outermost for loop of $k$,
  for $i = t$, we have that with high probability
  $E^{\supi{k}}_t = \emptyset$. This means that $J^{\supi{k}}_t$ consists of solely edges
  in $F^{\supi{k}}_t$. Thus by Lemma~\ref{lem:inductionk}, $(H^{\supi{k}}_t)^{\sq}$ is a
  $(1+\eps_2/(10^6t))^2$-spectral sparsifier of $(J^{\supi{k}}_t)^{\sq}$.
  Also by Lemma~\ref{lem:inductionk}, $(J^{\supi{k}}_t)^{\sq}$ is a $(1+\eps_2/(10^4 t))^t$-spectral sparsifier 
  of $(G^{\supi{k}})^{\sq}$. These combined imply that
  $(H^{\supi{k}})^{\sq}$ is a $(1 + \eps_2/1000)$-spectral sparsifier of $(G^{\supi{k}})^{\sq}$.
  Now applying an induction on $k$,
  we have that $(H^{\supi{t}})^{\sq}$ is a $(1 + \eps_2/1000)$-spectral sparsifier of $(G^{\supi{t}})^{\sq}$.
  Since $(G^{\supi{t}})^{\sq}$ is a $(1 + \eps_2/2)$-spectral sparsifier of $G^{\sq}$,
  $(H^{\supi{t}})^{\sq}$ is a $(1 + \eps_2)$-spectral sparsifier of $G^{\sq}$, as desired.
\end{proof}

\subsubsection{Recovery of heavy edges in \texorpdfstring{$G^{\sq}$}{}}\label{sec:shv}

\begin{figure}[ht]
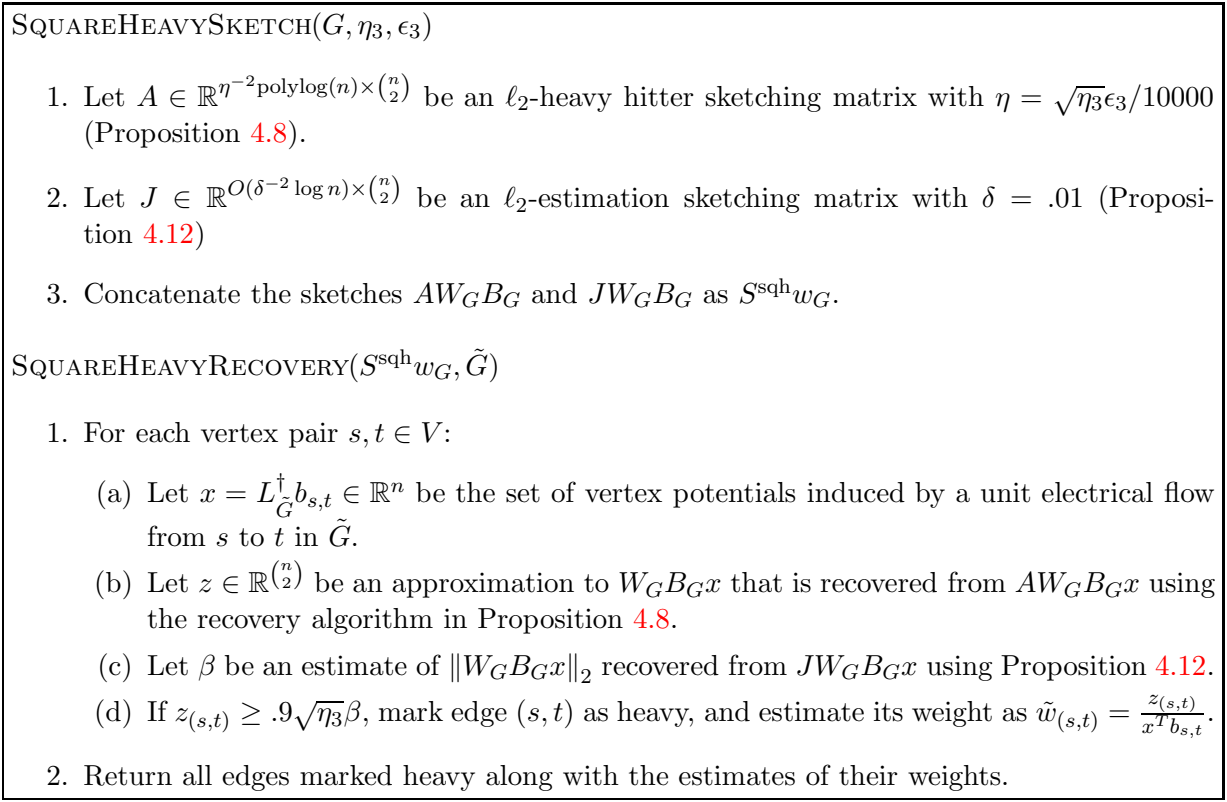

\begin{algbox}
  $\SVS(G,\omega_3,\eps_3)$

  \quad

  \begin{enumerate}
    \item Let $A\in\mathbb{R}^{\eta^{-2}\polylog(n)\times\binom{n}{2}}$ be an $\ell_2$-heavy hitter sketching
      matrix with $\eta = \sqrt{\omega_3} \eps_3 / 10000$ (Proposition~\ref{prop:heavyhitter}).
    \item Let $J\in\mathbb{R}^{O(\delta^{-2}\log n)\times \binom{n}{2}}$ be an $\ell_2$-estimation sketching matrix
      with $\delta = .01$ (Proposition~\ref{l2-norm})
    \item Concatenate the sketches $A W_G B_G$ and $J W_G B_G$ as $\SSV w_G$.
  \end{enumerate}

  \quad

  $\SVR(\SSV w_G, \Gtil)$

  \begin{enumerate}
    \item For each vertex pair $s,t\in V$:
      \begin{enumerate}
        \item Let $x = L_{\Gtil}^{\dag} b_{s,t} \in \mathbb{R}^{n}$ be the set of vertex potentials induced
          by a unit electrical flow from $s$ to $t$ in $\Gtil$.
        \item Let $z\in \mathbb{R}^{\binom{n}{2}}$ be an approximation to $W_G B_G x$ that is recovered
          from $A W_G B_G x$ using the recovery algorithm in Proposition~\ref{prop:heavyhitter}.
        \item Let $\beta$ be an estimate of $\norm{W_G B_G x}_2$ recovered from $J W_G B_G x$ using
          Proposition~\ref{l2-norm}.
        \item If $z_{(s,t)} \geq .9\sqrt{\omega_3} \beta$, mark edge $(s,t)$ as heavy,
          and estimate its weight as $\wtil_{(s,t)} = \frac{z_{(s,t)}}{x^T b_{s,t}}$.
      \end{enumerate}
    \item Return all edges marked heavy along with the estimates of their weights.
  \end{enumerate}
\end{algbox}
\caption{Linear sketch for recovering heavy edges in $G^{\sq}$.}
\label{fig:svs}
\end{figure}

We now give in Figure~\ref{fig:svs} a linear sketch for recovering heavy edges in $G^{\sq}$ and its analysis.
This linear sketch is essentially a direct application of $\ell_2$-heavy hitters.

The performance of the linear sketch is characterized in Lemma~\ref{lem:shv}:

\lemshv*

\begin{proof}[Proof of Lemma~\ref{lem:shv}]
  \textbf{Number of linear measurements.}{
    Notice that the sketches we compute satisfy
    $A W_G B_G \in \mathbb{R}^{\omega_3^{-1} \eps_3^{-2} \polylog(n)\times n}$
    and $J W_G B_G \in \mathbb{R}^{O(\log n)\times n}$.
    Therefore by concatenating them, the total number of linear measurements
    is bounded by $n \omega_3^{-1} \eps_3^{-2} \polylog(n)$.
  }

  \textbf{First guarantee.}{
    By the guarantee of the $\ell_2$-heavy hitter sketch, we have
    \begin{align}
      |\kh{ W_G B_G x }_{(s,t)} - z_{(s,t)}| \leq & \eta \norm{W_G B_G x}_2 \notag \\
      \leq & (\sqrt{\omega_3} \eps_3/10000) \norm{W_G B_G x}_2
      \qquad \text{(by the value of $\eta$)} \notag \\
      = & (\sqrt{\omega_3} \eps_3/10000) \sqrt{b_{s,t}^T L_{\Gtil}^{\dag} L_{G^{\sq}} L_{\Gtil}^{\dag} b_{s,t}}
      \label{eq:errz}
    \end{align}
    By the guarantee of the $\ell_2$-estimation sketch
    \begin{align}\label{eq:betaer}
      \beta \in \left[\frac{1}{1.01} \sqrt{b_{s,t}^T L_{\Gtil}^{\dag} L_{G^{\sq}} L_{\Gtil}^{\dag} b_{s,t}},
      1.01 \sqrt{b_{s,t}^T L_{\Gtil}^{\dag} L_{G^{\sq}} L_{\Gtil}^{\dag} b_{s,t}}\right].
    \end{align}
    Now fix any $e$ with
    \begin{align}\label{eq:heavy3}
      \frac{((w_e)_G b_e^T L_{\Gtil}^{\dag} b_e)^2}{ b_e^T L_{\Gtil}^{\dag} L_{G^{\sq}} L_{\Gtil}^{\dag} b_e }
      \geq \omega_3.
    \end{align}
    We then write
    \begin{align}
      | z_e - \kh{ W_G B_G x }_{e} | \leq & \frac{\eps_3}{10000}
      \sqrt{\omega_3 b_{e}^T L_{\Gtil}^{\dag} L_{G^{\sq}} L_{\Gtil}^{\dag} b_{e}}
      \qquad \text{by~(\ref{eq:errz})}
      \notag \\
      \leq & \frac{\eps_3}{10000} \sqrt{\omega_3} \sqrt{1/\omega_3} w_e b_{e}^T L_{\Gtil}^{\dag} b_{e}
      \qquad \text{(by~(\ref{eq:heavy3}))}
      \notag \\
      \leq & .001 \eps_3 w_e b_e^T L_{\Gtil}^{\dag} b_e \notag \\
      = & .001\eps_3 \kh{ W_G B_G x }_{e}. \label{eq:.01}
    \end{align}
    This implies that
    \begin{align*}
      z_e \geq & .999 w_e b_e^T L_{\Gtil}^{\dag} b_e \\
      \geq & .999 \sqrt{{\omega_3} b_{e}^T L_{\Gtil}^{\dag} L_{G^{\sq}} L_{\Gtil}^{\dag} b_{e}} 
      \qquad \text{(by~(\ref{eq:heavy3}))}
      \\
      \geq & .999 \sqrt{{\omega_3}} \frac{\beta}{1.01}
      \qquad \text{(by~(\ref{eq:betaer}))}
      \\
      \geq & .9 \sqrt{\omega_3} \beta.
    \end{align*}
    Therefore $e$ will be marked as heavy.
  }

  \textbf{Second guarantee.}{
    For $e$ such that $z_e \geq .9\sqrt{\omega_3}\beta$, we have
    \begin{align*}
      (W_G B_G x)_e \geq & z_e - \frac{\eps_3}{10000} \sqrt{\omega_3 b_{e}^T L_{\Gtil}^{\dag} L_{G^{\sq}} L_{\Gtil}^{\dag} b_{e}}
      \qquad \text{(by~(\ref{eq:errz}))}
      \\
      \geq & .9\sqrt{\omega_3} \frac{1}{1.01}\sqrt{b_{e}^T L_{\Gtil}^{\dag} L_{G^{\sq}} L_{\Gtil}^{\dag} b_{e}} -
      \frac{\eps_3}{10000}\sqrt{b_{e}^T L_{\Gtil}^{\dag} L_{G^{\sq}} L_{\Gtil}^{\dag} b_{e}}
      \qquad \text{(by~(\ref{eq:betaer}))}
      \\
      \geq & .87 \sqrt{\omega_3 b_{e}^T L_{\Gtil}^{\dag} L_{G^{\sq}} L_{\Gtil}^{\dag} b_{e}}
      \qquad \text{(by $\eps_3 < 1$).}
    \end{align*}
    Since $(W_G B_G x)_e = w_e b_e^T L_{\Gtil}^{\dag} b_e$,
    we have
    \begin{align}\label{eq:heavy4}
      \frac{((w_e)_G b_e^T L_{\Gtil}^{\dag} b_e)^2}{ b_e^T L_{\Gtil}^{\dag} L_{G^{\sq}} L_{\Gtil}^{\dag} b_e }
      \geq .7 \omega_3.
    \end{align}
    We then write
    \begin{align}
      | z_e - \kh{ W_G B_G x }_{e} | \leq & \frac{\eps_3}{10000}
      \sqrt{\omega_3 b_{s,t}^T L_{\Gtil}^{\dag} L_{G^{\sq}} L_{\Gtil}^{\dag} b_{s,t}}
      \qquad \text{(by~(\ref{eq:errz}))}
      \notag \\
      \leq & %
      \frac{\eps_3}{10000} \sqrt{\omega_3} \sqrt{1/(.7\omega_3)} w_e b_{e}^T L_{\Gtil}^{\dag} b_{e}
      \qquad \text{(by~(\ref{eq:heavy4}))}
      \notag \\
      \leq & .01 \eps_3 w_e b_e^T L_{\Gtil}^{\dag} b_e
      \notag \\
      = & .01\eps_3 \kh{ W_G B_G x }_{e}.
    \end{align}
    By dividing both sides by $x^T b_e$, we have
    \begin{align*}
      \sizeof{ \frac{z_e}{x^T b_e} - \frac{w_e b_e^T L_{\Gtil}^{\dag} b_e}{x^T b_e} } \leq
      .01\eps_3 \frac{w_e b_e^T L_{\Gtil}^{\dag} b_e}{x^T b_e}.
    \end{align*}
    Since
    ${x^T b_e} =
    {b_e^T L_{\Gtil}^{\dag} b_e}$,
    the above is equivalent to $\sizeof{\frac{z_e}{x^T b_e} - w_e} \leq .01\eps_3 w_e$,
    and thus we conclude
    $\frac{1}{1+\eps_3} w_e \leq \frac{w_e b_e^T L_{\Gtil}^{\dag} b_e}{x^T b_e} \leq (1 + \eps_3) w_e$,
    as desired.
  }
\end{proof}

\let\omega\oldomega

\section{Preliminaries on matrix-weighted graphs}\label{sec:hdg}

We consider an undirected, $k\times k$ matrix-weighted graph $G = (V,E)$ with $|V| = n$ and $|E| = m$,
where each edge $(u,v)$'s weight is given by $\phi_{uv} \phi_{uv}^T$ for some $\phi_{uv}\in\mathbb{R}^{k}$.
We will assume all $\phi_{uv}\neq 0$, since one could remove all zero weight edges from $E$.

\begin{remark}[Block matrices and vectors]
  Below, all $nk\times nk$ matrices
  (or $nk$-dimensional vectors) are written in block form with block size $k\times k$ (or $k\times 1$).
  All subscripts also refer to row/column block numbers as opposed to row/column numbers.
  For example, for an $nk\times nk$ matrix $M$, we write $M_{ij}$ to denote the $k\times k$ submatrix
  that is at the intersection of the $i^{\mathrm{th}}$ row block and the $j^{\mathrm{th}}$ column block.
\end{remark}

\begin{definition}[Degree matrices]
  For a vertex $u$, its degree is given by
  \begin{align*}
    D_u = \sum_{u\sim v} \phi_{uv} \phi_{uv}^T\in\mathbb{R}^{k\times k}.
  \end{align*}
  We then define the $nk\times nk$ degree matrix $D$ as a $k\times k$-block diagonal matrix:
  \begin{align*}
    D =
    \begin{pmatrix}
      D_1 &     &        &       \\
          & D_2 &        &       \\
          &     & \ddots &       \\
          &     &        & D_n
        \end{pmatrix}.
  \end{align*}
\end{definition}

\begin{definition}
  For each edge $(u,v)\in E$, define two $nk$-dimensional vectors
  $e_{u\la v}$ and $e_{v\la u}$
  \begin{align*}
    e_{u\la v} = 
    \begin{blockarray}{c@{}c@{}cl}
      \begin{block}{(c@{}c@{}c)l}
        & 0 \\
        & \vdots \\
        & \phi_{uv} & & \text{$u^{\mathrm{th}}$ block} \\
        & 0 \\
        & \vdots \\
        & \vdots \\
      \end{block}
    \end{blockarray}
    \qquad
    \text{and}
    \qquad
    e_{v\la u} =
    \begin{blockarray}{c@{}c@{}cl}
      \begin{block}{(c@{}c@{}c)l}
        & 0 \\
        & \vdots \\
        & \vdots \\
        & \phi_{uv} & & \text{$v^{\mathrm{th}}$ block} \\
        & 0 \\
        & \vdots \\
      \end{block}
    \end{blockarray}
  \end{align*}
\end{definition}

\begin{proposition}
  $D = \sum_{u\sim v} (e_{u\la v} e_{u\la v}^{T} + e_{v\la u} e_{v\la u}^T).$
\end{proposition}

When defining almost regular graphs, we will be interested in the quantities
$\phi_{uv}^T D_u^{\dag} \phi_{uv}$. We will call $\phi_{uv}^T D_u^{\dag} \phi_{uv}$ the leverage score of edge $(u,v)$
with respect to vertex $u$.

\begin{definition}[Leverage score of an edge w.r.t. a vertex]
  \label{def:lvgsc}
  We call $\phi_{uv}^T D_u^{\dag} \phi_{uv}$
  {\em the leverage score of edge $(u,v)$ w.r.t. vertex $u$}.
\end{definition}

\begin{proposition}
  For any edge $(u,v)$,
  $\phi_{uv}^T D_{u}^{\dag} \phi_{uv} = e_{u\la v}^T D^{\dag} e_{u\la v}$.
\end{proposition}

\begin{definition}[Adjacency matrices]
  The $nk\times nk$ adjacency matrix $A$ is given by $A_{uv} = \phi_{uv} \phi_{uv}^T$
  for $u\sim v$ and $A_{uv} = 0$ otherwise.
\end{definition}

In the following definitions, we
assume there is a (arbitrarily) fixed orientation of all edges.

\begin{definition}[Incidence vectors]
  For an edge $(u,v)$ oriented as $u\to v$,
  its $nk$-dimensional incidence vector $b_{uv}$ is given by
  \begin{align*}
    b_{uv} = 
    \begin{blockarray}{c@{}c@{}cl}
      \begin{block}{(c@{}c@{}c)l}
        & 0 \\
        & \vdots \\
        & \phi_{uv} & & \text{$u^{\mathrm{th}}$ block} \\
        & 0 \\
        & \vdots \\
        & -\phi_{uv} & & \text{$v^{\mathrm{th}}$ block} \\
        & 0 \\
        & \vdots \\
      \end{block}
    \end{blockarray}.
  \end{align*}
\end{definition}

\begin{definition}[Laplacian matrices]
  The Laplacian matrix is given by $L = D - A$.
\end{definition}
\begin{proposition}
  We have $L = \sum_{u\sim v} b_{uv} b_{uv}^T$.
\end{proposition}

\begin{proposition}\label{prop:quadraticL}
  For any vector $x\in \mathbb{R}^{nk}$, we have
  \begin{align*}
    x^T L x = \sum_{u\sim v} \ip{x_u - x_v}{\phi_{uv}}^2.
  \end{align*}
\end{proposition}

\begin{definition}[Projection matrices]
  For an $s$-dimensional
  subspace $\Scal\subseteq \mathbb{R}^{d}$ with orthonormal basis $b_1,\ldots,b_s\in\mathbb{R}^{d}$,
  we write
  \begin{align*}
    \Pi_{\Scal} = \sum_{i=1}^{s} b_i b_i^T \in \mathbb{R}^{d\times d}
  \end{align*}
  to denote the projection matrix onto $\Scal$.
  We will also write, for a symmetric matrix $M\in \mathbb{R}^{d\times d}$,
  $\Pi_{M}\in\mathbb{R}^{d\times d}$ to denote the projection matrix onto the range of $M$.
\end{definition}

\begin{definition}[Normalized Laplacians]
  The normalized Laplacian matrix is given by
  \begin{align*}
    N\defeq D^{\dag/2} L D^{\dag/2} = \Pi_D - D^{\dag/2} A D^{\dag/2}.
  \end{align*}
\end{definition}

\begin{proposition}
  $N = \sum_{u\sim v} (D^{\dag/2} b_{uv}) (D^{\dag/2} b_{uv})^T$.
\end{proposition}

\begin{proposition}\label{prop:quadraticN}
  For any vector $x\in \mathbb{R}^{nk}$, we have
  \begin{align*}
    x^T N x = \sum_{u\sim v} \kh{\ip{x_u}{D_u^{\dag/2} \phi_{uv}}- \ip{x_v}{D_v^{\dag/2}\phi_{uv}}}^2.
  \end{align*}
\end{proposition}

\begin{proposition}\label{prop:eigvalofN}
  The eigenvalues of $N$ is between $[0,2]$.
\end{proposition}

Sometimes, instead of using exact degrees, we will normalize the Laplacian matrix by approximate degrees.
Thus we also define $\Dtil$-normalized Laplacian matrix where
$\Dtil\in \mathbb{R}^{nk\times nk}$ is a semidefinite, $k\times k$-block diagonal matrix that has the same range as $D$.

\begin{definition}[$\Dtil$-normalized Laplacians]
  \label{def:dtilnL}
  For a semidefinite, $k\times k$-block diagonal matrix $\Dtil\in\mathbb{R}^{nk\times nk}$ whose
  range is the same as $D$,
  the $\Dtil$-normalized Laplacian matrix is given by
  \begin{align*}
    \Ntil\defeq \Dtil^{\dag/2} L \Dtil^{\dag/2}.
  \end{align*}
\end{definition}

\begin{proposition}
  $\Ntil = \sum_{u\sim v} (\Dtil^{\dag/2} b_{uv}) (\Dtil^{\dag/2} b_{uv})^T$.
\end{proposition}

\begin{proposition}
  For $\Dtil$ satisfying
  \begin{align*}
    \frac{1}{\kappa} D \pleq \Dtil \pleq \kappa D
  \end{align*}
  for some $\kappa > 1$,
  the eigenvalues of $\Ntil = \Dtil^{\dag/2} L \Dtil^{\dag/2}$ is between $[0,2\kappa]$.
\end{proposition}

Note that, since the degree matrix $D$ is not necessarily full-rank,
for either $L$, $N$, or $\Ntil$, there will always be some ``trivial'' zero eigenvalues
that correspond to the kernel of $D$. Therefore we shall focus only on the remaining ``nontrivial'' eigenvalues
and their eigenvectors.

\begin{definition}[Nontrivial eigenvalues and eigenvectors]
  \label{def:nontrivial}
  An eigenvalue $\lambda$ and its associated eigenvector $f\in\mathbb{R}^{nk}$ of either $L$, $N$, or $\Ntil$
  are said to be {\em nontrivial} if we have
  \begin{align*}
    f_u\,\bot\,\mathrm{ker}(D_u), \ \forall u\in V.
  \end{align*}
\end{definition}

\begin{proposition}
  The number of nontrivial eigenvalues (counted with multiplicity)
  of either $L$, $N$, or $\Ntil$ is exactly
  $\mathrm{rank}(D)$.
\end{proposition}

\section{Almost regular graphs have only few small eigenvalues}\label{sec:fewsmall}

We first give the definition of almost regularity in matrix-weighted graphs.

\begin{definition}[Almost regular graphs]
  \label{def:harg}
  Let $G = (V,E)$ be a $k\times k$ matrix-weighted graph with edge weights $\phi_{uv} \phi_{uv}^T$.
  For a $\gamma\geq 1$, %
  we say $G$ is $\gamma$-almost regular if for all
  $u\in V$ and $(u,v)\in E$, we have
  \begin{align}\label{eq:reg}
    \phi_{uv}^T D_u^{\dag} \phi_{uv} \leq \frac{\gamma\cdot k}{n}.
  \end{align}
\end{definition}

We then consider the spectrum of the $\Dtil$-normalized Laplacian (Definition~\ref{def:dtilnL}) of
a graph $G$ for some semidefinite, $k\times k$-block diagonal matrix
$\Dtil$ with the same range as $D$.
Specifically,
we show that when $G$ is almost regular and $\Dtil$ is close to $D$,
the number of small nontrivial eigenvalues (Definition~\ref{def:nontrivial}) is small.
\begin{theorem}\label{thm:fewsmall}
  Let $G = (V,E)$ be a $k\times k$ matrix-weighted, $\gamma$-almost regular graph.
  Suppose $\Dtil$ is a semidefinite, $k\times k$-block diagonal matrix satisfying that
  \begin{align*}
    \frac{1}{\kappa} D \pleq \Dtil \pleq \kappa D
  \end{align*}
  for some $\kappa \geq 1$.
  Then for any $\delta \in (0,1)$,
  the number of nontrivial eigenvalues of $\Ntil = \Dtil^{\dag/2} L \Dtil^{\dag/2}$
  that are at most $\frac{1}{\kappa} - \delta$
  is at most $\frac{\gamma \kappa^2 k^2}{\delta^2}$.
\end{theorem}
This theorem will be a consequence a lemma that
characterizes certain properties of the ``spectral embedding'' induced by the bottom eigenvectors of $\Ntil$.
We note that similar spectral embeddings were previously used to prove higher-order Cheeger inequalities
for scalar-weighted graphs~\cite{LouisRTV12,LeeGT14}.

\begin{definition}[Spectral embeddings]
  Given orthonormal vectors $f_1,\ldots,f_{\ell}\in\mathbb{R}^{nk}$,
  define an $\ell\times nk$ matrix $\Fcal$ whose rows are
  transposes of $f_1,\ldots,f_{\ell}$:
  \renewcommand{\arraystretch}{1.4}
  \begin{center}
    $\Fcal\,=\ $
    \begin{tabular}[h]{|x{4em}|x{7em}|x{4em}|x{7em}|x{4em}|}
      \hline
      $(f_1)_1^T $ & $\cdots$ & $(f_1)_u^T$ & $\cdots$ & $(f_1)_n^T$ \\
      \hline
      $(f_2)_1^T$ & $\cdots$ & $(f_2)_u^T$ & $\cdots$ & $(f_2)_n^T$ \\
      \hline
      \multicolumn{5}{|c|}{\xrowht{30pt}$\vdots$} \\
      \hline
      $(f_\ell)_1^T$ & $\cdots$ & $(f_\ell)_u^T$ & $\cdots$ & $(f_\ell)_n^T$ \\
      \hline
    \end{tabular}
    $\ \in \mathbb{R}^{\ell\times nk}$.
  \end{center}
  Then define an embedding $F : V \to \mathbb{R}^{\ell\times k}$ by
  letting $F(u)$ equal the $u^{\mathrm{th}}$ column block of $\Fcal$:
  \begin{center}
    $F(u)\,=\ $
    \begin{tabular}[h]{|x{4em}|}
      \hline
      $(f_1)_u^T$ \\
      \hline
      $(f_2)_u^T$ \\
      \hline\xrowht{30pt}$\vdots$ \\
      \hline
      $(f_\ell)_u^T$ \\
      \hline
    \end{tabular}
    $\ \in \mathbb{R}^{\ell\times k}$.
  \end{center}
  We call $F$ the spectral embedding induced by $f_1,\ldots,f_{\ell}$,
  and $\Fcal$ the embedding matrix induced by $f_1,\ldots,f_{\ell}$.
\end{definition}

Essentially, the following lemma says that the spectral embedding induced by the (nontrivial) bottom eigenvectors
has norm spread out across a large number of vertices.

\begin{lemma}\label{lem:fsu2}
  Let $G = (V,E)$ be a $k\times k$ matrix-weighted, $\gamma$-almost regular graph.
  Suppose $\Dtil$ is a semidefinite, $k\times k$-block diagonal matrix satisfying that
  \begin{align*}
    \frac{1}{\kappa} D \pleq \Dtil \pleq \kappa D
  \end{align*}
  for some $\kappa \geq 1$. %
  Fix a $\delta \in (0,1)$ and
  let
  $0\leq \lambdatil_1 \leq \ldots \leq \lambdatil_{\ell} \leq \frac{1}{\kappa}  - \delta$
  be all nontrivial eigenvalues of $\Ntil = \Dtil^{\dag/2} L \Dtil^{\dag/2}$ that are $\leq \frac{1}{\kappa} - \delta$.
  Let
  $\ftil_1,\ldots,\ftil_{\ell}$ be a corresponding set of orthonormal, nontrivial eigenvectors,
  which by the definition of nontriviality satisfies
  \begin{align*}
    (\ftil_i)_u\,\bot\,\mathrm{ker}(D_u), \ \forall i\in [\ell], u\in V.
  \end{align*}
  Let $\Ftil$ be the spectral embedding induced by $\ftil_1,\ldots,\ftil_{\ell}$.
  Then we have for all $u\in V$ %
  \begin{align}\label{eq:bsf2}
    \lambdamax\kh{ \Ftil(u)^T \Ftil(u) }
    \leq
    \frac{\gamma \kappa^2 k}{\delta^2 n},
  \end{align}
  where $\lambdamax(\cdot)$ denotes taking the largest eigenvalue.
\end{lemma}
\begin{proof}[Proof of Theorem~\ref{thm:fewsmall} using Lemma~\ref{lem:fsu2}]
  Let $\ell$ be the number of nontrivial eigenvalues of $\Ntil$ that are
  $\leq \frac{1}{\kappa} - \delta$ and let $\ftil_1,\ldots,\ftil_{\ell}$
  be a corresponding set of orthonormal, nontrivial eigenvectors.
  Let $\Ftil$ be the spectral embedding induced by $\ftil_1,\ldots,\ftil_{\ell}$.
  Since all $\ftil_i$'s are unit vectors, we have
  \begin{align*}
    \sum_{u} \trace{\Ftil(u)^T \Ftil(u) } =
    \sum_{u} \norm{\Ftil(u)}_F^2 =
    \sum_{u} \sum_{i=1}^{\ell} \norm{(\ftil_i)_u}^2 = \ell.
  \end{align*}
  Therefore there must exist a vertex $u$ with
  \begin{align*}
    \trace{ \Ftil(u)^T \Ftil(u) } \geq \frac{\ell}{n}.
  \end{align*}
  As $\Ftil(u)^T \Ftil(u)$ is a $k\times k$ positive semidefinite matrix,
  we have
  \begin{align*}
    \lambdamax \kh{ { \Ftil(u)^T \Ftil(u) } }  \geq
    \frac{1}{k}\cdot
    \trace{ \Ftil(u)^T \Ftil(u) } \geq \frac{\ell}{nk}.
  \end{align*}
  On the other hand, by Lemma~\ref{lem:fsu2},
  $\lambdamax \kh{ { \Ftil(u)^T \Ftil(u) } } \leq \frac{\gamma \kappa^2 k}{\delta^2 n}$.
  Thus we must have
  $\ell \leq \frac{\gamma \kappa^2 k^2}{\delta^2}$.
\end{proof}
\begin{proof}[Proof of Lemma~\ref{lem:fsu2}]
  Let $\Fcaltil$ be the embedding matrix induced by $\ftil_1,\ldots,\ftil_{\ell}$.
  Suppose for the sake of contradiction~(\ref{eq:bsf2}) is violated by some $u\in V$.
  That is, we have
  \begin{align}\label{eq:assumption}
    \lambdamax\kh{ \Ftil(u)^T \Ftil(u) } > 
    \frac{\gamma \kappa^2 k}{\delta^2 n}.
  \end{align}
  Define a diagonal matrix $\Lambdatil \in \mathbb{R}^{\ell\times \ell}$ by
  \begin{align*}
    \Lambdatil =
    \begin{pmatrix}
      \lambdatil_1 & & \\
      & \ddots & \\
      & & \lambdatil_{\ell}
    \end{pmatrix} \in \mathbb{R}^{\ell\times \ell}.
  \end{align*}
  Then, since $\lambdatil_1,\ldots,\lambdatil_{\ell}$ and
  $\ftil_1,\ldots,\ftil_{\ell}$ are corresponding eigenvalues and eigenvectors of $\Ntil$,
  we immediately have (recall that $\Fcaltil$'s rows are $\ftil_1^T,\ldots,\ftil_{\ell}^T$)
  \begin{align*}
    \Lambdatil \Fcaltil = \Fcaltil \Ntil,
  \end{align*}
  and in particular, by restricting to the $u^{\mathrm{th}}$ column block on both sides,
  \begin{align}\label{eq:lambdaF}
    \Lambdatil \Ftil(u)
    = \Ftil(u) \Dtil_{uu}^{\dag/2} D_{u} \Dtil_{uu}^{\dag/2} -
    \sum_{u\sim v}{ \Ftil(v) \Dtil_{vv}^{\dag/2} \phi_{uv} \phi_{uv}^T \Dtil_{uu}^{\dag/2} }.
  \end{align}
  Now consider the $\ell\times \ell$ matrix $\Ftil(u) \Ftil(u)^T$, whose largest eigenvalue is equal to
  $\lambdamax(\Ftil(u)^T \Ftil(u))$.
  Let $g\in\mathbb{R}^{\ell}$ be a unit eigenvector corresponding to $\lambdamax(\Ftil(u) \Ftil(u)^T)$.
  Since $\ftil_1,\ldots,\ftil_{\ell}$ are orthonormal, we have
  \begin{align}
    \sum_{v\in V} g^T \Ftil(v) \Ftil(v)^T g =
    & g^T \kh{ \sum_{v\in V} \Ftil(v) \Ftil(v)^T } g \notag \\
    = & g^T \kh{ \Fcaltil \Fcaltil^T } g
    \notag \\ = & g^T I g
    \notag \\
    = & \norm{g}^2 = 1. \label{eq:gffg}
  \end{align}
  We will then show a contradiction by arguing that~(\ref{eq:lambdaF})
  cannot be true given~(\ref{eq:assumption}).
  First by multiplying $g^T$ to the left and $\Ftil(u)^T g$ to the right to both sides of~(\ref{eq:lambdaF})
  we have
  \begin{align}\label{eq:contradict}
    g^T \Lambdatil \Ftil(u) \Ftil(u)^T g = g^T \Ftil(u) \Dtil_{uu}^{\dag/2} D_{u} \Dtil_{uu}^{\dag/2} \Ftil(u)^T g -
    \sum_{u\sim v}{ g^T \Ftil(v) \Dtil_{vv}^{\dag/2} \phi_{uv} \phi_{uv}^T \Dtil_{uu}^{\dag/2} \Ftil(u)^T g}.
  \end{align}
  Using the fact that $g$ is a unit eigenvector corresponding to $\lambdamax(\Ftil(u) \Ftil(u)^T)$,
  we can upper bound %
  the LHS of~(\ref{eq:contradict}) by
  \begin{align*}
    g^T \Lambdatil \Ftil(u) \Ftil(u)^T g  = & \lambdamax\kh{\Ftil(u) \Ftil(u)^T}\cdot g^T \Lambdatil g \\
    \leq & \kh{\frac{1}{\kappa} - \delta} \cdot \lambdamax\kh{\Ftil(u) \Ftil(u)^T},
  \end{align*}
  where the last inequality holds since all eigenvalues $\lambdatil_1,\ldots,\lambdatil_\ell \leq \frac{1}{\kappa} - \delta$.
  We can also lower bound %
  the first term on the RHS of~(\ref{eq:contradict}) by
  \begin{align*}
    g^T \Ftil(u) \Dtil_{uu}^{\dag/2} D_{u} \Dtil_{uu}^{\dag/2} \Ftil(u)^T g \geq
    \frac{1}{\kappa} \cdot g^T \Ftil(u) \Ftil(u)^T g = \frac{1}{\kappa}\cdot\lambdamax(\Ftil(u) \Ftil(u)^T),
  \end{align*}
  where the first inequality follows from $\Dtil \pleq \kappa D$.
  Therefore, in order to contradict~(\ref{eq:contradict}),
  it suffices to show that the second term on the RHS of~(\ref{eq:contradict}) satisfies
  \begin{align*}
    \sum_{u\sim v}{ g^T \Ftil(v) \Dtil_{vv}^{\dag/2} \phi_{uv} \phi_{uv}^T \Dtil_{uu}^{\dag/2} \Ftil(u)^T g}\,<\,
    \delta \cdot \lambdamax\kh{\Ftil(u) \Ftil(u)^T}.
  \end{align*}
  We then do so by applying the Cauchy-Schwarz inequality:
  \begin{align*}
    & \sum_{u\sim v} g^T \Ftil(v) \Dtil_{vv}^{\dag/2} \phi_{uv} \phi_{uv}^T \Dtil_{uu}^{\dag/2} \Ftil(u)^T g \\
    \leq &
    \sqrt{\sum_{u\sim v} g^T \Ftil(v) \Ftil(v)^T g }
    \sqrt{\sum_{u\sim v}
      g^T \Ftil(u) \Dtil_{uu}^{\dag/2} \phi_{uv} \phi_{uv}^T \Dtil_{vv}^{\dag} \phi_{uv} \phi_{uv}^T \Dtil_{uu}^{\dag/2}
      \Ftil(u)^T g
    } \\
    \leq & 1\cdot
    \sqrt{\sum_{u\sim v}
      g^T \Ftil(u) \Dtil_{uu}^{\dag/2} \phi_{uv} \phi_{uv}^T \Dtil_{vv}^{\dag} \phi_{uv} \phi_{uv}^T \Dtil_{uu}^{\dag/2}
      \Ftil(u)^T g
    } \qquad \text{(by~(\ref{eq:gffg}))} \\
    \leq &
    \sqrt{\frac{\gamma \kappa\cdot k}{n}}\cdot
    \sqrt{\sum_{u\sim v}
      g^T \Ftil(u) \Dtil_{uu}^{\dag/2} \phi_{uv}  \phi_{uv}^T \Dtil_{uu}^{\dag/2}
      \Ftil(u)^T g
    } \\
    & \qquad \text{(since $\phi_{uv}^T \Dtil_{vv}^{\dag} \phi_{uv} \leq \frac{\gamma \kappa\cdot k}{n}$
  by $\gamma$-almost regularity and $\Dtil \pgeq \frac{1}{\kappa} D$)}\\
    = & \sqrt{\frac{\gamma \kappa\cdot k}{n}}
    \sqrt{g^T \Ftil(u) \Dtil_{uu}^{\dag/2} \kh{\sum_{u\sim v} \phi_{uv} \phi_{uv}^T} \Dtil_{uu}^{\dag/2} \Ftil(u)^T g } \\
    = & \sqrt{\frac{\gamma \kappa\cdot k}{n}}
    \sqrt{ g^T \Ftil(u) \Dtil_{uu}^{\dag/2} D_u \Dtil_{uu}^{\dag/2} \Ftil(u)^T g } \\
    \leq & \sqrt{\frac{\gamma\kappa\cdot k}{n}} \sqrt{\kappa}
    \sqrt{ g^T \Ftil(u) \Ftil(u)^T g }
    \qquad \text{(by $\Dtil \pgeq \frac{1}{\kappa} D$)} \\
    = & \sqrt{\frac{\gamma \kappa^2 k}{n}}
    \sqrt{\lambdamax(\Ftil(u) \Ftil(u)^T)}
    \\
    < & \delta\cdot \lambdamax(\Ftil(u) \Ftil(u)^T)
    \qquad \text{
      (since by our assumption~(\ref{eq:assumption})
        $\lambdamax( \Ftil(u)^T \Ftil(u) ) > 
    \frac{\gamma \kappa^2 k}{\delta^2 n}$).}
  \end{align*}
  Therefore it must be the case that all $\lambdamax( \Ftil(u)^T \Ftil(u) ) \leq \frac{\gamma \kappa^2 k}{\delta^2 n}$,
  as desired
\end{proof}

\section{Almost regular graph decomposition}\label{sec:hargd}

\let\oldlambda\lambda
\renewcommand*{\lambda}{\mu}

In this section, we show that every matrix-weighted graph
that is sufficiently dense
can be made into an almost regular graph (in the sense of Definition~\ref{def:harg}) by downscaling a small number of edges.

Let us first introduce some notations for rescaled graphs.
For a $k\times k$ matrix-weighted graph $G = (V,E)$ with edge weights $\phi_{uv} \phi_{uv}^T$'s
and a scaling $s: E\to [0,1]$,
we will write $G^s$ to denote the graph obtained from $G$ by rescaling each edge $(u,v)$'s weight
to $(s_{uv} \phi_{uv}) (s_{uv} \phi_{uv})^T$.
For simplicity we will use the superscript $s$ when dealing with vectors and matrices associated with $G^s$.
For example, we have $\phi_{uv}^s = s_{uv} \phi_{uv}$,
$D_u^s = \sum_{u\sim v} (s_{uv} \phi_{uv}) (s_{uv} \phi_{uv})^T$, and
$L^s = \sum_{(u,v)\in E} s_{uv}^2 b_{uv} b_{uv}^T$.
Similarly, for a subset of edges $F\subseteq E$, we also use the superscript $F$
to denote matrices associated with the induced subgraph $G[F]$,
and thus, for example, $D_u^F = \sum_{(u,v)\in F} \phi_{uv} \phi_{uv}^T$.

Our main result in this section is a deterministic algorithm
for finding a large almost regular subgraph within any given graph that is sufficiently dense.

\begin{theorem}\label{thm:hargd}
  There is a deterministic algorithm $\HARGD$ that,
  given any $k\times k$ matrix-weighted graph $G = (V,E)$ with edge weights $\phi_{uv}\phi_{uv}^T$'s
  and any $\gamma \geq 1$, outputs a scaling
  $s: E \to [0,1]$ such that
  \begin{enumerate}
    \item The rescaled graph $G^{s}$ is $\gamma$-almost regular:
      \begin{align*}
        (\phi_{uv}^s)^T \kh{ D_{u}^s }^{\dag} \phi_{uv}^s
        \leq \frac{\gamma\cdot k}{n},
        \ \forall (u,v) \in E.
      \end{align*}
    \item The number of edges $(u,v)\in E$ with $s_{uv} \in (0,1)$ is at most $\frac{8 n^2}{\gamma}$.
    \item The number of edges $(u,v)\in E$ with $s_{uv} = 0$ is at most
      $\frac{8n}{\gamma \cdot k} (\rank{D} - \rank{D^s})$.
  \end{enumerate}
  The algorithm terminates in finite time.
\end{theorem}

Note that since $\rank{D} \leq nk$, the number of edges with $s_{uv} = 0$ is at most $\frac{8n^2}{\gamma}$.
Therefore, by setting $\gamma$ to be $\geq \frac{32n^2}{|E|}$,
the total fraction of edges with $s_{uv} < 1$ is no more than $1/2$.
Also note that our goal is to prove the existence of such a scaling for any given weights $\phi_{uv} \phi_{uv}^T$'s
that can potentially have {\em infinite} precision, so we only focus on designing an algorithm
that terminates in finite time, as opposed to giving an explicit bound on its running time.

Our approach for proving Theorem~\ref{thm:hargd} is motivated by the one in~\cite{CohenLMMPS15},
using which the authors showed that given a set of vectors, one can, by downscaling a small number of them,
make every vector have small leverage score compared to the average.
The main difference between our result and theirs is that we have an additional
bound on the number of completely deleted edges that is proportional to the rank reduction
(the third guarantee in Theorem~\ref{thm:hargd}),
which will be useful when we incorporate Theorem~\ref{thm:hargd} into
our expander decomposition later in Section~\ref{sec:hared}.

We now give the pseudocodes of our algorithm $\HARGD$ and its subroutine $\WAM$ below.
At a high level, in $\HARGD$, we first try to eliminate edges with high
leverage scores\footnote{Recall that in Definition~\ref{def:lvgsc}
  we call the quantity $(\phi_{uv})^T \kh{ D_{u} }^{\dag} \phi_{uv}$
the leverage of edge $(u,v)$ w.r.t. vertex $u$.},
by iteratively halving the weight of any such edge for some large number of iterations.
The latter iterative process is essentially what $\WAM$ does.
Then after $\WAM$ terminates, either (i) all edges have small leverage scores, in which case we are done, or
(ii) the edge weights have become sufficiently divergent so that we can decrease the rank of $D$ by deleting
a small number of edges.
If it is the second case, we then delete those edges and restart the process from beginning.
We will show that the total number of deleted/rescaled edges is small, so that we end up with a large almost regular graph.

\quad

\begin{algorithm}[H]
  \label{algo:wam}
  \caption{$\WAM(G,\gamma,T,s^0)$}
  \Input{
    $G$: a $k\times k$ matrix-weighted graph $G = (V,E)$ with edge weights $\phi_{uv} \phi_{uv}^T$. \\
    $\gamma$: an almost regularity parameter. \\
    $T$: an integer denoting the iteration count. \\
    $s^0$: an initial scaling, mapping from $E\to [0,1]$.
  }
  \Output{
    $s$: a scaling, mapping from $E\to [0,1]$.
  }
  Let $s\gets s^0$. \;
  \For{$t \gets 1\ \mathrm{to}\ T$}{
    \If{$\exists u\in V,(u,v) \in E$ {\rm with}
    $(\phi_{uv}^s)^T (D_{u}^{s})^{\dag} \phi_{uv}^s > \frac{\gamma\cdot k}{n}$}{
      Pick an arbitrary pair of such $u$ and $(u,v)$. \;
      $s_{uv} \gets s_{uv} / 2$.
    }
  }
  \Return $s$.
\end{algorithm}

\quad

\begin{algorithm}
  \label{algo:hargd}
  \caption{$\HARGD(G,\gamma)$}
  \Input{
    $G$: a $k\times k$ matrix-weighted graph $G = (V,E)$ with edge weights $\phi_{uv} \phi_{uv}^T$. \\
    $\gamma$: an almost regularity parameter.
  }
  \Output{
    $s$: a scaling, mapping from $E\to [0,1]$.
  }
  Initially, let $s_{uv} \gets 1$ for all $(u,v)\in E$. \;
  Let $\ell \gets \min_{(u,v)\in E} \norm{\phi_{uv}}^2$ and $r \gets \max_{(u,v)\in E} \norm{\phi_{uv}}^2$. \;
  Let $\alpha \gets \max\setof{8 n^4 k,\ 4nk\cdot r\cdot \max_{F\subseteq E} \lambdamax\kh{\kh{D^{F}}^{\dag}}}$.
  \label{line:alpha} \;
  Let $T \gets n^2 (4n^2 k \log \alpha + \log (r/\ell))$.\;
  \While{$\rank{D^s} > 0$}{
    $s\gets \WAM(G, \gamma, T, s)$.\;
    \If{$\forall u\in V,(u,v) \in E$ {\rm we have}
    $(\phi_{uv}^s)^T (D_{u}^{s})^{\dag} \phi_{uv}^s \leq \frac{\gamma\cdot k}{n}$}{
      Return $s$ and halt.
    }
    \Else{
      Compute all eigenvalues of $D^{s}$ and let
      $0 < \lambda_1\leq \ldots \leq \lambda_{p}$ be the positive ones.\;
      Let $\ell^s \gets \min_{(u,v)\in E, s_{uv} > 0} \norm{\phi_{uv}^s}^2$\; %
      Find the smallest $\rho \geq \ell^{s}$ such that
      all $\norm{\phi_{uv}^{s}}^2$ and $\lambda_i$ fall in $\mathbb{R}\setminus (\rho,\rho \cdot \alpha)$.
      \label{line:rho}
      \;
      Set $s_{uv}\gets 0$ for all $(u,v)\in E$ with $\norm{\phi_{uv}^s}^2 \leq \rho$.
      \label{line:delete}
      \;
      Set $s_{uv}\gets 1$ for all $(u,v)\in E$ with $s_{uv} > 0$.
      \label{line:reset}
    }
  }
  \Return $s$.
\end{algorithm}

\quad

\paragraph{Analysis of Algorithm~\ref{algo:wam}.}{
  We first show that after $\WAM$ terminates,
  all edges that have been downscaled have large leverage scores w.r.t. at least one of their endpoints.

  \begin{lemma}
    \label{lem:wamlvgsc}
    Suppose the initial scaling $s^0$ satisfies that
    $s^0_{uv} \in \setof{0,1}$ for all $(u,v)\in E$.
    Then after $\WAM$ terminates, for any $(u,v)\in E$ such that $s_{uv}\in (0,1)$,
    the leverage score of $(u,v)$ w.r.t. at least one of $u,v$ is $\geq \frac{\gamma\cdot k}{4n}$.
    That is, at least one of the following two statements is true:
    \begin{enumerate}
      \item $(\phi_{uv}^s)^T (D_{u}^{s})^{\dag} \phi_{uv}^s \geq \frac{\gamma\cdot k}{4n}$.
      \item $(\phi_{uv}^s)^T (D_{v}^{s})^{\dag} \phi_{uv}^s \geq \frac{\gamma\cdot k}{4n}$.
    \end{enumerate}
  \end{lemma}
  \begin{proof}
    Fix an edge $(u,v)\in E$ for which $s_{uv} \in (0,1)$ after $\WAM$ terminates.
    Since initially $s^0_{uv} \in \setof{0,1}$,
    it must be the case that $s^0_{uv} = 1$, and thus $s_{uv}$ must change at least once during the process.
    Let $t_{uv}\in[1,T]$ be the last for loop iteration in which $s_{uv}$ changes.
    Then we know that at the beginning of iteration $t_{uv}$,
    we have  either $(\phi_{uv}^s)^T (D_{u}^{s})^{\dag} \phi_{uv}^s > \frac{\gamma\cdot k}{n}$
    or $(\phi_{uv}^s)^T (D_{v}^{s})^{\dag} \phi_{uv}^s > \frac{\gamma\cdot k}{n}$.
    Suppose w.l.o.g. it is the first case.
    We first show that after we set $s_{uv}\gets s_{uv} / 2$, we have
    $(\phi_{uv}^s)^T (D_{u}^{s})^{\dag} \phi_{uv}^s \geq \frac{\gamma\cdot k}{4n}$.
    For clarity let us write $s$ and $s'$ to denote the scalings before and after the update in iteration $t_{uv}$
    respectively.
    Then we have
    \begin{align*}
      (\phi_{uv}^{s'})^T (D_{u}^{s'})^{\dag} \phi_{uv}^{s'} \geq
      & (\phi_{uv}^{s'})^T (D_{u}^{s})^{\dag} \phi_{uv}^{s'}
      \qquad
      \text{ ($D_{u}^{s'} = D_{u}^{s} - \frac{3s_{uv}^2}{4} \phi_{uv} \phi_{uv}^T \pleq D_u^{s}$) } \\
      = &
      \frac{1}{4} (\phi_{uv}^{s})^T (D_{u}^{s})^{\dag} \phi_{uv}^{s}
      \geq
      \frac{\gamma\cdot k}{4n} \qquad
      \text{ ($(\phi_{uv}^s)^T (D_{u}^{s})^{\dag} \phi_{uv}^s > \frac{\gamma\cdot k}{n}$). }
    \end{align*}
    After iteration $t_{uv}$, we only downscale the edge weights of other edges.
    Note that for any $s''$ that is pointwise smaller than $s'$, we have
    $D_u^{s''} \pleq D_u^{s'}$, and thus the leverage score of $(u,v)$ w.r.t. $u$
    can only increase afterwards. Therefore we have
    $(\phi_{uv}^{s})^T (D_{u}^{s})^{\dag} \phi_{uv}^{s} \geq \frac{\gamma\cdot k}{4n}$ in the end,
    as desired.
  \end{proof}

  As a consequence of Lemma~\ref{lem:wamlvgsc}, the total number of edges
  that have been downscaled during $\WAM$ is small.
  \begin{corollary}[of Lemma~\ref{lem:wamlvgsc}]
    \label{cor:nofs<1}
    Suppose the initial scaling $s^0$ satisfies that
    $s^0_{uv} \in \setof{0,1}$ for all $(u,v)\in E$.
    Then after $\WAM$ terminates, the number of edges $(u,v)\in E$
    with $s_{uv} \in (0,1)$ is at most
    $\frac{4 n \cdot \rank{D^{s}}}{\gamma\cdot k}$.
  \end{corollary}
  \begin{proof}
    Note that
    \begin{align*}
      \sum_{(u,v)\in E} \kh{ (\phi_{uv}^s)^T (D_{u}^{s})^{\dag} \phi_{uv}^s
      + (\phi_{uv}^s)^T (D_{v}^{s})^{\dag} (\phi_{uv}^s) } = &
      \sum_{u\in V} \sum_{u\sim v} (\phi_{uv}^s)^T (D_{u}^{s})^{\dag} \phi_{uv}^s \\
      = &
      \sum_{u\in V} \sum_{u\sim v} \trace{ (D_{u}^{s})^{\dag} (\phi_{uv}^s) (\phi_{uv}^s)^T } \\
      = &
      \sum_{u\in V} \trace{ (D_{u}^{s})^{\dag} \kh{ \sum_{u\sim v} (\phi_{uv}^s) (\phi_{uv}^s)^T } } \\
      = &
      \sum_{u\in V} \trace{ (D_{u}^{s})^{\dag} D_{u}^s } \\
      = & 
      \sum_{u\in V} \rank{ D_{u}^s }
      = \rank{D^s}.
    \end{align*}
    By Lemma~\ref{lem:wamlvgsc}, for edges with $s_{uv}\in(0,1)$,
    at least one of $(\phi_{uv}^s)^T (D_{u}^{s})^{\dag} \phi_{uv}^s$,
    $(\phi_{uv}^s)^T (D_{v}^{s})^{\dag} (\phi_{uv}^s)$ is $\geq \frac{\gamma\cdot k}{4n}$.
    Therefore there cannot be more than
    $\rank{D^{s}} / (\frac{\gamma\cdot k}{4n}) = \frac{4 n \cdot \rank{D^{s}}}{\gamma\cdot k}$ such
    edges.
  \end{proof}
}

\paragraph{Analysis of Algorithm~\ref{algo:hargd}.}{
  We first show that all edges that are deleted at Line~\ref{line:delete} have been downscaled during $\WAM$,
  and thus by Lemma~\ref{lem:wamlvgsc} have large leverage scores w.r.t. at least one of their endpoints.
  \begin{lemma}
    \label{lem:rhoproeprty}
    The $\rho$ found at Line~\ref{line:rho} has the property that for all edges $(u,v)\in E$
    such that $\norm{\phi_{uv}^s}^2 \leq \rho$ at the moment (that is, they will be deleted
    immediately at the next line), we have $s_{uv} < 1$.
  \end{lemma}
  \begin{proof}
    First note that before Line~\ref{line:rho} we have run $\WAM$ for
    $T = n^2 ( 4 n^2 k\log \alpha + \log( r/\ell) )$ iterations.
    Therefore there must be an edge $(u,v)$ with $s_{uv} \in (0,1)$
    whose weight has been halved at least $4 n^2 k \log \alpha + \log( r/\ell)$ times.
    Therefore for such an edge $(u,v)$ we have
    \begin{align*}
      \ell^s = \min_{(u',v')\in E, s_{u'v'} > 0} \norm{\phi_{u'v'}^s}^2 \leq
      & \norm{\phi_{uv}^{s}}^2 \\ \leq &
      \frac{1}{4^{4 n^2 k \log\alpha + \log(r/\ell)}}
      \norm{\phi_{uv}}^2  \\
      = & \frac{1}{\alpha^{8 n^2 k}} \cdot \frac{\ell^2}{r^2}\cdot \norm{\phi_{uv}}^2 \\
      \leq & \frac{1}{\alpha^{8n^2 k}}\cdot \min_{(u,v)\in E} \norm{\phi_{uv}}^2.
    \end{align*}
    Therefore for any $(u,v)$ whose $\norm{\phi_{uv}^s}^2$ falls in
    $[\ell^{s}, \ell^{s}\cdot \alpha^{8 n^2k})$, it must be the case that $s_{uv} < 1$.
    Also, note that the total number of different
    $\norm{\phi_{uv}^s}^2$'s and $\lambda_i$'s is at most $n^2 + nk \leq 2n^2 k$.
    Therefore, as long as $nk > 1$, there must be an interval
    $(\rho, \rho\cdot \alpha)\subseteq [\ell^{s}, \ell^{s} \cdot \alpha^{8 n^2k})$
    where no $\norm{\phi_{uv}^s}^2$ or $\lambda_i$ resides.
    Therefore the $\rho$ found at Line~\ref{line:rho} must be $< \ell^s\cdot \alpha^{8n^2 k}$.
    So for all edges $(u,v)\in E$
    such that $\norm{\phi_{uv}^s}^2 \leq \rho$ we must have $s_{uv} < 1$.
  \end{proof}

  We then show that the number of edges deleted at Line~\ref{line:delete} of $\HARGD$
  is small compared to the rank reduction.
  To this end let us fix an iteration of the while loop where
  we go to the ``else'' branch. Let $s$ denote
  the scaling returned by $\WAM$, and let $s'$ denote the scaling
  obtained after we delete the small weight edges at Line~\ref{line:delete}
  (but before we reset the weights of other edges at Line~\ref{line:reset}).

  \begin{lemma}
    \label{lem:nerd}
    The number of edges $(u,v)\in E$ with $\norm{\phi_{uv}^{s}}^2 \in (0, \rho]$ at Line~\ref{line:delete} is at most
    \begin{align*}
      \frac{8n}{\gamma\cdot k}\kh{\rank{D^s} - \rank{D^{s'}}}.
    \end{align*}
  \end{lemma}
  \begin{lemma}\label{lem:lmaxinv}
    We have
    \begin{align*}
      \lambdamax\kh{ \kh{D^{s'}}^{\dag} } \leq \frac{1}{4nk\cdot \rho}.
    \end{align*}
  \end{lemma}
  \begin{proof}
    Let $s'_{\min}  = \min \setof{ s'_{uv}: s'_{uv} \neq 0}$.
    Note that we have
    \begin{align*}
      (s'_{\min})^2 = \min_{s'_{uv} > 0} \frac{\norm{\phi_{uv}^{s'}}^2}{\norm{\phi_{uv}}^2}
      \geq \frac{\min_{s'_{uv} > 0} \norm{\phi_{uv}^{s'}}^2}{\max_{(u,v)\in E} \norm{\phi_{uv}}^2}
      \geq \frac{\rho \cdot \alpha}{r}.
    \end{align*}
    Let $F' = \setof{(u,v)\in E: s'_{uv} > 0}$. Then we have
    \begin{align*}
      D^{s'} \pgeq (s'_{\min})^2 D^F \pgeq
      \frac{\rho\cdot \alpha}{r}\cdot D^F,
    \end{align*}
    and as a result
    \begin{align}\label{eq:rra}
      \kh{ D^{s'} }^{\dag} \pleq \frac{r}{\rho\cdot \alpha}\cdot \kh{D^F}^{\dag}.
    \end{align}
    By Line~\ref{line:alpha}, we have
    \begin{align*}
      \alpha \geq 4nk\cdot r\cdot \max_{F'\subseteq E} \lambdamax\kh{ \kh{ D^{F'} }^{\dag} }
      \geq 4nk\cdot r\cdot \lambdamax \kh{ \kh{ D^F }^{\dag} }.
    \end{align*}
    Plugging this into~(\ref{eq:rra}) gives
    \begin{align*}
      \lambdamax\kh{ \kh{D^{s'}}^{\dag} } \leq \frac{1}{4nk\cdot \rho}.
    \end{align*}
    as desired.
  \end{proof}
  To prove Lemma~\ref{lem:nerd}, we first need
  to show that the rank of $D^{s}$ decreases by a certain amount after deleting small weight edges.
  \begin{lemma}\label{lem:ranki}
    Let $i$ be such that $\lambda_i \leq \rho < \rho \cdot \alpha \leq \lambda_{i+1}$.
    Then we have
    \begin{align*}
      \rank{D^s} - \rank{D^{s'}} \geq i.
    \end{align*}
  \end{lemma}
  \begin{proof}
    Let $f_1,\ldots,f_i \in \mathbb{R}^{nk}$ be a set of orthonormal eigenvectors corresponding to
    $\lambda_1,\ldots,\lambda_i$.
    Let $\Scal\subseteq \mathbb{R}^{nk}$ be the column space of $D^{s'}$.
    Then it suffices to show that after projecting $f_1,\ldots,f_i$ onto the subspace $\Scal^{\bot}$
    (the subspace orthogonal to $\Scal$),
    they are still linearly independent.
    To this end, we first note that for each $f_j$ where $j\leq i$, its projection onto $\Scal$ can be written as
    \begin{align*}
      \Pi_{\Scal} f_j = \kh{ D^{s'} }^{\dag/2} \kh{D^{s'}}^{1/2} f_j.
    \end{align*}
    Using the fact that $f_j$ is an eigenvector of $D^{s}$ corresponding to eigenvalue $\lambda_j \leq \rho$,
    we have
    \begin{align*}
      \norm{\kh{D^{s'}}^{1/2} f_j}^2 =
      & f_j^T D^{s'} f_j \\
      \leq & f_j^T D^{s} f_j
      \qquad \text{($D^{s'} \pleq D^{s}$)} \\
      = & \lambda_j \leq \rho.
    \end{align*}
    Then by Lemma~\ref{lem:lmaxinv},
    \begin{align}
      \norm{\kh{ D^{s'} }^{\dag/2} \kh{D^{s'}}^{1/2} f_j}^2 =
      & f_j^T \kh{D^{s'}}^{1/2} \kh{D^{s'}}^{\dag} \kh{D^{s'}}^{1/2} f_j \notag \\
      \leq & \lambdamax\kh{ \kh{D^{s'}}^{\dag} } f_j^T D^{s'} f_j \notag \\
      \leq & \frac{1}{4nk\cdot \rho}\cdot \rho = \frac{1}{4nk}.
      \label{eq:smallproj}
    \end{align}
    
    Let $d$ be the dimension of the subspace $\Scal^{\bot}$,
    and let $b_1,\ldots,b_d$ be an orthonormal basis of $\Scal^{\bot}$.
    Also, let $b_{d+1},\ldots,b_{nk}\in \mathbb{R}^{nk}$ be an orthonormal basis of $\Scal$.
    Then we can write each $f_j$ as a linear combination of $b_1,\ldots,b_{nk}$.
    That is, there exists a matrix $C\in\mathbb{R}^{i\times nk}$ such that
    \begin{align*}
      \begin{pmatrix}
        f_1^T \\ \vdots \\  f_i^T
      \end{pmatrix} =
      C
      \begin{pmatrix}
        b_1^T \\ \vdots \\ b_{nk}^T
      \end{pmatrix}.
    \end{align*}
    Since this is an orthogonal transformation, we have that the rows of $C$ are orthonormal,
    namely $C C^T = I$. If we write the $\tau^{\mathrm{th}}$ column of $C$ as $c_{\tau} \in \mathbb{R}^{i}$,
    then we have $\sum_{\tau=1}^{nk} c_{\tau} c_{\tau}^T = C C^T = I$.
    Moreover, we have
    \begin{align*}
      \sum_{\tau=d+1}^{nk} \norm{c_{\tau}}^2 = \sum_{j=1}^{i} \norm{\Pi_{\Scal} f_j}^2
      \leq i\cdot \frac{1}{4nk} \leq nk\cdot \frac{1}{4nk} = \frac{1}{4},
    \end{align*}
    where the first inequality follows from~(\ref{eq:smallproj}).
    Therefore
    \begin{align*}
      \trace{ \sum_{\tau = 1}^{d} c_\tau c_\tau^T } = & \trace{\sum_{\tau=1}^{nk} c_{\tau} c_{\tau}^T }
      - \trace{ \sum_{\tau=d+1}^{nk} c_{\tau} c_{\tau}^T } \geq i - \frac{1}{4}.
    \end{align*}
    On the other hand, we have
    \begin{align*}
      { \sum_{\tau = 1}^{d} c_\tau c_\tau^T } \pleq {\sum_{\tau=1}^{nk} c_{\tau} c_{\tau}^T },
    \end{align*}
    and thus all eigenvalues of $\sum_{\tau = 1}^{d} c_\tau c_\tau^T$ are at most $1$. These two imply that
    $\sum_{\tau = 1}^{d} c_\tau c_\tau^T$ must be of full rank $i$.
    As a result, $\Pi_{\Scal^{\bot}} f_1, \ldots, \Pi_{\Scal^{\bot}} f_i$ are linearly independent.
  \end{proof}

  \begin{proof}[Proof of Lemma~\ref{lem:nerd}]
    By Lemma~\ref{lem:ranki} we know that the rank reduction in $D^{s}$ is at least $i$,
    where $i$ is such that $\lambda_i \leq \rho < \rho\cdot \alpha \leq \lambda_{i+1}$.
    Let $f_1,\ldots,f_{p}$ be a set of orthonormal eigenvectors corresponding to $\lambda_1,\ldots,\lambda_{p}$.
    By Lemma~\ref{lem:rhoproeprty}, we know that for any $(u,v)\in E$ with $\norm{\phi_{uv}^{s}}^2 \in (0,\rho]$,
    we have $s_{uv} \in (0,1)$.
    Let $F \defeq \setof{(u,v)\in E: \norm{\phi_{uv}^{s}}^2 \in (0,\rho]}$.
    As the initial scaling $s^{0}$ that we send to $\WAM$ always has range $\setof{0,1}$,
    we have by Lemma~\ref{lem:wamlvgsc} that for each $e\in F$,
    its leverage score is at least $\frac{\gamma\cdot k}{4n}$ w.r.t. at least one of its endpoints.
    
    For each edge $(u,v)$, define two $nk$-dimensional vectors
    $x_{uv}^u$ and $x_{uv}^v$, such that $(x_{uv}^u)_u = \phi_{uv}^s$ and $(x_{uv}^{u})_w = 0$ for all $w\neq u$,
    and $(x_{uv}^v)_v = \phi_{uv}^s$ and $(x_{uv}^v)_w = 0$ for all $w\neq v$.
    Then for any $e\in F$ we have
    \begin{align*}
      (x^{u}_{uv})^T \kh{D^s}^{\dag} x^u_{uv} +
      (x^{v}_{uv})^T \kh{D^s}^{\dag} x^v_{uv} \geq \frac{\gamma\cdot k}{4n}.
    \end{align*}
    Now if we restrict to only the eigenvalues smaller than $\rho$, we have
    \begin{align}
      & (x^{u}_{uv})^T \kh{\sum_{j=1}^{i} \lambda_j f_j f_j^T}^{\dag} x^u_{uv} +
      (x^{v}_{uv})^T \kh{\sum_{j=1}^{i} \lambda_j f_j f_j^T}^{\dag} x^v_{uv} \notag \\
      = &
      (x^{u}_{uv})^T \kh{\kh{D^{s}}^{\dag}  - \sum_{j=i+1}^{p} \frac{1}{\lambda_j} f_j f_j^T} x^u_{uv} +
      (x^{v}_{uv})^T \kh{\kh{D^{s}}^{\dag}  - \sum_{j=i+1}^{p} \frac{1}{\lambda_j} f_j f_j^T} x^v_{uv}.
      \label{eq:lvgbot}
    \end{align}
    Notice that for all $e\in F$ we have
    $\norm{x_{uv}^u}^2 = \norm{x_{uv}^{v}}^2 = \norm{\phi_{uv}^{s}}^2 \leq \rho$. And thus
    for any $j\geq i+1$,
    \begin{align*}
      & \frac{1}{\lambda_j} \ip{x_{uv}^{u}}{f_j}^2 \leq \frac{\rho}{\lambda_j}
      \leq \frac{1}{\alpha}, \\
      & \frac{1}{\lambda_j} \ip{x_{uv}^{v}}{f_j}^2 \leq \frac{\rho}{\lambda_j}
      \leq \frac{1}{\alpha},
    \end{align*}
    where both second inequalities follow from $\lambda_{j}\geq \rho\cdot \alpha$ for $j\geq i+1$.
    Now by Line~\ref{line:alpha}, $\alpha \geq 8n^4 k$.
    Therefore combining this with~(\ref{eq:lvgbot}) we have
    \begin{align*}
      & \sum_{(u,v)\in F}(x^{u}_{uv})^T \kh{\sum_{j=1}^{i} \lambda_j f_j f_j^T}^{\dag} x^u_{uv} +
      (x^{v}_{uv})^T \kh{\sum_{j=1}^{i} \lambda_j f_j f_j^T}^{\dag} x^v_{uv} \\
      \geq &
      \sum_{(u,v)\in F}(x^{u}_{uv})^T \kh{D^{s}}^{\dag} x^u_{uv} +
      (x^{v}_{uv})^T \kh{D^{s}}^{\dag} x^v_{uv}
      - 2nk\cdot \frac{1}{8n^4 k} \\
      \geq & |F|\cdot \kh{ \frac{\gamma\cdot k}{4n} - \frac{1}{4n^3} }
      \geq |F|\cdot \frac{\gamma\cdot k}{8n}.
    \end{align*}
    On the other hand, we have
    \begin{align*}
      & \sum_{(u,v)\in F}(x^{u}_{uv})^T \kh{\sum_{j=1}^{i} \lambda_j f_j f_j^T}^{\dag} x^u_{uv} +
      (x^{v}_{uv})^T \kh{\sum_{j=1}^{i} \lambda_j f_j f_j^T}^{\dag} x^v_{uv} \\
      = & \trace{\kh{\sum_{j=1}^{i} \lambda_j f_j f_j^T}^{\dag/2}
        \kh{ \sum_{(u,v)\in F} x_{uv}^{u} \kh{ x_{uv}^{u} }^T +
          x_{uv}^{v} \kh{ x_{uv}^{v} }^T
        }
        \kh{\sum_{j=1}^{i} \lambda_j f_j f_j^T}^{\dag/2}
      } \\
      \leq  &
      \trace{\kh{\sum_{j=1}^{i} \lambda_j f_j f_j^T}^{\dag/2}
        \kh{ \sum_{(u,v)\in E} x_{uv}^{u} \kh{ x_{uv}^{u} }^T +
          x_{uv}^{v} \kh{ x_{uv}^{v} }^T
        }
        \kh{\sum_{j=1}^{i} \lambda_j f_j f_j^T}^{\dag/2}
      } \\
      = &
      \trace{\kh{\sum_{j=1}^{i} \lambda_j f_j f_j^T}^{\dag/2}
        D^{s}
        \kh{\sum_{j=1}^{i} \lambda_j f_j f_j^T}^{\dag/2}
      } \\
      = &
      \trace{\kh{\sum_{j=1}^{i} \lambda_j f_j f_j^T}^{\dag/2}
        \kh{ \sum_{j=1}^{p} \lambda_j f_j f_j^T }
        \kh{\sum_{j=1}^{i} \lambda_j f_j f_j^T}^{\dag/2}
      }
      \qquad \text{(spectral decomposition)}
      \\
      = &
      \rank{\sum_{j=1}^{i} \lambda_j f_j f_j^T} \leq i.
    \end{align*}
    Combining the above two inequalities we have
    \begin{align*}
      |F| \leq \frac{8n i}{\gamma\cdot k} \leq
      \frac{8n }{\gamma\cdot k}\kh{ \rank{D^{s}} - \rank{D^{s'}}},
    \end{align*}
    as desired.
  \end{proof}
}

\begin{proof}[Proof of Theorem~\ref{thm:hargd}]
  Since in each while loop iteration we zero out the weight of at least one edge,
  the while loop must terminate in $O(n^2)$ iterations, and thus
  the algorithm terminates in finite time.
  Also by the termination condition of the while loop,
  the resulting graph $G^s$ is $\gamma$-almost regular.
  Finally, Corollary~\ref{cor:nofs<1} and Lemma~\ref{lem:nerd} give
  the desired bounds on the number of downscaled/deleted edges.
\end{proof}

\let\lambda\oldlambda

\section{Almost regular expander decomposition}\label{sec:hared}

In this section, we show that every matrix-weighted graph that is sufficiently dense
can be made into an almost regular ``expander'' graph, by downscaling a small number of edges.
This can be seen as a matrix-weighted analog of the celebrated expander decomposition
of scalar-weighted graphs~\cite{GoldreichR99,KannanVV04,SpielmanT04}.

We first give the definition of matrix-weighted expanders.
Consider a $k\times k$ matrix-weighted graph $G = (V,E)$ with edge weights $\phi_{uv} \phi_{uv}^T$.
Let $\lambda_1 \leq \ldots \lambda_r$, where $r = \rank{D}$,
be the nontrivial eigenvalues (Definition~\ref{def:nontrivial}) of its normalized Laplacian
$N = D^{\dag/2} L D^{\dag/2}$.
Let $f_1,\ldots,f_{r}$ be a corresponding set of orthonormal, nontrivial eigenvectors,
which by definition of nontriviality satisfies
\begin{align*}
  (f_i)_u\,\bot\,\mathrm{ker}(D_u), \ \forall i\in [r], u\in V.
\end{align*}

\begin{definition}[Almost regular matrix-weighted expanders]
  \label{def:hde}
  For $\gamma\geq 1$, $\zeta\in (0,1)$, and $\psi\geq 1$,
  we say $G$ is a $(\gamma,\zeta,\psi)$-almost regular expander if
  \begin{enumerate}
    \item ($\gamma$-almost regularity) For every vertex $u$ and every incident edge $(u,v)\in E$, we have
      \begin{align*}
        \phi_{uv}^T D_u^{\dag} \phi_{uv} \leq \frac{\gamma\cdot k}{n}.
      \end{align*}
    \item ($(\zeta,\psi)$-expander) for every edge $(u,v)\in E$ we have
      \begin{align}\label{eq:explvg}
        \kh{D^{\dag/2} b_{uv} }^T
        \kh{ \sum_{\lambda_i\in(0,\zeta]} \frac{1}{\lambda_i} f_i f_i^T }
        \kh{ D^{\dag/2} b_{uv} } \leq \frac{\psi \cdot k^2}{n^2}
      \end{align}
  \end{enumerate}
\end{definition}

\paragraph{Notations for rescaled graphs.}{
  In describing our main result in this section,
  we will use similar
  notations for rescaled graphs as in Section~\ref{sec:hargd}.
  Namely, for a $k\times k$ matrix-weighted graph $G = (V,E)$ with edge weights $\phi_{uv} \phi_{uv}^T$'s
  and a scaling $s: E\to [0,1]$,
  we will write $G^s$ to denote the graph obtained from $G$ by rescaling each edge $(u,v)$'s weight
  to $(s_{uv} \phi_{uv}) (s_{uv} \phi_{uv})^T$.
  For ease of presentation we will use the superscript $s$ when dealing with vectors and matrices associated with $G^s$.
  For instance, we write $\phi_{uv}^s = s_{uv} \phi_{uv}$,
  $D^s = \sum_{u\sim v} s_{uv}^2 (e_{u\la v} e_{u\la v}^T + e_{v\la u} e_{v\la u}^T)$,
  and $L^s = \sum_{u\sim v} s_{uv}^2 b_{uv} b_{uv}^T$.
  Analogously, for a subset of edges $F\subseteq E$, we also use the superscript $F$
  when dealing with matrices associated with the induced subgraph $G[F]$,
  and thus, for instance, $D^F = \sum_{(u,v)\in F} (e_{u\la v} e_{u\la v}^T + e_{v\la u} e_{v\la u}^T)$.

  We will write $\lambda_1^s \leq \ldots \leq \lambda_{\rank{D^{s}}}^{s}$ to denote the
  nontrivial eigenvalues of the normalized Laplacian $N^{s} = (D^{s})^{\dag/2} L^{s} (D^{s})^{\dag/2}$ of $G^s$,
  and write $f_1^s,\ldots,f_{\rank{D^{s}}}^{s}$ to denote a corresponding set of orthonormal, nontrivial
  eigenvectors.
  We then define a function $H_{\zeta}^{s} : E \to \mathbb{R}$ by
  \begin{align*}
    H_{\zeta}^{s}(u,v) \defeq
    \kh{ \kh{D^s}^{\dag/2} b_{uv}^s }^T
    \kh{ \sum_{\lambda_i^s \in (0,\zeta]} \frac{1}{\lambda_i^s} f_i^s (f_i^s)^T }
    \kh{ \kh{D^s}^{\dag/2} b_{uv}^s }.
  \end{align*}
  We also define another function $R^s: E\to \mathbb{R}$ by
  \begin{align*}
    R^{s}(u,v) \defeq
    & (e_{u\la v}^{s})^T (D^s)^{\dag} e_{u\la v}^s + (e_{v\la u}^s)^T (D^s)^{\dag} e_{v\la u}^s \\ =
    & (\phi_{uv}^s)^T (D_u^s)^{\dag} \phi_{uv}^s + (\phi_{uv}^s)^T (D_v^s)^{\dag} \phi_{uv}^s.
  \end{align*}
  Then, for our goal, it suffices to find a scaling $s$ such that
  for all $(u,v)\in E$ we have
  \begin{align*}
    R^{s}(u,v) \leq \frac{\gamma\cdot k}{n} \qquad \text{and} \qquad
    H_{\zeta}^{s}(u,v) \leq \frac{\psi\cdot k^2}{n^2}.
  \end{align*}
}

Our main result here is a deterministic algorithm
for finding a large almost regular expander subgraph within any given graph that is sufficiently dense.

\begin{theorem}\label{thm:hared}
  There is a deterministic algorithm $\HED$ that,
  given any $k\times k$ matrix-weighted graph $G = (V,E)$ with edge weights
  $\phi_{uv} \phi_{uv}^T$'s, any $\gamma \geq 1$, $\zeta\in (0,1)$,
  and
  \begin{align*}
    \psi \geq \frac{1024 \gamma^2}{(1 - \zeta)^2},
  \end{align*}
  outputs a scaling $s: E \to [0,1]$ such that
  \begin{enumerate}
    \item The rescaled graph $G^s$ is a $(\gamma,\zeta,\psi)$-almost regular expander.
    \item The number of edges $(u,v)\in E$ with $s_{uv} < 1$ is at most
      \begin{align*}
        \frac{1000 n^2}{\gamma} + \frac{100000 n^2 \gamma^2 k^2}{\psi (1 - \zeta)^2}.
      \end{align*}
  \end{enumerate}
  The algorithm terminates in finite time.
\end{theorem}

Note that similar to Section~\ref{sec:hargd},
our goal is to prove the existence of such a scaling for any given weights $\phi_{uv} \phi_{uv}^T$'s
that may potentially have {\em infinite} precision, so we only focus on designing an algorithm
that terminates in finite time, as opposed to giving an explicit bound on its running time.

We give the pseudocodes of our algorithm $\HED$ in Algorithm~\ref{algo:hed}
and its subroutine $\WTM$ in Algorithm~\ref{algo:wam2}.
At a high level, in $\HED$, we first try to eliminate edges with large $H^s_{\zeta}(u,v)$ values by iteratively
halving the weight of any such edge for some large number of iterations,
while simultaneously maintaining the almost regularity of the graph,
by also iteratively halving the weight of any edge with large $R^{s}(u,v)$ value.
This iterative process is essentially achieved by $\WTM$.
Then after $\WTM$ terminates, either (i) all edges have small $H^s_{\zeta}(u,v)$ and $R^{s}(u,v)$ values,
in which case we are done, or (ii) we can strictly increase\footnote{This statement is a slight simplification of
our actual analysis, but helps with getting intuition.}
the number of nontrivial (Definition~\ref{def:nontrivial})
zero eigenvalues of the normalized Laplacian by removing a small number of edges
that have low weights after $\WTM$.
Since we maintain the almost regularity of the graph throughout,
by Theorem~\ref{thm:fewsmall}, the total number of nontrivial zero eigenvalues is small,
therefore we will only repeat (ii) for a limited number of times.

Since we want to ensure that the number of rescaled edges is small in $\WTM$,
we maintain the invariant
that any edge that has been downscaled satisfies that either $H^{s}_{\zeta}(u,v)$ is large
or $R^s(u,v)$ is large,
by iteratively {\em doubling} the weight of any edge with both values small.
In order to make sure that there is consistent progress in this process,
we set up two thresholds $\psi_1 < \psi_2$ for $H^{s}_{\zeta}(u,v)$
and also two thresholds $\gamma_1 < \gamma_2$ for $R^{s}(u,v)$.
Then we halve the weight of an edge $(u,v)$ if either $H^{s}_{\zeta}(u,v) > \frac{\psi_2 k^2}{n^2}$
or $R^{s}(u,v) > \frac{\gamma_2 k}{n}$, but only double its weight if {\em both}
$H^{s}_{\zeta}(u,v) < \frac{\psi_1 k^2}{n^2}$ {\em and} $R^{s}(u,v) < \frac{\gamma_1 k}{n}$.
To measure the progress in this elaborate process,
we will need to use some global quantities as our potential functions.
Specifically, we will design our potential functions based on
certain variants of determinant, which we define below.

\begin{definition}
  For a positive semidefinite matrix $M\in\mathbb{R}^{N\times N}$ with eigenvalues
  \begin{align*}
    0 = \lambda_1 = \dots = \lambda_{z} < \lambda_{z+1} \leq \ldots \leq \lambda_{N},
  \end{align*}
  define
  \begin{align*}
    \detp(M) \defeq \prod_{\lambda_i > 0} \lambda_i.
  \end{align*}
  For an integer $\ell$, define
  \begin{align*}
    \detl(M) \defeq \prod_{i=z+1}^{z+\ell} \lambda_i.
  \end{align*}
\end{definition}

We will need the following variational characterization of $\detl$,
whose proof is deferred to Appendix~\ref{sec:aphared}.

\begin{lemma}\label{lem:potentialfrac}
  Let $f_1,\ldots,f_{\ell}$ be a set of orthonormal eigenvectors corresponding to
  the $\ell$ smallest nonzero eigenvalues of $N$,
  and let $F =
  \begin{pmatrix}
    f_1 & \ldots & f_{\ell}
  \end{pmatrix}\in\mathbb{R}^{nk\times \ell}$.
  Let $\Xcal\subseteq \mathbb{R}^{nk\times \ell}$ be the set of all full-rank $nk\times \ell$ matrices
  whose columns are in the range of $N$. Then
  \begin{align*}
    \detl(N) = \min_{X\in\Xcal}
    \frac{ \det\kh{ X^T L X } }{ \det\kh{X^T D X} },
  \end{align*}
  and a minimizer can be obtained by letting
  $X^* = D^{\dag/2} F$.
\end{lemma}

The following matrix determinant lemma will be helpful in our analysis of the potential functions.
\begin{lemma}[\cite{Harville1998matrix}]
  For an invertible square matrix $M\in\mathbb{R}^{d\times d}$ and vectors $x,y\in\mathbb{R}^{d}$,
  we have
  \begin{align*}
    \det(M + xy^T) = (1 + y^T M^{-1} x) \det(M).
  \end{align*}
\end{lemma}

\quad

\begin{algorithm}[H]
  \label{algo:wam2}
  \caption{$\WTM(G,\gamma_1,\gamma_2, \zeta, \psi_1, \psi_2, \alpha ,s^0)$}
  \Input{
    $G$: a $k\times k$ matrix-weighted graph $G = (V,E)$ with edge weights $\phi_{uv} \phi_{uv}^T$. \\
    $\gamma_1 < \gamma_2, \zeta, \psi_1 < \psi_2$: parameters for almost regularity and expander. \\
    $\alpha > 0$: a gap parameter. \;
    $s^0$: an initial scaling, mapping from $E\to [0,1]$, s.t.
    $G^{s_0}$ is $\gamma_1/2$-almost regular.
  }
  \Output{
    $s$: a scaling, mapping from $E\to [0,1]$. \\
    $\rho$: a threshold.
  }
  Let $s\gets s^0$. \;
  \While{$\mathrm{true}$}{
    Compute the eigenvalues $\lambda_1 \leq \ldots \leq \lambda_{nk}$ of $(D^s)^{\dag/2} L^s (D^s)^{\dag/2}$. \;
    Compute the eigenvalues $\mu_1 \leq \ldots \leq \mu_{nk}$ of $D^s$.\;
    \If{{\rm no $\lambda_i$ is in the range $(0,\zeta]$}}{
      Return $s$ and $\rho$, and halt. \;
    }
    Let $\lambda_{\min} = \min_{\lambda_i \in (0, \zeta]} \lambda_i$.\;
    Find the smallest $\rho \geq \lambda_{\min}$ s.t.
    all %
    $(s_{uv}/s_{uv}^0)^2$, $\mu_i$, and
    $\lambda_i$ fall in $\mathbb{R}\setminus (\rho,\rho\cdot \alpha)$.
    \label{line:rhofind}
    \;
    \If{$\rho < 1$}{
      Return $s$ and $\rho$, and halt.\;
    }
    \ElseIf{$\nexists (u,v)\in E$ {\rm with}
    $H^{s}_{\zeta}(u,v) > \frac{\psi_2\cdot k^2}{n^2}$}{
      Return $s$ and $\rho$, and halt.\;
    }
    \Else{
      Let $s_{uv} \gets s_{uv} / 2$ for an arbitrary edge $(u,v)$ with $H^{s}_{\zeta}(u,v) > \frac{\psi_2\cdot k^2}{n^2}$.
      \label{line:wackH} \;

      \While{\label{line:while1s}$\exists (u,v) \in E$ {\rm with}
      $R^{s}(u,v) > \frac{\gamma_2 \cdot k}{n}$}{
        Let $s_{uv} \gets s_{uv} / 2$ for an arbitrary such edge $(u,v)$.
        \label{line:while1t}
      }
      \While{\label{line:while2s} $\exists (u,v): s_{uv} < s_{uv}^0$ {\rm satisfying both} $R^{s}(u,v) < \frac{\gamma_1 \cdot k}{n}$
      {\rm and} $H^{s}_{\zeta}(u,v) < \frac{\psi_1 \cdot k^2}{n^2}$}{
        Let $s_{uv}\gets s_{uv}\cdot 2$ for an arbitrary such edge $(u,v)$.
        \label{line:while2t}
      }
    }
  }
\end{algorithm}

\quad

\begin{algorithm}[H]
  \label{algo:hed}
  \caption{$\HED(G,\gamma, \zeta, \psi)$}
  \Input{
    $G$: a $k\times k$ matrix-weighted graph $G = (V,E)$ with edge weights $\phi_{uv} \phi_{uv}^T$. \\
    $\gamma$, $\zeta$, $\psi$: parameters for almost regularity and expander.
  }
  \Output{
    $s$: a scaling, mapping from $E\to [0,1]$.
  }
  $r_{\phi} \gets \max_{(u,v)\in E} \norm{\phi_{uv}}^2$. \;
  Let $\alpha_{1} \gets \max_{F\subseteq E} \lambdamax\kh{\kh{D^{F}}^{\dag}}$. \;
  Let $\alpha_{2} \gets \max_{F\subseteq E} \lambdamax\kh{\kh{D^{\dag/2} L^{F} D^{\dag/2}}^{\dag}}$. \;
  Initially, let $s_e\gets 1$ for all $e\in E$.\;
  \While{$\rank{D^s} > 0$}{
    Let $G' = (V,E')$ where $E' = \setof{e: s_e > 0}$. \;
    Let $s' \gets \HARGD(G',\gamma/32)$, and then $s_e\gets s'_e$ for all $e\in E'$.
    \label{line:argd}
    \;
    $s'_{\min} \gets \min_{s'_{uv}> 0} s'_{uv}$. \label{line:smin} \;
    Let $\alpha \gets \max\setof{8 n^4 k, 128nk^3 \gamma \cdot r_{\phi}\cdot \alpha_{1} / (s'_{\min})^2,
    128nk^3\gamma\cdot\alpha_2 / (s'_{\min})^2}$. \;
    $(s,\rho)\gets \WTM(G, \gamma/16, \gamma, \zeta, \psi / 1024, \psi, \alpha, s)$.
    \label{line:wtm}
    \;
    \If{$\forall (u,v) \in E$ {\rm we have}
    $R^{s}(u,v)\leq \frac{\gamma\cdot k}{n}$ and $H_{\zeta}^{s}(u,v)\leq \frac{\psi\cdot k^2}{n^2}$}{
      Return $s$ and halt.
    }
    \Else{
      Set $s_{uv}\gets 0$ for all $(u,v)\in E$ with $(s_{uv} / s'_{uv})^2 \leq \rho$.
      \label{line:delsmall}
      \;
      Set $s_{uv}\gets 1$ for all $(u,v)\in E$ with $s_{uv} > 0$. \label{line:resetw}
    }
  }
  \Return $s$.
\end{algorithm}

\quad

\subsection{Analysis of Algorithm~\ref{algo:wam2}}

For a fixed scaling $s: E\to [0,1]$, let $\ell$ be the number of eigenvalues
of the normalized Laplacian $N^{s}$ of $G^{s}$ that are in the range $(0,\zeta]$.
Let $\ell_0 = \ceil{ \frac{\gamma_2 k^2}{(1-\zeta)^2} }$, which, by Theorem~\ref{thm:fewsmall},
is an upper bound
on $\ell$ when the graph $G^{s}$ is $\gamma_2$-almost regular.
Then we consider the following potential function: %
\begin{align}\label{eq:potential}
  \Upsilon(s) \defeq
  \begin{cases}
    \zeta^{\ell_0 - \ell} \det\nolimits_{\ell}(N^{s}) & \ell\leq \ell_0 \\
    \det_{\ell_0}(N^{s}) & \ell > \ell_0.
  \end{cases}
\end{align}
Thus, $\Upsilon(s)$ is always a product of exactly $\ell_0$ numbers between $(0,\zeta]$.
It is not hard to see the following alternative form of $\Upsilon(s)$, which will be helpful for analyzing it.
\begin{proposition}\label{prop:altups}
  Let $\lambda_1,\ldots,\lambda_{\ell_0}$ be the smallest $\ell_0$ nonzero eigenvalues of $N^{s}$.
  Then
  \begin{align*}
    \Upsilon(s) = \prod_{i=1}^{\ell_0} \min\setof{\lambda_i,\zeta}.
  \end{align*}
\end{proposition}
\begin{corollary}[of Proposition~\ref{prop:altups}]
  \label{cor:altups}
  For any $\ell' \leq \ell_0$, we have
  $\Upsilon(s) \leq \det_{\ell'}(N^{s}) \zeta^{\ell_0 - \ell'}$.
\end{corollary}

Since we always maintain the $\gamma_2$-regularity,
we will also need to consider the determinant of the degree matrix,
$\detp(D)$, as a potential function.
We show that starting from an almost-regular graph,
$\detp(D)$ can only decrease at a limited speed when edges are downscaled.
The proof of the lemma below is deferred to Appendix~\ref{sec:aphared}.

\begin{lemma}\label{lem:decmost}
  Let $G = (V,E)$ be a $\gamma$-almost regular graph.
  Let $s: E\to (0,1]$ be a strictly positive scaling,
  and let $S \defeq \prod_{e\in E} \frac{1}{s_e}$.
  Then we have
  \begin{align*}
    \detp(D^s) \geq \kh{1 - \frac{2 \gamma k}{n}}^{2 \log S} \detp(D).
  \end{align*}
\end{lemma}

As a result of the above lemma,
we then show that the while loop at Lines~\ref{line:while1s}-\ref{line:while1t}
of $\WTM$ will terminate after a bounded number of iterations.
The proof of the following lemma is also deferred to Appendix~\ref{sec:aphared}.

\begin{lemma}\label{lem:tt1}
  Suppose the input to $\WTM$ satisfies that
  $G^{s_0}$ is $\gamma_1/2$-regular
  and $\gamma_2 \geq 16 \gamma_1$.
  Consider the first $t$ iterations of the outermost while loop.
  Then the total number of iterations executed so far by the inner while loop
  at Lines~\ref{line:while1s}-\ref{line:while1t} is at most $2t$.
\end{lemma}

We now prove our key lemma, which shows that our potential
$\Upsilon$ will decrease at least as fast as a geometric series with rate bounded away from $1$.
As a result, the smallest nonzero eigenvalue of the normalized Laplacian must also
decrease at a steady rate.
This implies that eventually we will be able to find a $\rho < 1$ at Line~\ref{line:rhofind},
and therefore terminate the outermost while loop in finite time.

\begin{lemma}
  Suppose the input to $\WTM$ satisfies that
  $G^{s_0}$ is $\gamma_1/2$-regular,
  $\gamma_2 \geq 16 \gamma_1$,
  $\psi_2 \geq 1024 \psi_1$,
  and
  \begin{align*}
    \psi_2 \geq \frac{1024 \gamma_2^2}{(1 - \zeta)^2}.
  \end{align*}
  Consider the first $t$ iterations of the outermost while loop,
  and let $s^1$ be the scaling obtained at the end of the $t^{\mathrm{th}}$ iteration.
  Then
  \begin{align*}
    \frac{\Upsilon(s^1)}{\Upsilon(s^0)}
    \leq
    \kh{ 1 - \frac{\psi_2 k^2}{16 n^2} }^t.
  \end{align*}
\end{lemma}
\begin{proof}
  We first show that the potential reduces by a certain amount at Line~\ref{line:wackH}.

  \begin{claim}\label{lem:line14}
    Let the scalings before and after one execution of Line~\ref{line:wackH} be $s$ and $s'$ respectively.
    Then we have
    \begin{align*}
      \frac{\Upsilon(s')}{\Upsilon(s)} \leq 1 - \frac{\psi_2 k^2}{4 n^2}.
    \end{align*}
  \end{claim}
  \begin{proof}[Proof of Claim~\ref{lem:line14}]
    Let $\ell$ be the number of eigenvalues of $N^s$ that are between $(0,\zeta]$.
    Since $G^{s}$ is $\gamma_2$-almost regular, we have $\ell \leq \ell_0$ by Theorem~\ref{thm:fewsmall}.
    Thus $\Upsilon(s) = \detl(N^s) \zeta^{\ell_0 - \ell}$.
    Also, by Corollary~\ref{cor:altups}, we have
    $\Upsilon(s') \leq \detl(N^{s'}) \zeta^{\ell_0 - \ell}$.
    Therefore, it suffices to show that
    \begin{align}
      \frac{\detl(N^{s'})}{\detl(N^{s})} \leq 1 - \frac{\psi_2 k^2}{4 n^2}.
    \end{align}

    We then do so by considering the variational characterization $\detl$
    from Lemma~\ref{lem:potentialfrac}.
    Let $X = (D^s)^{\dag/2} F$ and $X' = (D^{s'})^{\dag/2} F'$
    be the optimal matrix that minimizes the variational characterization 
    of $\detl(N^s)$ and $\detl(N^{s'})$ respectively.
    Here $F$ and $F'$ are both $nk\times \ell$ matrices whose columns
    are bottom nonzero eigenvectors of $N^{s}$ and $N^{s'}$ respectively.

    It then suffices to show that the fraction in the characterization
    reduces by much after the execution of Line~\ref{line:wackH},
    even if we do not switch from $X$ to $X'$,
    since switching to the latter can only decrease the potential function.
    Namely, since by the optimality of $X'$ we have
    \begin{align*}
      \frac{ \det\kh{ X^T L^{s'} X } }{ \det\kh{X^T D^{s'} X} } \geq
      \frac{ \det\kh{ (X')^T L^{s'} (X') } }{ \det\kh{(X')^T D^{s'} (X')} },
    \end{align*}
    it suffices to show
    \begin{align}\label{eq:sufficesss}
      \kh{ \frac{ \det\kh{ X^T L^{s'} X } }{ \det\kh{X^T D^{s'} X} } } \Big /
      \kh{ \frac{ \det\kh{ X^T L^{s} X } }{ \det\kh{X^T D^{s} X} } } \leq
      1 - \frac{\psi_2 k^2}{4 n^2}.
    \end{align}
    For the numerator we have by matrix determinant lemma
    \begin{align}
      \frac{ \det\kh{ X^T L^{s'} X } }
      { \det\kh{ X^T L^{s} X } }
      = & 1 - \frac{3}{4}\cdot (b_{uv}^s)^T X \kh{X^T L^s X}^{-1} X^T (b_{uv}^s) \notag \\
      = & 1 - \frac{3}{4}\cdot (b_{uv}^s)^T \kh{D^{s}}^{\dag/2} F
      \kh{ F^T \kh{ D^{s} }^{\dag/2} L^s \kh{D^{s}}^{\dag/2} F}^{-1} F^T \kh{D^{s}}^{\dag/2} b_{uv}^s \notag \\
      = & 1 - \frac{3}{4}\cdot H_{\zeta}^{s}(u,v)
      \leq 1 - \frac{3\psi_2 k^2}{4n^2},
      \label{eq:detnum}
    \end{align}
    where the last equality follows from that
    $F$'s columns are eigenvectors corresponding to all eigenvalues of $N^{s}$ that are between $(0,\zeta]$.
    For the denominator, we have, again by matrix determinant lemma,
    \begin{align*}
      \frac{ \det\kh{ X^T D^{s'} X } }
      { \det\kh{ X^T D^{s} X } } \geq &
      { 1 - \frac{3}{4}(e_{u\la v}^{s})^T X (X^T D^s X)^{-1} X^T e_{u\la v}^{s} -
      \frac{3}{4}(e_{v\la u}^{s})^T X (X^T D^s X)^{-1} X^T e_{v\la u}^{s} } \\ = &
      { 1 - \frac{3}{4}(e_{u\la v}^{s})^T X  X^T e_{u\la v}^{s} -
      \frac{3}{4}(e_{v\la u}^{s})^T X  X^T e_{v\la u}^{s} }
      \qquad
      \text{(as $X^T D^{s} X = F^T F = I$)} \\ = &
      1 - \frac{3}{4}(e_{u\la v}^{s})^T (D^{s})^{\dag/2} F F^T (D^{s})^{\dag/2} (e_{u\la v}^{s}) -
      \frac{3}{4}(e_{v\la u}^{s})^T (D^{s})^{\dag/2} F F^T (D^{s})^{\dag/2} (e_{v\la u}^{s}) \\ = &
      1 - \frac{3}{4}\norm{ F^T(u) (D^{s}_u)^{\dag/2} \phi_{uv} }^2 -
      \frac{3}{4}\norm{ F^T(v) (D^{s}_v)^{\dag/2} \phi_{uv} }^2,
    \end{align*}
    where in the last equality we let $F(u)$ be the $u^{th}$ row block of $F$.
    Since $G^{s}$ is $\gamma_2$-almost regular,
    we have
    $\norm{(D^s_u)^{\dag/2} \phi_{uv}}^2 \leq \frac{\gamma_2\cdot k}{n}$ and
    $\norm{(D^s_v)^{\dag/2} \phi_{uv}}^2 \leq \frac{\gamma_2\cdot k}{n}$.
    By Lemma~\ref{lem:fsu2}, we have
    $\lambdamax(F(u) F(u)^T) \leq \frac{\gamma_2 k}{(1 - \zeta)^2 n}$, and
    $\lambdamax(F(v) F(v)^T) \leq \frac{\gamma_2 k}{(1 - \zeta)^2 n}$, and therefore
    \begin{align*}
      & \norm{F^T(u) (D^s_u)^{\dag/2} \phi_{uv}}^2 \leq \frac{\gamma_2^2\cdot k^2}{(1 - \zeta)^2 n^2} \\
      & \norm{F^T(v) (D^s_v)^{\dag/2} \phi_{uv}}^2 \leq \frac{\gamma_2^2\cdot k^2}{(1 - \zeta)^2 n^2}.
    \end{align*}
    This give us
    \begin{align}\label{eq:detdom}
      \frac{ \det\kh{ X^T D^{s'} X } }
      { \det\kh{ X^T D^{s} X } } \geq & 1 - \frac{3\gamma_2^2\cdot k^2}{2(1 - \zeta)^2 n^2}.
    \end{align}
    (\ref{eq:detnum}),(\ref{eq:detdom}) coupled with $\psi_2 \geq \frac{1024 \gamma_2^2}{(1 - \zeta)^2}$
    imply~(\ref{eq:sufficesss}), which finishes the proof of the claim.
  \end{proof}

  We next show that during the first while loop at Lines~\ref{line:while1s}-\ref{line:while1t},
  the potential function cannot increase much.

  \begin{claim}\label{claim:firstwhile}
    Consider a fixed iteration of the outermost while loop.
    Suppose in this iteration of the outer while loop, the total number of iterations executed by the first inner while loop
    at Lines~\ref{line:while1s}-\ref{line:while1t} is $t_1$.
    Let $s,s'$ be the scalings
    before and after the while loop respectively.
    Then
    \begin{align*}
      \frac{\Upsilon(s')}{\Upsilon(s)}
      \leq
      \kh{ 1 - \frac{96 \gamma_2^2 k^2}{(1-\zeta)^2 n^2} }^{-t_1}.
    \end{align*}
  \end{claim}
  \begin{proof}[Proof of Claim~\ref{claim:firstwhile}]
    We know that before Line~\ref{line:wackH}, the graph is $\gamma_2$-almost regular.
    Then at Line~\ref{line:wackH} we halve the scale of a single edge,
    so the graph $G^{s}$ is $4\gamma_2$-almost regular afterwards.
    We also know that after the while loop terminates at Line~\ref{line:while1t},
    the graph $G^{s'}$ is $\gamma_2$-almost regular, by the termination condition of the while loop.
    Now consider the following process for obtaining $s'$ from $s$.
    \begin{enumerate}
      \item While $s\neq s'$:
        \begin{enumerate}
          \item For each $(u,v)\in E$ such that $s_{uv} > s'_{uv}$,
            let $s_{uv} \gets s_{uv} / 2$.
        \end{enumerate}
    \end{enumerate}
    We first argue, by induction, that at the end of each iteration of the while loop of the above process,
    $G^s$ is $4\gamma_2$-almost regular.
    As noted above, initially, the graph $G^s$ is $4\gamma_2$-almost regular.
    For the induction step,
    consider a fixed iteration, and let $F$ be the edges $(u,v)$ for which $s_{uv} > s'_{uv}$.
    Since we decrease the weights of all edges in $F$ by a same factor,
    from the point view of leverage scores, it is equivalent
    to increase the weights of all other edges by a same multiple.
    Therefore the leverage scores of edges in $F$ can only decrease,
    and thus can be at most $\frac{4\gamma_2 k}{n}$ after this iteration.
    As for edges $(u,v)$ not in $F$, they satisfy $s_{uv} = s'_{uv}$.
    We know that in $G^{s'}$ their leverage scores are at most $\frac{4\gamma_2 k}{n}$,
    as $G^{s'}$ is $\gamma_2$-almost regular.
    Since $G^{s}$'s weights always dominate those of $G^s$, their leverage scores in $G^{s}$
    can only be smaller, and thus at most $\frac{4\gamma_2 k}{n}$ as well.

    We then argue that at any point of the process, the graph is $16\gamma_2$-almost regular.
    This follows by noting that at any point of the algorithm, $G^{s}$'s edge weights
    are within a factor $4$ of those at the end of previous iteration, and those
    at the end of the current iteration.

    We now show that in this process, each time we let $s_{uv}\gets s_{uv}/2$,
    the potential function increases by at most $(1 - \frac{400\gamma_2^2 k^2}{(1 - \zeta)^2 n^2})^{-1}$,
    which implies the statement of this claim.
    Let $q,q'$ be the scalings before and after one execution of $s_{uv}\gets s_{uv}/2$.
    Let $\ell$ be the number of nonzero eigenvalues of $N^q$ that are between $(0,\zeta]$,
    and let $\ell_1 = \min(\ell,\ell_0)$.
    Then we have $\Upsilon(q) = \detn_{\ell_1}(N^{q}) \zeta^{\ell_0 - \ell_1}$,
    and, by Corollary~\ref{cor:altups},
    $\Upsilon(q') \leq \detn_{\ell_1}(N^{q'}) \zeta^{\ell_0 - \ell_1}$.
    Thus it suffices to show
    \begin{align}\label{eq:32}
      \frac{\detn_{\ell_1}(N^{q'})}{\detn_{\ell_1}(N^{q})} \leq
      \kh{ 1 - \frac{400\gamma_2^2 k^2}{(1 - \zeta)^2n^2} }^{-1}.
    \end{align}
    As in our proof of Claim~\ref{lem:line14}, we once again consider the variational characterization
    in Lemma~\ref{lem:potentialfrac}.
    Let $X = (D^{q})^{\dag/2} F$ where $F\in\mathbb{R}^{nk\times \ell_1}$'s columns are
    nonzero bottom eigenvectors of $N^{q}$.
    The numerator of the characterization can only decrease.
    For the denominator, we have by matrix determinant lemma
    \begin{align*}
      \frac{ \det\kh{ X^T D^{q'} X } }
      { \det\kh{ X^T D^{q} X } } \geq &
      { 1 - \frac{3}{4}(e_{u\la v}^{q})^T X (X^T D^q X)^{-1} X^T e_{u\la v}^{q} -
      \frac{3}{4}(e_{v\la u}^{q})^T X (X^T D^q X)^{-1} X^T e_{v\la u}^{q} } \\ = &
      { 1 - \frac{3}{4}(e_{u\la v}^{q})^T X  X^T e_{u\la v}^{q} -
      \frac{3}{4}(e_{v\la u}^{q})^T X  X^T e_{v\la u}^{q} }
      \qquad
      \text{(as $X^T D^{q} X = F^T F = I$)} \\ = &
      1 - \frac{3}{4}(e_{u\la v}^{q})^T (D^{q})^{\dag/2} F F^T (D^{q})^{\dag/2} (e_{u\la v}^{q}) -
      \frac{3}{4}(e_{v\la u}^{q})^T (D^{q})^{\dag/2} F F^T (D^{q})^{\dag/2} (e_{v\la u}^{q}) \\ = &
      1 - \frac{3}{4}\norm{ F^T(u) (D^{q}_u)^{\dag/2} \phi_{uv} }^2 -
      \frac{3}{4}\norm{ F^T(v) (D^{q}_v)^{\dag/2} \phi_{uv} }^2,
    \end{align*}
    where in the last equality we let $F(u)$ be the $u^{th}$ row block of $F$.
    Since $G^{q}$ is $16\gamma_2$-almost regular,
    we have
    $\norm{(D^q_u)^{\dag/2} \phi_{uv}}^2 \leq \frac{16\gamma_2\cdot k}{n}$ and
    $\norm{(D^q_v)^{\dag/2} \phi_{uv}}^2 \leq \frac{16\gamma_2\cdot k}{n}$.
    By Lemma~\ref{lem:fsu2}, we have
    $\lambdamax(F(u) F(u)^T) \leq \frac{16\gamma_2 k}{(1 - \zeta)^2 n}$, and
    $\lambdamax(F(v) F(v)^T) \leq \frac{16\gamma_2 k}{(1 - \zeta)^2 n}$, and therefore
    \begin{align*}
      & \norm{F^T(u) (D^q_u)^{\dag/2} \phi_{uv}}^2 \leq \frac{256\gamma_2^2\cdot k^2}{(1 - \zeta)^2 n^2} \\
      & \norm{F^T(v) (D^q_v)^{\dag/2} \phi_{uv}}^2 \leq \frac{256\gamma_2^2\cdot k^2}{(1 - \zeta)^2 n^2}.
    \end{align*}
    This give us
    \begin{align}\label{eq:detdom1}
      \frac{ \det\kh{ X^T D^{q'} X } }
      { \det\kh{ X^T D^{q} X } } \geq & 1 - \frac{400 \gamma_2^2\cdot k^2}{(1 - \zeta)^2 n^2}.
    \end{align}
    This then implies~(\ref{eq:32}) and finishes the proof of the claim.
  \end{proof}

  We next show that during the second while loop at Lines~\ref{line:while2s}-\ref{line:while2t},
  the potential function cannot increase much either.

  \begin{claim}\label{claim:secondwhile}
    Consider a fixed iteration of the outermost while loop.
    Suppose in this iteration of the outer while loop, the total number of iterations executed by the first inner while loop
    at Lines~\ref{line:while2s}-\ref{line:while2t} is $t_2$.
    Let $s,s'$ be the scalings
    before and after the while loop respectively.
    Then
    \begin{align*}
      \frac{\Upsilon(s')}{\Upsilon(s)}
      \leq
      \kh{ 1 + \frac{3\psi_1 k^2}{n^2} }^{t_2}.
    \end{align*}
  \end{claim}
  \begin{proof}[Proof of Claim~\ref{claim:secondwhile}]
    Let $q,q'$ be the scalings before and after one execution of $s_{uv}\gets s_{uv}\cdot 2$.
    Let $\ell$ be the number of nonzero eigenvalues of $N^q$ between $(0,\zeta]$,
    and let $\ell_2 = \min(\ell,\ell_0)$.
    Then we have $\Upsilon(q) = \detn_{\ell_2}(N^{q}) \zeta^{\ell_0 - \ell_2}$,
    and, by Corollary~\ref{cor:altups},
    $\Upsilon(q') \leq \detn_{\ell_2}(N^{q'}) \zeta^{\ell_0 - \ell_2}$.
    Thus it suffices to show
    \begin{align}\label{eq:54}
      \frac{\detn_{\ell_2}(N^{q'})}{\detn_{\ell_2}(N^{q})} \leq 1 + \frac{3\psi_1 k^2}{n^2}.
    \end{align}
    As in our proofs of Claims~\ref{lem:line14},~\ref{claim:firstwhile},
    we also consider the variational characterization
    in Lemma~\ref{lem:potentialfrac}.
    Let $X = (D^{q})^{\dag/2} F$ where $F\in\mathbb{R}^{nk\times \ell_2}$'s columns are
    nonzero bottom eigenvectors of $N^{q}$.
    The denominator of the characterization can only increase.
    For the numerator we have by matrix determinant lemma
    \begin{align}
      \frac{ \det\kh{ X^T L^{q'} X } }
      { \det\kh{ X^T L^{q} X } }
      = & 1 +  3 (b_{uv}^q)^T X \kh{X^T L^q X}^{-1} X^T (b_{uv}^q) \notag \\
      = & 1 +  3 (b_{uv}^q)^T \kh{D^{q}}^{\dag/2} F
      \kh{ F^T \kh{ D^{q} }^{\dag/2} L^q \kh{D^{q}}^{\dag/2} F}^{-1} F^T \kh{D^{q}}^{\dag/2} b_{uv}^q \notag \\
      \leq & 1 +  3 H_{\zeta}^{q}(u,v)
      \leq 1 + \frac{3 \psi_1 k^2}{n^2}.
      \label{eq:det54}
    \end{align}
    This implies~(\ref{eq:54}) and finishes the proof the claim.
  \end{proof}
  For the first $t$ iterations of the outermost while loop,
  let $t_1$ be the total number of iterations executed by the first inner while loop,
  and $t_2$ be the total number of iterations executed by the second inner while loop.
  Then we have
  $t_1 \leq 2t$ by Lemma~\ref{lem:tt1},
  and $t_2 \leq t + t_1 \leq 3t$.
  Therefore, by Claims~\ref{lem:line14},~\ref{claim:firstwhile},~\ref{claim:secondwhile},
  we have
  \begin{align*}
    \frac{\Upsilon(s^1)}{\Upsilon(s^{0})}
    \leq \kh{1 - \frac{\psi_2 k^2}{4 n^2}}^{t}
    \kh{ 1 - \frac{400 \gamma_2^2 k^2}{(1-\zeta)^2 n^2} }^{-2t}
    \kh{ 1 + \frac{3\psi_1 k^2}{n^2} }^{3t}
    \leq \kh{ 1 - \frac{\psi_2 k^2}{16 n^2} }^t,
  \end{align*}
  where the last inequality follows from
  $\psi_2 \geq 1024 \psi_1$ and
  $\psi_2 \geq \frac{1024 \gamma_2^2}{(1 - \zeta)^2}$.
\end{proof}

Notice that by design, after $\WTM$ terminates,
each edge $(u,v)$ with $s_{uv} < s_{uv}^0$ satisfies either
$R^s(u,v) \geq \frac{\gamma_1 k}{n}$ or
$H^s_{\zeta}(u,v) \geq \frac{\psi_1 k^2}{n^2}$.
We show that the total number of such edges is small.

\begin{lemma}\label{lem:ne1}
  After $\WTM$ terminates, the number of edges $(u,v)$ with $s_{uv} < s_{uv}^0$
  such that $R^s(u,v) \geq \frac{\gamma_1 k}{n}$ is at most
  $\frac{ n\cdot \rank{D^{s}} }{\gamma_1 k}$.
\end{lemma}
\begin{proof}
  Since $R^{s}(u,v)$'s are leverage scores, we have
  \begin{align*}
    \sum_{(u,v)\in E} R^{s}(u,v) = \rank{D_s}.
  \end{align*}
  Then the desired bound follows.
\end{proof}

\begin{lemma}\label{lem:ne2}
  After $\WTM$ terminates, the number of edges $(u,v)$ with $s_{uv} < s_{uv}^0$
  such that $H^s_{\zeta}(u,v) \geq \frac{\psi_1 k^2}{n^2}$ is at most
  $\frac{ n^2 \gamma_2 }{\psi_1 (1 - \zeta)^2}$.
\end{lemma}
\begin{proof}
  Let $\ell$ be the number of eigenvalues of $N^{s}$ between $(0,\zeta]$.
  Let $\lambda_1,\ldots,\lambda_{\ell}$ be all eigenvalues between $(0,\zeta]$,
  and let $f_1,\ldots,f_{\ell}$ be a set of orthonormal eigenvectors.
  By Theorem~\ref{thm:fewsmall}, $\ell \leq \frac{\gamma_2 k^2}{(1 - \zeta)^2}$.
  Also we have by Proposition~\ref{prop:ellavg} that
  \begin{align*}
    \sum_{(u,v)\in E} H^{s}_{\zeta}(u,v) = \ell.
  \end{align*}
  Thus our desired bound follows.
\end{proof}

\subsection{Analysis of Algorithm~\ref{algo:hed}}

Our analysis of Algorithm~\ref{algo:hed} will mostly focus on
bounding the total number of deleted edges at Line~\ref{line:delsmall}.
Since we only delete edges with $(s_{uv}/s'_{uv})^2 \leq \rho < 1$,
by the termination condition of the second inner while loop of $\WTM$ we know that
each deleted edge satisfies either $R^s(u,v)\geq \frac{\gamma_1 k}{n}$ or
$H^s_{\zeta}(u,v) \geq \frac{\psi_1 k^2}{n^2}$,
where $\gamma_1 = \gamma / 16$ and $\psi_1 = \psi/1024$.
Thus we will show that the numbers of both types of edges are small.

Consider fixing a while loop iteration of $\HED$ where we go the ``else'' branch.
Let $s'$ be the scaling we obtain after we invoke $\HARGD$ at Line~\ref{line:argd},
$s$ be the scaling returned by $\WTM$, and $\hat{s}$ be the scaling obtained
after we delete the small weight edges at Line~\ref{line:delsmall}
(but before we reset the edge weights at Line~\ref{line:resetw}).
Then by an (almost) identical proof to that of Lemma~\ref{lem:nerd},
we have:
\begin{lemma}\label{lem:sdddd}
  The number of edges $(u,v)\in E$ with $(s_{uv}/s'_{uv})^2 \leq \rho$
  such that $R^{s}(u,v)\geq \frac{\gamma_1\cdot k}{n}$ is at most
  \begin{align*}
    \frac{2n}{\gamma_1 k} \kh{ \rank{D^{s'}} - \rank{D^{\shat}} }.
  \end{align*}
\end{lemma}
This means that we can charge the number of deleted edges at Line~\ref{line:delsmall} that
satisfy $R^s(u,v)\geq \frac{\gamma_1\cdot k}{n}$ to the rank change of $D$.
We will then bound the number of deleted edges that satisfy
$H^{s}_{\zeta}(u,v)\geq \frac{\psi_1 k^2}{n^2}$ by considering the number of nontrivial (Definition~\ref{def:nontrivial})
zero eigenvalues of the normalized Laplacian,
which,
by Theorem~\ref{thm:fewsmall}, is at most
$2 \gamma k^2$ when the graph is $\gamma$-almost regular.

To that end,
for a normalized Laplacian matrix $N$,
let $\eta(N)$ denote the number of nontrivial 
zero eigenvalues of $N$.
Additionally, let $s''$ denote the scaling we obtain after we invoke $\HARGD$ at Line~\ref{line:argd} in the {\em next}
iteration of the while loop.
The following lemma characterizes how $\eta(N)$ changes after a while loop iteration.
\begin{lemma}\label{lem:etachange}
  We have
  \begin{align}\label{eq:etachange}
    \eta(N^{s''}) \geq \eta(N^{s'}) + 1 -
    \floor{ \frac{2 \gamma k}{n} \kh{\rank{D^{s'}} - \rank{D^{s''}}} + \sqrt{\frac{1}{16nk}}}.
  \end{align}
\end{lemma}

Before proving the lemma, we first give the proof of Theorem~\ref{thm:hared}.

\begin{proof}[Proof of Theorem~\ref{thm:hared}]
  Since we always maintain $\gamma$-almost regularity,
  the number of nontrivial zero eigenvalues can be at most
  $2\gamma k^2$. Therefore by Lemma~\ref{lem:etachange}
  the total number of iterations of the while loop is at most
  $6\gamma k^2$.
  Thus, the algorithm terminates in finite time.
  Also, by the termination condition, the resulting graph $G^s$ must be
  a $(\gamma,\zeta,\psi)$-almost regular expander.

  By Lemma~\ref{lem:ne2}, the total number of deleted edges
  with $H^s_{\zeta}(u,v) \geq \frac{\psi_1 k^2}{n^2}$ is at most
  \begin{align*}
    \frac{n^2 \gamma}{\psi_1 (1 - \zeta)^2}\cdot 6\gamma k^2 \leq
    \frac{10000 n^2 \gamma^2 k^2}{\psi (1 - \zeta)^2}.
  \end{align*}
  By Lemma~\ref{lem:sdddd} and Theorem~\ref{thm:hargd},
  the total number of deleted edges with $R^{s}(u,v)\geq \frac{\gamma_1\cdot k}{n}$,
  plus those deleted by $\HARGD$, is at most
  \begin{align*}
    \frac{8n}{(\gamma/32) k}\cdot nk = \frac{256 n^2}{\gamma}.
  \end{align*}
  These two bounds coupled with Lemmas~\ref{lem:ne1},~\ref{lem:ne2} imply
  the desired bound on the number of rescaled edges, and thus finish the proof. %
\end{proof}

It then remains to proof Lemma~\ref{lem:etachange}, for which we need the following lemma.

\begin{lemma}\label{lem:231}
  We have
  \begin{align*}
    \lambdamax\kh{ \kh{D^{\dag/2} L^{\hat{s}} D^{\dag/2}}^{\dag} } \leq \frac{1}{128nk^3 \gamma \cdot \rho}.
  \end{align*}
\end{lemma}
\begin{proof}
  Let $\shat_{\min} = \min\setof{\shat_{uv}: \shat_{uv}\neq 0}$.
  Since $\shat$ is obtained from $s$ by zeroing out the edges $(u,v)$ with $(s_{uv}/s'_{uv})^2\leq \rho$,
  and no $(s_{uv}/s'_{uv})^2$ is in the range $(\rho,\rho\cdot \alpha)$,
  we have
  \begin{align*}
    \shat_{\min}^2 \geq \rho\cdot \alpha\cdot (s'_{\min})^2 \geq 128nk^3\gamma \alpha_2 \rho,
  \end{align*}
  where $s'_{\min} = \min_{s'_{uv} > 0} s'_{uv}$ is defined at Line~\ref{line:smin},
  and the last inequality follows from $\alpha \geq 128nk^3\gamma\cdot \alpha_2 / (s'_{\min})^2$.
  Let $F = \setof{(u,v)\in E: \shat_{uv} > 0}$. Then we have
  \begin{align*}
    D^{\dag/2} L^{\shat} D^{\dag/2} \pgeq \shat_{\min}^2 D^{\dag/2} L^{F} D^{\dag/2}
    \pgeq 128nk^3\gamma \alpha_2 \rho D^{\dag/2} L^{F} D^{\dag/2},
  \end{align*}
  and therefore
  \begin{align}\label{eq:4nka2}
    \kh{ D^{\dag/2} L^{\shat} D^{\dag/2} }^{\dag} \pleq 
    \frac{1}{128nk^3\gamma\alpha_2 \rho} \kh{ D^{\dag/2} L^{F} D^{\dag/2} }^{\dag}.
  \end{align}
  By definition,
  \begin{align*}
    \alpha_{2} = \max_{F'\subseteq E} \lambdamax\kh{\kh{D^{\dag/2} L^{F'} D^{\dag/2}}^{\dag}}
    \geq \lambdamax\kh{ \kh{ D^{\dag/2} L^{F} D^{\dag/2} }^{\dag} }.
  \end{align*}
  This coupled with~(\ref{eq:4nka2}) implies that
  \begin{align*}
    \kh{ D^{\dag/2} L^{\shat} D^{\dag/2} }^{\dag} \pleq 
    \frac{1}{128nk^3\gamma \rho} I
  \end{align*}
  as desired.
\end{proof}
\begin{proof}[Proof of Lemma~\ref{lem:etachange}]
  Since $s$ and $s'$ have the same support, we have
  $\eta(N^{s'}) = \eta(N^s)$.
  Let $z = \eta(N^s)$.
  Let $0 = \lambda_1 = \ldots = \lambda_{z} < \lambda_{z+1} \leq \rho$ be the smallest $z+1$ nontrivial eigenvalues
  of $N^s$, and let $f_1,\ldots,f_{z+1}$ be a corresponding set of orthonormal, nontrivial eigenvectors.
  Since $G^{\hat{s}}$ is a subgraph of $G^{s}$, $f_1,\ldots,f_{z}$ are also zero eigenvectors of
  $(D^s)^{\dag/2} L^{\hat{s}} (D^s)^{\dag/2}$.
  We first show that $f_{z+1}$ is not in the range of $(D^s)^{\dag/2} L^{\hat{s}} (D^s)^{\dag/2}$.
  In particular, we will show that the projection of $f_{z+1}$ onto the range of 
  $(D^s)^{\dag/2} L^{\hat{s}} (D^s)^{\dag/2}$, i.e.
  \begin{align*}
    \kh{ (D^s)^{\dag/2} L^{\hat{s}} (D^s)^{\dag/2} }^{\dag/2}
    \kh{ (D^s)^{\dag/2} L^{\hat{s}} (D^s)^{\dag/2} }^{1/2} f_{z+1},
  \end{align*}
  has small norm.
  First note that
  \begin{align}
    \norm{\kh{ (D^s)^{\dag/2} L^{\hat{s}} (D^s)^{\dag/2} }^{1/2} f_{z+1}}^2 =
    & f_{z+1}^T (D^s)^{\dag/2} L^{\hat{s}} (D^s)^{\dag/2} f_{z+1} \notag \\ \leq
    & f_{z+1}^T (D^s)^{\dag/2} L^{s} (D^s)^{\dag/2} f_{z+1} \notag \\ =
    & \lambda_{z+1} \leq \rho.
    \label{eq:232}
  \end{align}
  Then
  \begin{align}
    & \norm{ \kh{ (D^s)^{\dag/2} L^{\hat{s}} (D^s)^{\dag/2} }^{\dag/2}
    \kh{ (D^s)^{\dag/2} L^{\hat{s}} (D^s)^{\dag/2} }^{1/2} f_{z+1} }^2 \notag \\ =
    & f_{z+1}^T \kh{ (D^s)^{\dag/2} L^{\hat{s}} (D^s)^{\dag/2} }^{1/2}  
    \kh{ (D^s)^{\dag/2} L^{\hat{s}} (D^s)^{\dag/2} }^{\dag}
    \kh{ (D^s)^{\dag/2} L^{\hat{s}} (D^s)^{\dag/2} }^{1/2} f_{z+1} \notag \\
    \leq & 
    \lambdamax\kh{ \kh{ (D^s)^{\dag/2} L^{\hat{s}} (D^s)^{\dag/2} }^{\dag} }
    \norm{ \kh{ (D^s)^{\dag/2} L^{\hat{s}} (D^s)^{\dag/2} }^{1/2} f_{z+1} }^2 \notag \\
    \leq &
    \lambdamax\kh{ \kh{ D^{\dag/2} L^{\hat{s}} D^{\dag/2} }^{\dag} }
    \norm{ \kh{ (D^s)^{\dag/2} L^{\hat{s}} (D^s)^{\dag/2} }^{1/2} f_{z+1} }^2 \notag \\
    \leq &
    \frac{1}{128nk^3\gamma\rho} \cdot \rho = \frac{1}{128nk^3\gamma},
    \label{eq:firstproj}
  \end{align}
  where the last inequality follows from Lemma~\ref{lem:231} and~(\ref{eq:232}).
  
  Define a matrix $\Fcal\in \mathbb{R}^{(z+1)\times nk}$ by
  \begin{align*}
    \Fcal :=
    \begin{pmatrix}
      f_1^T \\
      \vdots \\
      f_{z+1}^T
    \end{pmatrix} \in \mathbb{R}^{(z+1)\times nk}
  \end{align*}
  and a function $F: V\to \mathbb{R}^{(z+1)\times k}$ by
  \begin{align*}
    F(u) =
    \begin{pmatrix}
      (f_1)_u^T \\
      \vdots \\
      (f_{z+1})_u^T
    \end{pmatrix} \in \mathbb{R}^{(z+1)\times nk}
  \end{align*}
  Since $G^{s}$ is $\gamma$-almost regular,
  we have by Lemma~\ref{lem:fsu2} that $\lambdamax(F(u) F(u)^T)\leq\frac{2\gamma k}{n}$
  and by Theorem~\ref{thm:fewsmall} that $z+1 \leq 2\gamma k^2$.

  Now consider projecting the rows of $\Fcal$ twice: (i) project each row onto
  the null space of $(D^{s})^{\dag/2} L^{\shat} (D^{s})^{\dag/2}$, and then (ii)
  project each row onto the range of $(D^{s})^{\dag/2} D^{s''}(D^{s})^{\dag/2}$.
  Specifically, let $\Pi_1,\Pi_2\in\mathbb{R}^{nk\times nk}$ be the projection matrix
  onto the range of $(D^{s})^{\dag/2} L^{\shat} (D^{s})^{\dag/2}$ and $(D^{s})^{\dag/2} D^{s''}(D^{s})^{\dag/2}$
  respectively,
  and let $\Pi_1^\bot,\Pi_2^\bot$ be the projection matrix onto the maximal subspace orthogonal to
  the range of $\Pi_1$ and $\Pi_2$ respectively.
  Then we consider the matrix $\Fcal':= \Fcal \Pi_1^{\bot} \Pi_2$.

  In order to prove the lemma, it suffices to show that
  the rank of $\Fcal'$ is at least
  \begin{align}\label{eq:rankp}
    z + 1 -
    \floor{ \frac{2 \gamma k}{n} \kh{\rank{D^{s'}} - \rank{D^{s''}}} + \sqrt{\frac{1}{16nk}}},
  \end{align}
  because this will imply that the number of zero eigenvalues
  of $(D^{s})^{\dag/2} L^{s''} (D^{s})^{\dag/2}$ (and hence of $N^{s''}$) is at least
  $nk - \rank{D^{s''}} + (\ref{eq:rankp})$.

  Since
  $$\Fcal' (\Fcal')^{T} = \Fcal \Pi_1^{\bot} \Pi_2 \Pi_1^{\bot} \Fcal^T \pleq
  \Fcal \Pi_1^{\bot} \Pi_1^{\bot} \Fcal^T
  \pleq \Fcal \Fcal^T = I_{(z+1)\times (z+1)},$$
  we know that all singular values of $\Fcal'$ is at most $1$.
  Thus, it suffices to show that
  \begin{align*}
    \trace{\Fcal' (\Fcal')^T} \geq
    z + 1 -
    \kh{ \frac{2 \gamma k}{n} \kh{\rank{D^{s'}} - \rank{D^{s''}}} + \sqrt{\frac{1}{16nk}} },
  \end{align*}
  which will imply that the number of nonzero singular values of $\Fcal'$
  is at least~(\ref{eq:rankp}).
  To this end, let us first write $\trace{\Fcal \Pi_2 \Fcal^T}$ as
  \begin{align}
    & \trace{\Fcal \Pi_2 \Fcal^T} \notag \\ =
    & \trace{\Fcal (\Pi_1^{\bot} + \Pi_1) \Pi_2 (\Pi_1^{\bot} + \Pi_1)\Fcal^T} \notag \\
    = & \trace{\Fcal \Pi_1^{\bot} \Pi_2 \Pi_1^{\bot} \Fcal^T}
    + \trace{\Fcal \Pi_1 \Pi_2 \Fcal^T} +
    \trace{\Fcal\Pi_1^{\bot} \Pi_2 \Pi_1 \Fcal^T} \notag \\
    \leq & \trace{\Fcal \Pi_1^{\bot} \Pi_2 \Pi_1^{\bot} \Fcal^T} \notag\\
    + & \sqrt{ \trace{\Fcal \Pi_1 \Fcal^T} \trace{\Fcal \Pi_2 \Fcal^T} }
    + \sqrt{\trace{\Fcal\Pi_1^{\bot} \Pi_2 \Pi_1^{\bot} \Fcal^T} \trace{\Fcal \Pi_1 \Fcal^T}},
    \label{eq:trineq}
  \end{align}
  where the last equality follows from expanding,
  and the last inequality follows from Cauchy-Schwarz.
  This combined with~(\ref{eq:firstproj}) gives that
  \begin{align}
    \trace{\Fcal' (\Fcal')^T} =
    & \trace{\Fcal \Pi_1^{\bot} \Pi_2 \Pi_1^{\bot} \Fcal^T} \notag \\ \geq
    & \trace{\Fcal \Pi_2 \Fcal^T} - 2 \sqrt{\frac{z+1}{128nk^3\gamma}} \notag\\ \geq
    & \trace{\Fcal \Pi_2 \Fcal^T} - 2 \sqrt{\frac{2\gamma k^2}{128nk^3\gamma}} \notag \\ =
    & \trace{\Fcal \Pi_2 \Fcal^T} - \sqrt{\frac{1}{16nk}}. \label{eq:pi1}
  \end{align}
  Using the fact that $\lambdamax(F(u) F(u)^T) \leq \frac{2\gamma k}{n}$,
  we have
  \begin{align*}
    \trace{\Fcal \Pi_2^{\bot} \Fcal^T}
    \leq \frac{2 \gamma k}{n} \kh{\rank{D^{s'}} - \rank{D^{s''}}},
  \end{align*}
  and hence
  \begin{align}
    \trace{\Fcal \Pi_2 \Fcal^T} =
    & \trace{\Fcal \Fcal^{T}} - \trace{\Fcal \Pi_2^{\bot} \Fcal^T} \notag\\ \geq
    & z+1 - \frac{2 \gamma k}{n} \kh{\rank{D^{s'}} - \rank{D^{s''}}}. \label{eq:pi2}
  \end{align}
  Combining~(\ref{eq:pi1}) and~(\ref{eq:pi2}) finishes the proof of the lemma.
\end{proof}

\section{Expanders are preserved under vertex sampling}\label{sec:hdevs}

\subsection{Warm-up: ordinary expanders are preserved under vertex sampling}\label{sec:expwarmup}

As a warm-up, we start with ordinary unweighted graphs, and prove that
an expander with large minimum degree still has good expansion after uniform vertex sampling.
We will use the algebraic definition of expansion, which could be translated
to combinatorial expansion using Cheeger's inequality~\cite{AlonM85}.

\begin{definition}[Expanders]
  An unweighted graph $G = (V,E)$ is a $\zeta$-expander for some $0 < \zeta\leq 1$ if
  the second smallest eigenvalue of the normalized Laplacian $N$ is $\lambda_2 \geq \zeta$.
\end{definition}

Note that any ordinary, unweighted graph $G=(V,E)$ with
minimum degree $d_{\min}$
can be seen as a $1\times 1$-matrix weighted,
$\frac{n}{d_{\min}}$-almost regular graph.
Therefore we can apply Lemma~\ref{lem:fsu2} and obtain similar properties of the spectral embedding induced
by the bottom eigenvectors of the normalized Laplacian of $G$.

\begin{definition}[Spectral embeddings of scalar-weighted graphs]
  Given orthonormal vectors $f_1,\ldots,f_{\ell}\in\mathbb{R}^{n}$,
  define an $\ell\times n$ matrix $\Fcal$ whose rows are
  transposes of $f_1,\ldots,f_{\ell}$:
  \renewcommand{\arraystretch}{1.4}
  \begin{center}
    $\Fcal\,=\ $
    \begin{tabular}[h]{|x{4em}|x{7em}|x{4em}|x{7em}|x{4em}|}
      \hline
      $(f_1)_1 $ & $\cdots$ & $(f_1)_u$ & $\cdots$ & $(f_1)_n$ \\
      \hline
      $(f_2)_1$ & $\cdots$ & $(f_2)_u$ & $\cdots$ & $(f_2)_n$ \\
      \hline
      \multicolumn{5}{|c|}{\xrowht{30pt}$\vdots$} \\
      \hline
      $(f_\ell)_1$ & $\cdots$ & $(f_\ell)_u$ & $\cdots$ & $(f_\ell)_n$ \\
      \hline
    \end{tabular}
    $\ \in \mathbb{R}^{\ell\times n}$.
  \end{center}
  Then define an embedding $F : V \to \mathbb{R}^{\ell}$ by
  letting $F(u)$ equal the $u^{\mathrm{th}}$ column of $\Fcal$:
  \begin{center}
    $F(u)\,=\ $
    \begin{tabular}[h]{|x{4em}|}
      \hline
      $(f_1)_u$ \\
      \hline
      $(f_2)_u$ \\
      \hline\xrowht{30pt}$\vdots$ \\
      \hline
      $(f_\ell)_u$ \\
      \hline
    \end{tabular}
    $\ \in \mathbb{R}^{\ell}$.
  \end{center}
  We call $F$ the spectral embedding induced by $f_1,\ldots,f_{\ell}$,
  and $\Fcal$ the embedding matrix induced by $f_1,\ldots,f_{\ell}$.
\end{definition}

By Lemma~\ref{lem:fsu2}, we immediately have:

\begin{corollary}[of Lemma~\ref{lem:fsu2}]\label{cor:fsu2}
  Let $G = (V,E)$ be an ordinary, unweighted graph with minimum degree $d_{\min}$.
  Fix a $\delta \in (0,1)$ and
  let
  $0\leq \lambda_1 \leq \ldots \leq \lambda_{\ell} \leq 1  - \delta$
  be all eigenvalues of $N = D^{-1/2} L D^{-1/2}$ that are $\leq 1 - \delta$.
  Let
  $f_1,\ldots,f_{\ell}$ be a corresponding set of orthonormal eigenvectors.
  Let $F$ be the spectral embedding induced by $f_1,\ldots,f_{\ell}$.
  Then we have for all $u\in V$ %
  \begin{align}\label{eq:cbsf2}
    \norm{F(u)}^2
    \leq
    \frac{1}{\delta^2 d_{\min}}.
  \end{align}
\end{corollary}

By Theorem~\ref{thm:fewsmall} we also have a bound on the number of small eigenvalues:

\begin{corollary}[of Theorem~\ref{thm:fewsmall}]\label{cor:fewsmall}
  Let $G = (V,E)$ be an ordinary, unweighted graph with minimum degree $d_{\min}$.
  Then for any $\delta \in (0,1)$,
  the number of eigenvalues of $N = D^{-1/2} L D^{-1/2}$
  that are at most $1 - \delta$
  is at most $\frac{n}{\delta^2 d_{\min}}$.
\end{corollary}

\begin{theorem}\label{thm:ordinary}
  Let $G = (V,E)$ be a $\zeta$-expander
  with minimum degree $d_{\min} \geq 4\cdot 10^6 \cdot \zeta^{-1} n^{\frac{32}{\log\log n}}$
  for some $\zeta \leq \frac{1}{\log n}$.
  For an $s\geq \frac{4\cdot 10^6 \cdot \zeta^{-1} n^{\frac{32}{\log\log n}}}{d_{\min}}\cdot n$,
  let $C$ be a uniformly random vertex subset of size $s$.
  Then with probability $1 - n^{-7}$,
  the induced subgraph $G[C]$ is a $\zeta / n^{\frac{1024}{\log\log n}}$-expander
  with minimum degree at least $\frac{s}{2n}\cdot d_{\min}$.
\end{theorem}

This theorem is consequence of applying the following lemma $O(\frac{\log n}{\log \log n})$ times.

\begin{restatable}{lemma}{lemsample}
  \label{lem:sample}
  Let $G = (V,E)$ be a $\zeta$-expander
  with minimum degree $d_{\min} \geq  2\cdot 10^6 \cdot \zeta^{-1}\log^{10} n$
  for some $\zeta \leq \frac{1}{\log n}$.
  For an $s\geq \frac{n}{\log n}$, let $C$ be a uniformly random vertex subset of size $s$.
  Then with probability $1 - n^{-8}$,
  the induced subgraph $G[C]$ is a $\zeta/16$-expander
  with minimum degree at least $(1 - \frac{1}{2\log n})\cdot \frac{s}{n}\cdot d_{\min}$.
\end{restatable}

We will prove Lemma~\ref{lem:sample} via a matrix Martingale argument, for which end we will need to set up the
notions of Cholesky factorization and Schur complements.
The following definitions and facts are from~\cite{KyngS16,DurfeeKPRS17}.

\paragraph{Schur complements.}{
  Let $L\in\mathbb{R}^{n\times n}$
  be the Laplacian matrix of an $n$-vertex, connected graph $G$,
  and let $L(:,u)$ denote the column of $L$ corresponding to vertex $u$.
  For a vertex $u_1$,
  we define the {\em Schur complement} of $L$ with respect to $u_1$ as
  \begin{align}\label{eq:defsc1}
    S^{(1)} = L - \frac{1}{L_{u_1u_1}} L(:,u_1) L(:,u_1)^T\in\mathbb{R}^{n\times n}.
  \end{align}
  It is straightforward to see:
  \begin{fact}
    The entries on row $u_1$ of $S^{(1)}$ and the entries
    on column $u_1$ of $S^{(1)}$ are all zero.
    \label{fact:i10}
  \end{fact}
  The following fact is less straightforward, but well known:
  \begin{fact}\label{fact:lapli1}
    $S^{(1)}$ is a Laplacian matrix of a graph supported on $V\setminus \setof{u_1}$.
  \end{fact}
  We call the operation of subtracting $\frac{1}{L_{u_1u_1}} L(:,u_1) L(:,u_1)^T$ from $L$ the
  {\em elimination of vertex $u_1$}.
  Suppose we perform a sequence of eliminations of vertices $u_1,u_2,\ldots,u_{t}$ for some $1\leq t < n$, and define
  for each $i\in [t]$
  \begin{align*}
    \alpha_i &= S^{(i-1)}_{u_iu_i} \in \mathbb{R} \\
    c_i &= S^{(i-1)}(:,u_i) \in \mathbb{R}^{n} \\
    S^{(i)} &= S^{(i-1)} - \frac{1}{\alpha_i} c_i c_i^T \in \mathbb{R}^{n\times n}
  \end{align*}
  where $S^{(0)} := L$. %
  We call $S^{(t)}$ the {\em Schur complement} of $L$ with respect to $u_1,\ldots,u_t$.
  Then similar to Facts~\ref{fact:i10} and~\ref{fact:lapli1}, we have:
  \begin{fact}
    The entries on rows $u_1,\ldots,u_t$ of $S^{(t)}$ and the entries
    on columns $u_1,\ldots,u_t$ of $S^{(t)}$ are all zero.
    Moreover,
    $S^{(t)}$ is a Laplacian matrix of a graph supported on $V\setminus \setof{u_1,\ldots,u_t}$.
    \label{fact:it0}
  \end{fact}
  It is also known that: %
  \begin{fact}
    Changing the order in which we eliminate $u_1,\ldots,u_t$ does not change
    the resulting Schur complement $S^{(t)}$.
  \end{fact}
  Let $C = V\setminus \setof{u_1,\ldots,u_t}$.
  We define
  \begin{align}\label{eq:defsc2}
    \mathrm{SC}(L,C) := S^{(t)}_{CC} \in \mathbb{R}^{|C|\times |C|}
  \end{align}
  as the {\em Schur complement of $L$ onto $C$},
  where $S^{(t)}_{CC}$ is $S^{(t)}$ restricted to rows and columns in $C$.
  Note that by Fact~\ref{fact:it0}, $\mathrm{SC}(L,C)$ contains all nonzero entries
  of $S^{(t)}$.
  We also write $SC(G,C) = SC(L,C)$.
  To connect Schur complements to the Laplacian $L$, we need to introduce
  partial Cholesky factorization.
}

\paragraph{Partial Cholesky factorization.}{
  Suppose we eliminate a sequence of vertices $u_1,\ldots,u_t$ as above.
  Let $\Lcal$ be the $n\times t$ matrix
  whose $i^{\mathrm{th}}$ column is $c_i$,
  and let $\Dcal$ be the $t\times t$ diagonal matrix with $\Dcal_{ii} = \alpha_i$.
  Then by adding the matrices subtracted in the elimination steps back to $S^{(t)}$, we have
  \begin{align}\label{eq:choleskyadd}
    L = S^{(t)} + \sum_{i=1}^{t} \alpha_i c_i c_i^T = S^{(t)} + \Lcal \Dcal \Lcal^T.
  \end{align}
  Let us define $F = \setof{u_1,\ldots,u_t}$ and $C = V\setminus F$.
  Let $S = \mathrm{SC}(L,C)$, where we recall that the latter is $S^{(t)}$ restricted to rows and columns in $C$.
  Since $S$ contains all nonzero entries of $S^{(t)}$, by writing
  $\Lcal$ in block form as
  $\Lcal = \begin{pmatrix} \Lcal_{FF} \\ \Lcal_{CF} \end{pmatrix}$, we get
  \begin{align}
    L = \begin{pmatrix} \Lcal_{FF} \\ \Lcal_{CF} \end{pmatrix} \Dcal
    \begin{pmatrix} \Lcal_{FF} \\ \Lcal_{CF} \end{pmatrix}^T +
    \begin{pmatrix}
      0_{FF} & 0_{FC} \\
      0_{CF} & S
    \end{pmatrix} =
    \begin{pmatrix}
      \Lcal_{FF} & 0 \\
      \Lcal_{CF} & I_{CC}
    \end{pmatrix}
    \begin{pmatrix}
      \Dcal & 0 \\
      0 & S
    \end{pmatrix}
    \begin{pmatrix}
      \Lcal_{FF} & 0 \\
      \Lcal_{CF} & I_{CC}
    \end{pmatrix}^T.
    \label{eq:choleskymul}
  \end{align}
  We call (\ref{eq:choleskymul}) a {\em partial Cholesky factorization} of $L$.
}

We now state a fact about
matrix factorizations of the form in~(\ref{eq:choleskymul}).
For a set of vectors $x_1,\ldots,x_s$,
we will write $\Pi_{x_1,\ldots,x_s}$ to denote the projection matrix
onto their linear span, and $\Pi_{x_1,\ldots,x_s}^{\bot}$ to denote the projection matrix onto
the maximal subspace orthogonal to their linear span.

\begin{fact}\label{lem:lowertri}
  Given an $n\times n$ positive semidefinite matrix $M$ that can be factorized as
  \begin{align*}
    M = \begin{pmatrix} \Lcal_{FF} \\ \Lcal_{CF} \end{pmatrix} \Dcal
    \begin{pmatrix} \Lcal_{FF} \\ \Lcal_{CF} \end{pmatrix}^T +
    \begin{pmatrix}
      0_{FF} & 0_{FC} \\
      0_{CF} & S
    \end{pmatrix} =
    \begin{pmatrix}
      \Lcal_{FF} & 0 \\
      \Lcal_{CF} & I_{CC}
    \end{pmatrix}
    \begin{pmatrix}
      \Dcal & 0 \\
      0 & S
    \end{pmatrix}
    \begin{pmatrix}
      \Lcal_{FF} & 0 \\
      \Lcal_{CF} & I_{CC}
    \end{pmatrix}^T
  \end{align*}
  for some bi-partition $(C,F)$ of $[n]$,
  such that (i) $\Lcal_{FF}\in\mathbb{R}^{|F|\times |F|}$ is a full-rank matrix;
  (ii) $\Lcal_{CF}\in\mathbb{R}^{|C|\times |F|}$;
  (iii) $\Dcal$ is a diagonal matrix of non-negative entries.
  Let $f_1,\ldots,f_z$ be a basis of the null space of $M$.
  Then
  \begin{align*}
    S^{\dag} = \Pi_{(f_1)_C,\ldots,(f_z)_C}^{\bot} (M^{\dag})_{CC} \Pi_{(f_1)_C,\ldots,(f_z)_C}^{\bot},
  \end{align*}
  where $(M^{\dag})_{CC}$ is $M^{\dag}$ restricted to rows and columns in $C$.
  \label{fact:msc}
\end{fact}
As a direct corollary of the above fact, we have:
\begin{fact}[Corollary of Fact~\ref{lem:lowertri}]\label{lem:factsc}
  Let $f_1\in\mathbb{R}^{n}$ be the all-one vector.
  For any $C\subseteq V$,
  \begin{align*}
    \SC(L,C) = \Pi_{(f_1)_C}^{\bot} (L^{\dag})_{CC} \Pi_{(f_1)_C}^{\bot}.
  \end{align*}
\end{fact}

Below we will need to use the following lemma,
which is a direct consequence of the Matrix Chernoff bound (Theorem~\ref{thm:matrixchernoff}).

\begin{lemma}\label{lem:lsample}
  Let $f_1,\ldots,f_{\ell}\in\mathbb{R}^{n}$ be any $\ell$ orthonormal vectors.
  Define $F: [n]\to \mathbb{R}^{\ell}$ by
  letting $F(i) = ( (f_1)_1, \ldots, (f_\ell)_i )^T\in\mathbb{R}^{\ell}$.
  Suppose for any $i\in [n]$,
  \begin{align*}
    \norm{F(i)}^2 \leq \rho %
  \end{align*}
  for some $\rho$.
  For an $s \geq 100 \rho n \log n$,
  let $C\subseteq[n]$ be a uniformly random subset of indices of size $s$.
  Then we have with probability $1 - n^{-10}$ that
  \begin{align*}
    \frac{1}{2} I_{\ell\times \ell} \pleq
    \frac{n}{s} \kh{ \sum_{i\in C} F(i) F(i)^T } \pleq
    2 I_{\ell\times \ell}
  \end{align*}
  and therefore
  \begin{align*}
    \sum_{j=1}^{\ell} (f_j)_{C} (f_j)_{C}^T \pleq \frac{2s}{n} I_{n\times n}.
  \end{align*}
\end{lemma}

The following lemma shows that the Schur complement onto a random vertex
subset is a better expander than the original graph.
Here we use $\Pi_{x}^{\bot}$ to denote the projection matrix onto the maximal subspace
orthogonal to vector $x$.

\begin{lemma}\label{lem:samplephi}
  Let $G = (V,E)$ be a $\zeta$-expander with
  minimum degree $d_{\min} \geq 2\cdot 10^6 \log^5 n$
  where $\zeta\leq \frac{1}{\log n}$.
  Let the eigenvalues of $N_G$ be $0 = \lambda_1 < \zeta \leq \lambda_2 \leq \ldots \leq \lambda_n$
  and let $f_1,\ldots,f_n$ be a set of corresponding orthonormal eigenvectors.
  For an $s\geq \frac{n}{\log n}$, let $C$ be a uniformly random vertex subset of size $s$.
  Then with probability $1 - n^{-10}$
  \begin{align*}
    D^{-1/2}_{CC} \mathrm{SC}(L_G,C) D^{-1/2}_{CC} \pgeq
    \frac{n}{4s} \cdot \zeta\cdot \Pi_{(f_1)_{C}}^{\bot}.
  \end{align*}
\end{lemma}
\begin{proof}
  By eliminating vertices outside of $C$,
  we get a partial Cholesky factorization of $L_G$:
  \begin{align*}
    L_G =
    \begin{pmatrix}
      \Lcal_{FF} & 0 \\
      \Lcal_{CF} & I_{CC}
    \end{pmatrix}
    \begin{pmatrix}
      \Dcal & 0 \\
      0 & \SC(L_G,C)
    \end{pmatrix}
    \begin{pmatrix}
      \Lcal_{FF} & 0 \\
      \Lcal_{CF} & I_{CC}
    \end{pmatrix}^T.  %
  \end{align*}
  By multiplying $D^{-1/2}$ on both sides, we then get a factorization of $N_G$:
  \begin{align*}
    N_G = &
    \begin{pmatrix}
      D^{-1/2}_{FF} \Lcal_{FF} & 0 \\
      D^{-1/2}_{CC} \Lcal_{CF} & D^{-1/2}_{CC}
    \end{pmatrix}
    \begin{pmatrix}
      \Dcal & 0 \\
      0 & \SC(L_G,C)
    \end{pmatrix}
    \begin{pmatrix}
      D^{-1/2}_{FF} \Lcal_{FF} & 0 \\
      D^{-1/2}_{CC} \Lcal_{CF} & D^{-1/2}_{CC}
    \end{pmatrix}^T \\ = &
    \begin{pmatrix}
      D^{-1/2}_{FF} \Lcal_{FF} & 0 \\
      D^{-1/2}_{CC} \Lcal_{CF} & I_{CC}
    \end{pmatrix}
    \begin{pmatrix}
      \Dcal & 0 \\
      0 & D_{CC}^{-1/2}\SC(L_G,C) D_{CC}^{-1/2}
    \end{pmatrix}
    \begin{pmatrix}
      D^{-1/2}_{FF} \Lcal_{FF} & 0 \\
      D^{-1/2}_{CC} \Lcal_{CF} & I_{CC}
    \end{pmatrix}^T.  %
  \end{align*}
  Then by Fact~\ref{fact:msc},
  \begin{align*}
    \kh{ D^{-1/2}_{CC} \mathrm{SC}(L_G,C) D^{-1/2}_{CC} }^{\dag} =
    \Pi_{(f_1)_{C}}^{\bot} \kh{ \sum_{i=2}^{n} \frac{1}{\lambda_i} (f_i)_C (f_i)_C^T } \Pi_{(f_1)_{C}}^{\bot}.
  \end{align*}
  Let $\ell$ be such that $\lambda_{\ell} \leq \max\setof{\frac{n}{2s},1} \cdot \zeta < \lambda_{\ell+1}$. 
  Define the embedding $F: V\to\mathbb{R}^{\ell}$ as
  \begin{align*}
    F(u) = \begin{pmatrix}
      (f_1)_u \\
      (f_2)_u \\
      \vdots \\
      (f_\ell)_u
    \end{pmatrix} \in \mathbb{R}^{\ell}.
  \end{align*}
  By $s\geq n/\log n$ and $\zeta\leq 1/\log n$, we have
  $\frac{n}{2s} \cdot \zeta \leq 1/2$,
  and thus by Corollary~\ref{cor:fsu2}
  we have that for each $u\in V$
  \begin{align*}
    \norm{F(u)}^2 \leq \frac{4}{d_{\min}} \leq 10^{-4} \cdot \log^{-3} n.
  \end{align*}
  Then we have %
  \begin{align*}
    \kh{ D^{-1/2}_{CC} \mathrm{SC}(L_G,C) D^{-1/2}_{CC} }^{\dag} = &
    \Pi_{(f_1)_{C}}^{\bot}
    \kh{ \sum_{i=2}^{n} \frac{1}{\lambda_i} (f_i)_C (f_i)_C^T}
    \Pi_{(f_1)_{C}}^{\bot} \\ \pleq
    &
    \Pi_{(f_1)_{C}}^{\bot}
    \kh{ \sum_{i = 2}^{\ell} \frac{1}{\zeta} (f_i)_C (f_i)_C^T +
    \sum_{i=\ell + 1}^{n} \frac{2s}{n}\cdot \frac{1}{\zeta} (f_i)_C (f_i)_C^T }
    \Pi_{(f_1)_{C}}^{\bot} \\
    \pleq &
    \Pi_{(f_1)_{C}}^{\bot}
    \kh{ \sum_{i = 2}^{\ell} \frac{1}{\zeta} (f_i)_C (f_i)_C^T +
    \frac{2s}{n}\cdot \frac{1}{\zeta} I }
    \Pi_{(f_1)_{C}}^{\bot} \\
    \pleq &
    \Pi_{(f_1)_{C}}^{\bot}
    \kh{ \frac{2s}{n}\cdot \frac{1}{\zeta} I +
    \frac{2s}{n}\cdot \frac{1}{\zeta} I }
    \Pi_{(f_1)_{C}}^{\bot} \\
    = &
    \frac{4s}{n}\cdot \frac{1}{\zeta} \Pi_{(f_1)_{C}}^{\bot},
  \end{align*}
  where
  the third line follows from
  $\lambdamax(\sum_{i=\ell+1}^{n} (f_i)_C (f_i)_C^T)\leq \lambdamax(\sum_{i=1}^{n} f_i f_i^T) = 1$,
  and the second to last line holds with probability $1 - n^{-10}$ by Lemma~\ref{lem:lsample}.
  By inverting both sides, we then get
  \begin{align*}
    { D^{-1/2}_{CC} \mathrm{SC}(L_G,C) D^{-1/2}_{CC} } \pgeq \frac{n}{4s}\cdot \zeta \cdot \Pi_{(f_1)_C}^{\bot},
  \end{align*}
  as desired.
\end{proof}

We now define the notion of graph squaring, which will be useful in setting up our martingale.
\begin{definition}[Graph squaring]
  For a graph $G = (V,E)$ with degree matrix $D$ and adjacency matrix $A$, define the square of $G$ by
  \begin{align*}
    L_{G^2} \defeq D - A D^{-1} A.
  \end{align*}
\end{definition}

\begin{fact}[\cite{ChengCLPT15,JindalKPS17}]\label{fact:gsq}
  $L_{G^2} \pleq 2 L_G$ for any graph $G$.
\end{fact}

Below, we will write $\SCN(L,C)$ or $\SCN(G,C)$ to denote an $n\times n$ matrix
obtained by augmenting $\SC(L,C)$ to $n\times n$ by adding zeros on rows and columns in $V\setminus C$.
We will also abuse the notation a bit and write $L_{G[C]}$ to denote an $n\times n$ matrix
obtained by augmenting $L_{G[C]}$ to $n\times n$ by adding zeros on rows and columns in $V\setminus C$.

\begin{proposition}\label{prop:sumsc}
  $\sum_{u\in V} \SCN(L_G,V\setminus \setof{u}) =
  L_{G^2} + (n-2) L_G$ for any graph $G$.
\end{proposition}
\begin{proof}
  Notice that by definition
  \begin{align*}
    \sum_{u\in V} \SCN(L,V\setminus \setof{u}) =
    & \sum_{u\in V} \kh{ L_G - \frac{1}{L_{uu}} L_G(:,u) L_G(:,u)^T } \\ =
    & n L_G - L_G D^{-1} L_G \\ =
    & n(D - A) - (D - A) D^{-1} (D - A) \\ =
    & n D - n A - (D -2 A + A D^{-1} A) \\ =
    & (n-2) (D - A) + D - A D^{-1} A = (n-2) L_G + L_{G^2}
  \end{align*}
  as desired.
\end{proof}

Let $v_1,\ldots,v_n$ be a uniformly random permutation of vertices in $V$.
Define
\begin{align*}
  V_i = \setof{v_1,\ldots,v_i}.
\end{align*}
Consider the following sequence of matrices:
\begin{align*}
  X_0 & = L_G \\
  X_{1} & = \frac{1}{d_{v_1}} L_G(:,v_1) L_G(:,v_1)^T +
  \kh{1 + \frac{2}{n-2}} { L_{G[V - V_1]} + \frac{1}{n} \kh{ L_{G^2} - 2L_G } } \\
  X_{2} & = \frac{1}{d_{v_1}} L_G(:,v_1) L_G(:,v_1)^T + \frac{1}{n} \kh{ L_{G^2} - 2 L_G } + \kh{1 + \frac{2}{n-2}}\cdot  \\
  &\quad  \kh{ \frac{1}{d^{G[V-V_1]}_{v_2}} L_{G[V-V_1]}(:,v_2) L_{G[V-V_1]}(:,v_2)^T +
  \kh{1 + \frac{2}{n-3}} L_{G[V - V_2]} + \frac{1}{n-1} \kh{ L_{G[V - V_1]^2} - 2L_{G[V - V_1]} } }
\end{align*}
$X_{i+1}$ is obtained by replacing the $L_{G[V-V_i]}$ in $X_i$ by
\begin{align}\label{eq:rpl}
  \frac{1}{d^{G[V-V_i]}_{v_{i+1}}} L_{G[V-V_i]}(:,v_{i+1}) L_{G[V-V_i]}(:,v_{i+1})^T +
  \kh{1 + \frac{2}{n-2-i}} L_{G[V-V_{i+1}]} + \frac{1}{n-i} \kh{ L_{G[V-V_i]^2} - 2L_{G[V-V_i]} }.
\end{align}
Here, we have used $d^{H}_{v}$ to denote the degree of vertex $v$ in graph $H$.

\begin{lemma}
  $X_0,X_1,X_2,\ldots$ is a matrix-valued martingale.
\end{lemma}
\begin{proof}
  It suffices to prove that
  \begin{align*}
    \expec{v_{i+1}}{(\ref{eq:rpl})\ |\ V_{i}} = L_{G[V - V_i]}.
  \end{align*}
  Let us calculate the LHS term by term.
  \begin{align*}
    & \expec{v_{i+1}}{\frac{1}{d^{G[V-V_i]}_{v_{i+1}}} L_{G[V-V_i]}(:,v_{i+1}) L_{G[V-V_i]}(:,v_{i+1})^T\ |\ V_i} \\
    = & \frac{1}{n-i} L_{G[V-V_i]} D_{G[V-V_i]}^{-1} L_{G[V-V_i]} \\
    = & \frac{1}{n-i} \kh{ D_{G[V-V_i]} - 2 A_{G[V-V_i]} + A_{G[V-V_i]} D_{G[V-V_i]}^{-1} A_{G[V-V_i]} }
  \end{align*}
  \begin{align*}
    \expec{v_{i+1}}{\kh{1 + \frac{2}{n-2-i}} L_{G[V-V_{i+1}]}\ |\ V_i} = \frac{n - 2 - i}{n-i} L_{G[V-V_{i}]}
    + \frac{2}{n-i} L_{G[V-V_{i}]} =
    L_{G[V-V_i]}
  \end{align*}
  \begin{align*}
    & \expec{v_{i+1}}{\frac{1}{n-i} \kh{ L_{G[V-V_i]^2} - 2L_{G[V-V_i]} }\ |\ V_i} \\
    = & \frac{1}{n-i} \kh{ L_{G[V-V_i]^2} - 2L_{G[V-V_i]} } \\
    = & \frac{1}{n-i} \kh{ D_{G[V-V_i]} - A_{G[V-V_i]} D_{G[V-V_i]}^{-1} A_{G[V-V_i]} } -
    \frac{2}{n-i} L_{G[V-V_i]}.
  \end{align*}
  One can verify that these three add up to $L_{G[V-V_{i}]}$.
\end{proof}

To analyze this matrix-valued martingale, we will resort to the
following theorem:

\begin{theorem}[Matrix Freedman Inequality~\cite{Tropp11}]\label{thm:tropp}
  Consider a zero-mean matrix martingale $\setof{Y_i: i = 0,1,\ldots}$ whose values are symmetric matrices
  with dimensional $n$, and let $\setof{Z_i: i = 0,1,\ldots}$ be the difference sequence such that
  $Z_i = Y_{i+1} - Y_{i}$.
  Assume the difference sequence is uniformly bounded in the sense that for any $i=0,1,\ldots$
  \begin{align*}
    \lambdamax(Z_i) \leq R,
  \end{align*}
  where $\lambdamax(Z_i)$ is the maximum absolute value of any eigenvalue of $Z_i$.
  Define the predictable quadratic variation process of the martingale as
  \begin{align*}
    W_i \defeq \sum_{j=0}^{i} \ex{Z_j^2\ |\ Y_1,\ldots,Y_{j}}.
  \end{align*}
  Then for all $i\geq 0$ and $\sigma^2 > 0$,
  \begin{align*}
    \pr{ \exists i\geq 0: \lambdamax(Y_i)\geq t\quad \text{and}\quad \norm{W_i}\leq \sigma^2 }
    \leq d\cdot \exp\setof{ - \frac{t^2/2}{\sigma^2 + Rt/3} }.
  \end{align*}
\end{theorem}

We now prove Lemma~\ref{lem:sample}.

\lemsample*

\begin{proof}
  The minimum degree guarantee follows from a direct application of Chernoff bounds.
  We then focus on showing that $G[C]$ is a $\zeta/4$-expander.

  Define $Y_i = L_G^{\dag/2} X_i L_G^{\dag/2} - L_G^{\dag/2} L_G L_G^{\dag/2}$.
  Since $X_0,X_1,\ldots$ is a martingale, so is $Y_0,Y_1,\ldots$, by linearity of expectation.
  Moreover, since $X_0 = L_G$, $Y_0,Y_1,\ldots$ has zero mean.
  Consider the first $n - s + 1$ terms
  $Y_0,Y_1,Y_2,\ldots,Y_{n-s}$. First let us calculate the difference
  sequence $Z_i$:
  \begin{align*}
    Z_i = & Y_{i+1} - Y_i \\
    = & \prod_{j=1}^{i} \kh{1 + \frac{2}{n-1-j}}\cdot L_G^{\dag/2} \\
    & \kh{
      - \SCN \kh{G[V-V_i], V - V_{i+1}} +
      \kh{1 + \frac{2}{n-2-i}} L_{G[V-V_{i+1}]} +
      \frac{1}{n-i} \kh{ L_{G[V-V_i]^2} - 2L_{G[V-V_i]} }
    } L_G^{\dag/2},
  \end{align*}
  where we have
  used Definitions~(\ref{eq:defsc1}),(\ref{eq:defsc2}), and recall that
  $\SCN(G,C)$ is obtained by
  augmenting $\SC(G,C)$ to $n\times n$ by adding zeros.
  We then consider to bound the maximum (in absolute value) eigenvalue of $Z_i$.
  For the last term, we have
  \begin{align}\label{eq:lastterm}
    \lambdamax\kh{ L_G^{\dag/2}
      \frac{1}{n-i} \kh{ L_{G[V-V_i]^2} - 2L_{G[V-V_i]} }
      L_G^{\dag/2} 
    }\leq \frac{4}{n-i} \leq \frac{4}{s} \leq \frac{4\log n}{d_{\min}} \leq \frac{4}{10^6 \log^4 n},
  \end{align}
  where the first inequality follows from $L_{G[V-V_i]^2}\pleq 2 L_{G[V-V_i]} \pleq 2 L_{G}$ by
  Fact~\ref{fact:gsq}.
  Now to bound the first two terms, we first rearrange them as
  \begin{align}\label{eq:rearrange}
    \kh{ - \SCN \kh{G[V-V_i], V - V_{i+1}} + L_{G[V-V_{i+1}]} }
    + {\frac{2}{n-2-i}} L_{G[V-V_{i+1}]}.
  \end{align}
  Then for the second term of~(\ref{eq:rearrange}), we have
  $L_{G[V-V_{i+1}]}\pleq L_G$, and hence
  \begin{align}\label{eq:secondterm}
    \lambdamax\kh{ L_G^{\dag/2} \frac{2}{n-2-i}L_{G[V - V_{i+1}]} L_G^{\dag/2} } \leq
    & \frac{2}{n-2-i} \leq \frac{2}{ s - 2 }
    \leq \frac{4}{s} \leq \frac{4\log n}{d_{\min}} \leq \frac{4}{10^6 \log^4 n}.
  \end{align}
  We then claim that the first term of~(\ref{eq:rearrange}) can be bounded by
  \begin{align}\label{eq:firstterm}
    \lambdamax \kh{ L_G^{\dag/2}
    \kh{ \SCN \kh{G[V-V_i], V - V_{i+1}} - L_{G[V - V_{i+1}]} } L^{\dag/2}_G } \leq
    \frac{1}{d_{\min}\zeta} \leq \frac{1}{8 \cdot 10^6 \log^5 n}.
  \end{align}
  To see why this is the case,
  let us define an edge set $E_{i+1} = \setof{ (v_{i+1}, w) \in G[V - V_i] }$ %
  and write $L_{E_{i+1}}$ to denote the Laplacian of the subgraph of $G$ induced by $E_{i+1}$.
  Notice that $$L_{E_{i+1}}(:,v_{i+1}) = L_{G[V-V_i]}(:,v_{i+1}).$$
  Therefore, we have
  \begin{align*}
    \SCN \kh{G[V-V_i], V - V_{i+1}} - L_{G[V - V_{i+1}]} =
    \SCN \kh{L_{E_{i+1}}, V - V_{i+1}}.
  \end{align*}
  Therefore to prove~(\ref{eq:firstterm}), it suffices to show that
  for any $x$ such that $x\in \mathrm{range}(\SCN(L_{E_{i+1}},V-V_{i+1}))$, we have
  \begin{align}\label{eq:smallflow}
    x^T L^{\dag} x \leq \frac{1}{\zeta d_{\min}} \cdot
    x^T \SCN \kh{L_{E_{i+1}}, V - V_{i+1}}^{\dag} x
    = \frac{1}{\zeta d_{\min}}\cdot x^T L_{E_{i+1}}^{\dag} x,
  \end{align}
  where the equality follows from Fact~\ref{lem:factsc}.
  Since $G$ is a $\zeta$-expander with minimum degree $\dmin$,
  we know that $x^T L^{\dag} x \leq \frac{1}{\zeta d_{\min}} \norm{x}^2$.
  On the other hand, as $G[E_{k+1}]$ is a star graph,
  and $x\in \mathrm{range}(\SCN(L_{E_{i+1}},V-V_{i+1}))$ implies that
  $x$ is supported on $V-V_{i+1}$,
  we have $x^T L_{E_{i+1}}^{\dag} x = \norm{x}^2$. This
  proves our desired inequality~(\ref{eq:smallflow}).

  We also note that
  \begin{align*}
    \prod_{j=1}^{i} \kh{1 + \frac{2}{n-1-j}}= \frac{n}{n-2} \frac{n-1}{n-3} \ldots \frac{n+1-i}{n-1-i}
    = \frac{n(n-1)}{(n-i)(n-1-i)} \leq \kh{\frac{n}{s}}^2 \leq \log^2n.
  \end{align*}
  Therefore we have
  \begin{align}\label{eq:mtglR}
    \lambdamax(Z_i) \leq \frac{1}{1000\cdot \log^{2} n}.
  \end{align}
  Fixing a $j$,
  we then consider bounding the spectral norm of $\ex{Z_j^2\ |\ Y_1,\ldots,Y_{j-1}}$.
  \begin{align*}
    & \ex{Z_j^2\ |\ Y_1,\ldots,Y_{j-1}} \\ \pleq
    & \frac{2}{n-j} \cdot \kh{\frac{n}{n-1-j}}^4 \sum_{v \in V - V_{j}}
    \kh{ L_G^{\dag/2}
    \kh{ \SCN(G[V- V_j], V - V_j - \setof{v}) - L_{G[V - V_j - \setof{v}]} }
  L_G^{\dag/2} }^2 + \\
  & \frac{2}{n-j}\cdot \kh{ \frac{n}{n-1-j} }^4 \sum_{v\in V - V_{j}} \kh{ L_G^{\dag/2}
    \kh{
      \frac{2}{n-2-j} L_{G[V-V_{j} - \setof{v}]} +
  \frac{1}{n-j} \kh{ L_{G[V-V_j]^2} + 2L_{G[V-V_j]} } }
  L_G^{\dag/2}
  }^2 \\ \pleq
  & \frac{2 \log^5 n}{n-j} \cdot\frac{1}{\zeta d_{\min}} \sum_{v\in V-V_j} L_G^{\dag/2}
    \kh{ \SCN(G[V- V_j], V - V_j - \setof{v}) - L_{G[V - V_j - \setof{v}]} }
    L_G^{\dag/2} + \\
    & \frac{2\log^5 n}{n-j}\cdot
    \sizeof{V-V_j}\cdot \kh{ \frac{8}{s} }^2 I_{n\times n}
    \qquad \text{(by~(\ref{eq:firstterm}),(\ref{eq:secondterm}),(\ref{eq:lastterm}))}
    \\
    \pleq & \frac{2\log^6 n}{n \zeta d_{\min}} %
    L_G^{\dag/2} L_{G[V-V_j]^2} L_G^{\dag/2} +
    \frac{10000\log^7 n}{n^2} I_{n\times n}
    \\
    & \qquad \qquad
    \text{(by Proposition~\ref{prop:sumsc} and $n-j\geq s\geq n/\log n$)}
    \\
    \pleq & \frac{64}{2\cdot 10^6\cdot n \log^3 n} I_{n\times n} +
    \frac{10000\log^7 n}{n^2} I_{n\times n}
    \qquad \text{(by Fact~\ref{fact:gsq} and the lower bound on $\dmin$)}
    \\
    \pleq & \frac{1}{1000 n \log^{2} n} I_{n\times n}.
  \end{align*}
  Therefore we have
  \begin{align*}
    \normm{ \ex{Z_j^2\ |\ Y_1,\ldots,Y_{j-1}} } \leq \frac{1}{1000 n \log^2 n}
  \end{align*}
  and
  \begin{align}\label{eq:mtgls}
    \normm{ \sum_{j=0}^{n - s - 1} \ex{Z_j^2\ |\ Y_1,\ldots,Y_{j-1} } } \leq
    & \sum_{j=0}^{n - s - 1} \normm{ \ex{Z_j^2\ |\ Y_1,\ldots,Y_{j-1} } } \notag \\ \leq
    & n\cdot \frac{1}{1000 n \log^2 n} \notag \\
    = & \frac{1}{1000 \log^2 n}.
  \end{align}
  We now invoke Theorem~\ref{thm:tropp} with $R = \frac{1}{1000\log^{2}n}$,
  $\sigma^2 = \frac{1}{1000\log^{2} n}$, and
  $t = 1/2$, and get
  \begin{align*}
    \pr{\exists i\in [0,n-s]: \lambdamax(Y_i) \geq \frac{1}{10\log n}} \leq
    n\cdot \exp\kh{- \frac{\frac{1}{8}}{\frac{1}{1000\log^{2} n} +
    \frac{1}{6000 \log^2 n} } } \leq n^{-10}.
  \end{align*}
  This coupled with Fact~\ref{lem:lowertri} implies that
  \begin{align*}
    L_{G[C]} \pgeq \frac{1}{2} \cdot \frac{s^2}{n^2} \cdot \mathrm{SC}(L_G, C).
  \end{align*}
  Then we have
  \begin{align*}
    \lambda_2\kh{ D_{CC}^{-1/2} L_{G[C]} D_{CC}^{-1/2} } \geq
    & \frac{1}{2} \cdot \frac{s^2}{n^2} \cdot
    \lambda_2\kh{ D_{CC}^{-1/2} \mathrm{SC}(L_G,C) D_{CC}^{-1/2} } %
    \\ \geq
    & \frac{1}{8} \cdot \frac{s}{n}\cdot \zeta,
  \end{align*}
  where the last inequality holds with probability $1 - n^{-10}$ by Lemma~\ref{lem:samplephi}.
  Note that, by a Chernoff bound, we have with probability $1 - n^{-10}$ that
  \begin{align*}
    \frac{1}{2} D_{CC} \pleq
    \frac{n}{s} D_{G[C]} \pleq
    2 D_{CC}.
  \end{align*}
  These two combined imply that
  $G[C]$ is a $\zeta/16$-expander with probability $1 - 3\cdot n^{-10}$.
\end{proof}

\subsection{Matrix-weighted expanders are preserved under vertex sampling}\label{sec:mtexpsamp}

We now introduce Schur complements and Cholesky factorization for Laplacian matrices of $k\times k$
matrix-weighted graphs.

\paragraph{Schur complements.}{
  Let $L\in\mathbb{R}^{nk\times nk}$ be the Laplacian of a $k\times k$ matrix-weighted graph $G$,
  and let $L(:,u)\in\mathbb{R}^{nk\times k}$ to denote the column block of $L$ corresponding to vertex $u$.
  For a vertex $u_1$, we define the {\em Schur complement} of $L$ with respect to $u_1$ as
  \begin{align}\label{eq:defsck1}
    S^{(1)} = L -  L(:,u_1) L_{u_1u_1}^{\dag} L(:,u_1)^T.
  \end{align}
  It is straightforward to see:
  \begin{fact}\label{fact:i10k}
    The entries on row block $u_1$ of $S^{(1)}$ and the entries on column block $u_1$ of $S^{(1)}$ are all zero.
  \end{fact}
  We call the operation of subtracting $L(:,u_1) L_{u_1u_1}^{\dag} L(:,u_1)^T$ from $L$
  the {\em elimination of vertex $u_1$}.
  Unlike the ordinary graph case,
  if we eliminate a sequence of vertices one by one,
  the resulting matrix is not necessarily the Laplacian
  of another $k\times k$ matrix-weighted graph.
  So we alternatively define the {\em Schur complement with respect to a vertex set $F$} as
  \begin{align}\label{eq:defsck2}
    \SC(L,C) = L_{CC} - L_{CF} L_{FF}^{\dag} L_{FC},
  \end{align}
  where $C:= V\setminus F$. We also call $\SC(L,C)$ the {\em Schur complement of $L$ (or $G$) onto $C$}.
  Note that when $F = \setof{u_1}$ consists of only a single vertex,
  we have $S^{(1)}_{CC} = \SC(L,C)$.
  Also note that when $k=1$, this definition matches the alternative definition of Schur complements in
  ordinary graphs~\cite{KyngLPSS16}.
  We then connect $\SC(L,C)$ to the Laplacian $L$ by introducing partial Cholesky factorization.
}

\paragraph{Partial Cholesky factorization.}{
  We show that $L$ can be factorized in a similar way as in~\cite{KyngLPSS16}.
  The proof of the following proposition is deferred to Appendix~\ref{sec:apexp}.
  \begin{proposition}\label{prop:choleskyk}
    We have
    \begin{align}\label{eq:choleskyk}
      L =
      \begin{pmatrix}
        I_{FF} & 0 \\
        L_{CF} L_{FF}^{\dag} & I_{CC}
      \end{pmatrix}
      \begin{pmatrix}
        L_{FF} & 0 \\
        0 & \SC(L,C)
      \end{pmatrix}
      \begin{pmatrix}
        I_{FF} & 0 \\
        L_{CF} L_{FF}^{\dag} & I_{CC}
      \end{pmatrix}^T.
    \end{align}
  \end{proposition}
  We call~(\ref{eq:choleskyk}) a {\em partial Cholesky factorization} of $L$.
  We now state a fact about matrix factorizations of the form in~(\ref{eq:choleskyk}).
  As in the previous subsection, for a set of vectors $x_1,\ldots,x_s$,
  we will write $\Pi_{x_1,\ldots,x_s}$ to denote the projection matrix
  onto their linear span, and $\Pi_{x_1,\ldots,x_s}^{\bot}$ to denote the projection matrix onto
  the maximal subspace orthogonal to their linear span.
  \begin{fact}[See e.g.~\cite{KyngLPSS16}]
    Given an $nk\times nk$ positive semidefinite matrix $M$ divided into $k\times k$ blocks
    that can be factorized as
    \begin{align*}
      M =
      \begin{pmatrix}
        I_{FF} & 0 \\
        \Lcal_{CF} & I_{CC}
      \end{pmatrix}
      \begin{pmatrix}
        \Lcal_{FF} & 0 \\
        0 & S
      \end{pmatrix}
      \begin{pmatrix}
        I_{FF} & 0 \\
        \Lcal_{CF} & I_{CC}
      \end{pmatrix}^T
    \end{align*}
    for some bi-partition $(C,F)$ of $[n]$.
    Let $f_1,\ldots,f_z\in\mathbb{R}^{nk}$ be a basis of the null space of $M$.
  Then
  \begin{align*}
    S^{\dag} = \Pi_{(f_1)_C,\ldots,(f_z)_C}^{\bot} (M^{\dag})_{CC} \Pi_{(f_1)_C,\ldots,(f_z)_C}^{\bot},
  \end{align*}
  where $(M^{\dag})_{CC}$ is $M^{\dag}$ restricted to row and column blocks in $C$.
  \label{fact:msck}
  \end{fact}
}

As a direct corollary of the above fact, we have:
\begin{fact}[Corollary of Fact~\ref{lem:lowertri}]\label{lem:factsck}
  Let $f_1,\ldots,f_z\in\mathbb{R}^{nk}$ be a basis of the null space of $L$.
  For any $C\subseteq V$,
  \begin{align*}
    \SC(L,C) = \Pi_{(f_1)_C,\ldots,(f_z)_C}^{\bot} (L^{\dag})_{CC} \Pi_{(f_1)_C,\ldots,(f_z)_C}^{\bot}.
  \end{align*}
\end{fact}

Below we will need to use the following lemma,
which is a direct consequence of the Matrix Chernoff bound (Theorem~\ref{thm:matrixchernoff}).

\begin{lemma}\label{lem:lsamplek}
  Let $f_1,\ldots,f_{\ell}\in\mathbb{R}^{nk}$ be any $\ell$ orthonormal vectors.
  Define $F:[n]\to \mathbb{R}^{\ell\times k}$ by
  \begin{align*}
    F(u) = 
    \begin{pmatrix}
      (f_1)_u^T \\
      \vdots \\
      (f_\ell)_u^T
    \end{pmatrix} \in \mathbb{R}^{\ell\times k}.
  \end{align*}
  Suppose for any $u\in [n]$
  \begin{align*}
    \lambdamax\kh{ F(u)^T F(u) } \leq R.
  \end{align*}
  For an $s\geq 100 R n \log n$,
  let $C\subseteq [n]$ be a uniformly random subset of indices of
  size $s$.
  Then we have with probability $1 - n^{-10}$ that
  \begin{align*}
    \frac{1}{2} I_{\ell\times \ell} \pleq
    \frac{n}{s} \kh{ \sum_{u\in C} F(u) F(u)^T } \pleq
    2 I_{\ell\times \ell}
  \end{align*}
  and therefore
  \begin{align*}
    \sum_{j=1}^{\ell} (f_j)_{C} (f_j)_C^T \pleq \frac{2s}{n} I_{nk\times nk}.
  \end{align*}
\end{lemma}

\begin{definition}[Subgraph preservation]
  Let $G = (V,E)$ be a $k\times k$ matrix-weighted graph. %
  Let $\lambda_1,\ldots,\lambda_{nk}$ be eigenvalues of the normalized Laplacian $N = D^{\dag/2} L D^{\dag/2}$
  of $G$ and let $f_1,\ldots,f_{nk}$ be a corresponding set of orthonormal eigenvectors.
  For a vertex subset $U\subseteq V$ of size $t$, we say the vertex-induced subgraph $G[U]$
  $(\alpha,\beta,\zeta)$-preserves $G$ for some $\alpha,\beta\geq 1$, $\zeta\in(0,1)$ iff
  \begin{enumerate}
    \item The null space of $D^{\dag/2}_{UU} L_{G[U]} D^{\dag/2}_{UU}$ is exactly the linear span
      of $\setof{(f_i)_U : \lambda_i = 0}$.
    \item For all vectors $x\in\mathbb{R}^{|U|k}$ such that
      $x^T (f_i)_U = 0,\forall i: \lambda_i = 0$,
      \begin{align*}
        x^T \kh{ D^{\dag/2}_{UU} L_{G[U]} D^{\dag/2}_{UU} }^{\dag} x
        \leq \cdot x^T \kh{\alpha\cdot
          \frac{n^2}{t^2} \sum_{i: \lambda_i\in (0,\zeta]} \frac{1}{\lambda_i} (f_i)_U (f_i)_U^T +
          \beta\cdot \frac{n}{t} \cdot \frac{1}{\zeta} I
        } x.
      \end{align*}
  \end{enumerate}
\end{definition}

Therefore,
$G$ $(1,1,\zeta)$-preserves itself for any $\zeta$.

\begin{theorem}\label{thm:hexppreserve}
  Let $G = (V,E)$ be a $k\times k$ matrix-weighted $(\gamma,\zeta,\psi)$-almost regular expander
  with $\zeta \leq 1/\log n$.
  For an
  \begin{align*}
    s\geq 2\cdot 10^6\cdot \gamma \psi\zeta^{-1} k^2 n^{\frac{50000}{\log \log n}},
  \end{align*}
  let $C\subseteq V$ be a uniformly random vertex subset of size $s$. Then
  with probability at least $1 - n^{-7}$,
  the induced subgraph $G[C]$ $(2,n^{\frac{4096}{\log\log n}},\zeta)$-preserves $G$.
\end{theorem}

The theorem is a consequence of applying the following lemma $O(\frac{\log n}{\log\log n})$ times.

\begin{lemma}\label{lem:subpreserve}
  Let $G = (V,E)$ be a $k\times k$ matrix-weighted $(\gamma,\zeta,\psi)$-almost regular expander
  with $\zeta \leq 1/\log n$.
  Suppose there exist an $\alpha\in [1,2]$, a $\beta \geq 1$, and a $\delta\in(0,1)$, and a
  \begin{align*}
    t\geq 2\cdot 10^6\cdot \alpha \beta \gamma \psi\zeta^{-1} k^2 \log^{10} n,
  \end{align*}
  such that,
  a subgraph $G[U]$ induced by a random vertex subset $U$ of size $t$
  $(\alpha,\beta,\zeta)$-preserves $G$ with probability at least $1 - \delta$.

  For an $s \in [t/\log n,t]$, let $C\subseteq V$ be a uniformly random subset of size $s$.
  Then with probability at least $1 - \delta - n^{-8}$, the induced subgraph $G[C]$
  $((1 + \frac{1}{\log n}) \alpha, 64 \beta, \zeta)$-preserves 
  $G$.
\end{lemma}

\begin{proof}[Proof of Lemma~\ref{lem:subpreserve}]
  Let $\lambda_1\leq\ldots\leq\lambda_{nk}$ be eigenvalues of the normalized Laplacian $N = D^{\dag/2} L D^{\dag/2}$
  of $G$ and let $f_1,\ldots,f_{nk}$ be a corresponding set of orthonormal eigenvectors.
  Let $z$ be such that $\lambda_z = 0 < \lambda_{z+1}$.

  Consider generating $C$ by first randomly sampling a subset $U\subseteq V$ of size $t$,
  and then sub-sampling a subset $C\subseteq U$ of size $s$ from $U$.
  We first prove a claim about
  the Schur complement of $L_{G[U]}$ onto $C$.
  \begin{claim}\label{claim:betterexp}
    With probability at least $1 - \delta - 2 n^{-10}$, we have
    for all $x\in\mathbb{R}^{|C|k}$ such that
    $x^T (f_i)_C = 0, \forall i: \lambda_i = 0$,
    \begin{align*}
      x^T \kh{
        D^{\dag/2}_{CC} 
        \SC\kh{L_{G[U]}, C}
      D^{\dag/2}_{CC} }^{\dag} x
      \leq  x^T \kh{\alpha\cdot 
        \frac{n^2}{t^2} \sum_{i: \lambda_i\in (0,\zeta]} \frac{1}{\lambda_i} (f_i)_C (f_i)_C^T +
        \beta\cdot
        \frac{16s}{t} \cdot \frac{n}{t}\cdot \frac{1}{\zeta} I
      } x.
    \end{align*}
  \end{claim}
  \begin{proof}[Proof of Claim~\ref{claim:betterexp}]
    Let $F := U\setminus C$.
    By Proposition~\ref{prop:choleskyk},
    we can factorize $L_{G[U]}$ as
    \begin{align*}%
      L_{G[U]} =
      \begin{pmatrix}
        I_{FF} & 0 \\
        (L_{G[U]})_{CF} (L_{G[U]})_{FF}^{\dag} & I_{CC}
      \end{pmatrix}
      \begin{pmatrix}
        (L_{G[U]})_{FF} & 0 \\
        0 & \SC(L_{G[U]},C)
      \end{pmatrix}
      \begin{pmatrix}
        I_{FF} & 0 \\
        (L_{G[U]})_{CF} (L_{G[U]})_{FF}^{\dag} & I_{CC}
      \end{pmatrix}^T.
    \end{align*}
    By multiplying $D_{UU}^{\dag/2}$ on both sides, we get a factorization of $D^{\dag/2}_{UU} L_{G[U]} D^{\dag/2}_{UU}$:
    \begin{align*}
      & D_{UU}^{\dag/2}
      L_{G[U]} D_{UU}^{\dag/2} \\
      = & \begin{pmatrix}
        D_{FF}^{\dag/2} & 0 \\
        D_{CC}^{\dag/2} (L_{G[U]})_{CF} (L_{G[U]})_{FF}^{\dag} & D_{CC}^{\dag/2}
      \end{pmatrix}
      \begin{pmatrix}
        (L_{G[U]})_{FF} & 0 \\
        0 & \SC(L_{G[U]},C)
      \end{pmatrix}
      \begin{pmatrix}
        D_{FF}^{\dag/2} & 0 \\
        D_{CC}^{\dag/2} (L_{G[U]})_{CF} (L_{G[U]})_{FF}^{\dag} & D_{CC}^{\dag/2}
      \end{pmatrix}^T \\ =
      & \begin{pmatrix}
         I_{FF} & 0 \\
         D_{CC}^{\dag/2} (L_{G[U]})_{CF} (L_{G[U]})_{FF}^{\dag} D_{FF}^{1/2} & I_{CC}
      \end{pmatrix}
      \begin{pmatrix}
        D_{FF}^{\dag/2}(L_{G[U]})_{FF} D_{FF}^{\dag/2}& 0 \\
        0 & D_{CC}^{\dag/2}\SC(L_{G[U]},C)D_{CC}^{\dag/2}
      \end{pmatrix} \\
      & \begin{pmatrix}
        I_{FF} & 0 \\
        D_{CC}^{\dag/2} (L_{G[U]})_{CF} (L_{G[U]})_{FF}^{\dag} D_{FF}^{1/2}& I_{CC}
      \end{pmatrix}^T,
    \end{align*}
    where in the last inequality we have used that the rows of $(L_{G[U]})_{FF}$
    are all in the range of $D_{FF}$.
    In the case that $G[U]$ $(\alpha,\beta,\zeta)$-preserves $G$, we have
    that the null space of $D^{\dag/2}_{UU} L_{G[U]} D^{\dag/2}_{UU}$ is exactly
    the linear span of $\setof{(f_i)_U : 1\leq i\leq z}$.
    Then by Fact~\ref{fact:msck},
    \begin{align*}
      \kh{ D_{CC}^{\dag/2}\SC(L_{G[U]},C)D_{CC}^{\dag/2} }^{\dag} =
      \Pi_{(f_1)_C,\ldots,(f_z)_C}^{\bot}
      \kh{ \kh{ D_{UU}^{\dag/2} L_{G[U]} D_{UU}^{\dag/2} }^{\dag} }_{CC}
      \Pi_{(f_1)_C,\ldots,(f_z)_C}^{\bot}.
    \end{align*}
    Let $\lambdatil_1,\ldots,\lambdatil_{\ell}$ be all eigenvalues of
    $D^{\dag/2}_{UU} L_{G[U]} D^{\dag/2}_{UU}$ that are in the range
    $\left(0, \frac{1}{2}\cdot\frac{t}{n}\right]$. %
    Let $\ftil_1,\ldots,\ftil_{\ell}$ be a set of orthonormal eigenvectors,
    and define the embedding $\Ftil: V\to R^{\ell\times k}$ by
    \begin{align*}
      \Ftil(u) =
      \begin{pmatrix}
        (\ftil_1)_u^T \\
        \vdots \\
        (\ftil_{\ell})_u^T
      \end{pmatrix} \in \mathbb{R}^{\ell\times k}.
    \end{align*}
    Since $s\in [t/\log n,t]$ and $\zeta\leq 1/\log n$,
    we have $\frac{t}{2s}\cdot\frac{t}{n}\cdot \zeta \leq
    \frac{1}{2}\cdot \frac{t}{n}$.
    Then %
    we have
    \begin{multline}\label{eq:util}
      \kh{ D^{\dag/2}_{UU} L_{G[U]} D^{\dag/2}_{UU} }^{\dag} \pleq \\
      \Pi_{(f_1)_U,\ldots,(f_z)_U}^{\bot}
      \kh{\alpha\cdot
        \frac{n^2}{t^2} \sum_{i: \lambda_i\in (0,\zeta]} \frac{1}{\lambda_i} (f_i)_U (f_i)_U^T +
        \beta\cdot \frac{4n}{t} \sum_{j=1}^{\ell} \frac{1}{\zeta} \ftil_j \ftil_j^T +
        \beta\cdot \frac{4}{\frac{t}{2s}\cdot\frac{t}{n}\cdot \zeta}  I
      } \Pi_{(f_1)_U,\ldots,(f_z)_U}^{\bot}.
    \end{multline}
    Using the matrix Chernoff bound and that $t\geq 2\cdot 10^6 \gamma k \log^5 n$,
    we have with probability $1 - n^{-10}$ that
    \begin{align}
      \frac{1}{2} D_{G[U]} \pleq \frac{t}{n} D_{UU} \pleq
      2 D_{G[U]}.
    \end{align}
    As a result, $G[U]$ is $2\gamma$-almost regular.
    Therefore by applying Lemma~\ref{lem:fsu2} to $G[U]$
    with $\Dtil = \frac{t}{n} D_{UU}$ with $\kappa = 2$
    and $\delta = \frac{1}{2}$, we have for each vertex $u$
    \begin{align*}
      \lambdamax\kh{\Ftil(u)^T \Ftil(u)} \leq  \frac{32 \gamma k}{n}.
    \end{align*}
    By Lemma~\ref{lem:lsamplek}, with probability $1 - n^{-10}$,
    \begin{align}\label{eq:plug}
      \sum_{j=1}^{\ell} (\ftil_j)_{C} (\ftil_j)_C^T
      \pleq \frac{2s}{t} I.
    \end{align}
    Now by restricting both sides of~(\ref{eq:util}) to vertices in $C$
    and then plugging in~(\ref{eq:plug}), we have our desired claim.
  \end{proof}

  We now set up a matrix-valued martingale.
  To that end, we also need the notion of graph squaring as we did in the ordinary graph case.
  \begin{definition}[Graph squaring]
    For a graph $G = (V,E)$, define the square of $G$ by
    \begin{align*}
      L_{G^2} \defeq D - A D^{\dag} A.
    \end{align*}
  \end{definition}
  The proof of the following claim is deferred to Appendix~\ref{sec:apexp}.
  \begin{claim}\label{claim:gsq}
    $L_{G^2} \pleq 2 L_G$ for any graph $G$.
  \end{claim}
  Below, we will write $\SCN(L,C)$ or $\SCN(G,C)$ to denote an $nk\times nk$ matrix
  obtained by augmenting $\SC(L,C)$ to $nk\times nk$ by adding zeros on row and column blocks in $V\setminus C$.
  We will also abuse the notation a bit and write $L_{G[C]}$ to denote an $nk\times nk$ matrix
  obtained by augmenting $L_{G[C]}$ to $nk\times nk$ by adding zeros on row and column blocks in $V\setminus C$.
  The proof of the following proposition is also deferred to Appendix~\ref{sec:apexp}.
  \begin{proposition}\label{prop:sumsck}
    $\sum_{u\in V} \SCN(L_G,V\setminus \setof{u}) =
    L_{G^2} + (n-2) L_G$ for any graph $G$.
  \end{proposition}
  Let $v_1,\ldots,v_t$ be a uniformly random permutation of vertices in $U$.
  Define
  \begin{align*}
    V_i = \setof{v_1,\ldots,v_i}.
  \end{align*}
  Let $H = G[U]$.
  Consider the following sequence of matrices:
  \begin{align*}
    X_0 & = L_H \\
    X_{1} & = L_H(:,v_1) D_{v_1}^{\dag} L_H(:,v_1)^T +
    \kh{1 + \frac{2}{t-2}} { L_{H[U - V_1]} + \frac{1}{t} \kh{ L_{H^2} - 2L_H } } \\
    X_{2} & = L_H(:,v_1) D_{v_1}^{\dag} L_H(:,v_1)^T + \frac{1}{t} \kh{ L_{H^2} - 2 L_H } + \kh{1 + \frac{2}{t-2}}\cdot  \\
    &\quad  \bigg( L_{H[U-V_1]}(:,v_2) \kh{D^{H[U-V_1]}_{v_2}}^{\dag} L_{H[U-V_1]}(:,v_2)^T +
    \kh{1 + \frac{2}{t-3}} L_{H[U - V_2]} \\
  &\quad + \frac{1}{t-1} \kh{ L_{H[U - V_1]^2} - 2L_{H[U - V_1]} } \bigg)
  \end{align*}
  $X_{i+1}$ is obtained by replacing the $L_{H[U-V_i]}$ in $X_i$ by
  \begin{align}\label{eq:rplk}
    & L_{H[U-V_i]}(:,v_{i+1}) \kh{D^{H[U-V_i]}_{v_{i+1}}}^{\dag} L_{H[U-V_i]}(:,v_{i+1})^T +
    \kh{1 + \frac{2}{t-2-i}} L_{H[U-V_{i+1}]} \notag \\
    & \quad + \frac{1}{t-i} \kh{ L_{H[U-V_i]^2} - 2L_{H[U-V_i]} }.
  \end{align}
  Here, we have used $D^H_v$ to denote the degree of vertex $v$ in graph $H$.
  The proof of the following claim is also deferred to Appendix~\ref{sec:apexp}.
  \begin{claim}\label{claim:mart}
    $X_0,X_1,X_2,\ldots$ is a matrix-valued martingale.
  \end{claim}

  Define $Y_i = L_H^{\dag/2} X_i L_H^{\dag/2} - L_H^{\dag/2} L_H L_H^{\dag/2}$.
  As $X_0,X_1,\ldots$ is a martingale, so is $Y_0,Y_1,\ldots$, by the linearity of expectation.
  Additionally, since $X_0 = L_H$, $Y_0,Y_1,\ldots$ has zero expectation.
  We focus on the first $t - s + 1$ terms
  $Y_0,Y_1,Y_2,\ldots,Y_{t-s}$. First we calculate the difference
  sequence $Z_i = Y_{i+1} - Y_i$:
  \begin{align*}
    Z_i = & \prod_{j=1}^{i} \kh{1 + \frac{2}{t-1-j}}\cdot \\
    & L_H^{\dag/2} \bigg(
      - \SCN \kh{H[U-V_i], U - V_{i+1}} + \kh{1 + \frac{2}{t-2-i}} L_{H[U-V_{i+1}]} \\
      & \quad
      - \frac{1}{t-i} \kh{ L_{H[U-V_i]^2} - 2L_{H[U-V_i]} }
    \bigg) L_H^{\dag/2},
  \end{align*}
  where we have used the definition of Schur complements.
  We then consider to bound the maximum (in absolute value) eigenvalue of $Z_i$.
  For the last term, we have
  \begin{align}\label{eq:lasttermk}
    \lambdamax\kh{ L_H^{\dag/2}
      \frac{1}{t-i} \kh{ L_{H[U-V_i]^2} - 2L_{H[U-V_i]} }
      L_H^{\dag/2} 
    }\leq \frac{4}{t-i} \leq \frac{4}{s} \leq \frac{4\log n}{t} \leq \frac{4}{10^6 \log^6 n},
  \end{align}
  where the first inequality follows from $L_{H[U-V_i]^2} \pleq 2L_{H[U-V_i]} \pleq 2L_H$ by
  Claim~\ref{claim:gsq}.
  Now to bound the first two terms, we first rearrange them as
  \begin{align}\label{eq:rearrangek}
    \kh{ - \SCN \kh{H[U-V_i], U - V_{i+1}} + L_{H[U-V_{i+1}]} } + \frac{2}{t-2-i} L_{H[U-V_{i+1}]}.
  \end{align}
  Then for the second term of~(\ref{eq:rearrangek}),
  we have $L_{H[U - V_{i+1}]} \pleq L_H$, and hence
  \begin{align}\label{eq:secondtermk}
    \lambdamax\kh{ L_H^{\dag/2} \frac{2}{t-2-i}L_{H[U - V_{i+1}]} L_H^{\dag/2} } \leq
    & \frac{2}{t-2-i} \leq \frac{2}{ s - 2 }
    \leq \frac{4}{s} \leq \frac{4\log n}{t} \leq \frac{4}{10^6 \log^6 n}.
  \end{align}
  We then claim that the first term of~(\ref{eq:rearrangek}) can be bounded by
  \begin{align}\label{eq:firsttermk}
    \lambdamax \kh{ L^{\dag/2}_{H}
    \kh{ \SCN \kh{H[U-V_i], U - V_{i+1}} - L_{H[U - V_{i+1}]} } L^{\dag/2}_H } \leq
    { \alpha \frac{\psi k^2}{t} + \beta\frac{\gamma k}{\zeta t} } \leq \frac{1}{8 \cdot 10^6 \log^6 n}.
  \end{align}
  To see why this is the case,
  let us define an edge set $E_{i+1} = \setof{ (v_{i+1}, w) \in E : w \in U - V_{i+1} }$,
  and write $L_{E_{i+1}}$ to denote the Laplacian of the subgraph of $H$ induced by $E_{i+1}$.
  Notice that
  \begin{align*}
    L_{E_{i+1}}(:,v_{i+1}) = L_{H[U-V_{i}]}(:,v_{i+1}).
  \end{align*}
  Therefore, we have
  \begin{align*}
    \mathrm{SC} \kh{H[U-V_i], U - V_{i+1}} - L_{H[U - V_{i+1}]} =
    \mathrm{SC} \kh{L_{E_{i+1}}, U - V_{i+1}}.
  \end{align*}
  Therefore to prove~(\ref{eq:firsttermk}), it suffices to show that
  for any vector $x$ such that $x\in \mathrm{range}(\SCN(L_{E_{i+1}}, U - V_{i+1}))$, we have
  \begin{align}\label{eq:smallflowk}
    x^T L^{\dag}_H x \leq \kh{ \alpha \frac{\psi k^2}{t} + \beta \frac{\gamma k}{\zeta t} } \cdot
    x^T \mathrm{SC} \kh{L_{E_{i+1}}, U - V_{i+1}}^{\dag} x
    = \kh{ \alpha \frac{\psi k^2}{t} + \beta \frac{\gamma k}{\zeta t} }\cdot x^T L_{E_{i+1}}^{\dag} x.
  \end{align}
  Since $x\in \mathrm{range}(\SCN(L_{E_{i+1}}, U - V_{i+1}))$,
  we can write $x$ as a liner combination of the incidence vectors of the edges in $E_{i+1}$:
  \begin{align*}
    x = \sum_{e\in E_{i+1}} c_e b_e
  \end{align*}
  for some coefficients $c_e$'s,
  where we have $x_{v_{i+1}} = \sum_{e\in E_{i+1}} c_e (b_e)_{v_{i+1}} = 0$.
  Then since $G[E_{i+1}]$ is a star graph, we have
  \begin{align*}
    x^T L_{E_{i+1}}^{\dag} x = \sum_{e\in E_{i+1}} c_e^2.
  \end{align*}
  We now calculate $x^T L^{\dag}_H x$ by
  \begin{align*}
    x^T L^{\dag}_H x = & x^T D^{\dag/2}_{UU} \kh{ D^{\dag/2}_{UU} L_H D^{\dag/2}_{UU} }^{\dag} D^{\dag/2}_{UU} x \\
    \leq & x^T D^{\dag/2}_{UU}
    \kh{\alpha \cdot \frac{n^2}{t^2} \sum_{i: \lambda_i \in (0,\zeta]}\frac{1}{\lambda_i} (f_i)_U (f_i)_U^T
    + \beta\cdot \frac{n}{t}\cdot \frac{1}{\zeta} I} D^{\dag/2}_{UU} x \\
    & \qquad \text{(since $H = G[U]$ $(\alpha,\beta,\zeta)$-preserves $G$)} \\
    \leq & |E_{i+1}| \cdot \alpha \cdot \frac{n^2}{t^2} \sum_{e\in E_{i+1}} c_e^2 \frac{\psi k^2}{n^2}
    + \beta \frac{n}{t}\cdot \frac{1}{\zeta} \norm{ \sum_{e\in E_{i+1}} c_e D_{UU}^{\dag/2} b_e }^2 \\
    & \qquad \text{(by expanding and that $G$ is a $(\gamma,\zeta,\psi)$-almost regular expander)} \\
    \leq & { \alpha \frac{\psi k^2}{t} }
    \kh{ \sum_{e\in E_{i+1}} c_e^2 } +
    \beta \frac{n}{t}\cdot \frac{1}{\zeta} \kh{ \sum_{(w,v_{i+1})\in E_{i+1}}
    c_e^2 \norm{ D_{w}^{\dag/2} \phi_{v_{i+1}w} }^2 }
    \\
    & \qquad \text{(by $|E_{i+1}| \leq t$ and that $\sum_{e\in E_{i+1}} c_e D_{UU}^{\dag/2} b_e$
      is zero on $v_{i+1}$)} \\
    \leq & \kh{ \alpha \frac{\psi k^2}{t}  + \beta \frac{\gamma k}{\zeta t} }
    \kh{ \sum_{e\in E_{i+1}} c_e^2 } \\
    & \qquad \text{(by $\gamma$-almost regularity of $G$).}
  \end{align*}
  This proves our desired inequality~(\ref{eq:smallflowk}).

  We also note that
  \begin{align*}
    \prod_{j=1}^{i} \kh{1 + \frac{2}{t-1-j}}= \frac{t}{t-2} \frac{t-1}{t-3} \ldots \frac{t+1-i}{t-1-i}
    = \frac{t(t-1)}{(t-i)(t-1-i)} \leq \kh{\frac{t}{s}}^2 \leq \log^2n.
  \end{align*}
  Therefore
  \begin{align}\label{eq:mtglR2}
    \lambdamax(Z_i) \leq \frac{1}{1000\cdot \log^{4} n}.
  \end{align}
  Fixing a $j$,
  we then consider bounding the spectral norm of $\ex{Z_j^2\ |\ Y_1,\ldots,Y_{j-1}}$.
  \begin{align*}
    & \ex{Z_j^2\ |\ Y_1,\ldots,Y_{j-1}} \\ \pleq
    & \frac{2}{t-j} \cdot \kh{\frac{t}{t-1-j}}^4 \sum_{v \in U - V_{j}}
    \kh{ L_H^{\dag/2}
    \kh{ \SCN(H[U- V_j], U - V_j - \setof{v}) - L_{H[U - V_j - \setof{v}]} }
  L_H^{\dag/2} }^2 + \\
  & \frac{2}{t-j}\cdot \kh{ \frac{t}{t-1-j} }^4 \sum_{v\in U - V_{j}} \kh{ L_H^{\dag/2}
    \kh{
      \frac{2}{t-2-j} L_{H[U-V_{j} - \setof{v}]} +
  \frac{1}{t-j} \kh{ L_{H[U-V_j]^2} + 2L_{H[U-V_j]} } }
  L_H^{\dag/2}
  }^2 \\ \pleq
  & \frac{2 \log^5 n}{t-j} \cdot\kh{ \alpha \frac{\psi k^2}{t}
  + \beta \frac{\gamma k}{\zeta t} } \sum_{v\in U-V_j} L_H^{\dag/2}
    \kh{ \SCN(H[U- V_j], U - V_j - \setof{v}) - L_{H[U - V_j - \setof{v}]} }
    L_H^{\dag/2} + \\
    & \frac{2 \log^5 n}{t-j}\cdot
    \sizeof{U-V_j}\cdot \kh{ \frac{8}{s} }^2 I_{tk\times tk}
    \qquad \text{(by~(\ref{eq:firsttermk}),(\ref{eq:secondtermk}),(\ref{eq:lasttermk}))}
    \\
    \pleq & \frac{2 \log^6 n}{t} \kh{ \alpha \frac{\psi k^2}{t} + \beta \frac{\gamma k}{\zeta t} } %
    L_H^{\dag/2} L_{H[U-V_j]^2} L_H^{\dag/2} +
    \frac{10000\log^7n}{t^2} I_{tk\times tk} \\
    & \qquad \text{(by Proposition~\ref{prop:sumsck} and $t-j\geq s\geq t/\log n$)}
    \\
    \pleq & \frac{64}{2\cdot 10000 \cdot t \log^4 n} I_{tk\times tk} +
    \frac{10000\log^7n}{t^2} I_{tk\times tk}
    \qquad \text{(by Claim~\ref{claim:gsq} and the values of $\gamma,\zeta,\psi$)}
    \\
    \pleq & \frac{1}{1000 t \log^{4} n} I_{tk\times tk}.
  \end{align*}
  Therefore we have
  \begin{align*}
    \normm{ \ex{Z_j^2\ |\ Y_1,\ldots,Y_{j-1}} } \leq \frac{1}{1000 t \log^4 n}
  \end{align*}
  and
  \begin{align}\label{eq:mtgls2}
    \normm{ \sum_{j=0}^{t - s - 1} \ex{Z_j^2\ |\ Y_1,\ldots,Y_{j-1} } } \leq
    & \sum_{j=0}^{t - s - 1} \normm{ \ex{Z_j^2\ |\ Y_1,\ldots,Y_{j-1} } } \notag \\ \leq
    & t\cdot \frac{1}{1000 t \log^4 n} \notag \\
    = & \frac{1}{1000 \log^4 n}.
  \end{align}
  We now invoke Theorem~\ref{thm:tropp} with $R = \frac{1}{1000\log^{4}n}$,
  $\sigma^2 = \frac{1}{1000\log^{4} n}$, and
  $t = \frac{1}{10\log n}$, and get
  \begin{align*}
    \pr{\exists i\in [0,n-s]: \lambdamax(Y_i) \geq 1/2} \leq
    n\cdot \exp\kh{- \frac{\frac{1}{200 \log^2 n}}{\frac{1}{1000\log^{4} n} +
    \frac{1}{30000 \log^5 n} } } \leq n^{-10}.
  \end{align*}
  This coupled with Fact~\ref{fact:msck} implies that
  \begin{align}\label{eq:prsv1}
    L_{G[C]} \pgeq \kh{ 1 - \frac{1}{5\log n} } \cdot \frac{s^2}{t^2} \cdot \mathrm{SC}(L_{G[U]}, C).
  \end{align}
  Then using Claim~\ref{claim:betterexp},
  for all $x\in\mathbb{R}^{|C|k}$ such that
  $x^T (f_i)_C = 0, \forall i: \lambda_i = 0$,
  \begin{align}
    & x^T \kh{ D_{CC}^{\dag/2} L_{G[C]} D_{CC}^{\dag/2} }^{\dag} x \\
    \leq & \kh{1 + \frac{1}{2\log n}} \cdot\frac{t^2}{s^2}\cdot
    x^T \kh{
      \alpha\cdot \frac{n^2}{t^2} \sum_{i: \lambda_i\in (0,\zeta]} \frac{1}{\lambda_i} (f_i)_C (f_i)_C^T +
      \beta\cdot\frac{16s}{t} \cdot \frac{n}{t}\cdot \frac{1}{\zeta} I
    } x \notag\\
    \leq &
    x^T  \kh{
      \kh{1 + \frac{1}{\log n}} \alpha\cdot\frac{n^2}{s^2} \sum_{i: \lambda_i\in (0,\zeta]} 
      \frac{1}{\lambda_i} (f_i)_C (f_i)_C^T +
      64 \beta\cdot \frac{n}{s}\cdot \frac{1}{\zeta} I
    } x. \label{eq:prsv2}
  \end{align}
  (\ref{eq:prsv1}),(\ref{eq:prsv2}) together imply that
  $G[C]$ $((1+\frac{1}{\log n}) \alpha, 64\beta,\zeta)$-preserves $G$, as desired.
\end{proof}

\section{A lower bound for weighted spectral sparsification}\label{sec:lbspectral}

In this section, we prove a superlinear lower bound of $n^{21/20-o(1)}$ for computing
some $O(1)$-spectral sparsifier.

\spectrallb*

By Proposition~\ref{prop:B},
in order to prove Theorem~\ref{thm:spectrallb},
it suffices to prove Theorem~\ref{thm:tvdsmall}.
Recall that $B^w\in\mathbb{R}^{\binom{n}{2}\times n}$ is the weighted signed edge-vertex incidence matrix
of the input graph generated from the distribution specified in Section~\ref{sec:ovspectrallb}.
\tvdsmall*

\begin{proof}[Proof of Theorem~\ref{thm:tvdsmall}]
  Similar to Section~\ref{sec:onerow}, we can assume w.l.o.g.
  the number of $(u,v)\in E$ for which $\phi_{uv}\neq 0$
  is at least $\Omega(n^2)$, since otherwise 
  \begin{align*}
    \expec{\pi}{d_{\tv}\kh{ (\Phi B^w)_{\pi,\yes}, (\Phi B^w)_{\pi,\no} }} \leq
    O(1)\cdot \prob{\pi}{\phi_{\pi(1)\pi(n/2+1)} \neq 0} \leq o(1),
  \end{align*}
  and we would already have our desired result.
  Let us then define the $k\times k$ matrix-weighted graph $H_{\phi} = (V,E_{\phi})$,
  where $E_{\phi}$ contains {\em all} edges $(u,v)$ whose $\phi_{uv}\neq 0$
  (including the ones not present in the input graph),
  and each edge $(u,v)$ has matrix weight $\phi_{uv} \phi_{uv}^T$.
  Thus by the above assumption we have $|E_{\phi}| \geq \Omega(n^2)$.

  By Theorem~\ref{thm:edintro}, there exists a scaling $s:E_{\phi}\to [0,1]$ s.t.
  $H_{\phi}^{s}$ is a $(\gamma,\zeta,\psi)$-almost regular expander,
  where $(\gamma,\zeta,\psi) = (8\log n, 1/\log n, 16 k^2 \log^3 n)$,
  and $|(u,v)\in E_{\phi}: s_{uv} < 1| \leq o(n^2)$.
  Let $\lambda_1 \leq \ldots \leq \lambda_{nk}$ be the eigenvalues of the normalized Laplacian matrix
  of $H_{\phi}^{s}$, and let $f_1,\ldots,f_{nk}\in\mathbb{R}^{nk}$ be a corresponding set of orthonormal
  eigenvectors.

  For $i = 0,\ldots, n^{4/5}-1$, let $B_i$ denote the vertices in the $i^{\mathrm{th}}$ block
  of the input graph:
  \begin{align*}
    B_i \defeq \setof{ \pi(n^{1/5} i + 1),\ldots,\pi(n^{1/5} i + n^{1/5}) }.
  \end{align*}

  Since $k\leq n^{1/20 - \eps}$, we have
  $2\cdot 10^6 \gamma \psi \zeta^{-1} k^2 n^{\eps} \leq n^{1/5}$.
  Thus by Theorem~\ref{thm:expsampintro}, %
  with probability $1 - 1/n^4$,
  all vertex-induced subgraphs $H_{\phi}^{s}[B_i\union B_{i+1}]$ preserve $H_{\phi}^{s}$
  in the sense of~(\ref{eq:preservance}).
  Using this fact, we prove the following claim:
  \begin{claim}\label{claim:hdsmaller}
    For each edge $(x,y)$ with $s_{xy} = 1$, conditioned on $(x,y)$ being the crossing edge
    (i.e. $\setof{\pi(1),\pi(n/2+1)} = \setof{x,y}$),
    with probability at least $1 - 1/n^2$ over $\pi$, we have
    that $b_{xy}$ is in the range of $L_{\pi}$, and
    \begin{align}\label{eq:claim:er}
      b_{xy}^{T} L_{\pi}^{\dag} b_{xy} \leq \frac{k^2}{n^{2/5 - o(1)}}.
    \end{align}
  \end{claim}
  Once again by $k\leq n^{1/20-\eps}$,
  the RHS of~(\ref{eq:claim:er}) is $o(1)$.
  Since $s_{uv} = 1$ holds for a $1-o(1)$ fraction of the edges,
  we have, by Proposition~\ref{prop:tver},
  $\expec{\pi}{d_{\tv}\kh{ (\Phi B^w)_{\pi,\yes}, (\Phi B^w)_{\pi,\no} }} = o(1)$.
\end{proof}

\begin{proof}[Proof of Claim~\ref{claim:hdsmaller}]
  Consider drawing $\pi$ from the conditional distribution
  on $\pi(1) = x$ and $\pi(n/2+1) = y$.
  Suppose w.l.o.g. $(x,y)$ is oriented $x\to y$, so $(b_{xy})_x = \phi_{xy}$ and $(b_{xy})_y = -\phi_{xy}$.
  Let $E_{\pi}$ be the set of edges of $\Hcal_{\pi}$,
  where we recall that $\Hcal_{\pi}$ consists of the non-crossing edges,
  such that each edge $(u,v)$ has matrix weight $n^{4/5}\log^{-1} n \phi_{uv} \phi_{uv}^T \in \mathbb{R}^{k\times k}$.
  Let us call a function $\fbf : E_{\pi} \to\mathbb{R}$ a {\em flow} in the graph $\Hcal_{\pi}$,
  and say $\fbf$ routes a {\em demand} $d\in\mathbb{R}^{nk}$ if $\sum_{e\in E_{\pi}} \fbf_e b_e = d$.

  By Fact~\ref{thm:dual}, to prove~(\ref{eq:claim:er}),
  it suffices to show that there exists a flow $\fbf$ in $\Hcal_{\pi}$ that routes demand $n^{-2/5} b_{xy}$
  satisfying $\norm{\fbf}_2^2 \leq \frac{k^2}{n^{2/5 - o(1)}}$.
  By scaling, it then suffices to show that there is a flow $\fbf$ that routes demand $b_{xy}$
  satisfying $\norm{\fbf}_2^2 \leq k^2 n^{2/5 + o(1)}$.
  To construct such a flow, we consider first breaking the demand $b_{xy}$
  into demands $d_{0},d_1,\ldots,d_{n^{4/5}/2-1}\in\mathbb{R}^{nk}$ such that
  \begin{enumerate}
    \item $\sum_{0\leq i < n^{4/5}/2} d_i = b_{xy}$.
    \item $d_i$ is supported on $B_i\union B_{i+1}$.
  \end{enumerate}
  We will then route each $d_i$ in the vertex induced subgraph $H_{\phi}[B_i\union B_{i+1}]$ using
  a flow $\fbf_i$, and finally obtain a flow $\fbf = \sum_{0\leq i < n^{4/5}/2} \fbf_i$ that routes $b_{xy}$
  and satisfies $\norm{\fbf}_2^2 \leq 2\sum_{0\leq i < n^{4/5}/2} \norm{\fbf_i}_2^2$.

  To show how these $d_i$'s are constructed, let us first define
  vectors $\dtil_0,\ldots,\dtil_{n^{4/5}/2}\in\mathbb{R}^{nk}$
  where $\dtil_i$ is supported on $B_i$.
  We will later define each $d_i$ by letting $d_i = \dtil_i - \dtil_{i+1}$.
  Let $D\in\mathbb{R}^{nk\times nk}$ be the degree matrix of $H_{\phi}^s$.
  For now we would like $\dtil_i$ to satisfy
  \begin{align}\label{eq:dtili}
    \forall j: 0\leq \lambda_j \leq \zeta,\quad
    f_j^T D^{\dag/2} \dtil_i = (1 - \mu_i) (f_j)_x^T D_{xx}^{\dag/2} \phi_{xy}
    + \mu_i (f_j)_y^T D_{yy}^{\dag/2} \phi_{xy}.
  \end{align}
  where we define
  \begin{align*}
    \mu_i := \frac{i}{n^{4/5}/2}.
  \end{align*}
  In particular, we would like $f_j^T D^{\dag/2} \dtil_0 = (f_j)_x^T D^{\dag/2}_{xx} \phi_{xy}$ and
  $f_j^T D^{\dag/2} \dtil_{n^{4/5}/2} = (f_j)_y^T D^{\dag/2}_{yy} \phi_{xy}$
  for all $j$ such that $0\leq \lambda_j\leq \zeta$.
  Thus, we let %
  $\dtil_0$ be such that
  \begin{align*}
    (\dtil_0)_u =
    \begin{cases}
      \phi_{xy} & u = x \\
      0 & \text{otherwise}
    \end{cases}
  \end{align*}
  and let $\dtil_{n^{4/5}/2}$ be such that
  \begin{align*}
    (\dtil_{n^{4/5}/2})_u =
    \begin{cases}
      \phi_{xy} & u = y \\
      0 & \text{otherwise.}
    \end{cases}
  \end{align*}
  We would then like to construct the remaining $\dtil_i$'s such that each $\norm{D^{\dag/2} \dtil_i}_2$ is small.
  First let us write the linear equations~(\ref{eq:dtili}) in matrix form.
  Let $\ell$ be s.t. $\lambda_{\ell} \leq \zeta < \lambda_{\ell+1}$.
  Define a function $F: V\to \mathbb{R}^{\ell\times k}$ such that
  \begin{align*}
    F(u) =
    \begin{pmatrix}
      (f_1)^T_u \\
      \vdots \\
      (f_{\ell})^T_u
    \end{pmatrix}
  \end{align*}
  and define another function $G : \setof{B_0,\ldots,B_{n^{4/5}/2}} \to \mathbb{R}^{\ell\times n^{1/5} k}$ such that
  \begin{align*}
    G(B_i) =
    \begin{pmatrix}
      (f_1)^T_{B_i} \\
      \vdots \\
      (f_{\ell})^T_{B_i}
    \end{pmatrix}
  \end{align*}
  Then we can write~(\ref{eq:dtili}) as
  \begin{align}\label{eq:dtili2}
    G(B_i) D_{B_iB_i}^{\dag/2} (\dtil_i)_{B_i} = (1 - \mu_i) F(x) D_{xx}^{\dag/2} \phi_{xy}
    + \mu_i F(y) D_{yy}^{\dag/2} \phi_{xy}
  \end{align}
  Let us write the RHS as $r_i := (1 - \mu_i) F(x) D_{xx}^{\dag/2} \phi_{xy} + \mu_i F(y) D_{yy}^{\dag/2} \phi_{xy}$.
  Then by Fact~\ref{thm:dual}, we can find a $(\dtil_i)_{B_i}$ satisfying~(\ref{eq:dtili2})
  such that $\norm{D_{B_iB_i}^{\dag/2} (\dtil_i)_{B_i}}_2^2 = r_i^T (G(B_i) G(B_i)^T)^{-1} r_i$. Now notice
  \begin{align*}
    G(B_i) G(B_i)^T = \sum_{u\in B_i} F(u) F(u)^T.
  \end{align*}
  Since $f_1,\ldots,f_{\ell}$ are orthonormal, we have
  $\sum_{u\in V} F(u) F(u)^T = I_{\ell\times \ell}$. By Lemma~\ref{lem:fsu2}, we have for all $u\in V$ that
  $\lambdamax(F(u) F(u)^T) \leq \frac{\gamma k}{(1 - \zeta)^2 n} \leq \frac{16k \log n}{n}$.
  Since the vertices in $B_i$ are chosen uniformly at random, by a Matrix Chernoff bound
  (Theorem~\ref{thm:matrixchernoff}),
  we have with probability $\geq 1 - n^{-5}$
  that $\frac{1}{2} I \pleq n^{4/5} G(B_i) G(B_i)^T \pleq 2 I$, which implies
  $r_i^T (G(B_i) G(B_i)^T)^{-1} r_i \leq 2 n^{4/5} \norm{r_i}^2$.
  As a result, we can find $\dtil_1,\ldots,\dtil_{n^{4/5}/2-1}$ such that
  \begin{align*}
    \norm{ D^{\dag/2} \dtil_i }_2^2 \leq 4 n^{4/5} \kh{ \norm{F(x) D_{xx}^{\dag/2} \phi_{xy}}_2^2
    + \norm{F(y) D_{yy}^{\dag/2} \phi_{xy}}_2^2 }.
  \end{align*}
  For simplicity let $C_i := B_i\union B_{i+1}$.
  We then construct each $d_i$ for $0\leq i < n^{4/5}/2$ by letting $d_i := \dtil_i -\dtil_{i+1}$.
  Thus $d_i$ is indeed supported on $C_i$,
  and we have $\sum_{0\leq i < n^{4/5}/2} d_i = b_{xy}$.
  Now we consider how to route each $d_i$ in $H_{\phi}[B_i\union B_{i+1}]$.
  Once again by Fact~\ref{thm:dual}, there is a flow $\fbf_i$ with $\ell_2^2$-norm
  $(d_i)_{C_i}^T L_{H_{\phi}[C_i]}^{\dag} (d_i)_{C_i}
  \leq (d_i)_{C_i}^T L_{H_{\phi}^s[C_i]}^{\dag} (d_i)_{C_i}$. We then write
  \begin{align*}
    & (d_i)_{C_i}^T L_{H_{\phi}^s[C_i]}^{\dag} (d_i)_{C_i} %
    \\
    =
    & (d_i)_{C_i}^T D^{\dag/2}_{C_iC_i} \kh{ D^{\dag/2}_{C_iC_i} L_{H_{\phi}^s[C_i]} D^{\dag/2}_{C_iC_i} }^{\dag}
    D^{\dag/2}_{C_iC_i} (d_i)_{C_i}
    \\
    \leq
    & n^{o(1)} (d_i)_{C_i}^T D^{\dag/2}_{C_iC_i}
    \kh{ \frac{n^2}{n^{2/5}} \sum_{j: \lambda_j\in(0,\zeta]} \frac{1}{\lambda_j}
    (f_j)_{C_i} (f_j)_{C_i}^T + \frac{n}{n^{1/5}} (\log n) I }
    D^{\dag/2}_{C_iC_i} (d_i)_{C_i} \\
    \leq &
    n^{8/5+o(1)} 
    \sum_{j: \lambda_j\in(0,\zeta]} \frac{1}{\lambda_j}
    \kh{ f_j^T D^{\dag/2} (\dtil_i - \dtil_{i+1}) }^2
    + n^{4/5+o(1)} d_i^T D^{\dag} d_i
    \\
    = &
    n^{8/5+o(1)} 
    \sum_{j: \lambda_j\in(0,\zeta]} \frac{1}{\lambda_j}
    \kh{ \frac{1}{n^{4/5}/2} f_j^T D^{\dag/2} b_{xy} }^2
    + n^{4/5+o(1)} \kh{ \dtil_i^T D^{\dag} \dtil_i + \dtil_{i+1}^T D^{\dag} \dtil_{i+1} }
    \\
    = & 
    n^{o(1)} b_{xy}^T D^{\dag/2} \kh{ \sum_{j: \lambda_j\in (0,\zeta]} \frac{1}{\lambda_j} f_j f_j^T }
    D^{\dag/2} b_{xy} +
    n^{4/5+o(1)} \kh{ \dtil_i^T D^{\dag} \dtil_i + \dtil_{i+1}^T D^{\dag} \dtil_{i+1} }.
  \end{align*}
  Notice that the first term in the above is at most
  $\frac{n^{o(1)} \psi k^2}{n^2} \leq \frac{n^{o(1)} k^4}{n^2}$ by the expander property
  and $s_{xy} = 1$.
  The second term for each $0 < i < n^{4/5}/2-1$ is at most
  \begin{align*}
    n^{8/5+o(1)}\kh{ \norm{F(x) D_{xx}^{\dag/2} \phi_{xy}}_2^2 + \norm{F(y) D_{yy}^{\dag/2} \phi_{xy}}_2^2 }
    \leq \frac{k^2}{n^{2/5-o(1)}},
  \end{align*}
  where the inequality follows from $\gamma$-regularity and that $\lambdamax(F(u)^T F(u))\leq \frac{16 k \log n}{n}$.
  For $i = 0$ or $i = n^{4/5}/2 - 1$,
  the second term is at most
  \begin{multline*}
    n^{4/5+o(1)} \kh{ \norm{D^{\dag/2}_{xx} \phi_{xy}}_2^2 + \norm{D^{\dag/2}_{yy} \phi_{xy} }_2^2 } + \\
    n^{8/5+o(1)}\kh{ \norm{F(x) D_{xx}^{\dag/2} \phi_{xy}}_2^2 + \norm{F(y) D_{yy}^{\dag/2} \phi_{xy}}_2^2 }
    \leq \frac{k}{n^{1/5-o(1)}}.
  \end{multline*}

  Finally by letting $\fbf = \sum_{0\leq i < n^{4/5}/2} \fbf_i$,
  we have $\norm{\fbf}_2^2 \leq k^2 n^{2/5+o(1)}$ as desired.
\end{proof}

\section{A lower bound for weighted spanner computation}\label{sec:spanner}

In this section, we prove a superlinear lower bound of $n^{1+\alpha-o(1)}$ for computing
an $O(n^{\frac{2}{3}(1 - \alpha)})$-spanner for any constant $\alpha \in (0,1/10)$.

\spannerlb*

Our proof will be very similar to that of our lower bound for spectral sparsifiers in
Sections~\ref{sec:ovspectrallb} and~\ref{sec:lbspectral}.
In our proof,
we will also use the block cycle graph as our hard instance as in Section~\ref{sec:harddist}.
However, we make the following modifications.
\begin{enumerate}
  \item We let the number of blocks be $\ell = n^{\frac{2}{3} (1 - \alpha)}$,
    and the number of vertices in each block be $s = n^{\frac{1}{3} + \frac{2}{3}\alpha}$.
  \item We draw the weights of all non-crossing edges independently from $\Ncal(8\log n, 1)$.
    and the weight of the crossing edge from $\Ncal(0, \log^2 n)$.
\end{enumerate}

Similarly, if the crossing edge has negative weight, we say the input is invalid,
and accept any sketch as a valid sketch.

Similar to Propositions~\ref{prop:erbig} and~\ref{prop:0.6},
we have:

\begin{proposition}
  With probability at least $1 - 1/n$,
  all non-crossing edges have weights in the range $[0.1\log n, 10\log n]$.
  If the crossing edge is present,
  then its weight is between $[0.1\log n, 10\log n]$ with probability
  at least $0.41$.
\end{proposition}

\begin{proposition}
  Any linear sketch that can compute,
  with probability $0.9$,
  a $\frac{1}{1000} n^{\frac{2}{3} (1 - \alpha)}$-spanner
  in a graph where the edge weights are within a factor of $100$ of each other,
  can distinguish between the $\sfy$ and $\sfn$ distributions
  with probability at least $0.6$.
\end{proposition}

We will prove the following theorem,
which by Proposition~\ref{prop:B}
implies Theorem~\ref{thm:spannerlb}.
Recall that $B^w\in\mathbb{R}^{\binom{n}{2}\times n}$ is the weighted signed edge-vertex incidence matrix
of the input graph, as defined in Section~\ref{sec:tver}.

\begin{theorem}\label{thm:tvdsmalls}
  For any fixed sketching matrix
  $\Phi\in\mathbb{R}^{k\times \binom{n}{2}}$ where
  $k \leq n^{\alpha - \eps}$ for some constant $\eps > 0$,
  we have
  \begin{align*}
    \expec{\pi}{d_{\tv}\kh{ (\Phi B^w)_{\pi,\yes}, (\Phi B^w)_{\pi,\no} } } \leq o(1).
  \end{align*}
\end{theorem}

We also have the following proposition similar to Proposition~\ref{prop:tver}.
\begin{proposition}\label{prop:tvers}
  For any permutation $\pi$ such that
  $b_{\pi(1)\pi(n/2+1)}$ is in the range of $L_{\pi}$,
  \begin{align*}
    d_{\tv}((\Phi B^w)_{\pi,\mathrm{yes}},(\Phi B^w)_{\pi,\mathrm{no}}) \leq O(1)\cdot
    \min\setof{1, (\log^2 n) b_{\pi(1)\pi(n/2+1)} L_{\pi}^{\dag} b_{\pi(1)\pi(n/2+1)}},
  \end{align*}
  where $L_{\pi} = \sum_{\text{non-crossing $(u,v)$}} b_{uv} b_{uv}^T$,
  and $b_{uv}$'s are defined in~(\ref{eq:defbb}).
\end{proposition}

\begin{proof}[Proof of Theorem~\ref{thm:tvdsmalls}]
  We can assume w.l.o.g. that
  the number of $(u,v)\in E$ with $\phi_{uv}\neq 0$
  is $\Omega(n^2)$, as otherwise we would already have
  \begin{align*}
    \expec{\pi}{d_{\tv}\kh{ (\Phi B^w)_{\pi,\yes}, (\Phi B^w)_{\pi,\no} }} \leq
    O(1)\cdot \prob{\pi}{\phi_{\pi(1)\pi(n/2+1)} \neq 0} \leq o(1).
  \end{align*}
  Then we define the $k\times k$ matrix-weighted graph $H_{\phi} = (V,E_{\phi})$,
  where $E_{\phi}$ consists of {\em all} edges $(u,v)$ whose $\phi_{uv}\neq 0$
  (including those not present in the input graph),
  and each edge $(u,v)$ has matrix weight $\phi_{uv} \phi_{uv}^T$.
  Therefore by the assumption above we have $|E_{\phi}| \geq \Omega(n^2)$.

  By Theorem~\ref{thm:edintro}, there exists a scaling $s:E_{\phi}\to [0,1]$ s.t.
  $H_{\phi}^{s}$ is a $(\gamma,\zeta,\psi)$-almost regular expander,
  where $(\gamma,\zeta,\psi) = (8\log n, 1/\log n, 16 k^2 \log^3 n)$,
  and $|(u,v)\in E_{\phi}: s_{uv} < 1| \leq o(n^2)$.
  Let $\lambda_1 \leq \ldots \leq \lambda_{nk}$ be the eigenvalues of the normalized Laplacian matrix
  of $H_{\phi}^{s}$, and let $f_1,\ldots,f_{nk}\in\mathbb{R}^{nk}$ be a corresponding set of orthonormal
  eigenvectors.

  For $i = 0,\ldots, n^{\frac{2}{3}(1 - \alpha)}-1$, let $B_i$ denote the vertices in the $i^{\mathrm{th}}$ block
  of the input graph:
  \begin{align*}
    B_i \defeq \setof{ \pi(n^{\frac{1}{3}+\frac{2}{3}\alpha} i + 1),\ldots,\pi(n^{\frac{1}{3}+\frac{2}{3}\alpha} i
    + n^{\frac{1}{3}+\frac{2}{3}\alpha}) }.
  \end{align*}

  Since $k\leq n^{\alpha - \eps}$ and $\alpha < 1/10$, we have
  $2\cdot 10^6 \gamma \psi \zeta^{-1} k^2 n^{\eps} \leq n^{\frac{1}{3}+\frac{2}{3}\alpha}$.
  Thus by invoking Theorem~\ref{thm:expsampintro}, %
  with probability $1 - 1/n^4$,
  all vertex-induced subgraphs $H_{\phi}^{s}[B_i\union B_{i+1}]$ preserve $H_{\phi}^{s}$
  in the sense of~(\ref{eq:preservance}).
  Using this fact, we prove the following claim:
  \begin{claim}\label{claim:hdsmallers}
    For each edge $(x,y)$ with $s_{xy} = 1$, conditioned on $(x,y)$ being the crossing edge
    (i.e. $\setof{\pi(1),\pi(n/2+1)} = \setof{x,y}$),
    with probability at least $1 - 1/n^2$ over $\pi$, we have
    that $b_{xy}$ is in the range of $L_{\pi}$, and
    \begin{align}\label{eq:claim:ers}
      b_{xy}^{T} L_{\pi}^{\dag} b_{xy} \leq %
      { \frac{k^2}{n^{2\alpha - o(1)}} }.
    \end{align}
  \end{claim}
  Once again by $k\leq n^{\alpha-\eps}$,
  the RHS of~(\ref{eq:claim:ers}) is $o(1)$.
  Since $s_{uv} = 1$ holds for a $1-o(1)$ fraction of the edges,
  we have, by Proposition~\ref{prop:tvers},
  $\expec{\pi}{d_{\tv}\kh{ (\Phi B^w)_{\pi,\yes}, (\Phi B^w)_{\pi,\no} }} = o(1)$.
\end{proof}

\begin{proof}[Proof of Claim~\ref{claim:hdsmallers}]
  Consider drawing $\pi$ from the conditional distribution
  on $\pi(1) = x$ and $\pi(n/2+1) = y$.
  Suppose w.l.o.g. $(x,y)$ is oriented $x\to y$, so $(b_{xy})_x = \phi_{xy}$ and $(b_{xy})_y = -\phi_{xy}$.
  Let $E_{\pi}$ be the set of edges of $\Hcal_{\pi}$,
  where we let $\Hcal_{\pi}$ consist of the non-crossing edges,
  such that each edge $(u,v)$ has matrix weight $\phi_{uv} \phi_{uv}^T \in \mathbb{R}^{k\times k}$.
  Let us call a function $\mathbf{f} : E_{\pi} \to\mathbb{R}$ a {\em flow} in the graph $\Hcal_{\pi}$,
  and say $\mathbf{f}$ routes a {\em demand} $d\in\mathbb{R}^{nk}$ if $\sum_{e\in E_{\pi}} \mathbf{f}_e b_e = d$.

  By Fact~\ref{thm:dual}, to prove~(\ref{eq:claim:ers}),
  it suffices to show that there exists a flow $\fbf$ in $\Hcal_{\pi}$ that routes demand $b_{xy}$
  satisfying $\norm{\fbf}_2^2 \leq \frac{k^2}{n^{2\alpha - o(1)}}$.
  To construct such a flow, we consider first breaking the demand $b_{xy}$
  into demands $d_{0},d_1,\ldots,d_{n^{\frac{2}{3}(1 - \alpha)}/2-1}\in\mathbb{R}^{nk}$ such that
  \begin{enumerate}
    \item $\sum_{0\leq i < n^{\frac{2}{3}(1 - \alpha)}/2} d_i = b_{xy}$.
    \item $d_i$ is supported on $B_i\union B_{i+1}$.
  \end{enumerate}
  We will then route each $d_i$ in the vertex induced subgraph $H_{\phi}[B_i\union B_{i+1}]$ using
  a flow $\fbf_i$, and finally obtain a flow
  $\fbf = \sum_{0\leq i < n^{\frac{2}{3}(1 - \alpha)}/2} \fbf_i$ that routes $b_{xy}$
  and satisfies $\norm{\fbf}_2^2 \leq 2\sum_{0\leq i < n^{\frac{2}{3}(1 - \alpha)}/2} \norm{\fbf_i}_2^2$.

  To show how these $d_i$'s are constructed, let us first define
  vectors $\dtil_0,\ldots,\dtil_{n^{\frac{2}{3}(1- \alpha)}/2}\in\mathbb{R}^{nk}$
  where $\dtil_i$ is supported on $B_i$.
  We will later define each $d_i$ by letting $d_i = \dtil_i - \dtil_{i+1}$.
  Let $D\in\mathbb{R}^{nk\times nk}$ be the degree matrix of $H_{\phi}^s$.
  For now we would like $\dtil_i$ to satisfy
  \begin{align}\label{eq:dtilis}
    \forall j: 0\leq \lambda_j \leq \zeta,\quad
    f_j^T D^{\dag/2} \dtil_i = (1 - \mu_i) (f_j)_x^T D_{xx}^{\dag/2} \phi_{xy}
    + \mu_i (f_j)_y^T D_{yy}^{\dag/2} \phi_{xy}.
  \end{align}
  where we define
  \begin{align*}
    \mu_i := \frac{i}{n^{\frac{2}{3}(1 - \alpha)}/2}.
  \end{align*}
  In particular, we would like $f_j^T D^{\dag/2} \dtil_0 = (f_j)_x^T D^{\dag/2}_{xx} \phi_{xy}$ and
  $f_j^T D^{\dag/2} \dtil_{n^{\frac{2}{3}(1 - \alpha)}/2} = (f_j)_y^T D^{\dag/2}_{yy} \phi_{xy}$
  for all $j$ such that $0\leq \lambda_j\leq \zeta$.
  Thus, we let %
  $\dtil_0$ be such that
  \begin{align*}
    (\dtil_0)_u =
    \begin{cases}
      \phi_{xy} & u = x \\
      0 & \text{otherwise}
    \end{cases}
  \end{align*}
  and let $\dtil_{n^{\frac{2}{3}(1 - \alpha)}/2}$ be such that
  \begin{align*}
    \kh{\dtil_{n^{\frac{2}{3}(1 - \alpha)}/2}}_u =
    \begin{cases}
      \phi_{xy} & u = y \\
      0 & \text{otherwise.}
    \end{cases}
  \end{align*}
  We would then like to construct the remaining $\dtil_i$'s such that each $\norm{D^{\dag/2} \dtil_i}_2$ is small.
  First let us write the linear equations~(\ref{eq:dtilis}) in matrix form.
  Let $\ell$ be s.t. $\lambda_{\ell} \leq \zeta < \lambda_{\ell+1}$.
  Define a function $F: V\to \mathbb{R}^{\ell\times k}$ such that
  \begin{align*}
    F(u) =
    \begin{pmatrix}
      (f_1)^T_u \\
      \vdots \\
      (f_{\ell})^T_u
    \end{pmatrix}
  \end{align*}
  and define another function $G : \setof{B_0,\ldots,B_{n^{\frac{2}{3}(1 - \alpha)}/2}} \to
  \mathbb{R}^{\ell\times n^{\frac{1}{3} + \frac{2}{3} \alpha} k}$ such that
  \begin{align*}
    G(B_i) =
    \begin{pmatrix}
      (f_1)^T_{B_i} \\
      \vdots \\
      (f_{\ell})^T_{B_i}
    \end{pmatrix}
  \end{align*}
  Then we can write~(\ref{eq:dtilis}) as
  \begin{align}\label{eq:dtili2s}
    G(B_i) D_{B_iB_i}^{\dag/2} (\dtil_i)_{B_i} = (1 - \mu_i) F(x) D_{xx}^{\dag/2} \phi_{xy}
    + \mu_i F(y) D_{yy}^{\dag/2} \phi_{xy}.
  \end{align}
  Let us write the RHS as $r_i := (1 - \mu_i) F(x) D_{xx}^{\dag/2} \phi_{xy} + \mu_i F(y) D_{yy}^{\dag/2} \phi_{xy}$.
  Then by Fact~\ref{thm:dual}, we can find a $(\dtil_i)_{B_i}$ satisfying~(\ref{eq:dtili2s})
  such that $\norm{D_{B_iB_i}^{\dag/2} (\dtil_i)_{B_i}}_2^2 = r_i^T (G(B_i) G(B_i)^T)^{-1} r_i$. Now notice
  \begin{align*}
    G(B_i) G(B_i)^T = \sum_{u\in B_i} F(u) F(u)^T.
  \end{align*}
  Since $f_1,\ldots,f_{\ell}$ are orthonormal, we have
  $\sum_{u\in V} F(u) F(u)^T = I_{\ell\times \ell}$. By Lemma~\ref{lem:fsu2}, we have for all $u\in V$ that
  $\lambdamax(F(u) F(u)^T) \leq \frac{\gamma k}{(1 - \zeta)^2 n} \leq \frac{16k \log n}{n}$.
  Since the vertices in $B_i$ are chosen uniformly at random, by a Matrix Chernoff bound
  (Theorem~\ref{thm:matrixchernoff}),
  we have with probability $\geq 1 - n^{-5}$
  that $\frac{1}{2} I \pleq n^{\frac{2}{3}(1 - \alpha)} G(B_i) G(B_i)^T \pleq 2 I$, which implies
  $r_i^T (G(B_i) G(B_i)^T)^{-1} r_i \leq 2 n^{\frac{2}{3}(1 - \alpha)} \norm{r_i}^2$.
  As a result, we can find $\dtil_1,\ldots,\dtil_{n^{\frac{2}{3}(1 - \alpha)}/2-1}$ such that
  \begin{align*}
    \norm{ D^{\dag/2} \dtil_i }_2^2 \leq 4 n^{\frac{2}{3}(1 - \alpha)} \kh{ \norm{F(x) D_{xx}^{\dag/2} \phi_{xy}}_2^2
    + \norm{F(y) D_{yy}^{\dag/2} \phi_{xy}}_2^2 }.
  \end{align*}
  For simplicity let $C_i := B_i\union B_{i+1}$.
  We then construct each $d_i$ for $0\leq i < n^{\frac{2}{3}(1 - \alpha)}/2$ by letting $d_i := \dtil_i -\dtil_{i+1}$.
  Thus $d_i$ is indeed supported on $C_i$,
  and we have $\sum_{0\leq i < n^{\frac{2}{3}(1 - \alpha)}/2} d_i = b_{xy}$.
  Now we consider how to route each $d_i$ in $H_{\phi}[B_i\union B_{i+1}]$.
  Once again by Fact~\ref{thm:dual}, there is a flow $\fbf_i$ with $\ell_2^2$-norm
  $(d_i)_{C_i}^T L_{H_{\phi}[C_i]}^{\dag} (d_i)_{C_i}
  \leq (d_i)_{C_i}^T L_{H_{\phi}^s[C_i]}^{\dag} (d_i)_{C_i}$. We then write
  \begin{align*}
    & (d_i)_{C_i}^T L_{H_{\phi}^s[C_i]}^{\dag} (d_i)_{C_i} %
    \\
    =
    & (d_i)_{C_i}^T D^{\dag/2}_{C_iC_i} \kh{ D^{\dag/2}_{C_iC_i} L_{H_{\phi}^s[C_i]} D^{\dag/2}_{C_iC_i} }^{\dag}
    D^{\dag/2}_{C_iC_i} (d_i)_{C_i}
    \\
    \leq
    & n^{o(1)} (d_i)_{C_i}^T D^{\dag/2}_{C_iC_i}
    \kh{ n^{\frac{4}{3}(1 - \alpha)} \sum_{j: \lambda_j\in(0,\zeta]} \frac{1}{\lambda_j}
    (f_j)_{C_i} (f_j)_{C_i}^T + n^{\frac{2}{3}(1 - \alpha)} (\log n) I }
    D^{\dag/2}_{C_iC_i} (d_i)_{C_i} \\
    \leq &
    n^{\frac{4}{3}(1 - \alpha) + {o(1)}} 
    \sum_{j: \lambda_j\in(0,\zeta]} \frac{1}{\lambda_j}
    \kh{ f_j^T D^{\dag/2} (\dtil_i - \dtil_{i+1}) }^2
    + n^{\frac{2}{3}(1 - \alpha)+o(1)} d_i^T D^{\dag} d_i
    \\
    = &
    n^{\frac{4}{3}(1 - \alpha)+o(1)} 
    \sum_{j: \lambda_j\in(0,\zeta]} \frac{1}{\lambda_j}
    \kh{ \frac{1}{n^{\frac{2}{3}(1 - \alpha)}/2} f_j^T D^{\dag/2} b_{xy} }^2
    + n^{\frac{2}{3}(1 - \alpha)+o(1)} \kh{ \dtil_i^T D^{\dag} \dtil_i + \dtil_{i+1}^T D^{\dag} \dtil_{i+1} }
    \\
    = & 
    n^{o(1)} b_{xy}^T D^{\dag/2} \kh{ \sum_{j: \lambda_j\in (0,\zeta]} \frac{1}{\lambda_j} f_j f_j^T }
    D^{\dag/2} b_{xy} +
    n^{\frac{2}{3}(1 - \alpha)+o(1)} \kh{ \dtil_i^T D^{\dag} \dtil_i + \dtil_{i+1}^T D^{\dag} \dtil_{i+1} }.
  \end{align*}
  Notice that the first term in the above is at most
  $\frac{n^{o(1)} \psi k^2}{n^2} \leq \frac{n^{o(1)} k^4}{n^2}$ by the expander property
  and $s_{xy} = 1$.
  The second term for each $0 < i < n^{\frac{2}{3}(1 - \alpha)}/2-1$ is at most
  \begin{align*}
    n^{\frac{4}{3}(1 - \alpha)+o(1)}\kh{ \norm{F(x) D_{xx}^{\dag/2} \phi_{xy}}_2^2 + \norm{F(y) D_{yy}^{\dag/2} \phi_{xy}}_2^2 }
    \leq \frac{k^2 n^{\frac{4}{3}(1 - \alpha) + o(1)}}{n^{2}},
  \end{align*}
  where the inequality follows from $\gamma$-regularity and that $\lambdamax(F(u)^T F(u))\leq \frac{16 k \log n}{n}$.
  For $i = 0$ or $i = n^{\frac{2}{3}(1- \alpha)}/2 - 1$,
  the second term is at most
  \begin{multline*}
    n^{\frac{2}{3}(1 - \alpha)+o(1)}
    \kh{ \norm{D^{\dag/2}_{xx} \phi_{xy}}_2^2 + \norm{D^{\dag/2}_{yy} \phi_{xy} }_2^2 } + \\
    n^{\frac{4}{3}(1 - \alpha)+o(1)}\kh{ \norm{F(x) D_{xx}^{\dag/2} \phi_{xy}}_2^2 + \norm{F(y) D_{yy}^{\dag/2} \phi_{xy}}_2^2 }
    \leq \frac{k n^{\frac{2}{3}(1 - \alpha) + o(1)}}{n}.
  \end{multline*}
  Finally by letting $\fbf = \sum_{0\leq i < n^{\frac{2}{3}(1 - \alpha)}/2} \fbf_i$,
  we have $\norm{\fbf}_2^2 \leq {k^2 n^{-2\alpha+o(1)}}$ as desired.
\end{proof}

\appendix

\section{Proof of Proposition~\ref{prop:spanner}}\label{sec:appendspanner}

\begin{proof}[Proof of Proposition~\ref{prop:spanner}]
  Consider generating a weighted graph as follows.
  Let $0 < a < b$ be some given parameters.
  First, we pick a uniformly random edge slot $e\in \binom{V}{2}$.
  Then we let the weight of each edge $f\neq e$ be an independent
  $\Ncal(b,b^2/(100\log n))$, a Gaussian distribution with mean $b$
  and variance $b^2/(100\log n)$.
  Finally, we include edge $e$ in the graph with probability $1/2$,
  and if we do include it,
  we draw its weight from $\Ncal(0,a^2)$.
  If $e$ is present has negative weight, we say the input is {\em invalid},
  and accept any sketch as a valid sketch.
  We will call the conditional distribution on the presence of $e$
  the \sfy distribution, and the conditional distribution on the absence of $e$
  the \sfn distribution.
  
  We then show that
  any linear sketch that computes an $o(\wmax/\wmin)$-spanner must
  detect the presence or absence of $e$ with 
  the edge slot $e$ with good probability.
  \begin{claim}\label{claim:.6}
    Any linear sketch that can compute an $o(\wmax/\wmin)$-spanner with probability $.9$
    can distinguish between the \sfy and \sfn distributions with probability $.6$.
  \end{claim}
  \begin{proof}
    By Chernoff bounds (Theorem~\ref{thm:GaussianChernoff}),
    with probability $1-1/n$, all edges other than $e$ have weights in $[b/100,100b]$.
    If we do add $e$, it has weight $\geq a/100$ with probability at least $.4$,
    by standard properties of Gaussian distributions.
    Then conditioned on both of these happening,
    one can detect $e$ by looking at an $o(\wmax/\wmin) = o(b/a)$ spanner
    of the graph. Therefore, if a linear sketch
    can compute an $o(\wmax/\wmin)$-spanner with probability $0.9$,
    it can detect if $e$ is present with probability at least
    \begin{align*}
      \geq (1 - 1/n)(.5+.5\cdot .4) .9 \geq .6.
    \end{align*}
  \end{proof}
  Suppose for the sake of contradiction,
  there is a sketching matrix $\Phi\in\mathbb{R}^{k\times \binom{n}{2}}$
  where $\phi_e$ is the $e^{\mathrm{th}}$ column of $\Phi$
  and $k = o(n^2)$, such that one can recover from $\Phi w$ an $o(\wmax/\wmin)$-spanner with probability $.9$.
  Let $(\Phi w)_{\yes}$ and $(\Phi w)_{\no}$ be the sketches obtained
  conditioned on the presence/absence of $e$.
  By Theorem~\ref{thm:tvd}, conditioned on a fixed choice of the edge slot $e$,
  if $\phi_e$ is in the linear span of $\setof{\phi_f: f\neq e}$,
  then the total variation distance between $(\Phi w)_{\yes}$ and $(\Phi w)_{\no}$ is bounded by
  \begin{align*}
    \dtv((\Phi w)_{\yes} | e,(\Phi w)_{\no} | e)
    \leq O(1)\cdot a^2 \phi_e^T \kh{ \sum_{e\neq f\in \binom{V}{2}} \frac{b^2}{\log n} \phi_f \phi_f^T }^{\dag} \phi_e,
  \end{align*}
  Since $k = o(n^2)$, for at least $(1 - o(1))\binom{n}{2}$ edge slots $e$,
  we have $\phi_e^T \kh{ \sum_{f\in \binom{V}{2}} \phi_f \phi_f^T }^{\dag} \phi_e\leq o(1)$,
  which implies that $\phi_e$ is in the linear span of $\setof{\phi_f: f\neq e}$.
  Let $E^*$ with $|E|\geq (1 - o(1))\binom{n}{2}$ denote the set of such edges.
  Thus we have
  \begin{align*}
    \dtv((\Phi w)_{\yes},(\Phi w)_{\no}) \leq
    & \frac{1}{\binom{n}{2}} \kh{ \sum_{e\in E^*} {\dtv((\Phi w)_{\yes} | e,(\Phi w)_{\no} | e)} +
    \sum_{e\notin E^*} 1}\\ \leq
    & O(1/n^2)\cdot a^2 \sum_{e\in E^*}
    \phi_e^T \kh{ \sum_{f\in \binom{V}{2}} \frac{b^2}{\log n} \phi_f \phi_f^T }^{\dag} \phi_e + o(1) \\ \leq
    & O(\log n/n^2) \frac{a^2}{b^2}\cdot \trace{ \kh{ \sum_{e\in\binom{V}{2}} \phi_e \phi_e^T }
    \kh{ \sum_{f\in \binom{V}{2}} \phi_f \phi_f^T }^{\dag} } + o(1) \\ =
    & O(\log n/n^2) \frac{a^2}{b^2} \rank{ \sum_{e\in\binom{V}{2}} \phi_e \phi_e^T } + o(1) \\ \leq
    & O\kh{ \frac{k a^2 \log n}{b^2 n^2} } + o(1).
  \end{align*}
  Thus, whenever $b^2 \geq a^2 \log n$, this total variation distance is $o(1)$,
  contradicting Claim~\ref{claim:.6}.
\end{proof}

\section{A hard instance for decomposing matrix-weighted graphs into large subgraphs
without small nonzero eigenvalues}\label{sec:aphd}

The proof of the following proposition appears in Appendix~\ref{sec:aphdg}
(in ``Proof of Proposition~\ref{prop:quadraticL}'').

\begin{proposition}
  For a vector $x \in \mathbb{R}^{nk}$, we have
  \begin{align*}
    x^T L x = \sum_{u\sim v} (x_u - x_v)^T \phi_{uv} \phi_{uv}^T (x_u - x_v).
  \end{align*}
\end{proposition}

By the above proposition, we have:

\begin{proposition}\label{prop:nullL}
  For any $nk$-dimensional vector $x$ such that $x_u = x_v \in \mathbb{R}^{k}$ for any $u,v\in V$,
  we have $x^T L x = 0$.
  As a result, the null space of $L$ has dimension at least $k$.
\end{proposition}

{
  First consider the following $n$-vertex $2\times 2$ matrix-weighted graph, call it $R$,
  whose vertices can be embedded into
  points that are uniformly distributed along
  a unit 2D-circle. Specifically, for each vertex $u\in \setof{1..n}$, assign to it a coordinate
  \begin{align*}
    c_{u} \defeq (\cos \frac{2 u \pi}{n}, \sin \frac{2 u \pi}{n})^T.
  \end{align*}
  Then for every $u\neq v$, connect them by an edge weighted by (the outer product of)
  the segment connecting $c_u, c_v$. That is, add an edge $(u,v)$ with weight
  $(c_u - c_v) (c_u - c_v)^T$.
  Then one can see that the Laplacian matrix (and thus also the normalized Laplacian matrix) of graph $R$
  has three zero eigenvalues --- apart from the two trivial zero eigenvectors mentioned in Proposition~\ref{prop:nullL},
  one can also construct another zero eigenvector $x$ where
  $x_u$ equals $c_u$ rotated by $90$ degrees counter-clockwise around the origin, namely
  $x_u = (-\sin \frac{2 u \pi}{n}, \cos \frac{2u\pi}{n})^T$. Then we have for any $u\neq v$ that
  $x_u - x_v$ is orthogonal to $\phi_{uv}$, and hence
  \begin{align*}
    x^T L_R x = \sum_{\setof{u,v}} (x_u - x_v)^T \phi_{uv} \phi_{uv}^T (x_u - x_v) = 0.
  \end{align*}
  Now consider creating a noisy version of the graph $R$, call it $\Rcal$, as follows.
  Between every pair of vertices $u,v$, we add a small stochastic noise on the weight of the edge connecting them.
  Specifically, let $\phitil_{uv}$'s be drawn independently from the 2-dimensional Gaussian
  $\Ncal(0, \eps I)$, for some sufficiently small $\eps$ (say $o(1/\poly(n))$).
  Then we let the edge weight of $(u,v)$ be
  \begin{align*}
    (\phi_{uv} + \phitil_{uv}) (\phi_{uv} + \phitil_{uv})^T.
  \end{align*}
  Now for the vector $x$ with $x_u = (-\sin \frac{2 u \pi}{n}, \cos \frac{2u\pi}{n})^T$,
  we have
  \begin{align*}
    \expec{\phitil_{uv}}{x^T L_{\Rcal} x} = & \sum_{\setof{u,v}}
    \expec{\phitil_{uv}}{(x_u - x_v)^T (\phi_{uv} + \phitil_{uv})
    (\phi_{uv} + \phitil_{uv})^T (x_u - x_v) } \\
    = &
    \sum_{\setof{u,v}}
    \expec{\phitil_{uv}}{(x_u - x_v)^T \phitil_{uv} \phitil_{uv}^T (x_u - x_v)}
    = \Theta(\eps n^2).
  \end{align*}
  Also note that $\norm{x}^2 = n$, and
  $x$ is orthogonal to the two trivial zero eigenvectors of $L_{\Rcal}$
  as $\sum_{u} x_u = 0$.
  Therefore $L_{\Rcal}$'s smallest nonzero eigenvalue is at most $o(1/\poly(n))$.
  One can verify that in $\Rcal$ the degree of each vertex $u$ has
  a constant ratio between its largest and smallest eigenvalues,
  and as a result the normalized Laplacian $N_{\Rcal}$ also has a nonzero eigenvalue bounded by $o(1/\poly(n))$.
  One can similarly show the same property for every large enough subgraph of $\Rcal$ (say, e.g.,
  with $\Omega(n^2)$ edges).
}

\section{Missing proofs from Section~\ref{sec:ovspectrallb}}\label{sec:apovlb}

\begin{proof}[Proof of proposition~\ref{prop:0.6}]
  By Proposition~\ref{prop:erbig}, with probability $1-1/n$,
  the effective resistance between $\pi(1)$ and $\pi(n/2+1)$ is at least $1/48$.
  Thus whenever the crossing edge 
  has positive weight at least $\frac{1}{4}$,
  one can tell if the crossing edge is present or not by looking at a $1.0001$-spectral sparsifier of the graph.
  On the other hand, if we do add the crossing edge,
  it has weight $\geq \frac{1}{4}$ with probability
  at least $0.4$, by properties of a standard Gaussian.
  Therefore if a linear sketch
  can compute a $1.0001$-spectral sparsifier with probability
  $0.9$, then it can decide if
  the crossing edge is present with probability
  \begin{align*}
    \geq (1 - 1/n)(0.5 + 0.5\cdot 0.4) 0.9 \geq 0.6.
  \end{align*}
\end{proof}

\begin{proof}[Proof of Proposition~\ref{prop:B}]
  For the $i^{\mathrm{th}}$ row of the sketching matrix $\Phi$,
  if its support is on edges incident on vertex $u$, we say the $i^{th}$ row belongs to $u$.
  Let $U\subseteq V$ be the subset of vertices $u$ such that the number of rows of $\Phi$ that belong
  to $u$ is at most $\frac{1000 N}{n}$. By Markov's inequality we have $|U| \geq 0.999 n$.
  Let $F = \binom{U}{2}$ be the edge slots within $U$.
  Then $|F|\geq 0.998\binom{n}{2}$.

  Let $(\Phi w)_{\pi,\yes}$ be $\Phi w$ conditioned on $\pi$ and the presence of the crossing edge;
  similarly 
  let $(\Phi w)_{\pi,\no}$ be $\Phi w$ conditioned on $\pi$ and the absence of the crossing edge.
  Let $\phi_{uv}\in\mathbb{R}^{N}$ be the column of $\Phi$ corresponding to edge slot $(u,v)$.
  By Theorem~\ref{thm:tvd}, for any $\pi$ we have %
  \begin{multline}
    d_{\tv}\kh{ (\Phi w)_{\pi,\yes}, (\Phi w)_{\pi,\no} }
    \leq \\ O(1)\cdot \min\setof{1,
      \phi_{\pi(1)\pi(n/2+1)}^T
      \kh{ \sum_{\text{non-crossing }(u,v)} n^{4/5}\log^{-1} n \phi_{uv} \phi_{uv}^T }^{\dag}
      \phi_{\pi(1)\pi(n/2+1)}
    }.
    \label{eq:dtvap}
  \end{multline}
  Note that we have slightly abused the notation $\dag$ in the second term on the RHS of~(\ref{eq:dtvap}):
  in the case that $\phi_{\pi(1)\pi(n/2+1)}$ is not in the range of the matrix inside the summation,
  the term should be understood as infinity.

  By restricting the summation on the RHS to non-crossing edges in $F$,
  we have, by Fact~\ref{thm:dual},
  for any $\pi$ %
  \begin{multline}\label{eq:abuse}
    \phi_{\pi(1)\pi(n/2+1)}^T
    \kh{ \sum_{\text{non-crossing }(u,v)} n^{4/5}\log^{-1} n \phi_{uv} \phi_{uv}^T }^{\dag}
    \phi_{\pi(1)\pi(n/2+1)} \leq \\
    \phi_{\pi(1)\pi(n/2+1)}^T
    \kh{ \sum_{\text{non-crossing }(u,v)\in F} n^{4/5}\log^{-1} n \phi_{uv} \phi_{uv}^T }^{\dag}
    \phi_{\pi(1)\pi(n/2+1)}.
  \end{multline}

  Now consider a sketching matrix $\Phi''\in\mathbb{R}^{N\times \binom{n}{2}}$ that is obtained from
  $\Phi$ by zeroing out the columns corresponding to edge slots {\em not} in $F$.
  Then we have by~(\ref{eq:abuse}) that for any $\pi$ such that $(\pi(1),\pi(n/2+1))\in F$
  (thus the column of $\Phi''$ corresponding to the crossing edge is {\em not} zeroed out)
  \begin{align*}
    d_{\tv}\kh{ (\Phi w)_{\pi,\yes}, (\Phi w)_{\pi,\no} } \leq
    d_{\tv}\kh{ (\Phi'' w)_{\pi,\yes}, (\Phi'' w)_{\pi,\no} }.
  \end{align*}
  Using the fact that $\prob{\pi}{(\pi(1),\pi(n/2+1))\in F} = |F|/\binom{n}{2} \geq 0.998$,
  we have
  \begin{align*}
    \expec{\pi}{d_{\tv}\kh{ (\Phi'' w)_{\pi,\yes}, (\Phi'' w)_{\pi,\no} }}
    \geq
    \expec{\pi}{d_{\tv}\kh{ (\Phi w)_{\pi,\yes}, (\Phi w)_{\pi,\no} }} - 0.002,
  \end{align*}
  and as a result,
  there exists an incidence sketch using $\Phi''$ that can distinguish between the \sfy and \sfn distributions
  with probability $0.55$.
  We then show that we can simulate such an incidence sketch that uses $\Phi''$ by a
  signed sketch with a sketching matrix $\Phi'\in\mathbb{R}^{k\times \binom{n}{2}}$
  such that $k\leq O(1)\cdot\max\setof{1,\frac{N\log n}{n}}$.
  
  Let $R_{u} \subseteq \setof{1,\ldots,N}$ denote the indices of the rows that belong to $u$.
  Then we know $|R_u| \leq 1000 N / n$ for all $u\in U$.
  It suffices to construct a $\Phi'$ from which we can recover,
  for each vertex $u \in U$, the sketch
  $\sum_{v\in U\setminus\setof{u}} w_{uv}\cdot (\phi_{uv})_{R_u} \in \mathbb{R}^{|R_u|}$.
  Now consider sketching matrices $\Phi'_{1},\ldots,\Phi'_{10\log n}\in\mathbb{R}^{\frac{1000 N}{n} \times \binom{n}{2}}$,
  which are initially set to be all-zero.
  We then assign values to these matrices as follows.
  For each $i = 1,\ldots,10\log n$, we generate a subset of vertices $S_i$
  by including each vertex independently in $S_i$ with probability $1/2$.
  By Chernoff bound, we know that with high probability, we have that
  for every vertex $u$ and every incident edge $(u,v)$,
  $u\in S_i, v\in V-S_i$ holds for at least one $i$.
  Let $i_{uv}$ be the smallest $i$ such that $u\in S_i, v\in V-S_i$.
  Then for each vertex $u$ and each incident edge $(u,v)\in F$,
  we let the column of $\Phi'_{i_{uv}}$ corresponding to $(u,v)$ be
  \begin{align*}
    \Phi'_{i_{uv}}(:,(u,v)) =
    \begin{cases}
      (\phi_{uv})_{R_u} & \text{$u$ is $(u,v)$'s head} \\
      -(\phi_{uv})_{R_u} & \text{$u$ is $(u,v)$'s tail},
    \end{cases}
  \end{align*}
  where we  augment $(\phi_{uv})_{R_u}$ by $0$'s if $|R_u| < \frac{1000N}{n}$.
  Then for each vertex $u\in S_i$, we have
  \begin{align*}
    \kh{ \Phi'_{i} B^w }(:,u) = \sum_{v\in U: i_{uv} = i} w_{uv}\cdot (\phi_{uv})_{R_u},
  \end{align*}
  and therefore
  \begin{align*}
    \sum_{i=1}^{10\log n} \kh{ \Phi'_{i} B^w }(:,u) = \sum_{v\in U\setminus\setof{u}} w_{uv}\cdot (\phi_{uv})_{R_u}.
  \end{align*}
  This implies that we can construct a desired $\Phi'$ by stacking up the $\Phi'_i$'s:
  \begin{align*}
    \Phi' =
    \begin{pmatrix}
      \Phi'_1 \\
      \vdots \\
      \Phi'_{10\log n}
    \end{pmatrix}\in\mathbb{R}^{\frac{10000N\log n}{n}\times \binom{n}{2}}.
  \end{align*}
\end{proof}

\begin{proof}[Proof of Proposition~\ref{prop:tver}]
  Consider fixing {a} sketching matrix
  $\Phi\in\mathbb{R}^{k\times \binom{n}{2}}$ and a permutation $\pi:\setof{1..n}\to \setof{1..n}$.
  We are interested in the total variation distance between $(\Phi B^w)_{\pi,\yes}$ and
  $(\Phi B^w)_{\pi,\no}$.
  Let us also fix a vertex $u\in \setof{1..n}$ and consider the $u$-th column
  $(\Phi B^w)(:,u)\in\mathbb{R}^{k}$ of the sketch obtained.
  We have (recall that $\phi_e$ is the $e$-th column of $\Phi$)
  \begin{align}
    (\Phi B^w)(:,u) = \sum_{e} B^w_{eu} \phi_{e}.
  \end{align}
  If $e$ is not present in the graph, or if $e$ is not incident on $u$,
  then the entry $B^w_{eu}$ is zero.
  Otherwise ($e$ present and incident on $u$),
  we know that $B^w_{eu}$ follows a univariate Gaussian distribution $\Ncal(\mu_{e,u},\sigma_{e,u}^2)$,
  where for non-crossing $e$
  \begin{align}
    \mu_{e,u} =
    \begin{cases}
      8n^{2/5} & \text{$u$ is $e$'s head} \\
      -8n^{2/5} & \text{$u$ is $e$'s tail}
    \end{cases}
    \qquad \text{and} \qquad
    \sigma_{e,u}^2 = n^{4/5} \log^{-1} n
  \end{align}
  and for the crossing edge $e$
  \begin{align}
    \mu_{e,u} = 0
    \qquad \text{and} \qquad
    \sigma_{e,u}^2 = 1.
  \end{align}
  Therefore, %
  $B^w_{eu} \phi_e$ follows a $k$-dimensional Gaussian distribution:
  \begin{align}
    B^w_{eu} \phi_e \sim \Ncal(\mu_{e,u} \phi_e, \sigma_{e,u}^2 \phi_e \phi_e^T).
  \end{align}
  Also notice that the distributions of $B^w_{eu}$'s are independent for different edges.
  Thus by Fact~\ref{thm:sumgaussian}
  \begin{align}
    (\Phi B^w) (:,u) \sim \Ncal\kh{ \sum_{e\sim u} \mu_{e,u} \phi_e, \sum_{e\sim u} \sigma_{e,u}^{2} \phi_e\phi_e^T  }
  \end{align}
  As a result, the sketch $\Phi B^w$ can be seen as an $nk$-dimensional Gaussian.

  We next consider the correlations between different columns of $\Phi B^w$.
  For two vertices $u,v$, if there is no edge between them, then the distributions
  of $(\Phi B^w)(:,u)$ and $(\Phi B^w)(:,v)$ are independent, so the correlation between them is zero.
  Otherwise (if the edge $e = (u,v)$ is present), the only correlation between $(\Phi B^w)(:,u)$ and $(\Phi B^w)(:,v)$
  is the one between $B^w_{eu}\phi_e$ and $B^w_{ev}\phi_e$, who are negations of each other.
  Therefore the correlation between $(\Phi B^w)(:,u)$ and $(\Phi B^w)(:,v)$ is just the negative covariance of $B^w_{eu} \phi_e$:
  \begin{align}
    \ex{\ \kh{\, (\Phi B^w)(:,u) - \ex{ (\Phi B^w)(:,u) } \,} \kh{\, (\Phi B^w)(:,v) - \ex{ (\Phi B^w)(:,v) } \,}^T \ } =
    - \sigma_{e,u}^2 \phi_e \phi_e^T.
  \end{align}

  We are now ready to write down the covariance matrix of the sketch $\Phi B^w$.
  First consider the covariance matrix of $(\Phi B^w)_{\pi,\no}$,
  call it $\Sigma_{(\Phi B^w)_{\pi,\no}}\in\mathbb{R}^{nk\times nk}$.
  We can write
  $\Sigma_{(\Phi B^w)_{\pi,\no}}$ as an $n\times n$ block matrix (with block size $k\times k$) by
  \begin{align*}%
    \kh{ \Sigma_{(\Phi B^w)_{\pi,\no}} }_{uv} =
    \begin{cases}
      \sum_{\text{non-crossing } e\sim u} n^{4/5} \log^{-1} n \phi_e \phi_e^T & u = v \\
      - n^{4/5} \log^{-1} n \phi_e \phi_e^T & u\neq v \text{ and $e=(u,v)$ is non-crossing} \\
      0 & \text{otherwise.}
    \end{cases}
  \end{align*}
  One can also verify that
  $\Sigma_{(\Phi B^w)_{\pi,\no}} = \sum_{\text{non-crossing} (u,v)} n^{4/5}\log^{-1} n b_{uv} b_{uv}^T$,
  where $b_{uv}$'s are defined in~(\ref{eq:defbb}).
  That is $\Sigma_{(\Phi B^w)_{\pi,\no}}$ is exactly the matrix $L_{\pi}$.

  If the crossing edge is present,
  using the fact that its weight follows a standard Gaussian distribution,
  the covariance matrix can be written as
  \begin{align}
    \Sigma_{(\Phi B^w)_{\pi,\yes}} = \Sigma_{(\Phi B^w)_{\pi,\no}} + b_{\pi(1)\pi(n/2+1)} b_{\pi(1)\pi(n/2+1)}^T
  \end{align}
  Notice that, since the crossing edge's weight has zero mean,
  $(\Phi B^w)_{\pi,\yes}$ and $(\Phi B^w)_{\pi,\no}$
  have the same mean.
  Therefore,
  we can invoke Theorem~\ref{thm:tvd} and obtain an upper bound (up to a constant factor)
  on the total variation distance between the two distributions as
  \begin{align}\label{eq:dtver}
    O(1)\cdot \min\setof{1, b_{\pi(1)\pi(n/2+1)}^T \Sigma_{(\Phi B^w)_{\pi,\no}}^{\dag} b_{\pi(1)\pi(n/2+1)}},
  \end{align}
  as desired.
\end{proof}

\begin{proof}[Proof of Claim~\ref{claim:ue}]
  Let $n' = |U| \geq \dmin \geq \frac{|E_{\phi}|}{8n}$ be the number of vertices in $I$.
  Notice that by conditioning on an $f\in I$ being the crossing edge,
  we have $\pi(1),\pi(n/2+1)\in U$ both.
  Let $U_i$ denote the vertices in $U$ that are in the $i^{\mathrm{th}}$ block of the cycle.
  We have with high probability over $\pi$ that each $|U_i| \in [\frac{n'}{2n^{4/5}}, \frac{2n'}{n^{4/5}}]$.
  By Lemma~\ref{lem:edvs}, with probability at least $1 - 1/n^4$ over $\pi$,
  each $I[U_i \union U_{i+1}]$ is a $\frac{1}{n^{o(1)}}$-expander
  with minimum degree $\geq \frac{|E_{\phi}|}{16 n^{9/5}}$.
  Now consider constructing a graph $\Itil$ supported on $U$ by,
  for each $0\leq i < \ell$,
  adding a clique of weight $\frac{|E_{\phi}|}{n'n}$ on $U_i\union U_{i+1}$
  (whose vertex degrees are all $\Theta(\frac{|E_{\phi}|}{n^{9/5}})$).
  One can show that in the graph $\Itil$,
  the effective resistance between $\pi(1),\pi(n/2+1)$
  is maximized when we set $n'$ to be its minimum $\Theta(\frac{|E_{\phi}|}{n})$,
  in which case the effective resistance is $\Theta(n^{4/5})\cdot \kh{\frac{n^{9/5}}{|E_{\phi}|}}^2$.
  Then by the expander property, the effective resistance between $\pi(1),\pi(n/2+1)$
  in $H_{\pi}$ is at most $\Theta(n^{4/5+o(1)})\cdot \kh{\frac{n^{9/5}}{|E_{\phi}|}}^2$.
  Finally, since $\Hcal_{\pi}$ is $n^{4/5} \log^{-1} n H_{\pi}$,
  we have that the effective resistance between $\pi(1),\pi(n/2+1)$
  in $\Hcal_{\pi}$ is at most $\Theta(n^{o(1)})\cdot \kh{\frac{n^{9/5}}{|E_{\phi}|}}^2$,
  which is bounded by $u_f$.
\end{proof}

\begin{proof}[Proof of Proposition~\ref{prop:ellavg}]
  Suppose w.l.o.g. $\lambda_1,\ldots,\lambda_{\ell}$ are all eigenvalues between $(0,\zeta]$. Then
  \begin{align*}
    & \sum_{(u,v)\in E}
    \kh{D^{\dag/2} b_{uv} }^T
    \kh{ \sum_{i=1}^{\ell} \lambda_i f_i f_i^T }^{\dag} D^{\dag/2} b_{uv} \\ =
    & \sum_{(u,v)\in E}
    \trace{\kh{ \sum_{i=1}^{\ell} \lambda_i f_i f_i^T }^{\dag} D^{\dag/2} b_{uv} b_{uv}^T D^{\dag/2} } \\
    =
    & \trace{\kh{ \sum_{i=1}^{\ell} \lambda_i f_i f_i^t }^{\dag}
    \kh{ \sum_{(u,v)\in E} D^{\dag/2} b_{uv} b_{uv}^T D^{\dag/2} } }
    \\
    = & \trace{ \kh{ \sum_{i=1}^{\ell} \lambda_i f_i f_i^t }^{\dag}  N_G } \\
    = & \trace{ \kh{ \sum_{i=1}^{\ell} \lambda_i f_i f_i^t }^{\dag} \kh{ \sum_{i=1}^{n} \lambda_i f_i f_i^T } } \\
    = & \trace{ \kh{ \sum_{i=1}^{\ell} \lambda_i f_i f_i^t }^{\dag} \kh{ \sum_{i=1}^{\ell} \lambda_i f_i f_i^T } }
    = \ell,
  \end{align*}
  as desired.
\end{proof}

\section{Missing proofs from Section~\ref{sec:hdg}}\label{sec:aphdg}

\begin{proof}[Proof of Proposition~\ref{prop:quadraticL}]
  The claim follows by
  \begin{align*}
    x^T L x = & x^T \kh{ \sum_{u\sim v} b_{uv} b_{uv}^T } x \\
    = & \sum_{u\sim v} \ip{x}{b_{uv}}^2 \\
    = & \sum_{u\sim v} \kh{ \ip{x_u}{\phi_{uv}} - \ip{x_v}{\phi_{uv}} }^2 \\
    = & \sum_{u\sim v} \ip{x_u - x_v}{\phi_{uv}}^2.
  \end{align*}
\end{proof}

\begin{proof}[Proof of Proposition~\ref{prop:quadraticN}]
  The claim follows by
  \begin{align*}
    x^T N x = & x^T \kh{ \sum_{u\sim v} (D^{\dag/2} b_{uv}) (D^{\dag/2} b_{uv})^T } x \\
    = & \sum_{u\sim v} \ip{x}{D^{\dag/2} b_{uv}}^2 \\
    = & \sum_{u\sim v} \kh{\ip{x_u}{D_u^{\dag/2} \phi_{uv}}- \ip{x_v}{D_v^{\dag/2}\phi_{uv}}}^2.
  \end{align*}
\end{proof}

\begin{proof}[Proof of Proposition~\ref{prop:eigvalofN}]
  The lower bound of $0$ follows from that $N$ is positive semi-definite. We then prove the upper
  bound by showing that for any $x\in \mathbb{R}^{nk}$, its Rayleigh quotient $R(x) = \frac{x^T N x}{x^T x}$ is at most $2$.
  It is then equivalent to show that $x^T N x \leq 2 \norm{x}^2$:
  \begin{align*}
    x^T N x
    = & \sum_{u\sim v} \kh{\ip{x_u}{D_u^{\dag/2} \phi_{uv}}- \ip{x_v}{D_v^{\dag/2}\phi_{uv}}}^2 \\
    \leq & \sum_{u\sim v} \kh{ 2\ip{x_u}{D_u^{\dag/2} \phi_{uv}}^2 + 2 \ip{x_v}{D_v^{\dag/2}\phi_{uv}}^2 } \\
    = & 2 \sum_{u} \sum_{v\sim u} \ip{x_u}{D_u^{\dag/2} \phi_{uv}}^2 \\
    = & 2 \sum_{u} \sum_{v\sim u} x_u^T D_u^{\dag/2} \phi_{uv} \phi_{uv}^T D_u^{\dag/2} x_u \\
    = & 2 \sum_{u} x_u^T D_u^{\dag/2} \kh{ \sum_{v\sim u} \phi_{uv} \phi_{uv}^T } D_u^{\dag/2} x_u \\
    = & 2 \sum_{u} x_u^T D_u^{\dag/2} D_u D_u^{\dag/2} x_u \\
    \leq & 2 \sum_{u} x_u^T x_u = 2 \sum_u \norm{x_u}^2 = 2 \norm{x}^2.
  \end{align*}
  Here the last inequality follows from that $D_u^{\dag/2} D_u D_u^{\dag/2} = \Pi_{D_u} \pleq I$.
\end{proof}

\section{Missing proofs from Section~\ref{sec:hared}}\label{sec:aphared}

\begin{proof}[Proof of Lemma~\ref{lem:potentialfrac}]
  Define $Y = D^{1/2} X$.
  Then
  \begin{align}\label{eq:fractionxy}
    \min_{X\in\Xcal} \frac{ \det\kh{ X^T L X } }{ \det\kh{X^T D X} }
    = & \min_{D^{\dag/2} Y \in \Xcal}
    \frac{ \det\kh{ Y^T D^{\dag/2} L D^{\dag/2} Y }}{\det\kh{Y^T Y}}.
  \end{align}
  Let the eigenvalues of $Y^T Y$ be $\nu_1,\ldots,\nu_{\ell}$
  and let $g_1,\ldots,g_{\ell}\in\mathbb{R}^{\ell}$ be a corresponding set of orthonormal eigenvectors.
  We also know that $Y Y^T$ has eigenvalues $\nu_1,\ldots,\nu_{\ell},0,\ldots,0$.
  Let $h_1,\ldots,h_{nk}\in\mathbb{R}^{nk}$ be a corresponding set of orthonormal
  eigenvectors of $Y Y^T$.
  Then
  \begin{align*}
    \det(Y^T Y) = \Pi_{i=1}^{\ell} \nu_i
  \end{align*}
  and
  \begin{align*}
    \det\kh{ Y^T D^{\dag/2} L D^{\dag/2} Y } =
    & \detp\kh{ L^{1/2} D^{\dag/2} Y Y^T D^{\dag/2} L^{1/2} } \\ =
    & \detp\kh{ L^{1/2} D^{\dag/2} \kh{ \sum_{i=1}^{\ell} \nu_i h_i h_i^T } D^{\dag/2} L^{1/2} } \\ =
    & \detp\kh{ L^{1/2} D^{\dag/2} H V H^T D^{\dag/2} L^{1/2} } \\ =
    & \det\kh{ V^{1/2} H^T D^{\dag/2} L D^{\dag/2} H V^{1/2} } \\ =
    & \kh{ \Pi_{i=1}^{\ell} \nu_i }
    \det\kh{ H^T D^{\dag/2} L D^{\dag/2} H },
  \end{align*}
  where we define
  $H =
  \begin{pmatrix}
    h_1 & \ldots & h_{\ell}
  \end{pmatrix}\in\mathbb{R}^{nk\times \ell}$
  and
  $V = \mathrm{diag}(\nu_1,\ldots,\nu_{\ell})\in\mathbb{R}^{\ell\times \ell}$.
  The claim in the lemma then follows by noting that
  $\det\kh{ H^T D^{\dag/2} L D^{\dag/2} H }$ is minimized (over all $H$'s with orthonormal columns
  in the range of $N$)
  when $H$'s columns are bottom nonzero eigenvectors of $N = D^{\dag/2} L D^{\dag/2}$,
  a result of Cauchy interlacing.
  Therefore~(\ref{eq:fractionxy}) equals $\detl(N)$.
\end{proof}

\begin{proof}[Proof of Lemma~\ref{lem:decmost}]
  Consider the following process for transitioning $G$ to $G^s$,
  where we use $t$ to denote the current scaling, which is all one in the beginning.
  \begin{enumerate}
    \item Initially, let $t_e \gets 1$ for all $e\in E$.
    \item While $t \neq s$:
      \begin{enumerate}
        \item Let $F\gets \setof{e: t_e > s_e}$,
          and let $\eta \gets \min_{e\in F} t_e / s_e$.
        \item For each $e\in F$,
          let $t_e \gets t_e / \eta$. \label{line:teta}
      \end{enumerate}
  \end{enumerate}
  Note that this process terminates in finite time as in each loop we make $t_e = s_e$
  for at least one extra $e$.
  \begin{claim}
    At the beginning of every while loop, we have
    for any $e$ such that $s_e < t_e$
    \begin{align*}
      R^t(e) \leq \frac{2\gamma k}{n}.
    \end{align*}
  \end{claim}
  \begin{proof}
    We prove this by induction.
    In the beginning of the process, by $\gamma$-regularity, we have
    $R^t(e) \leq \frac{2\gamma k}{n}$ for all $e\in E$.
    Then for the induction step, note that in each while loop,
    we downscale the weights of {\em all} edges with $s_e < t_e$ by a same amount.
    From the point view of leverage scores, this is equivalent to
    increasing the weights of all other edges by a same multiple.
    As a result, the leverage scores of {\em all} edges with $s_e < t_e$ can only decrease
    (by Fact~\ref{thm:dual}),
    and thus we have $R^t(e)\leq \frac{2\gamma k}{n}$ for these edges throughout.
  \end{proof}
  \begin{claim}\label{claim:decspeed}
    Consider fixing an iteration of the while loop
    and letting $t$ and $t'$ be the scaling before and after the execution of Line~\ref{line:teta} respectively.
    Then we have
    \begin{align*}
      \detp(D^{t'}) \geq \kh{1 - \frac{2\gamma k}{n}}^{2|F| \log \eta} \detp(D^{t}).
    \end{align*}
  \end{claim}
  \begin{proof}
    For an $x\in [1,\ln \eta]$, define $t^x : E\to [0,1]$ by
    \begin{align*}
      t^x_e =
      \begin{cases}
        e^{-x} \cdot t_e & e\in F \\
        t_e & \text{o.w.}
      \end{cases}
    \end{align*}
    Then we have $t' =  t^{\ln \eta}$. Consider differentiating $\ln \detp (D^{t^{x}})$:
    \begin{align*}
      \frac{\dr \ln \detp(D^{t^x})}{\dr x} =
      & \trace{ \kh{D^{t^x}}^{\dag} \frac{\dr D^{t^x}}{\dr x} } \\ = &
      -
      2\trace{ \kh{D^{t^x}}^{\dag} \kh{D^{t^x}}^{F} } \\ = &
      - 2\trace{ \kh{D^{t^x}}^{\dag} \kh{ \sum_{e\in F} \kh{e^{t^x}_{u\la v} \kh{e^{t^x}_{u\la v}}^T +
            e^{t^x}_{v\la u} \kh{e^{t^x}_{v\la u}}^T
      } } } \\ = - &
      \sum_{e\in F} 2R^{t^x}(e)
      \geq - 2|F|\cdot \frac{2\gamma k}{n}.
    \end{align*}
    Therefore
    \begin{align*}
      \ln \detp(D^{t'}) - \ln \detp(D^{t}) =
      \int_{0}^{\ln\eta} \frac{\dr \ln \detp(D^{t^x})}{\dr x} \dr x
      \geq  -|F|\cdot \frac{4\gamma k}{n}\cdot \ln\eta.
    \end{align*}
    Thus we have
    \begin{align*}
      \frac{\detp(D^{t'})}{\detp(D^{t})} \geq
      e^{-\frac{2\gamma k}{n} \cdot 2|F|\cdot \ln \eta}
      \geq \kh{1 - \frac{2\gamma k}{n}}^{2|F|\cdot \ln \eta}
      \geq \kh{1 - \frac{2\gamma k}{n}}^{2|F|\cdot \log \eta}
    \end{align*}
    as desired.
  \end{proof}
  The lemma is then a direct consequence of the above claim.
\end{proof}

\begin{proof}[Proof of Lemma~\ref{lem:tt1}]
  Let $s'$ be the scaling obtained at the end of the $t^{\mathrm{th}}$ iteration of the outermost while loop.
  Let $t_1$ be the total number of iterations executed so far by the first inner while loop
  (at Lines~\ref{line:while1s}-\ref{line:while1t}).
  Then by matrix determinant lemma
  \begin{align*}
    \frac{\detp(D^{s'})}{\detp(D^{s^0})} \leq
    & \kh{ 1 - \frac{3\gamma_2\cdot k}{8n} }^{2t_1}
    \kh{ 1 + \frac{3 \gamma_1\cdot k}{2 n} }^{2(t + t_1)} \\ \leq
    & \kh{ 1 - \frac{6 \gamma_1\cdot k}{n} }^{2t_1}
    \kh{ 1 + \frac{3 \gamma_1\cdot k}{2 n} }^{2(t + t_1)} \\ \leq
    & \kh{1 - \frac{4\gamma_1\cdot k}{n}}^{2t_1}
    \kh{1 + \frac{3 \gamma_1\cdot k}{2n}}^{2t}
  \end{align*}
  On the other hand, by Lemma~\ref{lem:decmost},
  \begin{align*}
    \frac{\detp(D^{s'})}{\detp(D^{s^0})} \geq
    \kh{ 1 - \frac{\gamma_1 k}{n} }^{2(t+t_1)}.
  \end{align*}
  These two together imply that $t_1 \leq 2t$.
\end{proof}

\section{Missing proofs from Section~\ref{sec:hdevs}}\label{sec:apexp}

\begin{proof}[Proof of Proposition~\ref{prop:choleskyk}]
  Multiplying the RHS out gives
  \begin{align}\label{eq:multout}
    \begin{pmatrix}
      L_{FF} & L_{FF} L_{FF}^{\dag} L_{FC} \\
      L_{CF} L_{FF}^{\dag} L_{FF} &
      L_{CF} L_{FF}^{\dag} L_{FF} L_{FF}^{\dag} L_{FC} +
      L_{CC} - L_{CF} L_{FF}^{\dag} L_{FC}
    \end{pmatrix}.
  \end{align}
  Then it suffices to show that 
  $L_{CF} L_{FF}^{\dag} L_{FF} = L_{CF}$, which will imply that~(\ref{eq:multout}) is equal to $L$.
  Notice that $L_{CF}$ contains the edges between $C$ and $F$.
  Moreover, we have
  $L_{FF} = L_{G[F]} + \sum_{(u,v)\in E\intersect (F\times C)} (e_{u\la v})_{F} (e_{u\la v})_{F}^T$,
  and thus all rows of $L_{CF}$ are in the range of $L_{FF}$, which gives
  that $L_{CF} L_{FF}^{\dag} L_{FF} = L_{CF}$.
\end{proof}

\begin{proof}[Proof of Claim~\ref{claim:gsq}]
  It suffices to prove that $D_G^{\dag/2} L_{G^2} D_G^{\dag/2} \pleq 2 D_G^{\dag/2} L_{G} D_G^{\dag/2}$.
  Notice that
  \begin{align*}
    D_G^{\dag/2} L_{G} D_G^{\dag/2} = I - D_G^{\dag/2} A_G D_G^{\dag/2}
  \end{align*}
  and
  \begin{align*}
    D_G^{\dag/2} L_{G^2} D_G^{\dag/2} = I - D_G^{\dag/2} A_G D_G^{\dag} A_G D_G^{\dag/2} =
    I - \kh{ D_G^{\dag/2} A_G D_G^{\dag/2} }^2.
  \end{align*}
  Let $\lambda_1,\ldots,\lambda_n$ be the eigenvalues of $D_G^{\dag/2} L_G D_G^{\dag/2}$,
  and let $f_1,\dots,f_n$ be a set of orthonormal eigenvectors. Then we have
  \begin{align*}
    D_G^{\dag/2} L_{G} D_G^{\dag/2} &= \sum_{i=1}^{n} \lambda_i f_i f_i^T \\
    D_G^{\dag/2} L_{G^2} D_G^{\dag/2} &= \sum_{i=1}^{n} \kh{ 1 - (1 - \lambda_i)^2 } f_i f_i^T =
    \sum_{i=1}^{n} (2 - \lambda_i)\lambda_i f_i f_i^T.
  \end{align*}
  Since $\lambda_i\in[0,2]$, we have our desired result.
\end{proof}

\begin{proof}[Proof of Proposition~\ref{prop:sumsck}]
  Notice that by definition
  \begin{align*}
    \sum_{u\in V} \SCN(L,V\setminus \setof{u}) =
    & \sum_{u\in V} \kh{ L_G - L_G(:,u) L_{uu}^{\dag} L_G(:,u)^T } \\ =
    & n L_G - L_G D^{\dag} L_G \\ =
    & n(D - A) - (D - A) D^{\dag} (D - A) \\ =
    & n D - n A - (D -2 A + A D^{\dag} A) \\ =
    & (n-2) (D - A) + D - A D^{\dag} A = (n-2) L_G + L_{G^2}
  \end{align*}
  as desired.
\end{proof}

\begin{proof}[Proof of Claim~\ref{claim:mart}]
  It suffices to prove that
  \begin{align*}
    \expec{v_{i+1}}{(\ref{eq:rpl})\ |\ V_{i}} = L_{H[V - V_i]}.
  \end{align*}
  Let us calculate the LHS term by term.
  \begin{align*}
    & \expec{v_{i+1}}{ L_{H[V-V_i]}(:,v_{i+1}) \kh{D^{H[V-V_i]}_{v_{i+1}}}^{\dag} L_{H[V-V_i]}(:,v_{i+1})^T\ |\ V_i} \\
    = & \frac{1}{t-i} L_{H[V-V_i]} D_{H[V-V_i]}^{\dag} L_{H[V-V_i]} \\
    = & \frac{1}{t-i} \kh{ D_{H[V-V_i]} - 2 A_{H[V-V_i]} + A_{H[V-V_i]} D_{H[V-V_i]}^{\dag} A_{H[V-V_i]} }
  \end{align*}
  \begin{align*}
    \expec{v_{i+1}}{\kh{1 + \frac{2}{t-2-i}} L_{H[V-V_{i+1}]}\ |\ V_i} = \frac{t - 2 - i}{t-i} L_{H[V-V_{i}]}
    + \frac{2}{t-i} L_{H[V-V_{i}]} =
    L_{H[V-V_i]}
  \end{align*}
  \begin{align*}
    & \expec{v_{i+1}}{\frac{1}{t-i} \kh{ L_{H[V-V_i]^2} - 2L_{H[V-V_i]} }\ |\ V_i} \\
    = & \frac{1}{t-i} \kh{ L_{H[V-V_i]^2} - 2L_{H[V-V_i]} } \\
    = & \frac{1}{t-i} \kh{ D_{H[V-V_i]} - A_{H[V-V_i]} D_{H[V-V_i]}^{\dag} A_{H[V-V_i]} } -
    \frac{2}{t-i} L_{H[V-V_i]}.
  \end{align*}
  One can verify that these three add up to $L_{H[V-V_{i}]}$.
\end{proof}

\bibliographystyle{alpha}
\bibliography{ref}

\end{document}